\documentclass[a4paper, 12pt]{report}
\usepackage[utf8]{inputenc}
\usepackage{graphicx}
\usepackage[english]{babel}
\usepackage{fancyhdr}
\usepackage{lettrine}

\usepackage{geometry}
\usepackage{minitoc}         %for ToC per chapter, 3 compilations needed
\newcommand{\mychapter}[3]{
	\chapter[#1]{#2}  %1: short title for ToC ; 2 : long title at the beginning of chapter 
	\label{#3}
	\minitoc
	\cleardoublepage
	}

\usepackage{natbib}
\usepackage{stmaryrd}
\usepackage{amsmath}
\usepackage{amsfonts}
\usepackage{amssymb,latexsym}
\usepackage{mathcomp}
\usepackage{dsfont}
\usepackage[usenames]{color}
\usepackage{array}
\usepackage{url}
\usepackage{pjournal} %Journals abbreviations
\usepackage{xspace}

\usepackage{tikz}
\usetikzlibrary{decorations.pathreplacing}
\newcommand{\tikzmark}[1]{\tikz[overlay,remember picture,baseline=(#1.base)]
  \node (#1) {\strut};}
  
\begin{document} 

%shortcuts
\newcommand{Ž}{\'{e}}
\newcommand{}{\`{e}}
\newcommand{}{\^{e}}
\newcommand{'}{\¬{e}}
\newcommand{ˆ}{\`{a}}
\newcommand{‰}{\^{a}}
\newcommand{}{\`{u}}
\newcommand{ž}{\^{u}}
\newcommand{Ÿ}{\¬{u}}
\newcommand{"}{\^{i}}
\newcommand{}{\c{c}}
\def\be{\begin{equation}}
\def\ee{\end{equation}}
\def\ben{\begin{equation*}}
\def\een{\end{equation*}}
\def\bea{\begin{eqnarray}}
\def\eea{\end{eqnarray}}
\def\bean{\begin{eqnarray*}}
\def\eean{\end{eqnarray*}}
\def\ba#1\ea{\begin{align}#1\end{align}}
\allowdisplaybreaks[2]
\def\mr{\mathrm}
\def\cl{$C_\ell$ }
\def\fnl{f_\mathrm{NL}}
\def\bps{b_\mathrm{PS}}
\def\air{A_\mathrm{IR}}
\def\bl{b_{\ell_1 \ell_2 \ell_3}}
\def\mbl{\mathbf{b}_{123}}
\def\lu{\ell_1}
\def\ld{\ell_2}
\def\lt{\ell_3}
\def\dd{\mathrm{d}}
\def\nn{\mathbf{n}}
\def\xx{\mathbf{x}}
\def\kk{\mathbf{k}}
\def\hn{\hat{n}}
\def\ci{\perp\!\!\!\perp}
\def\planck{\textit{Planck}\xspace}
\def\lbra{\left\langle}
\def\rbra{\right\rangle}
\newcommand{\threej}[6]{\begin{pmatrix}
#1 & #2 & #3 \\ #4 & #5 & #6
\end{pmatrix}}
\newcommand{\threejzero}[3]{\begin{pmatrix}
#1 & #2 & #3 \\ 0 & 0 & 0
\end{pmatrix}}
\newcommand{\degree}{ \ensuremath{^\circ} }
\def\Msun{M_{\odot}}
\newcommand{\leftexp}[2]{{\vphantom{#2}}^{#1}{#2}}

\definecolor{orange}{RGB}{255,127,0}
%\definecolor{darkgreen}{rgb}{0,0.9,0.5}
\definecolor{darkgreen}{RGB}{0,150,0}

\renewcommand{\topfraction}{0.9}	% max fraction of floats at top
\renewcommand{\bottomfraction}{0.9}	% max fraction of floats at bottom
\renewcommand{\floatpagefraction}{0.9}
\renewcommand{\textfraction}{0.01}
 %User-defined shortcuts & new commands, required packages etc

\pagestyle{empty} %no numbering of pages

\fancypagestyle{plain}
{
    \fancyhead{}
    \fancyfoot{}
    \renewcommand{\headrulewidth}{0pt}
}	% clear header and footer of plain page because of Table of Contents

%Page de titre
%special geometry for this page only
\newgeometry{textwidth=450pt, textheight=710pt,lmargin=70pt,tmargin=60pt}

%Upper bar with logo
\noindent\begin{minipage}{0.14\linewidth}
\includegraphics[width=\linewidth]{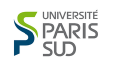}
\includegraphics[width=\linewidth]{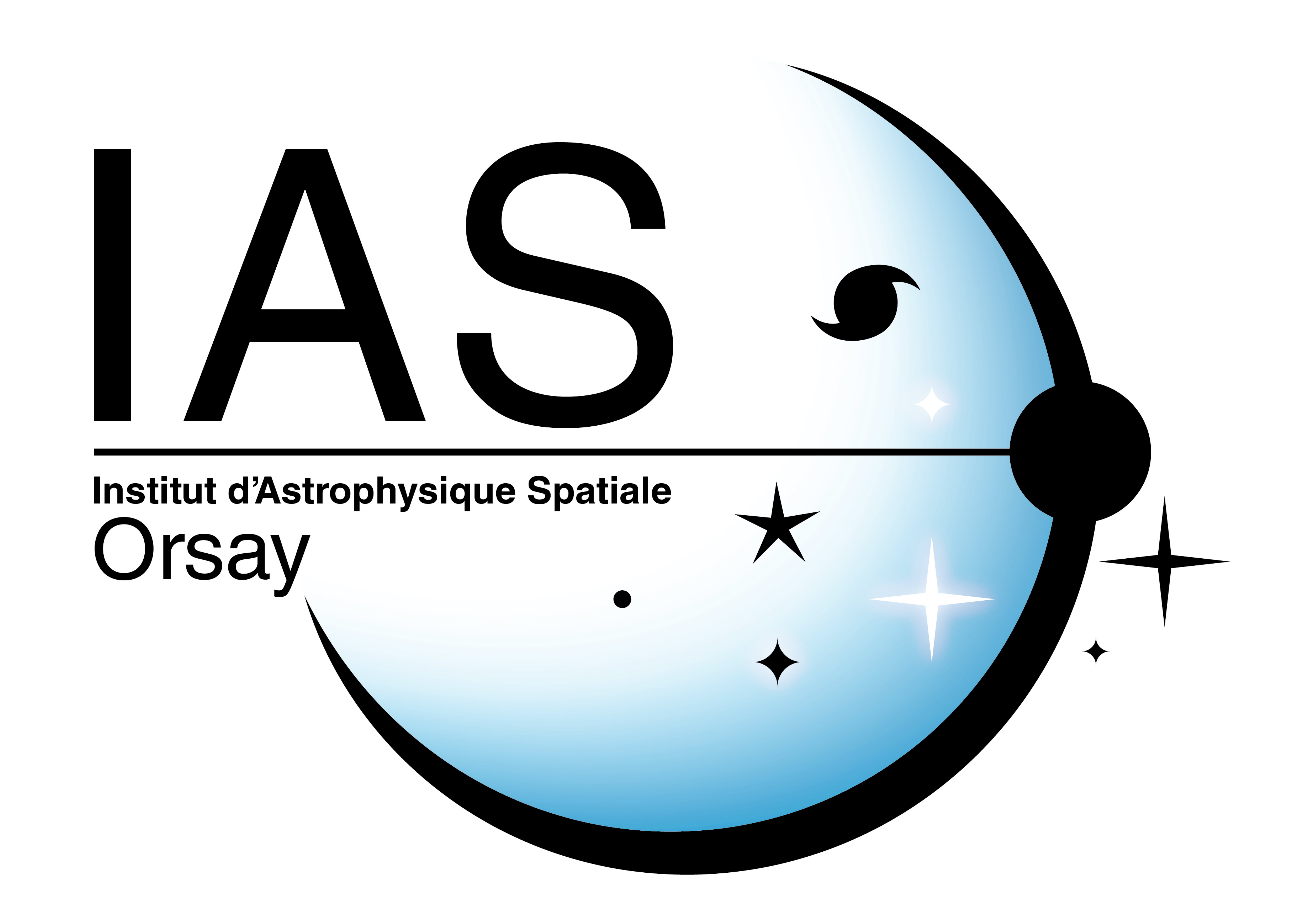}
\end{minipage}\hfill
\begin{minipage}{0.85\linewidth}
{\footnotesize {\bf Institut d'Astrophysique Spatiale} 
\hfill {\bf Mati\`{e}re Interstellaire et Cosmologie}
\rule{\linewidth}{1pt}
\begin{flushright} {\bf \'{E}cole doctorale Astronomie et Astrophysique d'\^{I}le-de-France} \end{flushright}}
\end{minipage}

%Rest of page
\ \\\ \\\ \\\
\begin{center}

%Title
{\bf {\fontsize{23pt}{30pt}\selectfont Non-Gaussianity and extragalactic foregrounds to the Cosmic Microwave Background}}

\ \\\ \\\

%{\bf {\fontsize{18pt}{25pt}\selectfont TH\`{E}SE DE DOCTORAT}}
%
%\ \\\
%
%pr\'{e}sent\'{e}e pour l'obtention du grade de
%
%\ \\\
%
%{\bf {\fontsize{16pt}{22pt}\selectfont Docteur de l'Universit\'{e} Paris-Sud}}
%\\\
%
%{\bf {\fontsize{12pt}{15pt}\selectfont Sp\'{e}cialit\'{e} Astrophysique et Instrumentations Associ\'{e}es}}
%
%\ \\\

{\bf {\fontsize{16pt}{22pt}\selectfont Universit\'{e} Paris-Sud}}

\ \\\

{\bf {\fontsize{15pt}{25pt}\selectfont TH\`{E}SE DE DOCTORAT}}

\ \\\

soutenue le 23/9/2013

\ \\\

par

\ \\\

{\fontsize{22pt}{30pt}\selectfont Fabien Lacasa}

\end{center}

\ \\\ \\\ \\\

{\small
\begin{tabular}{llr}
\!\!\!\!{\bf Directrice de th\`{e}se : } & Nabila Aghanim & \qquad \ Institut d'Astrophysique Spatiale, Orsay
\end{tabular}
}

%Jury
\noindent {\bf Composition du Jury}
\\\

{\small
\begin{tabular}{llr}
\!\!\!\!\textit{Pr\'{e}sident : } & Jean-Loup Puget & Institut d'Astrophysique Spatiale, Orsay \\
\!\!\!\!\textit{Rapporteurs : } & Olivier Dor\'{e} & Jet Propulsion Laboratory \\
 & Jean-Philippe Uzan & Institut d'Astrophysique de Paris \\
\!\!\!\!\textit{Examinateurs : } & Francis Bernardeau & Commissariat ˆ l'Žnergie atomique \\
& Bartjan van Tent & Laboratoire de Physique Th\'{e}orique, Orsay \\
\!\!\!\!\textit{Membres invit\'{e}s : } & Martin Kunz & Universit\'{e} de Gen\`{e}ve \\
\phantom{aaaaaaaaaaaaaaaaaaaa} & Guilaine Lagache & Institut d'Astrophysique Spatiale, Orsay
\end{tabular}

%\textit{Membres invit\'{e}s : } Martin Kunz \hfill Universit\'{e} de Gen\`{e}ve\\
%\phantom{\textit{Membres invit\'{e}s : }} Guilaine Lagache \hfill Institut d'Astrophysique Spatiale, Orsay
}
\ \\\

\noindent \rule{\linewidth}{1pt}
\begin{center} {\bf Universit\'{e} de Paris-Sud XI} \end{center}

%Back to previous geometry
\restoregeometry

\newpage
\null
\newpage

%Page de remerciements
%\input{thanks}

%\newpage
%\null
%\newpage

%Resume en francais
%\renewcommand{\abstractname}{R\'{e}sum\'{e}}
%\abstract{\input{resume}}

%\newpage
%\null

%Abstract en anglais
\renewcommand{\abstractname}{Abstract}
\abstract{This PhD thesis, written in english, is interested in the non-Gaussianity of extragalactic foregrounds to the Cosmic Microwave Background (CMB). Indeed, the CMB is on of the golden observables of contemporary cosmology, allowing in particular to highly constrain the cosmological parameters describing the universe. In the last decade has emerged the interest of looking for deviations of the CMB statistics to the Gaussian law. Indeed, the standard models for the generation of primordial perturbations (inflation) predict a close to Gaussian statistic, where the deviations to this law would in fact allow to discriminate the different models. However the measurements of the CMB, in particular by the \planck satellite for which I worked, can be contaminated by several emissions also known as foregrounds, depending on the observing frequency. I got interested in particular in some of these foregrounds due to extragalactic emissions tracing the large scale structure of the universe, specifically to radio and infrared point-sources (the latter constituting the Cosmic Infrared Background, CIB) and to the thermal Sunyaev-Zel'dovich effect (tSZ).
\newline

In this thesis, I hence describe the tools to characterise a random field and its deviation to the Gaussian law. The harmonic space is well adapted for theoretical and numerical computations and I thus define polyspectra. This thesis is particularly interested in the third order polyspectrum, the bispectrum, as it is the lowest order indicator of non-Gaussianity, with potentially the highest signal to noise ratio. I describe how the bispectrum can be estimated on data, accounting for the complexity due to partial sky coverage (because of a galactic mask for example) and instrumental resolution and noise. I then propose a method to visualise the bispectrum, which is more adapted than the already existing ones as it allows to probe the scale and configuration dependence while accounting for the invariance under the permutation of multipoles. I then describe the error bars/covariance of a polyspectrum measurement, a method to generate non-Gaussian simulations (useful to test the effect of diverse instrumental complications on a measurement), and how the statistic of a 3D field projects onto the sphere when the observations are line-of-sight integrals.

I then motivate my research from the cosmological point of view, by describing the generation of density perturbations \footnote{called adiabatics, the possibility of isocurvature modes being neglected here.} by the standard inflation model and their possible non-Gaussianity, how they yield the Cosmic Microwave Background anisotropies and grow to form the large scale structure of today's universe. To describe this large scale structure, I present the halo model and propose a diagrammatic method allowing to compute the polyspectra of the galaxy density field and to have a simple and powerful representation of the involved terms.

I then describe the different foregrounds to the Cosmic Microwave Background, galactic as well as extragalactic. I briefly describe the physics of the thermal Sunyaev-Zel'dovich effect and explain how its spatial distribution can be described with the halo model. I then describe the extragalactic point-sources and present a prescription for the non-Gaussianity of clustered sources and its generalisation to the case of several populations of sources, clustered or not. For the CIB I introduce a physical modeling with the halo model, and the diagrammatic method proves to be particularly useful in this case. I implement numerically the model to compute the 3D galaxy bispectrum and to produce the first theoretical prediction of the CIB angular bispectrum. I then explore the model results : I show the contributions of the different terms and the temporal evolution of the galaxy bispectrum. For the CIB angular bispectrum, I show its different terms and its scale and configuration dependence. I also show how the bispectrum varies with model parameters and which constraints a measurement would bring to these parameters. In the considered case, the bispectrum allows very good constraints, as it either breaks degeneracies present at the power spectrum level or either these constraints are better than those coming from the power spectrum.

Finally, I describe my work on measuring non-Gaussianity. I first explain the estimator that I introduced for the amplitude of the CIB bispectrum, and how this estimator can be combined with similar ones for radio sources and the CMB to produce a joint and robust constraint of the different sources of non-Gaussianity. Then, I quantify the contamination that extragalactic point-sources can produce to the estimation of primordial NG\,; in the \planck case I show that this contamination is negligible for the frequencies where the CMB is dominant. I then describe my measurement of the CIB bispectrum on \planck data~: the different difficulties encountered and the obtained results. The bispectrum is very significantly detected at 217, 353 and 545 GHz with signal to noise ratios ranging from 5.8 to 28.7. Its shape is consistent between frequencies, as well as the intrinsic amplitude of NG, which garanties the results' robustness. Interestingly, the measured bispectrum is significantly steeper than the empirical prescription that I introduced, and than the physical model based on the halo model described precedently. I speculate that this difference indicates the necessity to complexify the infrared emission model, and that it can be explained by the most recent models which consider that the emissivity depends on the host halo mass and on the type of galaxy (central/satellite). Ultimately, I describe my measurement work on the thermal Sunayev-Zel'dovich bispectrum, on simulations and on Compton parameter maps estimated by \planck. For the \planck data, the robustness of the estimation is validated thanks to realist foreground simulations, which allow to quantify their contamination to the estimated tSZ map. The tSZ bispectrum is very significantly detected with a signal to noise ratio $\sim$200, and I show that its amplitude is consistent with the projected map of detected clusters and with the \planck Sky Model simulation\,; I also show that its scale and configuration dependence is consistent with that I found on tSZ simulations. Finally, I use this measurement to put a constraint on the cosmological parameters $\sigma_8$ and $\Omega_b\,$, obtaining $\sigma_8 \; (\Omega_b / 0.049)^{0.35} = 0.74 \pm 0.04$ in agreement with the estimations of $\sigma_8$ through other tSZ statistics.
\newline

This thesis lead to four articles : \cite{Lacasa2012} published by MNRAS, \cite{Lacasa2013a} accepted by Astronomy \& Astrophysics and being revised, and \cite{Lacasa2013b} \& \cite{Penin2013} accepted by MNRAS.\\
I also contributed to the following \planck articles : \cite{planck2013-SZmap}, \cite{planck2013-NG} and \cite{planck2013-CIB}.}

\null
\newpage

%Synthese en francais
%\input{synthese}

%\newpage
%\null

\setcounter{page}{0}

%Table des matieres
\dominitoc

\setcounter{tocdepth}{3} %Depth of Table of Contents %3=up to subsubsection
\setcounter{secnumdepth}{3} %Depth of numbering
\tableofcontents

\newpage
\null
\newpage

\setcounter{page}{1}

\fancypagestyle{plain}
{
    \fancyhead{}
    \fancyfoot[C]{\thepage}
    \renewcommand{\headrulewidth}{0pt}
}	% re-define plain page after the Table of Contents
\pagestyle{plain}  %back to plain style with numbering of pages

%%%%%%%%%%%%%%%%%%%%%%%%%%%%%%%%%%%%%%%%%%%%%%
%Corps du texte

\chapter*{Introduction}
\addstarredchapter{Introduction}

Sky observations are nowadays carried from radio waves to gamma rays, extending astronomy across the whole electromagnetic spectrum. In particular, space-based all-sky surveys in the microwave range have become available since RELIKT-1 \citep{Klypin1987}, and later on with COBE \citep{Boggess1992}, WMAP \citep{Bennett1997}, and now \planck \citep{planck2011-mission}. These missions were designed to map the Cosmic Microwave Background (CMB), a relic radiation from the early universe emitted when it was in a plasma state. The Cosmic Microwave Background is indeed one of the golden observables of contemporary cosmology, along with standard rulers such as type Ia supernovae \citep{Riess2009}, and Large scale Structure (LSS) tracers, e.g. allowing the measurement of Baryon Acoustic Oscillations \citep{Percival2010}.

Extraction of cosmological information from the observed CMB map involves the measurement of its spatial correlation functions, in particular at second order i.e. the power spectrum. For example, the \planck mission has put the stringest constraint to date on cosmological parameters using the measurement of this power spectrum \citep{planck2013-cosmoparams}. However, in the last decade interest has risen for the measurement of higher orders (non-Gaussianity studies) as it may provide information on the primordial process generating the cosmological perturbation, in particular it may discriminate inflation models which are degenerate at the power spectrum level \citep{Bartolo2004}.

Foreground signals are nevertheless present at microwave frequencies and contaminate the CMB observations. These foregrounds stem from the emission of LSS tracers and probe information of cosmological interest (e.g., the ionised gas in clusters or the star-formation history). High order measurements are also of interest for these LSS tracers. Indeed, they allow for a more complete statistical characterisation of the field, hence an extraction of more cosmological information.

In the literature, the study of foregrounds non-Gaussianity first focussed on the case of extragalactic radio point-sources, which are the dominant non-Gaussian signal outside the galactic plane at WMAP frequencies. These studies were used to quantify the contamination of point-sources to the CMB power spectrum \citep{Komatsu2003} or to the estimation of primordial non-Gaussianity \citep[e.g.][]{Argueso2003,Gonzalez-Nuevo2005,Babich2008}. The study of LSS non-Gaussianity has also emerged in the case of galaxy surveys. It has indeed been shown to be a powerful probe cosmological parameters as well as galaxy formations model, in a way complementary to the power spectrum \citep[see e.g. the thesis by][]{Sefusatti2005}. 

This thesis is interested in the study of high order correlation functions for different extragalactic foregrounds to the CMB, in particular at third order. This study has two main motivations. First, it is required for an unbiased measurement of the corresponding information for the primordial CMB. Second, the measurement of these high orders and its comparison with theoretical expectation allows a more complete statistical analysis of the foreground signals. This statistical analysis could in turn provide better constraints on the models of these foregrounds, and thus a better understanding of the processes they trace and of the LSS distribution.
\newline

The thesis is divided as follows~: in the first chapter, I introduce the statistical tools needed to study random fields.\\
The second chapter focuses on the case of random field on the sphere, such as sky observations, and introduces harmonic analysis. It discusses in particular the issue of the estimation of high orders, and the projection of these statistics from the 3-dimensional space, where models are naturally set, to the celestial sphere.\\
The third chapter introduces the basics of cosmology motivating this thesis, and shows how the primeval perturbations of the universe are generated during the inflation period. It further shows how these perturbations generate the CMB anisotropies and later on the Large Scale Structure distribution. In particular I describe a diagrammatic formalism I have developed, which allows the computation of the high order correlation functions of the galaxy density field, a standard tracer of the LSS.\\
The fourth chapter introduces the different extragalactic foregrounds to the CMB that trace the LSS, in particular the Cosmic Infrared Background (CIB) and the thermal Sunyaev-Zel'dovich (tSZ) signal. Furthermore it shows how their correlation functions can be modeled theoretically.\\
The last chapter describes the data analysis I have performed for \planck on these extragalactic foregrounds. It shows my estimations of how CMB high order studies are biased by extragalactic foregrounds, and how to account for this bias. It also shows the third order measurements that I have produced for the CIB and tSZ signals, and the constraints that I could derive from these measurements.\\
A glossary is given after the conclusion and the perspectives, listing abbreviations used and symbols that may not be standard.
\newline

This thesis lead to four articles : \cite{Lacasa2012} published by MNRAS, \cite{Lacasa2013a} accepted by Astronomy \& Astrophysics and being revised, and \cite{Lacasa2013b} \& \cite{Penin2013} submitted to MNRAS.\\
I furthermore contributed to the following \planck articles : \cite{planck2013-SZmap}, \cite{planck2013-NG} and \cite{planck2013-CIB}.

%{\bf
%Chaque partie principale doit comporter si possible :
%\begin{itemize}
%\item littŽrature, Žtat de l'art
%\item questions posŽes que je vais adresser
%\item Mon boulot dans le dŽtail
%\item rŽponse aux questions posŽes
%\item que reste-t-il ˆ faire / comment aller plus loin
%\end{itemize}
%}

\mychapter{Some statistics}{Some statistics}{chapt:stat}

%\lettrine{lettre}{}

%In this chapter, which is \emph{not} an introduction to statistics and has no pedagogical pretention, I will briefly introduce the statistical notions used in cosmology which will be needed to expose my work. I illustrate these notions on examples which will find applications later on in the thesis.
In this chapter, which has no pedagogical pretention, I describe the statistical characterisation of random objects. I begin with the simpler case of random variables, before generalising to random fields. The statistical notions described are illustrated on several examples of distributions of random variables or random fields, and these examples find applications later on in the thesis.

%%%%%%%%%%%%%%%%%%%%%%%%%%%%%%%%%%%%%%%%%%%%%%%%
\section{Random variables}
\subsection{Notations}\label{Sect:rvnot}
Let $X$ be a random element, valued on some space E. Its probability law is a positive measure on E, with total measure unity~:
\be
\int \dd P_X = 1
\ee
The probability of an event $A$ is :
\be
\mathbb{P}(X \in A) = \dd P_X\left(X^{-1}(A)\right)
\ee
For circumstances of interest in cosmology, the probability law derives from a \textit{probability density function} (p.d.f.)~: 
\be
\dd P_X = p(x) \,\dd x
\ee
with $p(x)$ a positive function (with integral unity) and $\dd x$ the Lebesgues measure on the space of interest, that is $\mathbb{R}^n$ or $\mathbb{C}^n$ throughout this thesis.\\
Any function $g(X)$ then has an expectation value\footnote{traditionally noted $\mathbb{E}[g(X)]$ in mathematics, but we use here lighter notations in agreement with cosmology's conventions.}~:
\be
\lbra g(X) \rbra = \int g(x) \; p(x) \;\! \dd x
\ee
if the integral converges.

Two random elements are said \textit{independent} (noted $X \ci Y$) iff\footnote{if and only if} for any functions $g$ and $h$~:
\be
\lbra g(X) \, h(Y) \rbra = \lbra g(X) \rbra \, \lbra h(Y) \rbra
\ee

A random variable is a scalar random element, i.e. it is valued in $\mathbb{R}$ or $\mathbb{C}$.

\subsection{Moments}
The $n$-th order moment of a random variable $X$ is defined, for $n \in \mathbb{N}$, as :
\be
m_n = \lbra X^n \rbra
\ee
if the involved integral converges.\\
In particular $m_1$ is called the mean (or average) of $X$. One may define the \textit{centered moments} by substracting this average :
\be
\mu_n = \lbra \tilde{X}^n \rbra \quad \mathrm{with} \quad \tilde{X} = X - \lbra X \rbra
\ee
In the following, we will always consider centered moments and loosely refer to them as just `moments' for concision.\\
We note in passing that moments may not be defined beyond a certain order, explicit examples being given in Sect.\ref{Sect:exmom}.
\newline

The \textit{moment-generating function} (m.g.f.) is defined, for $t\in\mathbb{R}$, as :
\be\label{Eq:defmgf}
M_X(t) = \lbra e^{t X} \rbra
\ee 
if the expectation value exists. In other terms, it is the Laplace transform of the p.d.f. It may be defined only at $t=0$ for some random variables (see Sect.\ref{Sect:exmom}). However, if it can be defined on $\mathbb{R}$ and is sufficiently regular, the p.d.f. can be recovered from the moment-generating function through an inverse Laplace transform, meaning that two random variables having the same moment-generating function have the same probability law.\\
The moment-generating function inherits its name from the following property : if it is defined at least on a neighbourhood of $t=0$, then~:
\be
m_n = \left.\frac{\dd^n M_X(t)}{\dd t^n}\right|_{t=0}
\ee
and, if the series converges,
\be\label{Eq:seriesmgf}
M_X(t) = \sum_{n=1}^{+\infty} \frac{m_n \, t^n}{n!}
\ee
This is an important result for statitics as it means that the study of a random variable can be reduced to the measurement of the hierarchy of its moments (if the series in Eq.\ref{Eq:seriesmgf} converges). So that the uncountably infinite problem of measuring the p.d.f. of a random variable is reduced to the countable (but still infinite) problem of measuring its moments.\\
In practice one may want to measure only a finite number of moments and truncate the series Eq.\ref{Eq:seriesmgf} at that order. This approach will be further developed and discussed in Sect.\ref{Sect:GramCharlier}.
%however this suffers from known problems as the infered p.d.f. may not be positive everywhere, and the \textit{tails} of the distribution (the points where the p.d.f. is small compared to its maximum) may be poorly reproduced.

The moment-generating function has the following property~:
\be\label{Eq:gmfindep}
X \ci Y \Leftrightarrow M_{X+Y} = M_X \, M_Y
\ee
which will be useful later, in this chapter.

\subsection{Cumulants}
For a random variable, the \textit{cumulant-generating function} (c.g.f.) is defined by~:
\be\label{Eq:defcgf}
g_X(t) = \log(M_X(t))
\ee
which defines the cumulants as~:
\be
\kappa_n = \left.\frac{\dd^n g_X(t)}{\dd t^n}\right|_{t=0}
\ee
so that~:
\be\label{Eq:cgfseries}
g_X(t) = \sum_{n=1}^{+\infty} \frac{\kappa_n \, t^n}{n!}
\ee
Under the previously mentionned regularity hypothesis, the cumulant-generating function characterises univoquely the probability law. Hence characterising a random variable can be achieved by studying its cumulants.\\
Through Eq.\ref{Eq:gmfindep} we have $g_{X+Y} = g_X + g_Y$ iff $X \ci Y$, hence cumulants have the following property~:
\be
X \ci Y \quad \Leftrightarrow \quad \kappa_n(X+Y) = \kappa_n(X) + \kappa_n(Y) \ \ \forall n
\ee
This property is not shared by (centered) moments, hence cumulants are more advantageous for theoretical or practical computations.
\newline

Cumulants are related univoquely and hierarchically to moments : the cumulant at a given order is a function of the moments up to that order, and vice-versa. The following equations give these relations up to order 6, as they will be useful later on~:
\bea
\kappa_1 &=& m_1  \label{Eq:momtocum1} \\
\kappa_2 &=& \mu_2\\
\kappa_3 &=& \mu_3\\
\kappa_4 &=& \mu_4 - 3\mu_2^2\\
\kappa_5 &=& \mu_5 - 10\mu_3\,\mu_2\\
\kappa_6 &=& \mu_6 - 15 \mu_4 \, \mu_2 -10 \mu_3^2 + 30 \mu_2^3 \label{Eq:momtocum6}
\eea
and conversely~:
\bea
m_1 &=& \kappa_1\label{Eq:cumtomom1} \\
\mu_2 &=& \kappa_2\\
\mu_3 &=& \kappa_3\\
\mu_4 &=& \kappa_4 + 3\kappa_2^2\\
\mu_5 &=& \kappa_5 + 10\kappa_3 \, \kappa_2\\
\mu_6 &=& \kappa_6 + 15 \kappa_4 \, \kappa_2 +10 \kappa_3^2 +15\kappa_2^3 \label{Eq:cumtomom6}
\eea

{\bf Terminology : } $\kappa_2$ is called the variance, often noted $\sigma^2$, and its square root $\sigma$ is called the standard deviation. It is sometimes used to normalise higher order cumulants, e.g. the skewness is defined as $\kappa_3/\sigma^3$, and the (excess) kurtosis is $\kappa_4/\sigma^4$. In this thesis, we will not use this normalisation unless explicitly mentioned, so we may loosely use the term skewness to refer to the third-order cumulant.

\subsection{Examples and limitations}\label{Sect:exmom}
\subsubsection{Gaussian law}
The Gaussian (or normal) law is a real distribution (i.e. $X \in \mathbb{R}$) which is ubiquitous in physics, particularly because of the Central Limit theorem\footnote{which essentially tells that the average of many independent random variables tends to be Gaussian, for example the power deposited by photons on a \planck detector, as there are many photons hitting the detector within its response time.}. It is characterised by two parameters which can be chosen as its mean and variance, it is noted $\mathcal{N}(\mu,\sigma^2)$, and has p.d.f.~:
\be
p(x) = \frac{1}{\sqrt{2\pi \sigma^2}}\exp\left(-\frac{(x-\mu)^2}{2\sigma^2}\right) \qquad \forall x \in \mathbb{R}
\ee
As the Laplace transform of a Gaussian is a Gaussian, the moment-generating function is calculated easily ~:
\be
M_X(t) = \exp\left(\mu t + \frac{\sigma^2 t^2}{2}\right)
\ee
Consequently the cumulant-generating function is :
\be
g_X(t) = \mu t + \frac{\sigma^2 t^2}{2}
\ee
and the cumulants are :
\bea
\kappa_1 &=& \mu \\
\kappa_2 &=& \sigma^2 \\
\kappa_n &=& 0 \quad \forall n\geq 3
\eea
The moments are more complicated and may e.g. be recovered through the formulas \ref{Eq:cumtomom1}-\ref{Eq:cumtomom6}. The simplicity of cumulants is one of their conceptual and practical advantage and, anticipating on the following, it shows that studying cumulants at order $\geq 3$ probes potential non-Gaussianity.

\subsubsection{Poisson law}\label{Sect:statPoisson}
The Poisson law is a discrete distribution, meaning that the outcome of the random variable is an integer. It arises when counting the number of independent events happening on a given period of time, for example the number of photons hitting a {\planck} detector within one second. It is characterised by a single parameter $\lambda$ being the mean, and the probability law is :
\be
P(X=k) = e^{-\lambda} \; \frac{\lambda^k}{k!}
\ee
The m.g.f. is~:
\be
M_X(t) = \exp\left(\lambda (e^t-1)\right)
\ee
so that the cumulants are :
\be
\kappa_n = \lambda \quad \forall n \in \mathbb{N}^*
\ee
In particular the dimensionless skewness is $\kappa_3/\sigma^3 = \lambda^{-1/2}$ and can be made arbitrarily important by taking $\lambda$ small enough. This property will be important in Sect.\ref{Sect:bispcouplmat_sim} to generate non-Gaussian simulations.

\subsubsection{Probability laws without moments nor cumulants}
The Cauchy-Lorentz \citep{Hazewinkel2001} law is a real-valued distribution with a p.d.f. given by~:
\be
p(x) = \frac{1/\pi}{1+x^2}
\ee
Because the p.d.f. decreases slowly at infinity, all the moments, and hence cumulants, are undefined. Indeed the involved integral does not converge. The moment-generating function is also undefined for $t\neq0$.

Furthermore, one may define a probability law such that the moments and cumulants are defined only up to a certain order $n$, e.g.~:
\be
p(x) \propto \frac{1}{1+|x|^{n+2}}
\ee
defines a p.d.f. up to a normalisation factor. Its moment-generating function is also undefined for $t\neq0$.

High order moments or cumulants are hence sensitive to the tail of the p.d.f., i.e. to rare events. A sufficient condition for the existence of cumulants is that the p.d.f. has bounded support or decreases faster than any rational fraction (e.g. as an exponential) at infinity.

\subsubsection{Probability law not uniquely defined by its moments or cumulants}
A random variable $X$ follows a \textit{log-normal} distribution if $Y=\log(X)$ follows a Gaussian distribution. It is noted $\ln \mathcal{N}(\mu,\sigma^2)$, with $\mu$ and $\sigma^2$ respectively the mean and variance of $Y$, and its p.d.f. is~:
\be
p(x)=\frac{1}{x \sqrt{2\pi \sigma^2}} \exp\left(-\frac{(\ln x -\mu)^2}{2\sigma^2}\right)
\ee
All the moments and cumulants are defined, namely the uncentered moments are :
\be
m_n = \exp\left(\mu n + \frac{\sigma^2 n^2}{2}\right)
\ee
while the centered moments and the cumulants have more complicated expressions.

However the moment-generating function is undefined for $t\neq0$ as the involved integral does not converge. Indeed the series in Eq.\ref{Eq:seriesmgf} diverges, as $\frac{m_n}{n!} \rightarrow +\infty$ when $n \rightarrow +\infty$.\\
Hence we are not in the position of deducing the p.d.f. from the moments, and indeed it can be shown that there are other distributions sharing exactly the same moments (see \cite{Stoyanov1987} Sect. 11.2 for explicit details). The log-normal distribution is called \textit{moment-indeterminate}, in other terms statistical information is escaping the hierarchy of moments. This problem arises because the distribution is ``fat-tailed''.

The inadequacy of the direct moment approach in this case has been proven of interest for cosmology by \cite{Carron2011}, as the matter density field approximately follows a log-normal distribution. However, taking a log  transform of the field suffices to make the moment approach valid again, as the variable then becomes Gaussian distributed. This `Gaussianization' process --i.e. a 1-point non-linear remapping such that the 1-point marginal becomes Gaussian or close to Gaussian-- was first proposed for Large Scale Structure studies by \cite{Weinberg1992} and further developed by \cite{Neyrinck2011a,Neyrinck2011b}. Note however that `Gaussianizing' a general random field may not yield a Gaussian field\footnote{For example, consider the p.d.f. with two points $p_\epsilon(x_1,x_2)\propto \mr{Gauss}(x_1,x_2) \left(1+\epsilon\sin(x_1 \,x_2)\right)$\\ $\forall\epsilon\in[-1,1]$, with `Gauss' any two-point symmetric Gaussian. This p.d.f. is isotropic and its 1-point marginal is Gaussian, however it does not represent a Gaussian field.}, so that the power spectrum of the Gaussianized field is not a complete statistical characterisation of the field. Finally, building on the same principle as the `Gaussianization', several articles have been published during the writing of this thesis on applying non-linear transformations to put back information into the moment's hierarchy of cosmological fields \citep{Carron2013,Leclercq2013,Simpson2013}.

%%%%%%%%%%%%%%%%%%%%%%%%%%%%%%%%%%%%%%%%%%%%%%%%
\section{Random fields}
\subsection{Definition}
A random field is a collection of random variables $\mathbf{X}=(X_i)_i$ with an index $i$ running on a continuous space. Examples of relevance for cosmology are :
\begin{itemize}
\item the intensity of a signal depending on the direction in the sky. The index $i$ is then running on the sphere and may be taken as the unit vector $\hn$.
\item the baryonic matter density (or velocity, pressure...) depending on the space-time point. The index $i$ is then running on the manifold and may be taken as $(\vec{x},t)$ given a global choice of coordinates.
\end{itemize}
In the following, we will consider that the index runs over a finite number of values, as this will avoid technicalities such as dealing with functional derivatives. This is also realistic for practical applications as sky observations do not produce continuous fields but pixelised maps, with a finite number of pixels.\\
In summary, $\mathbf{X}$ can be thought of as a large random vector, e.g. $X_i = T(\hn_i)$ the temperature of a sky signal in direction $\hn_i$ with $i=1\cdots n_\mathrm{pix}$.

The definition of independence in Sect.\ref{Sect:rvnot} also applies to random fields.

\subsection{Correlation functions}\label{Sect:correlfunct}
In the following sections, we will assume the index $i$ of $\mathbf{X}=(X_i)_i$ runs from 1 to $n$.\\
The correlation function (c.f. hereafter) of order $k$ is then :
\be
\mu_k(X_{i_1}, \cdots,X_{i_k}) = \lbra \tilde{X}_{i_1} \cdots \tilde{X}_{i_k} \rbra
\ee
with $\tilde{X_j}=X_j-\lbra X_j\rbra$, and with the usual hypothesis that the expectation value is well-defined.\\
The correlation functions are generated by the moment-generating function :
\be
M_{\tilde{\mathbf{X}}}(\mathbf{t}) = \lbra \exp\left(\mathbf{t} \cdot \tilde{\mathbf{X}} \right) \rbra
\ee
with $\mathbf{t} = (t_i)_i$ a vector, and the dot $\cdot$ the canonical scalar product.\\
Indeed, if the m.g.f. is defined, we have :
\be
M_{\tilde{\mathbf{X}}}(\mathbf{t}) = \sum_{k=1}^{+\infty} \ \sum_{i_1 \cdots i_k = 1}^n \frac{t_{i_1} \cdots t_{i_k}}{k!} \times \mu_k(X_{i_1}, \cdots,X_{i_k})
\ee
and~:
\be
\mu_k(X_{i_1}, \cdots,X_{i_k}) = \left.\frac{\partial^k M_{\tilde{\mathbf{X}}}(\mathbf{t}) }{\partial t_{i_1} \cdots \partial t_{i_k}}\right|_{\mathbf{t} = \mathbf{0}}
\ee
As for random variables, we have :
\be
\mathbf{X} \ci \mathbf{Y} \quad \Leftrightarrow \quad M_{\mathbf{X+Y}}(\mathbf{t}) = M_{\mathbf{X}}(\mathbf{t}) \times M_{\mathbf{Y}}(\mathbf{t})
\ee
%Also, if only independent variables are involved, the m.g.f. factorizes.\\
%{\bf First example :}\\
%If $X_1$ is independent of $X_2 \cdots X_n$, then
%\be
%M_{\mathbf{X}}(\mathbf{t}) = M_{X_1}(t_1) \times M_{\mathbf{X'}}\left(\mathbf{t'}\right) \quad \mathrm{with} \quad \mathbf{X'}=(X_i)_{i=2\cdots n} \quad \mathrm{and} \quad \mathbf{t'}=(t_i)_{i=2\cdots n}
%\ee
%{\bf Second example :}\\
%If $X_1 = \cdots = X_m = X$ and $X_{m+1} = \cdots = X_n = Y$ with $X \ci Y$, then
%\be
%M_{\mathbf{X}}(\mathbf{t}) = M_X(t_1+\cdots+t_m) \times M_Y(t_{m+1} + \cdots + t_n)
%\ee

\subsection{Ursell functions -- or connected correlation functions}\label{Sect:ursellfunct}
For random fields the cumulant-generating function is defined as for random variables :
\be
g_{\mathbf{X}}(\mathbf{t}) = \log\left(M_{\mathbf{X}}(\mathbf{t})\right)
\ee
It generates so-called \textit{connected correlation functions} or \textit{Ursell functions} through :
\be
\kappa_k(X_{i_1}, \cdots,X_{i_k}) =  \left.\frac{\partial^k g_{\mathbf{X}}(\mathbf{t}) }{\partial t_{i_1} \cdots \partial t_{i_k}}\right|_{\mathbf{t} = \mathbf{0}}
\ee
so that~:
\be
g_{\mathbf{X}}(\mathbf{t}) = \sum_{k=1}^{+\infty} \ \sum_{i_1 \cdots i_k=1}^n \frac{t_{i_1} \cdots t_{i_k}}{k!} \times \kappa_k(X_{i_1}, \cdots,X_{i_k})
\ee 
The connected correlation functions can also be noted $\lbra X_{i_1} \cdots X_{i_k} \rbra_c$.\\
They are related to the simple correlation functions through formulas analog to Eq.\ref{Eq:momtocum1}-\ref{Eq:momtocum6} linking cumulants and moments. In particular, both coincide up to order 3 and differ afterwards.

As for random variables, we have :
\be
\mathbf{X} \ci \mathbf{Y} \quad \Leftrightarrow \quad g_{\mathbf{X+Y}}(\mathbf{t}) = g_{\mathbf{X}}(\mathbf{t}) + g_{\mathbf{Y}}(\mathbf{t})
\ee
so that~:
\be\label{Eq:additivconnectcf}
\kappa_k\left(\mathbf{X+Y}\right) = \kappa_k\left(\mathbf{X}\right) + \kappa_k\left(\mathbf{Y}\right)\quad \forall k
\ee
This is an important property which gives conceptual and practical advantages to connected correlation functions over simple correlation functions.

A useful consequence is that if all $X_i$ involved are either equal or independent (and at least two of them are different), then the connected correlation function is zero.\\
For example let's consider two independent random variables $Y \ci Z$, and $\mathbf{X}$ such that
$X_1 = \cdots = X_m = Y$ and $X_{m+1} = \cdots = X_k = Z$. We define $\mathbf{Y}$ and $\mathbf{Z}$ by~:
 \bean
 Y_1 = \cdots = Y_m = Y \quad &\mathrm{and}& \quad Y_{m+1} = \cdots = Y_k = 0\\
 Z_1 = \cdots = Z_m = 0 \quad &\mathrm{and}& \quad Z_{m+1} = \cdots = Z_k = Z
 \eean
 where 0 is the random variable which returns zero everytime. We have $\mathbf{X}=\mathbf{Y}+\mathbf{Z}$ with $\mathbf{Y} \ci \mathbf{Z}$. Hence
\bea
\nonumber\kappa_k(X_1,\cdots,X_k) &=& \kappa_k(\mathbf{Y})+\kappa_k(\mathbf{Z})\\
&=& 0 + 0 = 0
\eea

\subsection{Gaussian random fields}
A Gaussian random field $\mathbf{X}$ is a continuous random field parametrised by its mean $\mathbf{m}$ and an $n\times n$ matrix $C$ :
\be
p(\mathbf{X}) = \frac{1}{\sqrt{(2\pi)^n |\det C|}} \exp\left[-\frac{1}{2} \, ^{T}\left(\mathbf{X}^*-\mathbf{m}^*\right) \cdot C^{-1} \cdot \left(\mathbf{X}-\mathbf{m}\right)\right]
\ee
with $^*$ the complex conjugation, $\mathbf{X} = (X_i)_{i=1\cdots n}$ considered as a vector, the dot $\cdot$ is the matrix multiplication. Furthermore, $C$ is the covariance matrix~:
\be
\forall i,j \quad C_{i,j} = \lbra \tilde{X}_i \tilde{X}_j \rbra = \lbra X_i X_j \rbra_c
\ee
or with vector notations $C = \lbra \mathbf{X} \, ^{T}\mathbf{X} \rbra_c$
\newline

Gaussian random fields appear frequently and have many applications in physics, signal processing etc. Out of their numerous properties, the following will be useful here~:
\begin{itemize}
\item If the covariance matrix is diagonal, $X_i$ are independent. In other words if two Gaussian random variables are uncorrelated, they are independent. This is the converse statement of the fact that independent random variables are uncorrelated, and generally does not hold if the variables are not Gaussian.
\item $\kappa_1(\mathbf{X})=\mathbf{m}$, $\kappa_2(\mathbf{X}) = C$, and all higher order connected correlation functions are zero. Moreover empirical measurement of the first and second order correlation functions is optimal, meaning that no further information on the parameters of the field can be extracted from the data (where data means empirical realisations of $\mathbf{X}$).
\end{itemize}

\subsection{White noise}\label{Sect:stat-whitenoise}
A white noise field is a collection of independent random variables, i.e. all $X_i$ are independent. If the $X_i$ are furthermore identically distributed (i.i.d.), the white noise is called \textit{homogeneous}.\\
Sometimes the term white noise is used for Gaussian white noise, but we do not make this confusion here, as we will use white noises with different underlying p.d.f. later on.
\newline

\noindent {\bf Example :} the instrumental noise in {\planck} maps is an inhomogeneous Gaussian white noise. Indeed each pixel is independent of the others (at least at first approximation), the instrumental noise is Gaussian, and some pixels are less noisy than others (because they have been observed more often due to the {\planck} scanning strategy).

Due to the properties described in Sect.\ref{Sect:ursellfunct}, the Ursell functions vanish for a white noise unless all $X_i$ involved are equal~:
\be
\kappa_k(X_{i_1},\cdots,X_{i_k}) = \kappa_k(X_{i_1}) \, \delta_{i_1,\cdots,i_k}
\ee
with~:
\be
\delta_{i_1,\cdots,i_k} =
\left\{\begin{array}{ll}
1 & \qquad \mathrm{if}\quad i_1 = \cdots = i_k \\
0 & \qquad \mathrm{otherwise} \\
\end{array}\right.
\ee
and $\kappa_k(X_i)$ is the k-th order cumulant of $X_i$ considered as a single random variable. Furthermore, if the white noise is homogeneous, $\kappa_k(X_i)$ does not depend on $X_i$.

\subsection{Gram-Charlier expansion}\label{Sect:GramCharlier}
As stated previously, full knowledge of the cumulants (/Ursell functions) of a random variable (/field) gives knowledge of the p.d.f. of the variable (/field), under some regularity condition. The Gram-Charlier expansion puts this idea in practice when only some of the cumulants are known. It is an expansion of the p.d.f. with respect to a fiducial p.d.f., in terms of the difference of the cumulants. Most often, and in this thesis in particular, the fiducial p.d.f. is the Gaussian one with same mean and variance.\\
By inputting Eq.\ref{Eq:cgfseries} into Eq.\ref{Eq:defmgf} via Eq.\ref{Eq:defcgf} and taking the inverse Laplace transform, one gets for a random variable $X$~:
\be
p(x) = \exp\left[\sum_{n=3}^{+\infty} \frac{(-1)^n \kappa_n}{n!} \left(\frac{\dd}{\dd x}\right)^n\right] \frac{1}{\sqrt{2\pi \sigma^2}}\exp\left[-\frac{(x-\mu)^2}{2\sigma^2}\right]
\ee
Developing up to the fourth order we get~:
\bea
\nonumber p(x) &\approx& \frac{1}{\sqrt{2\pi \sigma^2}}\exp\left[-\frac{(x-\mu)^2}{2\sigma^2}\right] \times \bigg(1 +\frac{\kappa_3/\sigma^3}{3!}H_3\left(\frac{x-\mu}{\sigma}\right)+\frac{\kappa_4/\sigma^4}{4!}H_4\left(\frac{x-\mu}{\sigma}\right) \\
&& \qquad \qquad \qquad \qquad \qquad \qquad \qquad \qquad+ \mathcal{O}(\mathrm{order \ 5})\bigg)
\eea
with $H_n$ the $n$-th order Hermite polynomial \citep{Hermite1864}.\\
For a random field, computations become more cumbersome but we find up to the third order~:
\bea
\nonumber p(x_1,\cdots,x_n) &=& \exp\left[\sum_{n=3}^{+\infty} \frac{(-1)^n}{n!}  \sum_{i_1 \cdots i_n} \kappa_n(X_{i_1},\cdots,X_{i_n}) \frac{\partial^n}{\partial x_{i_1}\cdots \partial x_{i_n}}\right] \\
\label{GC_theo_rf} && \frac{1}{\sqrt{(2\pi)^n |\det C|}} \exp\left[-\frac{1}{2} \, ^{T}\left(\mathbf{X}-\mathbf{m}\right) \cdot C^{-1} \cdot \left(\mathbf{X}-\mathbf{m}\right)\right]\\
%\nonumber&\approx& \exp\left[-\frac{1}{2} \, ^{T}\left(\mathbf{X}-\mathbf{m}\right) \cdot C^{-1} \cdot \left(\mathbf{X}-\mathbf{m}\right)\right] \bigg(1 +\\
%&& \frac{1}{3!}\sum_{i_1,i_2,i_3} \frac{\kappa_3(X_{i_1},X_{i_2},X_{i_3})}{\sigma_{i_1} \sigma_{i_2} \sigma_{i_3}} H_3\left(\frac{x_{i_1}-m_{i_1}}{\sigma_{i_1}},\frac{x_{i_2}-m_{i_2}}{\sigma_{i_2}},\frac{x_{i_3}-m_{i_3}}{\sigma_{i_3}}\right)+ \mathcal{O}(\mathrm{order \ 4})\bigg) \qquad
\nonumber&\approx& \frac{1}{\sqrt{(2\pi)^n |\det C|}} \exp\left[-\frac{1}{2} \, ^{T}\left(\mathbf{X}-\mathbf{m}\right) \cdot C^{-1} \cdot \left(\mathbf{X}-\mathbf{m}\right)\right] \Bigg(1-\\
\nonumber&& \frac{1}{3!}\sum_{i_1,i_2,i_3}\kappa_3(X_{i_1},X_{i_2},X_{i_3})\bigg(\sum_a C^{-1}_{i_1,i_2}C^{-1}_{i_3,a} \,\tilde{x}_a + C^{-1}_{i_1,i_3}C^{-1}_{i_2,a} \,\tilde{x}_a+ C^{-1}_{i_2,i_3}C^{-1}_{i_1,a} \,\tilde{x}_a \\
\label{Eq:GCexplong} && \qquad +\sum_{a,b,c} C^{-1}_{i_1,a } C^{-1}_{i_2,b} C^{-1}_{i_3,c} \,\tilde{x}_a \tilde{x}_b \tilde{x}_c \bigg)\Bigg)
\eea
with $\tilde{x}_i = x_i - m_i$.\\
For concision, we can compact the Gaussian part in a single notation, and use Einstein's summation convention and the symmetrization operator
$A_{[i_1\cdots i_n]} = \left(A_{i_1 \cdots i_n} + \mathrm{perm.}\right)/n!\,$. Eq.\ref{Eq:GCexplong} then shortens to~:
\be\label{Eq:GC_ijk}
p(x_1,\cdots,x_n) \approx \mr{Gauss}(\mathbf{X}) \Bigg(1-\frac{\kappa_3(X_i,X_j,X_k)}{3!}\left(3 C^{-1}_{[ij} C^{-1}_{k]a} \tilde{x}_a - C^{-1}_{i,a } C^{-1}_{j,b} C^{-1}_{k,c} \,\tilde{x}_a \tilde{x}_b \tilde{x}_c\right)\Bigg)
\ee
This shows that the third-order correlation function is the leading order indicator of non-Gaussianity of the field. \cite{Babich2005} and \cite{Creminelli2006} have indeed used the previous equation to devise an estimator for the primordial non-Gaussianity in the Cosmic Microwave Background, and we will use this expansion in Sect.\ref{Sect:linearterm_bisp} to define the optimal estimator of the third order c.f. in harmonic space (the bispectrum). 
\newline
\newline

%Conclusion of the chapter
In this chapter, I have thus described random variables/fields and how they can be characterised by cumulants/connected correlation functions. The measurement of these correlation functions hence allows to extract statistical information from realisations of the field. In particular, non-zero cumulants/connected correlation functions probe deviations from the Gaussian case.

\mychapter{Some spherical analysis}{Some spherical analysis}{chapt:sphere}

In this chapter, we introduce the harmonic analysis used in cosmology. We will work principally on the sphere where {\planck} observations are set, but also in 3D where theoretical models are naturally set. Building up on the previous chapter, we describe correlation functions in harmonic space, i.e. polyspectra : their structure, their estimation, their error bars and  their representation. We also illustrate these tools on examples of cosmological relevance.

%%%%%%%%%%%%%%%%%%%%%%%%%%%%%%%%%%%%%%%%%%%%%%%%
\section{Spherical harmonics}
\subsection{Definition}
An element of the 2D sphere $\mathbb{S}_2$ is a unit vector $\hn$. It can also be parametrised by the spherical coordinates angle $(\theta,\phi) \in [0,\pi]\times[0,2\pi[$, where $\theta$ defines the latitude with respect to the North pole and $\phi$ defines the longitude.\\
The area element $\dd^2 \hn$ is the infinitesimal solid angle~:
\be
\dd^2 \hn = \dd \Omega = \sin \theta \, \dd \theta \, \dd \phi
\ee
measured in steradians (sr) with a total sky area of $4\pi$ steradians $\simeq 41\, 000 \;\mathrm{deg}^2$.\\
The spherical harmonics $Y_{\ell m}$ are the eigenfunctions of the Laplacian on the sphere\footnote{with eigenvalue $\lambda_{\ell m} = \ell(\ell+1)-m^2$.}. As such, they define an orthogonal basis for the canonical scalar product\footnote{The Laplacian defines a positive-definite quadratic form through $\lbra f , g \rbra = \int f \Delta g = \int \nabla f \cdot \nabla g = \int g \Delta f$. It can hence be diagonalised and the eigenfunctions form a complete set.}. They are indexed by two integers : the multipole $\ell \in \mathbb{N}$ and the azimuthal parameter $m$ with $-\ell \leq m \leq \ell$~; alternatively they may be indexed by $s = \ell (\ell+1) + m +1$.\\
If we normalise them with respect to the canonical scalar product, the complex-valued spherical harmonics are~:
\be\label{Eq:defspherharm}
Y_{\ell m}(\hn) = \sqrt{\frac{2\ell+1}{4\pi}\frac{(\ell-m)!}{(\ell+m)!}} P_\ell^m(\cos \theta) e^{i m \phi}
\ee
with $P_\ell^m$ the associated Legendre polynomial \citep{AbramovitzStegun}.\\
One can define real-valued spherical harmonics, by replacing $e^{im\phi}$ in Eq.\ref{Eq:defspherharm} with $\cos (m\phi)$ for $m>0$ and $\sin (|m| \phi)$ for $m<0$. However the cosmology convention is to work with complex-valued $Y_{\ell m}\;\!$.\\

\subsection{Useful properties}
The orthonormality condition reads~:
\be\label{Eq:spherharmorth}
\int \dd^2\hn \; Y_{\ell m}(\hn) \, Y^*_{\ell' m'}(\hn) = \delta_{\ell \ell'} \, \delta_{m m'}
\ee
and the completeness (or closure) relation is~:
\be\label{Eq:Ylmclosurerelation}
\sum_{\ell m} Y_{\ell m}(\hn) \, Y^*_{\ell m}(\hn') = \delta^2(\hn-\hn')
\ee
They express the fact that spherical harmonics form an orthonormal basis.\\
Under complex conjugation we have :
\be\label{Eq:conjugYlm}
Y^*_{\ell m} = (-1)^m \, Y_{\ell, -m}
\ee
The addition theorem allows to perform sums over the azimuthal parameter :
\be
\sum_{m=-\ell}^\ell Y_{\ell m}(\hn) \, Y^*_{\ell m}(\hn') = \frac{2\ell+1}{4\pi} P_\ell(\hn\cdot\hn')
\ee
\newline

{\bf Notations :}\\
In the following, I may note for concision $i$ for a subscript $(\ell_i, m_i)$, e.g. $Y_1 = Y_{\ell_1 m_1}\,$. A subscript $i^*$ will correspond, through Eq.\ref{Eq:conjugYlm}, to a subscript $(\ell_i, -m_i)$ and a multiplicative factor $(-1)^{m_i}$.\\
Furthermore, I introduce the convention of using multiple indices to denote a product over it. E.g. for a quantity $X$, we have $X_{1\cdots n} \equiv \prod_{i=1}^n X_i$.\\
For example, with these conventions Eq.\ref{Eq:spherharmorth} implies~:
\be
\int \dd^2\hn \; Y_{12} = \delta_{12^*}
\ee

\subsection{Relation with angular momentum and Gaunt coefficients}
It can be shown \citep{Condon1935} that the eigenstate of the (spinless) electron in the hydrogen atom admits a wavefunction of the form~:
\be
\psi(\vec{r}) = R_{n,\ell}(r) \, Y_{\ell m}(\theta,\phi)
\ee
with $n\in \mathbb{N}^*$ defining the energy state of the electron, $\ell$ defining its angular momentum $L$ and $m$ defining the projection of the angular momentum on the vertical axis $L_z$.\\
 Spherical harmonics are the eigenvectors of the operators $L^2$ and $L_z$, with eigenvalue $\hbar \, \ell(\ell+1)$ and $\hbar \, m$ respectively, which hence gives physical interpretation to $\ell$ and $m$.

The sum of two angular momenta $(\ell_1,m_1)$ \& $(\ell_2,m_2)$ yields a linear mixture of states with total angular momentum $(\ell_3,m_3)$, the coefficients involved in this linear combination being the \textit{Clebsch-Gordan} coefficients~:
\be
\lbra \lu m_1 , \ld m_2 | \lt m_3 \rbra = (-1)^{\lu-\ld+m} \sqrt{2\lt+1} \threej{\lu}{\ld}{\lt}{m_1}{m_2}{m_3}
\ee
where the last term is the Wigner $3j$ symbol. It is zero unless the triangular conditions are met~:
%\footnote{also known as 3j symbol or Wigner coefficient}
\bea
\lu+\ld+\lt \ \mr{is \ even,}\\
|\lu-\ld| \leq \lt \leq \lu+\ld \, , \\
m_1+m_2+m_3 = 0.
\eea
Correspondingly, the product of two spherical harmonics decomposes onto the harmonic basis as~:
\be
Y_{\lu m_1} \, Y_{\ld m_2} = \sum_{\lt m_3} G_{m_1 m_2 m_3}^{\lu \ld \lt} Y^*_{\lt m_3}
\ee
where $G_{m_1 m_2 m_3}^{\lu \ld \lt}$ ($\in\mathbb{R}$) is the Gaunt coefficient\footnote{For simplicity, I note $G_{123} = G_{m_1 m_2 m_3}^{\lu \ld \lt}$ and $Y_i = Y_{\ell_i m_i}$ in the following.}~:
\bea
G_{123} &=& \int \dd^2\hn \; Y_{123}(n) \\
&=& \sqrt{\frac{(2\ell+1)_{123}}{4\pi}} \threej{\lu}{\ld}{\lt}{0}{0}{0} \threej{\lu}{\ld}{\lt}{m_1}{m_2}{m_3}
\eea

Consequently, one can transform any product of spherical harmonics into a sum of single spherical harmonics. The integral of such a product is hence a combination of Gaunt coefficients~:
\be
\int \dd^2\hn \; Y_{1\cdots n}(\hn) = \sum_{\ell^\mathrm{d}_1 \cdots \ell^\mathrm{d}_{n-3}}\sum_{m^\mathrm{d}_1 \cdots m^\mathrm{d}_{n-3}} G_{1,2,1^d} \times G_{1^{d*}, 3, 2^d} \,\cdots\, G_{(n-4)^{d*} , n-2 , (n-3)^d} \times G_{(n-3)^{d*} , (n-1) , n}
\ee

%%%%%%%%%%%%%%%%%%%%%%%%%%%%%%%%%%%%%%%%%%%%%%%%
\section{Harmonic coefficients and polyspectra}
\subsection{Harmonic coefficients}
As the spherical harmonics form a basis, any function $T(\hn)$ on the sphere can be decomposed over it~:
\be
T(\hn) = \sum_{\ell m} a_{\ell m} \, Y_{\ell m}(\hn)
\ee
Through the orthonormality property Eq.\ref{Eq:spherharmorth}, the harmonic coefficients $a_{\ell m}$ can be expressed as~:
\be
a_{\ell m} = \int \dd^2\hn \; T(\hn) \, Y^*_{\ell m}(\hn)
\ee
and, if $T(\hn) \in \mathbb{R}$, under complex conjugation we have :
\be
a^*_{\ell m} = (-1)^m \, a_{\ell, -m}
\ee
Hence the set of coefficients $\{ a_{\ell m} | m\geq 0 \}$ provides a complete description of $T(\hn)$, with $a_{\ell m} \in \mathbb{C}$ for $m>0$ and $a_{\ell 0} \in \mathbb{R}$.

\subsection{Isotropic polyspectra}
In the following we will consider statistically isotropic random fields on the sphere. This means that the probability law of the field is invariant under rotation and under parity~:
\be
\forall \mathcal{R} \in \mathcal{O}(3) \quad P\left[\left(T(\mathcal{R} \hn_i)\right)_{i=1\cdots n_\mathrm{pix}}\right] = P\left[\left(T(\hn_i)\right)_{i=1\cdots n_\mathrm{pix}}\right]
\ee
This implies that (connected) correlation functions are invariant under rotation, and in particular the 1-point c.f. is independent of the pixel : $\lbra T(\hn) \rbra = \lbra T \rbra \ \forall \hn$. In the following, we will assume that the fields have zero mean (or that it has been substracted), as this will simplify the expressions of real-space correlation functions.

Under the assumption of statistical isotropy, the harmonic coefficient average (1-point c.f.) are zero : $\lbra a_{\ell m} \rbra = 0$.\\ %(except possibly the monopole $\lbra a_{00} \rbra = \lbra T \rbra$, if the mean was not substracted).\\
Statistical isotropy also simplifies the harmonic 2-point c.f.~:
\be\label{Eq:defcl}
\lbra a_{\ell m} \, a^*_{\ell' m'} \rbra_c = C_\ell \; \delta_{\ell \ell'} \, \delta_{m m'}
\ee
In other terms, the covariance matrix is diagonal in harmonic space. This equation defines the angular power spectrum $C_\ell$ of the field, which possesses one degree of freedom (d.o.f.). \\
The harmonic 3-point c.f. is~:
\be
\lbra a_1 \, a_2\, a_3 \rbra_c =  b_{\lu \ld \lt} \; G_{123}
\ee
which defines the angular bispectrum $b_{123} =b_{\lu \ld \lt}$ with three degrees of freedom\footnote{$\ b_{\lu \ld \lt}$ is sometimes called the reduced bispectrum in the literature, where the bispectrum is defined through $\lbra a_1 \, a_2\, a_3 \rbra_c =  B_{\lu \ld \lt} \threej{\lu}{\ld}{\lt}{m_1}{m_2}{m_3}$. However the convention I have adopted is more natural as it is the analogous of the power spectrum at third order, it provides simpler expressions for theoretical bispectra, and reduces to the Fourier bispectrum in the flat-sky limit.}. The study of the bispectrum is the main subject of this thesis.\\
Generally, the harmonic $n$-point c.f. is~:
\be\label{Eq:defpolyspectrum}
\lbra a_{1\cdots n} \rbra_c = \sum_{\ell^\mathrm{d}_1 \cdots \ell^\mathrm{d}_{n-3}} \mathcal{P}^{(n)}(\ell_1 \cdots \ell_n,\ell^\mathrm{d}_1 \cdots \ell^\mathrm{d}_{n-3}) \times \mathcal{G}(\mathbf{1} \cdots \mathbf{n},\ell^\mathrm{d}_1 \cdots \ell^\mathrm{d}_{n-3})
\ee
with~:
\be\label{Eq:defG4polysp}
\mathcal{G}(\mathbf{1} \cdots \mathbf{n},\ell^\mathrm{d}_1 \cdots \ell^\mathrm{d}_{n-3}) = \sum_{m^\mathrm{d}_1 \cdots m^\mathrm{d}_{n-3}} G_{1,2,1^d} \; G_{1^{d*}, 3, 2^d} \,\cdots\, G_{(n-4)^{d*} , n-2 , (n-3)^d} \; G_{(n-3)^{d*} , (n-1) , n}
\ee
which defines the angular polyspectrum $\mathcal{P}^{(n)}$ of order $n$ with $2n-3$ degrees of freedom.\\
At orders $n\geq 4$ the degrees of freedom cannot be parametrised solely by the involved multipoles and additional \textit{diagonal} degrees of freedom $(\ell_1^d \cdots \ell^d_{n-3})$ are introduced. Geometrical interpretation of the polyspectra and of the diagonal d.o.f. is more obvious in the flat case (when we have a random field over $\mathbb{R}^n$ instead of the sphere) and will hence be explained in Sect.\ref{Sect:2Dpolysp}.\\
A particular case is when the polyspectrum does not vary with $(\ell^\mathrm{d}_1 \cdots \ell^\mathrm{d}_{n-3})$, it is then said \textit{diagonal-independent} and Eq.\ref{Eq:defpolyspectrum} takes the simpler form~:
\be
\lbra a_{1\cdots n} \rbra_c = \mathcal{P}^{(n)}(\ell_1 \cdots \ell_n) \, \mathcal{G}(\mathbf{1} \cdots \mathbf{n}) \quad \mr{with} \quad \mathcal{G}(\mathbf{1} \cdots \mathbf{n}) = \int \dd^2\hn \; Y_{1\cdots n}(\hn)
\ee

\subsection{Relation with real-space correlation functions}
The correlation functions in harmonic space are univoquely related to their real-space counterparts. Indeed we can write~:
\bea
\nonumber\lbra T(\hn_1) \cdots T(\hn_n) \rbra_c &=& \sum_{\ell_i m_i} \lbra a_{1 \cdots n} \rbra_c \left[Y_i(\hn_i)\right]_{i=1\cdots n}\\
\nonumber&=& \sum_{\ell_i \ell_j^d} \mathcal{P}^{(n)}(\ell_1 \cdots \ell_n,\ell^\mathrm{d}_1 \cdots \ell^\mathrm{d}_{n-3})\\
\nonumber && \times \sum_{m_i m_j^d} G_{1,2,1^d} \; G_{1^{d*}, 3, 2^d} \,\cdots\, G_{(n-4)^{d*} , n-2 , (n-3)^d} \; G_{(n-3)^{d*} , (n-1) , n} \left[Y_i(\hn_i)\right]_{i=1\cdots n} \qquad \\
&=& \sum_{\ell_i \ell_j^d} \mathcal{P}^{(n)}(\ell_i,\ell^\mathrm{d}_j) \times F_{\ell_i \ell^d_j}(\hn_1, \cdots, \hn_n)
\eea 
where $F_{\ell_i \ell^d_j}$ is the contribution to the real-space $n$-point Ursell function of a given polyspectrum configuration. It is a rotationally invariant function for which there is no simple expression on the sphere to my knowledge, other than the definition implied by the previous lines. We can note nevertheless that, through the spherical harmonic closure relation Eq.\ref{Eq:Ylmclosurerelation}, we have~:
\be\label{Eq:closurerelFlld}
\sum_{\ell_i \ell^d_j} F_{\ell_i \ell^d_j}(\hn_1, \cdots, \hn_n) = \delta^{(2)}(\hn_1 - \hn_2)\cdots\delta^{(2)}(\hn_{n-1} - \hn_n)  \equiv \delta^{(2n)}(\hn_1 = \cdots = \hn_n)
\ee
\newline

At second order, $F_{\ell_i \ell^d_j}$ simplifies greatly~:
\bea
\nonumber F_{\ell \ell'}(\hn,\hn') &=& \sum_{m,m'} Y_{\ell m}(\hn) \, Y_{\ell' m'}(\hn') \int \dd^2 \hn'' \; Y_{\ell m}(\hn'') \, Y_{\ell' m'}(\hn'')\\
&=&\delta_{\ell \ell'} \sum_m  Y_{\ell m}(\hn) \, Y^*_{\ell m}(\hn')
= \delta_{\ell \ell'} \, \frac{2\ell+1}{4\pi} \, P_\ell (\hn\cdot\hn')
\eea
Hence, with $\hn\cdot\hn' = \cos \theta$, we have~:
\be
C(\theta) \equiv \lbra T(\hn) \, T(\hn') \rbra = \sum_\ell \frac{2\ell+1}{4\pi} \, C_\ell \, P_\ell(\cos\theta)
\ee
In particular $\frac{\ell(2\ell+1)\,C_\ell}{4\pi}$ is the power (contribution to the variance) per logarithmic multipole bin.
\newline

At third order, the formula is more complicated~:
\bea
\nonumber F_{\lu \ld \lt}(\hn_1,\hn_2,\hn_3) &=& \sum_{m_{123}} G_{123} \; Y_1(\hn_1) Y_2(\hn_2) Y_3(\hn_3) \\
\nonumber &=& \int \dd^2\hn \sum_{m_{123}} Y^*_1(\hn) Y_1(\hn_1) \; Y^*_2(\hn) Y_2(\hn_2) \; Y^*_3(\hn) Y_3(\hn_3)\\
&=& \frac{(2\ell+1)_{123}}{(4\pi)^3} \int \dd^2\hn \, P_{\lu}(\hn\cdot\hn_1) \, P_{\ld}(\hn\cdot\hn_2) \, P_{\lt}(\hn\cdot\hn_3) \,
\eea
and will become even more complex at higher orders.

In the particular case where $\hn_1 = \hn_2 = \hn_3$, $F_{\lu \ld \lt}$ reduces to~:
\be\label{Fl123dirac}
F_{\lu \ld \lt}(\hn_1=\hn_2=\hn_3) = \frac{(2\ell+1)_{123}}{(4\pi)^2} \threej{\lu}{\ld}{\lt}{0}{0}{0}^2 = \frac{N_{123}}{4\pi}
\ee
where I introduce preemptively $N_{123} = \frac{(2\ell+1)_{123}}{4\pi} \threej{\lu}{\ld}{\lt}{0}{0}{0}^2$. $N_{123}$ is the number of bispectrum modes for the triplet $(\lu,\ld,\lt)$, and is the analogous to $2\ell+1$ for the power spectrum.

Hence we have~:
\be
\kappa_3 = \lbra T^3 \rbra = \sum_{\ell_{123}} \frac{N_{123}}{4\pi} \, b_{123}
\ee
where the sum runs over multipole triplets following the triangular condition, otherwise $N_{123}=0$. We can conclude that $\frac{\lu \ld \lt N_{123}}{4\pi} \, b_{123}$ is the contribution to the skewness per logarithmic multipole bin.

\subsection{Estimation of the power spectrum and bispectrum}\label{Sect:estim_specandbisp}
\subsubsection{Power spectrum}
From Eq.\ref{Eq:defcl} we have $\lbra | a_{\ell m}|^2 \rbra = C_\ell$. We can thus introduce an unbiased estimator $\hat{C}_\ell$ of the power spectrum as~:
\be\label{Eq:clestim}
\hat{C}_\ell = \frac{1}{2\ell+1} \sum_m | a_{\ell m}|^2
\ee
$\hat{C}_\ell$ can also be shown to possess the minimal variance among unbiased estimators\footnote{For example $|a_{\ell 0}|^2$ is an unbiased estimator of $C_\ell$, with a greater variance than $\hat{C}_\ell$}, i.e. it is optimal.\\
Numerically Eq.\ref{Eq:clestim} takes $\mathcal{O}(N^{3/2}_\mathrm{pix})$ operations to compute from a given map. The harmonic transform to compute the $a_{\ell m}$ is numerically the bottleneck as it scales as $\mathcal{O}(N^{3/2}_\mathrm{pix})$ while the sum over $m$ scales as $\mathcal{O}(N_\mathrm{pix})$.

Using the so-called scalemaps defined by \cite{Spergel1999}~:
\be
T_\ell(\hn) = \sum_m a_{\ell m} \, Y_{\ell m}(\hn)
\ee
which contains a single multipole $\ell$, we can formally rewrite Eq.\ref{Eq:clestim} as~:
\be
\hat{C}_\ell = \frac{1}{2\ell+1} \int \dd^2\hn \, T_\ell(\hn)^2
\ee
Numerically, this is an inefficient estimation method for the power spectrum, as the computation of all scalemaps up to $\ell_\mathrm{max} \propto N^{1/2}_\mathrm{pix}$ takes $\mathcal{O}(N^{2}_\mathrm{pix})$ operations.

A common complication is that we do not usually have a full-sky map of the signal to analyse, either because the instrument is not able to scan all the sky (e.g. ground-based telescope) or because some areas of the sky are too contaminated by other signals (typically the Milky Way emission). Hence we are analysing a masked map~:
\be
T_M(\hn) = T(\hn) \times M(\hn)
\ee
with $M(\hn)$ the mask function. $M(\hn)$ is typically 0 for masked pixels and 1 for observed pixels, but we may choose to smooth the edges (apodisation) to avoid ``ringing''\footnote{Abrupt features in real space create long-tailed oscillations in harmonic space, e.g. the Fourier transform of a step function is a sinc function} \citep[see e.g.][]{Ponthieu2011}.\\
This multiplication in real space translates into a convolution in harmonic space, so that the harmonic coefficient of the masked map takes the form~:
\be
a_1(T_M) = \sum_{23} a_2(T) \, a_3(M) \, G_{1^*,2,3}
\ee
where $a_{\ell m}(T)$ and $a_{\ell m}(M)$ are the harmonic coefficients respectively of the true sky signal and of the mask.\\
Direct inversion of the whole convolution is not possible because information has been lost by masking. As the convolution is linear, correlation functions are convolved order by order i.e. the $n^\mr{th}$ order masked correlation function is a convolved version of the $n^\mr{th}$ order full-sky correlation function. However, there is no information leakage between different orders.\\
For example, the covariance matrix becomes non-diagonal \citep[see e.g.][]{Hivon2002} contrary to Eq.\ref{Eq:defcl}, which was valid in full-sky\,; the best possible estimation of the power spectrum should use these off-diagonal terms. It has been however shown that an unbiased estimation of the power spectrum can be obtained considering the pseudo-spectrum of the masked map, and that the method is nearly optimal for high multipoles \citep[e.g.][]{Hivon2002}. Indeed, if we define the pseudo-spectrum of the masked map~:
\be
\tilde{C}_\ell(T_M) = \frac{1}{2\ell+1} \sum_m |a_{\ell m}(T_M)|^2
\ee
Some algebra shows that it is on average a convolved version of the true power spectrum~:
\be\label{Eq:defmixmatcl}
\lbra \tilde{C}_\ell(T_M) \rbra = \sum_{\ell'} \mathcal{M}_{\ell \ell'} \, C_{\ell'}(T)
\ee
where $\mathcal{M}$ is the coupling matrix~:
\be\label{Eq:mixmatcl}
\mathcal{M}_{\lu \ld} = \frac{2\ld+1}{4\pi} \sum_3 |a_3(M)|^2 \threej{\lu}{\ld}{\lt}{0}{0}{0}^2
\ee
If the available sky is sufficiently large, the coupling matrix can be inverted to yield an unbiased estimation of $C_{\ell}(T)$. For smaller fraction of observed sky, the pseudo-power spectrum can be binned and the binned coupling matrix becomes invertible if the binning is large enough \citep{Hivon2002}.\\
The observed sky fraction is given in term of the mask as~:
\be
f_\mr{SKY} = \frac{1}{4\pi} \int \dd^2\hn \, M(\hn)^2
\ee
If the true power spectrum is varying slowly compared to the coupling matrix, it can be factored out of the sum in Eq.\ref{Eq:defmixmatcl} and we get~:
\be
\lbra \tilde{C}_\ell(T_M) \rbra \approx C_{\ell}(T) \times \sum_{\ell'} \mathcal{M}_{\ell \ell'} = f_\mr{SKY} \, C_{\ell}(T)
\ee
This is the so-called $f_\mr{SKY}$ approximation and it is justified for large $f_\mr{SKY}$ and/or slow varying power spectrum \citep{Komatsu2002COBE}. In particular, it is exact if the power spectrum is constant, e.g. for a white noise, see Sect.\ref{Sect:whitenoise}.

The coupling matrix has $\ell_\mr{max}^2 \propto N_\mr{pix}$ elements, and after precomputation of Wigner symbols and of the harmonic transform of the mask, Eq.\ref{Eq:mixmatcl} involves $N_\mr{pix}$ operations. Thus the whole computation scales as $\mathcal{O}(N_\mr{pix}^2)$, which, although not fast, is numerically tractable for {\planck}-like data sets (50 millions pixels per map at full resolution).
\newline

\begin{figure}[htbp]
\begin{minipage}{0.65\linewidth}
\includegraphics[width=.9\linewidth]{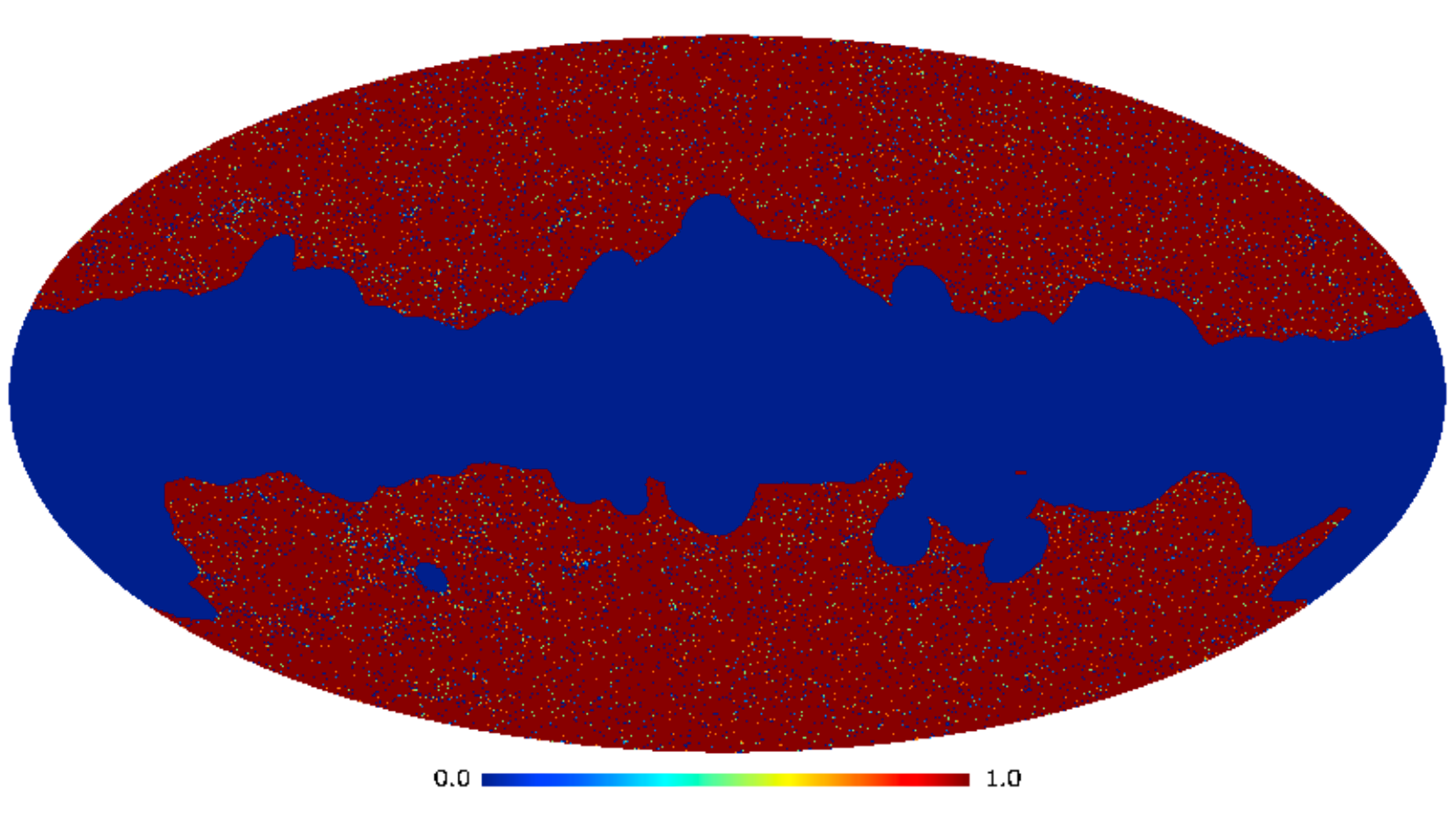}
\end{minipage}
\hfill
\begin{minipage}{0.35\linewidth}
\includegraphics[width=.9\linewidth]{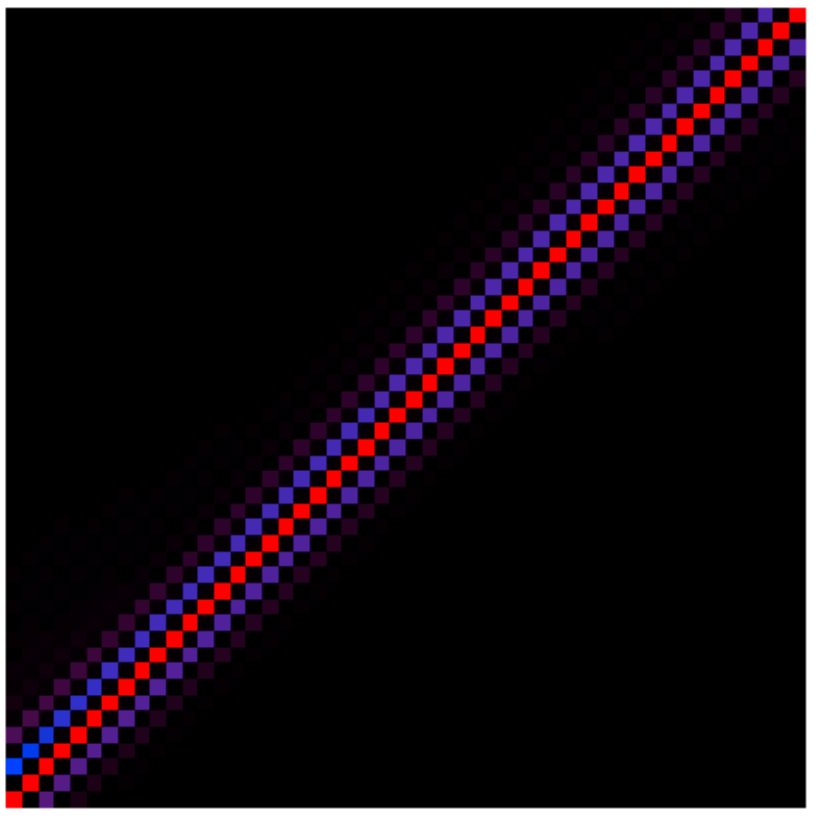}
\end{minipage}
\caption{\textit{Left :} galactic and point-source mask used in {\planck} Sunyaev-Zel'dovich analysis \citep{planck2013-SZmap}. \textit{Right :} power spectrum coupling matrix\,; the color code runs from black (zero) to red (maximum, here~$\sim 0.4$). The coupling matrix is dominated by a band around the diagonal, as multipoles tend to couple to neighboring multipoles.}
\label{Fig:mixmatcl}
\end{figure}

I have implemented a code to compute the coupling matrix and estimate power spectra on masked maps. Indeed, I wanted a flexible code written in IDL language, so that it could interface easily with all my other codes. And I wanted to be able to use it simply and quickly, for example for my work described in Sect.\ref{Sect:meas_CIB_NG}, without having to deal with the shortcomings of existing codes written in other programming languages.\\
As an illustration, Fig.\ref{Fig:mixmatcl} shows a mask used in {\planck} analysis and the corresponding coupling matrix for $\ell=1\cdots 50$.

Let us note that even-odd matrix elements (e.g. $(\ell,\ell+1)$ or $(\ell,\ell+3)$) are small but non-zero. They would be exactly zero if the mask were parity-invariant. Indeed in the latter case $a_{\ell m}(M)$ is zero if $\ell$ is odd, and through Eq.\ref{Eq:mixmatcl} and the parity condition for Wigner symbols ($\lu+\ld+\lt$ even), we see that $\mathcal{M}_{\lu \ld}$ is zero if $\lu$ and $\ld$ have different parity.

\subsubsection{Bispectrum}
Similarly to the power spectrum case, we can define an unbiased estimator of the bispectrum by~:
\be\label{Eq:blestim}
\hat{b}_{123} = \frac{1}{N_{123}} \sum_{m_{123}} G_{123} \; a_1 \, a_2 \, a_3
\ee
Indeed the properties of Wigner symbols lead to~:
\be
\sum_{m_{12}} G_{124} \, G_{125} = \frac{(2\ell+1)_{12}}{4\pi} \threej{\lu}{\ld}{\ell_4}{0}{0}{0}^2 \delta_{45}
\quad \Rightarrow \quad \sum_{m_{123}} G_{123} \, G_{123} = N_{123}
\ee
The estimator can also be shown to be optimal. It can be rewritten with scalemaps as \citep{Bucher2010,Lacasa2012}~:
\be
\hat{b}_{123} = \frac{1}{N_{123}} \int \dd^2\hn \, T_{\lu}(\hn) T_{\ld}(\hn) T_{\lt}(\hn)
\ee
which, once the scalemaps are computed, takes $\mathcal{O}(N_\mathrm{pix})$ operations for each multipole triplet. By contrast, Eq.\ref{Eq:blestim} runs over $\propto \ell_\mr{max}^2 \propto N_\mr{pix}$ elements but the Gaunt coefficients are costly to compute at high multipoles, so the scalemap method is computationally faster.

There are $\propto \ell_\mr{max}^3 \propto N_\mr{pix}^{3/2}$ triplets of multipoles, so the computation of a full bispectrum scales as $\mathcal{O}(N^{5/2}_\mathrm{pix})$ operations. For {\planck}-like data sets with millions of pixels, this can be numerically challenging as it takes several CPU-days. The resulting measurement is also challenging to use as it consists of $\mathcal{O}(N^{3/2}_\mathrm{pix})$ coefficients with low signal to noise ratio.

Several approaches to circumvent this problem have been proposed, and the one I have followed is the binned bispectrum method introduced by \cite{Bucher2010}. Cosmological signals have in general a smooth scale dependence so that polyspectra are smooth functions of the multipoles. Hence one can bin multipoles together, with little loss of information unless the bin-size becomes too large (compared to the scale of variation of the bispectrum).\\
If we define the binned scalemaps as~:
\be
T_{\Delta \ell}(\hn) = \sum_{\ell\in\Delta\ell, m} a_{\ell m} \, Y_{\ell m}(\hn)
\ee
and the binned number of triplets as~:
\be
N_{\Delta\lu \Delta\ld \Delta\lt} = \sum_{\ell_i \in \Delta\ell_i} N_{123}
\ee
we can define the binned bispectrum estimator as \citep{Lacasa2012}:
\be
\hat{b}_{\Delta\lu \Delta\ld \Delta\lt} = \frac{1}{N_{\Delta\lu \Delta\ld \Delta\lt}} \int \dd^2\hn \; T_{\Delta\lu}(\hn) T_{\Delta\ld}(\hn) T_{\Delta\lt}(\hn)
\ee
which estimates a weighted average of the bispectrum within the multipole bins. The binned bispectrum method drastically reduces the number of measured coefficients which goes as $\sim n_\mr{bin}^3$. It also increases their signal-to-noise ratio (SNR) so that the resulting measurement can be visualised and interpreted. The drawbacks are a loss of statistical information and of fine variations of the bispectrum, which are both mitigated if the binning is not too large and/or if the bispectrum is smooth.\\
For the signals of interests in chapters 4 and 5, a binning as large as $\Delta\ell=128$ can be used. Indeed for unclustered sources (a constant bispectrum, see chapter 4) binning does not lose any information, while for clustered sources I checked that the correlation between the binned bispectrum and the full bispectrum was higher than 94\% for $\Delta\ell=128$ and higher than 97\% for $\Delta\ell=64$.

\subsubsection{Effect of masking on the bispectrum : variance}\label{Sect:linearterm_bisp}

This section shows that a linear term is needed for the bispectrum estimator to lower its variance in the partial sky case. In fact, I will show that the linear term is needed for any estimation of a third order moment. This term however vanishes for the bispectrum in the isotropic case, which explains why it was not considered previously.

The computation of this section are based on the Gram-Charlier expansion (Sect.\ref{Sect:GramCharlier}) and in particular on Eq.\ref{Eq:GC_ijk}, which I will nevertheless need to develop to the next order $\mathcal{O}(\kappa_3^2)$. I will first illustrate the computation on the simpler case of the estimation of the third order cumulant of a random variable with independent samples, which will show the arising of the linear term. I will then generalise the result to the estimation of the bispectrum of a random field, in which case the algebra becomes cumbersome.
\newline

Let us first consider $X_1 \cdots X_n\,$, $n$ random variables i.i.d. with zero mean and variance $\sigma^2$. In that case the second and third order Ursell functions take the form~:
\bea
C_{ij} &=& \sigma^2 \delta_{ij} \\
\kappa_3(X_i,X_j,X_k) &=& \kappa_3 \, \delta_{ijk}
\eea
Following Eq.\ref{GC_theo_rf} and Eq.\ref{Eq:GC_ijk}, we have~:
\be
p(x_1\cdots x_n | \kappa_3) = \mr{Gauss}(\mathbf{X}) \times \bigg(1+\underbrace{\frac{\kappa_3}{3!}\sum_i \frac{x_i^3}{\sigma^6}-\frac{3 x_i}{\sigma^4}}_{=\frac{\kappa_3}{3!\sigma^3}\sum_i H_3\left(\frac{x_i}{\sigma}\right)}\bigg)+\frac{\kappa_3^2}{2 \, 3!^2} \sum_{ij} \frac{\partial^6}{\partial x_i^3 \partial x_j^3} \mr{Gauss}(\mathbf{X})
\ee
where $H_n$ denotes the $n$-th order Hermite polynomial, and I find~:
\be
\frac{\partial^6}{\partial x_i^3 \partial x_j^3} \mr{Gauss}(\mathbf{X}) = \mr{Gauss}(\mathbf{X}) \times \frac{1}{\sigma^6} \left[H_3\left(\frac{x_i}{\sigma}\right)H_3\left(\frac{x_j}{\sigma}\right)-\left(H_6\left(\frac{x_i}{\sigma}\right)+H_3\left(\frac{x_i}{\sigma}\right)^2\right) \delta_{ij}\right]
\ee
Following Baye's theorem \citep{Bayes1763}, the likelihood of the third order cumulant is $\mathcal{L}(\kappa_3) \propto p(x_1\cdots x_n | \kappa_3)$. Hence a Maximum Likelihood Estimator (MLE) is found by solving~:
\be
\frac{\partial p(x_1\cdots x_n | \kappa_3)}{\partial \kappa_3} = 0
\ee
This is equivalent to~:
\be
\frac{1}{3!^2} \frac{\kappa_3}{\sigma^6} \sum_i H_6\left(\frac{x_i}{\sigma}\right)+H_3\left(\frac{x_i}{\sigma}\right)^2 -\frac{1}{3!^2} \frac{\kappa_3}{\sigma^6} \sum_{i\neq j} H_3\left(\frac{x_i}{\sigma}\right)H_3\left(\frac{x_j}{\sigma}\right) = \frac{1}{3!\sigma^3}\sum_i H_3\left(\frac{x_i}{\sigma}\right)
\ee
so that the MLE takes the simple form~:
\be\label{Eq:hatk3full}
\hat{\kappa}_3 = \frac{3! \,\sigma^3 \sum_i H_3\left(\frac{x_i}{\sigma}\right)}{\sum_i H_6\left(\frac{x_i}{\sigma}\right)+H_3\left(\frac{x_i}{\sigma}\right)^2 - \sum_{i\neq j} H_3\left(\frac{x_i}{\sigma}\right)H_3\left(\frac{x_j}{\sigma}\right)}
\ee
Note that the Gram-Charlier expansion is valid if the p.d.f. is a perturbation of the Gaussian p.d.f., hence the computations developed above are valid in the weak non-Gaussianity (NG) limit. If the number $n$ of samples is sufficiently large, following \cite{Babich2005}, we can hence replace the denominator in Eq.\ref{Eq:hatk3full} by its Gaussian expectation value~:
\be
\lbra \sum_i H_6\left(\frac{x_i}{\sigma}\right)+H_3\left(\frac{x_i}{\sigma}\right)^2 - \sum_{i\neq j} H_3\left(\frac{x_i}{\sigma}\right)H_3\left(\frac{x_j}{\sigma}\right) \rbra_\mr{Gauss} = 0 + 6 - 0 = 6
\ee
Hence the MLE takes the form~:
\be
\hat{\kappa}_3 = \frac{1}{n} \sum_i x_i^3 -3\sigma^2 x_i
\ee
This estimator is obviously unbiased and contains an intuitive cubic part $\frac{1}{n} \sum_i x_i^3$, but also a linear part $\frac{1}{n} \sum_i 3\sigma^2 x_i\,$. Note that $X_i^3 -3\sigma^2 X_i = X_i^3 -3 \lbra X_i^2 \rbra X_i$ is the Wick product of $X_i$ \citep{Wick1950}.\\
The cubic term alone is an unbiased estimator : the linear term has zero expectation value. However the linear term allows the estimator to be optimal (in the weak NG limit). Indeed the Wick product of $X_i$ is the cubic polynomial\footnote{with unit leading coefficient} with the lowest variance in the Gaussian case \citep{Donzelli2012}.\\
To summarize, reducing the variance of the $\kappa_3$ estimator is achieved by replacing the cubic product in the intuitive estimator $\frac{1}{n} \sum_i x_i^3$  by the Wick product $x_i^3 -3\sigma^2 x_i$.
\newline

Let us now turn to the estimation of the bispectrum of a spherical random field. The full-sky estimator introduced previously is~:
\be
\hat{b}_{123} = \frac{1}{N_{123}} \sum_{m_{123}} G_{123} \, a_1 a_2 a_3
\ee
A straightforward generalisation inspired by the previous considerations is to replace the product of the harmonic coefficients by the corresponding Wick product~:
\be\label{Eq:bhat-wWickprod}
\hat{b}_{123} = \frac{1}{N_{123}} \sum_{m_{123}} G_{123} \, \left(a_1 a_2 a_3 -\lbra a_1 a_2\rbra a_3 -\lbra a_1 a_3\rbra a_2 -\lbra a_2 a_3\rbra a_1 \right)
\ee
However, if we start from the Gram-Charlier expansion for spherical harmonic coefficients and follow calculations similar to the $\kappa_3$ case above, we find that the situation is a bit more complex for the bispectrum~:
\begin{itemize}
\item in the partial sky case, the covariance matrix is no longer diagonal, hence the $C^{-1}_{ij}$ terms do not factor out of the sum in Eq.\ref{Eq:GC_ijk}. Hence the spherical harmonic coefficients have to be inverse-covariance filtered (also known as Wiener filtering). In theory, this just induces a different weighting of harmonic coefficients in the estimator Eq.\ref{Eq:bhat-wWickprod} and a corresponding change in the normalisation of the estimator. In practice, Wiener filtering maps is a computational challenge \footnote{e.g. the covariance matrix has $n_\mr{pix} \times n_\mr{pix}$ entries and is impossible to store nor invert for \planck-like data sets with $n_\mr{pix} = \mathcal{O}(10^6)$.
%A cunning algorithm was however designed by \cite{Elsner2013}.
}, and Wiener filtered maps were not available for the signals I have studied. I have hence neglected the Wiener filtering and focused on a \emph{pseudo-bispectrum} estimator, analogously to the pseudo-spectrum method.
\item in the full-sky case, by maximising the likelihood with respect to $b_{123}$ one obtains an equation for $b_{123}$ hence defining straightforwardly the MLE. On the contrary in the partial sky case, one obtains an equation involving all bispectrum coefficients. Hence, one has to solve a linear system, and the bispectrum estimates are coupled, as different estimates  may e.g. involve the same product $a_1 a_2 a_3$ (albeit with different weights). This situation, and the need to invert a coupling matrix, will be treated for the pseudo-bispectrum case in the following sections (Sect.\ref{Sect:analyt_bispcouplmat} \& \ref{Sect:bispcouplmat_sim}).
\end{itemize}

Thus, we see that Eq.\ref{Eq:bhat-wWickprod} defines a pseudo-bispectrum estimator which needs to be debiased (see Sect.\ref{Sect:analyt_bispcouplmat} \& \ref{Sect:bispcouplmat_sim}) and is not optimal (in all cases, the optimality would have been guaranteed only in the weak NG limit). The situation is analogous to the pseudo-spectrum method, which does not provide an optimal estimator but is however useful because it is fast and near-optimal at high multipoles \citep{Efstathiou2004}. The pseudo-bispectrum estimator Eq.\ref{Eq:bhat-wWickprod} can be rewritten with scale maps as~:
\be\label{Eq:bhat-wlin-scmap}
\hat{b}_{123} = \frac{1}{N_{123}} \int \dd^2\hn \, \left(T_{\lu} T_{\ld} T_{\lt} - \lbra T_{\lu} T_{\ld} \rbra T_{\lt} - \lbra T_{\lu} T_{\lt} \rbra T_{\ld} - \lbra T_{\ld} T_{\lt} \rbra T_{\lu}\right)
\ee
where $T_\ell = T_\ell(\hn)$ for clarity.\\
Note that the average $\lbra \rbra$ only involve 2-point correlations, hence they are in practice computed through average over Gaussian simulations. Moreover, in the full-sky case $\lbra T_\ell(\hn) T_{\ell'}(\hn) \rbra$ is either zero (if $\ell\neq\ell'$) or a constant (if $\ell = \ell'$), so that the linear term vanishes in the isotropic case.\\
Note also that Eq.\ref{Eq:bhat-wlin-scmap} is straightforwardly generalised when multipoles are binned. Furthermore, binning multipoles decreases the number of simulations needed for convergence of the linear term, because binning induces an averaging and reduces the coupling between bins.

\subsubsection{Effect of masking on the bispectrum : bias}\label{Sect:analyt_bispcouplmat}
As for the power spectrum, the estimation of the bispectrum is biased when part of the map is masked for any reason. I have worked on this subject, following the same idea as for the power spectrum. Indeed, if we compute the pseudo bispectrum of the masked map\footnote{Considering only the cubic term, as the linear term has zero average}~:
\be
\tilde{b}_{123}(T_M) = \frac{1}{N_{123}} \sum_{m_{123}} G_{123} \; a_1(T_M) \, a_2(T_M) \, a_3(T_M) 
\ee
The average $\tilde{b}_{123}(T_M)$ is a convolved version of the true bispectrum~:
\be\label{Eq:defmixmatbisp}
\lbra \tilde{b}_{\ell_{123}}(T_M) \rbra = \sum_{\ell'_{123}} \mathcal{M}_{\ell_{123},\ell'_{123}} \, b_{\ell'_{123}}(T)
\ee
I derived several possible formulae for the bispectrum coupling matrix~:
\bea
\nonumber \mathcal{M}_{\ell_{123},\ell'_{123}} & = & \frac{1}{N_{123}} \int \dd^2\hn_{123} \; M(\hn_1) \, M(\hn_2) \, M(\hn_3) \\
&& \qquad \qquad \times \; F_{\lu \ld \lt}(\hn_1,\hn_2,\hn_3) \, F_{\lu' \ld' \lt'}(\hn_1,\hn_2,\hn_3)\label{Eq:mixmatbispint}\\
& = & \frac{1}{N_{123}} \sum_{\substack{m_{123}\\m'_{123}\\1'' 2'' 3'' }} G_{123} \, G_{1'2'3'} \,\left(G_{i^*i'i''}\; a_{i''}(M)\right)_{i=123}\label{Eq:mixmatbispG}\\
&=& \frac{1}{N_{123}} \, \sum_{\substack{m_{123}\\m'_{123}}} G_{123} \, G_{1'2'3'} \; W_{1,1'} W_{2,2'} W_{3,3'} \label{Eq:mixmatbispW}
\eea
with $W_{i,i'} = W_{\ell_i m_i,\ell'_i m'_i}\,$, and~:
\be
W_{\ell m,\ell' m'} = \int \dd^2\hn \, M(\hn) \; Y_{\ell m}(\hn) \, Y^*_{\ell' m'}(\hn)
\ee
which measures the non-orthonormality of spherical harmonics on the masked sky.\\
To the best of my knowledge, the equations for the bispectrum coupling matrix do not simplify any further on the sphere. In all cases, the numerical computation of a full bispectrum coupling matrix is unconceivable as it has $\mathcal{O}(N^3_\mr{pix})$ elements.

As for the power spectrum, the equations simplify when the bispectrum varies slowly compared to the coupling matrix. Indeed, we then have~:
\be
\lbra \tilde{b}_{\ell_{123}}(T_M) \rbra \simeq b_{\ell_{123}}(T) \times \sum_{\ell'_{123}} \mathcal{M}_{\ell_{123},\ell'_{123}}
\ee
and through Eq.\ref{Eq:closurerelFlld}-\ref{Fl123dirac}-\ref{Eq:mixmatbispint} we have~:
\bean
\sum_{\ell'_{123}} \mathcal{M}_{\ell_{123},\ell'_{123}} &=& \frac{1}{N_{123}} \int \dd^2\hn \, M^3(\hn) \, F_{\lu \ld \lt}(\hn,\hn,\hn)\\
&=& \frac{1}{4\pi} \int \dd^2\hn \, M^3(\hn) = f_\mr{SKY}
\eean
Hence we find again the $f_\mr{SKY}$ approximation $\lbra \tilde{b}_{\ell_{123}}(T_M) \rbra \simeq f_\mr{SKY} \times b_{\ell_{123}}(T)$.

\subsubsection[Binned bispectrum coupling matrix evaluation]{Binned bispectrum coupling matrix evaluation through simulations}\label{Sect:bispcouplmat_sim}

For a binned bispectrum estimator, we would not need a full bispectrum coupling matrix but a binned coupling matrix, with $\mathcal{O}(n_\mr{bin}^6)$ elements instead of $\mathcal{O}(\ell_\mr{max}^6)$. Such matrix sizes are manageable, and hence I have embarked in the project to find a method to compute such a binned coupling matrix. The purpose is to have a tool allowing me to debias binned bispectrum estimations on a masked sky. As the direct computation of (binned) Eq.\ref{Eq:mixmatbispint}-\ref{Eq:mixmatbispW} is not possible, I have designed an alternative method : the evaluation of the coupling matrix through simulations.
%Although it is not published, the following material is original.

Let us note $n_\mr{config}$ the number of binned bispectrum configurations, and let us consider bispectra as vectors with length $n_\mr{config}$. Then Eq.\ref{Eq:defmixmatbisp} can be rewritten as~:
\be
\mathbf{b}^\mathrm{obs} = \mathcal{M} \cdot \mathbf{b}^\mathrm{true} + N
\ee
where $N$ is a noise vector and $\mathcal{M}$ is an $n_\mr{config}\times n_\mr{config}$ matrix. If we have $n_\mathrm{sim}$ simulations with different input bispectra, noting~:
\be
\mathcal{B}_o = \left( \mathbf{b}^\mathrm{obs}_1 \cdots \mathbf{b}^\mathrm{obs}_{n_\mathrm{sim}} \right) \qquad \mathrm{and} \qquad \mathcal{B}_t = \left( \mathbf{b}^\mathrm{true}_1 \cdots \mathbf{b}^\mathrm{true}_{n_\mathrm{sim}} \right)
\ee
we have~:
\be\label{Eq:BoMBtN}
\mathcal{B}_o = \mathcal{M} \cdot \mathcal{B}_t+N
\ee
Thus the idea is to generate maps with different bispectra such that $\mathbf{b}^\mathrm{true}_1 \cdots \mathbf{b}^\mathrm{true}_{n_\mathrm{sim}}$ form a spanning set of the bispectrum vector space, mask the maps and measure the corresponding masked pseudo-bispectra.\\
Then from the input $\mathcal{B}_t$ and the observable $\mathcal{B}_o\,$,  I should construct an estimator $\hat{\mathcal{M}}$ of $\mathcal{M}$. To this end, I choose an estimator of the form $\hat{\mathcal{M}} = \mathcal{B}_o \cdot W\,$, where $W$ is an $n_\mr{sim}\times n_\mr{config}$ matrix. I am looking for $W$ such that~:
\be
\left\{
\begin{array}{cc}
\mathcal{B}_t \, W = \mathrm{Id} \\
\lbra \mathrm{Tr}(^{T}\!\hat{\mathcal{M}}\, \hat{\mathcal{M}}) \rbra \ \mathrm{minimal}
\end{array}
\right.
\ee
These equations express the fact that I search for an unbiased estimator with minimal variance. The solution to this constrained minimization problem can be obtained through the method of Lagrange multipliers. After some algebra I find~:
\be
\hat{\mathcal{M}} = \mathcal{B}_o \, R_o^{-1} \, ^{T}\!\mathcal{B}_t \left(\mathcal{B}_t \, R_o^{-1} \, ^{T}\!\mathcal{B}_t \right)^{-1} \qquad \mr{with} \quad R_o = \lbra ^{T}\!\mathcal{B}_o \,\mathcal{B}_o \rbra
\ee
In the case where $n_\mathrm{sim} = n_\mr{config}$, all matrices are square ones and I get~:
\be
\hat{\mathcal{M}} = \mathcal{B}_o \, \mathcal{B}_t^{-1}
\ee
which is the obvious solution of Eq.\ref{Eq:BoMBtN} with square matrices and negligible noise. I have adopted this approach, simpler than with rectangular matrices, and decided to reduce the noise in the estimation by repeating the procedure and averaging over iterations. That is to say I first use $n_\mr{sim}=n_\mr{config}$ simulations and obtain a first estimation $\hat{\mathcal{M}}_1$, I repeat ... and my final estimation is $\hat{\mathcal{M}}_f = \sum_i \hat{\mathcal{M}}_i / n_\mr{iter}$.
\newline

The main question is how to generate simulations such that their bispectra form a basis, and that they have a high signal-to-noise ratio. Indeed, we want the smallest possible noise in the bispectra so that we have the smallest possible noise in the estimation of $\mathcal{M}$.\\
A bispectrum is a function of $(\lu,\ld,\lt)$ invariant under their permutations. As such it decomposes over symmetric polynomials~:
\bea
\sigma_1 &=& \lu + \ld + \lt \\
\sigma_2 &=& \lu\ld +\lu\lt +\ld\lt \\
\sigma_3 &=& \lu\ld\lt
\eea
So the aim is to generate any polynomial of $(\sigma_1,\sigma_2,\sigma_3)$, i.e.
\bean
&&1\\
&&\sigma_1, \sigma_2, \sigma_3 \\
&&\sigma_1^2, \sigma_1\sigma_2, \sigma_1\sigma_3, \sigma_2^2, \sigma_2\sigma_3, \sigma_3^2 \\
&&\cdots 
\eean
As will be shown in Sect.\ref{Sect:whitenoise}, I am able to quickly generate maps with a theoretical bispectrum of the form~:
\be\label{Eq:sepbisp4mixmat}
b_{123} = \alpha_{\lu} \, \alpha_{\ld} \, \alpha_{\lt}
\ee
with $\alpha_\ell$ an arbitrary function of $\ell$, and with each bispectrum coefficient having a high SNR.\\
At zeroth order we get easily $b_{123} = 1$ by taking $\alpha_\ell = 1$.\\
At first order we can recover the three symmetric polynomials using $\alpha_\ell = \ell - \tilde{\ell}_{(0)}\,$, $\alpha_\ell = \ell - \tilde{\ell}_{(1)}$ and $\alpha_\ell = \ell - \tilde{\ell}_{(2)}$ with $\tilde{\ell}_{(0)} \neq \tilde{\ell}_{(1)} \neq \tilde{\ell}_{(2)}$ arbitrary constants. Indeed~:
\be
(\ell_1 - \tilde{\ell}_{(i)})(\ell_2 - \tilde{\ell}_{(i)})(\ell_3 - \tilde{\ell}_{(i)}) = \tilde{\ell}_{(i)}^2 \, \sigma_1 - \tilde{\ell}_{(i)}\, \sigma_2 + \sigma_3
\ee
so we can invert\footnote{The determinant is that of Vandermonde and is non-zero iff $\tilde{\ell}_{(0)} \neq \tilde{\ell}_{(1)} \neq \tilde{\ell}_{(2)}$.} the linear system between the three bispectra and the three symmetric polynomials.

It can also be seen that the six symmetric second-order polynomials can be recovered with linear combinations of the six bispectra determined by $\alpha_\ell = (\ell - \tilde{\ell}_{(i)})(\ell - \tilde{\ell}_{(j)})$ with $(i,j) \in \llbracket 0,1,2 \rrbracket$. This is also true at higher orders.

For simplicity we may note an $\alpha_\ell$ with the indices of the $\tilde{\ell}_{(i)}$ involved, e.g. $(\ell~-~\tilde{\ell}_{(0)})(\ell~-~\tilde{\ell}_{(1)})$ can be noted 01. Then the 10 polynomials at third order can be noted in the form~:\\
000, 100, 110, 111, 200, 210, 211, 220, 221, 222 \\
and all $\alpha_\ell$ can be ordered through lexicographic order.
\newline

\begin{figure}[htbp]
\begin{center}
\begin{minipage}{0.6\linewidth}
\includegraphics[width=.9\linewidth]{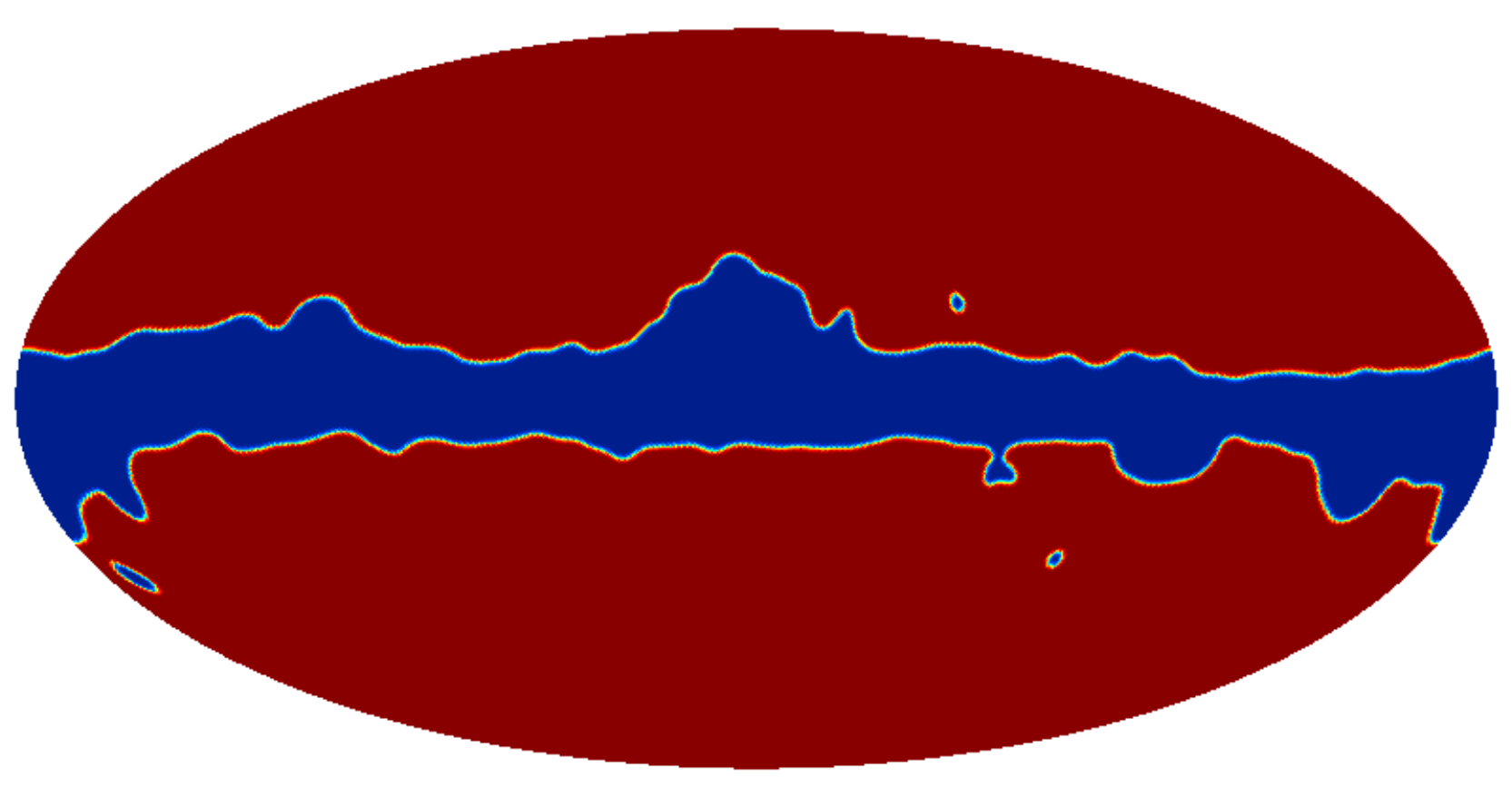}
\end{minipage}
\hfill
\begin{minipage}{0.35\linewidth}
\includegraphics[width=.9\linewidth]{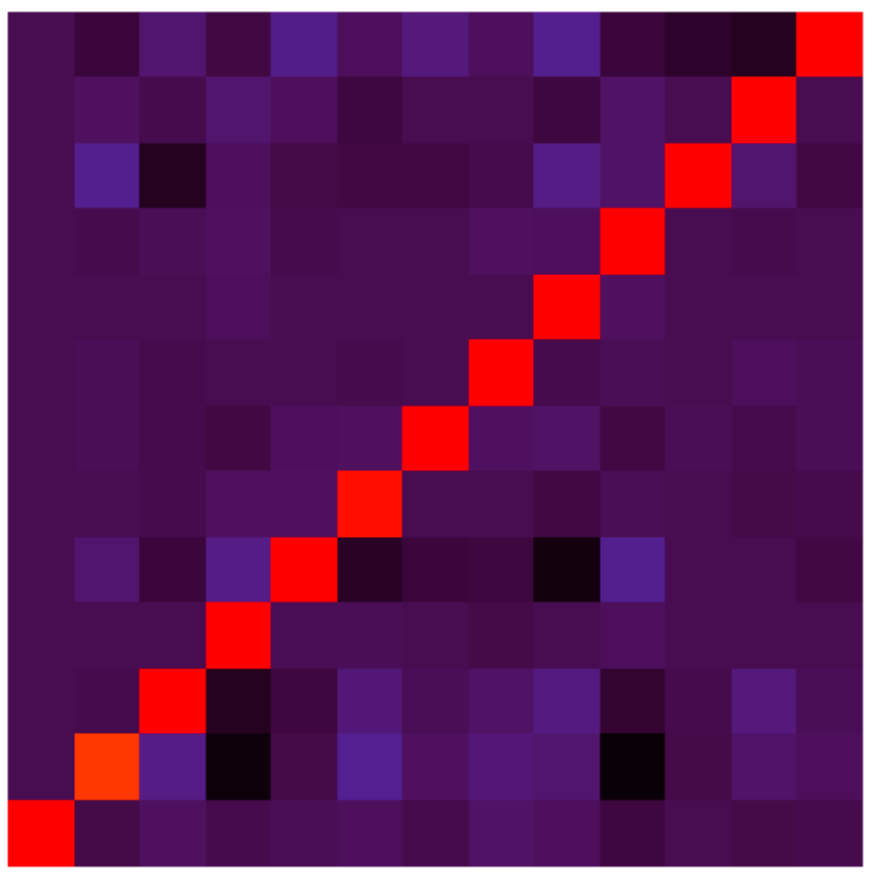}
\end{minipage}
\caption{\textit{Left :} 80\% apodized galactic mask. \textit{Right :} Corresponding binned bispectrum coupling matrix for $\Delta\ell=64$ $\ell_\mr{max}=256$ and 100 iterations ; the entries of the matrix are the bispectrum configurations.}
\label{Fig:mixmatbisp}
\end{center}
\end{figure}

I have developed this computation, then checked that the resulting bispectra indeed form a basis and that the coupling matrix is the identity for simulations without mask. However, the problem that I have encountered is that the matrices involved in the computation are poorly-conditioned\footnote{The condition number for a matrix $M$ and a norm $\mid\mid \!\cdot\! \mid\mid$ is cond(M)=$\mid\mid\! M \!\mid\mid \; \mid\mid\! M^{-1} \!\!\mid\mid$. It quantifies the sensitivity of the numerical inversion of the matrix. Indeed for a matrix with a large condition number, small errors on the matrix coefficients lead to large errors on the coefficient of the invert. In our case, the cosmic variance of the simulations produces noise in the $\mathcal{B}_0$ coefficients.} so that the noise gets amplified. For example with the $\alpha_\ell$ described previously and choosing $\tilde{\ell}_{(0)}=\ell_\mr{min}$ $\tilde{\ell}_{(1)}=\ell_\mr{mean}$ $\tilde{\ell}_{(2)}=\ell_\mr{max}\ $\footnote{So that the $\alpha_\ell$ are as different as possible.}, with multipole bins $\Delta\ell=64$ up to $\ell_\mr{max}=256$, the condition number of $\mathcal{B}_t$ is equal to $2\times 10^5$. Thus the recovered coupling matrices are quite noisy. Several iterations are needed to estimate this noise and decrease it through averaging. I have tried several changes to decrease the condition number, e.g. $\alpha_\ell = \sin(\ell - \tilde{\ell}_{(i)})$ or (many) other functional forms. For a fixed binning scheme $\Delta\ell=64$ up to $\ell_\mr{max}=256$, I have been able to reduce significantly the condition number, to $\mathcal{O}(100)$, and to obtain correct estimations of coupling matrices. For example, Fig.\ref{Fig:mixmatbisp} shows the matrix obtained with a 80\% galactic mask and this binning scheme, giving 4 multipole bins and 13 bispectrum configurations. Namely the center of the multipole bins are 32, 96, 160, 224 and the bispectrum configurations are : (32,32,32), (32,96,96), (32,160,160), (32,224,224), (96,96,96), (96,96,160), (96,160,160), (96,160,224), (96,224,224), (160,160,160), (160,160,224), (160,224,224) and (224,224,224)\,; these are the 13 entries of the matrix.

The matrix is dominated by its diagonal, and the off-diagonal terms are in fact dominated by noise. The diagonal terms are close to $f_\mr{SKY}$, as the binning size is large and we have a large sky fraction. Indeed, we saw for the power spectrum that the mask mostly couples neighboring multipoles, in the case of Fig.\ref{Fig:mixmatbisp} the correlation length is smaller than the bin size so that the coupling between different bins is negligible.
\newline

However, this method to compute numerically the bispectrum coupling matrix is not computationally conceivable for higher resolutions and/or numbers of multipole bins. Indeed, at higher resolution each bispectrum evaluation becomes slower. Moreover, if we increase the number of multipole bins, the number of configurations increases even more ($n_\mr{config} \propto n_\mr{bin}^3$) which increases the number of required simulations for each iteration. Furthermore, for binning schemes others than the previously described one, I have not been able to obtain condition number low enough ($\leq100$) to make computations numerically possible.

Although I have not been able to compute coupling matrices at high enough resolution to debias the bispectrum measurement which will be described in Chapt.\ref{chapt:NGmeas}, there is already valuable information in the results of Fig.\ref{Fig:mixmatbisp}. Indeed, we see that the coupling between different bins for this bin size is quite negligible until $\ell_\mr{max}=256$. Furthermore, we have seen for the power spectrum that the correlation length between multipoles does not increase at high multipoles, it even tends to decrease. Hence we can reasonably assume that for $\Delta\ell=64$, the correlation between bins will remain negligible at high resolution. Thus, for bispectra without steep variations, we expect the $f_\mr{SKY}$ approximation to produce a correct debiasing of binned bispectra for this bin size. I will describe in Chapt.\ref{chapt:NGmeas} how I dealt with the problem for measurement on \planck data.
\newline

Finally, one of the perspective of this project is to find novel ways to reduce the condition number. Indeed if we had condition number of order $\mathcal{O}(1)$, only a few iterations would be necessary per matrix. Thus the computation would become possible at higher resolution in some CPU-days/weeks, up to $\ell_\mr{max}=1024$ $N_\mr{side}=512$, which is a resolution high enough for interesting studies of the foregrounds that will be described later on (see Chapt.\ref{chapt:LSS} \& \ref{chapt:NGmeas}). Optimising the $\alpha_\ell$ to form a better basis, or other methods to generate simulations with a given bispectrum \citep[e.g.][]{Brown2013} are other possible tracks to solve the problem.

\subsubsection{Bispectrum representation : a new parametrisation}\label{Sect:bisp_param}

Once we have an optimal estimate of the bispectrum, we need a tool to visualise this complex three-dimensional quantity. Several ways of visualising the angular bispectrum have been proposed in the literature, e.g. isosurfaces in the ($\ell_1,\ell_2,\ell_3$) 3D space by \cite{Fergusson2010}, or slices of constant perimeter in the orthogonal transverse coordinate ($\ell _{\perp a},\ell _{\perp b}$)
space by \cite{Bucher2010}.

We know that $b_{\ell_1 \ell_2 \ell_3}$ is invariant under permutations of $\ell_1$, $\ell_2$ and $\ell_3$, i.e. it is a function of the shape and size of the triangle $(\ell_1, \ell_2, \ell_3)$ only. However the aforementioned visualisations do not account for this property and hence plot up to six times the same information. In \cite{Lacasa2012} I have proposed a parametrisation invariant under permutation of $\ell_1$, $\ell_2$, and $\ell_3$, as this will avoid redundancy of information and allow convenient visualisation and interpretation of data. Let us first denote $\overline{(\ell_1,\ell_2,\ell_3)}$ the equivalence class of the triplet under permutations. 

\begin{figure}[htbp]
\centering
\includegraphics[width=8.5cm]{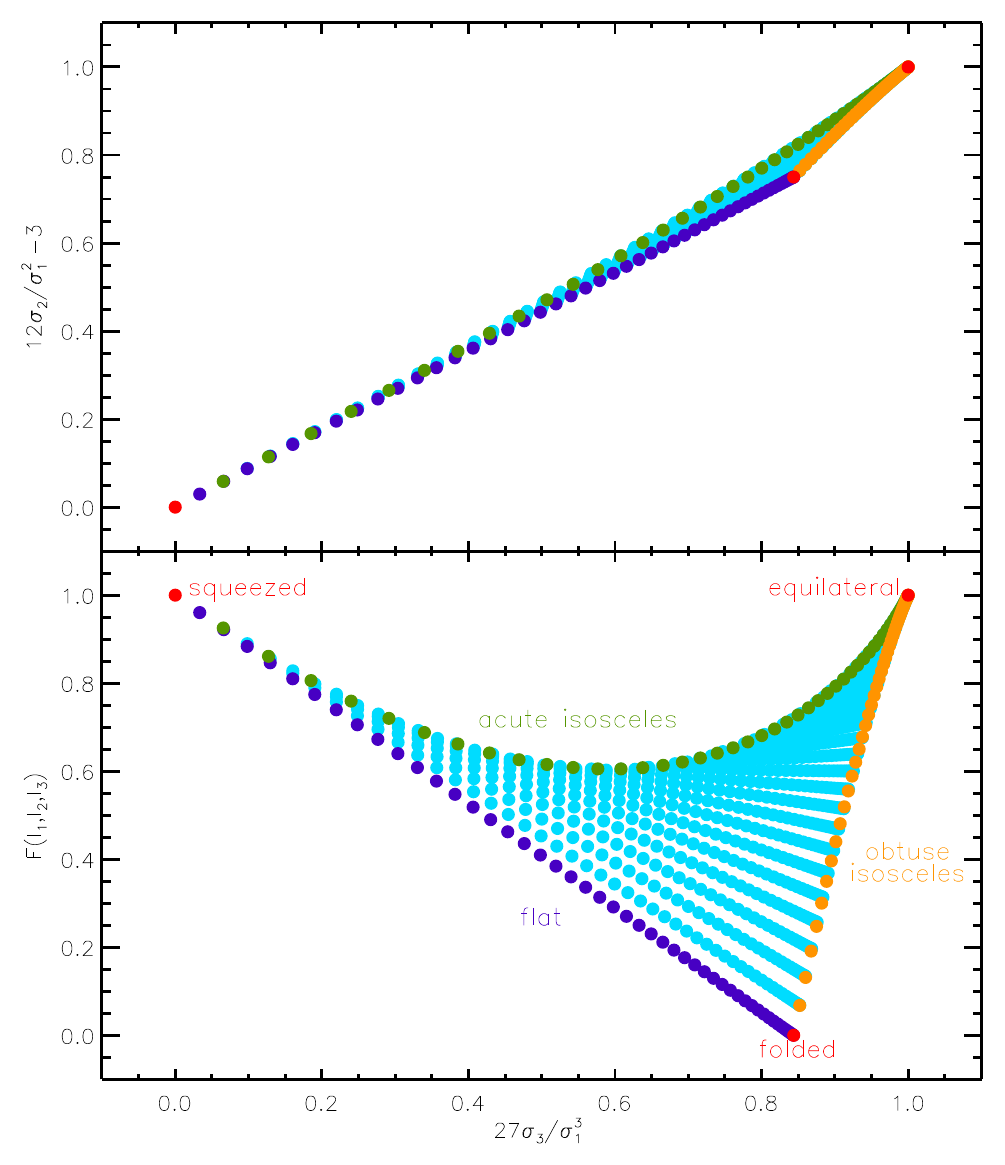}
\caption{\textit{Upper panel :} triangles of constant perimeter, $P$, in the parametrisation
  defined by the normalised symmetric polynomials. \textit{Bottom panel :} same triangles in
  the new parameter space ($F,S$) proposed by \cite{Lacasa2012} and defined in the text.} 
\label{Fig:triparam_both}
\end{figure}

The elementary symmetric polynomials ensure the invariance under permutations:
\begin{itemize}
\item $\sigma_1 = \ell_1+\ell_2+\ell_3 $
\item $\sigma_2 = \ell_1\ell_2+\ell_1\ell_3+\ell_2\ell_3 $
\item $\sigma_3 = \ell_1\ell_2\ell_3 $
\end{itemize}
Through Cardan's formula, there is a one-to-one correspondence between $\overline{(\ell_1,\ell_2,\ell_3)}$ and the triplet $(\sigma_1,\sigma_2,\sigma_3)$, as the multipoles are defined by the roots of the polynomial $X^3 - \sigma_1 X^2 + \sigma_2 X - \sigma_3$.\\
I further define the scale-invariant parameters $\tilde{\sigma}_2=12\sigma_2/\sigma_1^2-3$ and $\tilde{\sigma}_3=27\sigma_3/\sigma_1^3$ with coefficients chosen so that $\tilde{\sigma}_2$ and $\tilde{\sigma}_3$ vary in the range [0,1].  As illustrated in the upper panel of Fig. \ref{Fig:triparam_both}, this parametrisation does not allow us to discriminate efficiently between the different triangles.

I have introduced the parameters noted $(P,F,S)$~:
\begin{itemize}
\item $P=\sigma_1$ (perimeter)
\item $F=32(\tilde{\sigma}_2-\tilde{\sigma}_3)/3+1$
\item $S=\tilde{\sigma}_3$
\end{itemize}
which provide a clearer distinction of the triangles as is illustrated in the bottom panel of Fig. \ref{Fig:triparam_both}.

I have adopted this parametrisation to plot bispectra in the following way : defining perimeter bins between the minimal ($P_\mr{min} = 3 \ell_\mr{min}$) and the maximal ($P_\mr{max} = 3 \ell_\mr{max}$) perimeter, I plot all the triangles in a given perimeter bin in the $(F,S)$ space and code the value of the bispectrum with a color. In this thesis, bispectra will be plotted either with this new parametrisation or in some particular configurations (e.g., equilateral depending on $\ell=\lu=\ld=\lt$).

\subsection{(Co)variance}\label{Sect:polyspcov}

A measurement is useless without the associated error bar. Here I derive the (co)variance of a power spectrum or bispectrum measurement in the full-sky case (see \cite{Sefusatti2006} for a derivation in the 3D Fourier case and \cite{Joachimi2009} for the 2D flat sky limit). The partial sky case is more complex, and I describe an approximation in this case in Sect.\ref{Sect:Cov-polysp_fsky}.

\subsubsection{Power spectrum}
The power spectrum estimator $\hat{C}_\ell$ is~:
\be
\hat{C}_\ell = \frac{1}{2\ell+1} \sum_m a_{\ell m} \, a^*_{\ell m}
\ee
Hence we have~:
\begin{align*}
\lbra \hat{C}_\ell \, \hat{C}_{\ell'} \rbra &= \frac{1}{(2\ell+1)(2\ell'+1)} \sum_{m,m'} \lbra a_{\ell m} \, a^*_{\ell m} \, a_{\ell' m'} \, a^*_{\ell' m'} \rbra \\
&= \frac{1}{(2\ell+1)(2\ell'+1)} \sum_{m,m'} \left[\lbra a_{\ell m} \, a^*_{\ell m} \rbra \lbra a_{\ell' m'} \, a^*_{\ell' m'} \rbra
 + \lbra a_{\ell m} \, a_{\ell' m'} \rbra \lbra a^*_{\ell m} \, a^*_{\ell' m'} \rbra\right.\\
& \left.+ \lbra a_{\ell m} \, a^*_{\ell' m'} \rbra \lbra a_{\ell' m'} \, a^*_{\ell m} \rbra
 + \lbra a_{\ell m} \, a^*_{\ell m} \, a_{\ell' m'} \, a^*_{\ell' m'} \rbra_c \right]\\
 &= C_\ell \, C_{\ell'} + \delta_{\ell \ell'} \, \frac{2 C_\ell \, C_{\ell'}}{2\ell+1} + \frac{1}{(2\ell+1)(2\ell'+1)} \sum_{m,m'} (-1)^{m+m'} \\
 & \times \sum_{\ell^d} \mathcal{P}^{(4)}(\ell,\ell,\ell',\ell',\ell^d) \sum_{m^d} G_{\ell \ell \ell^d}^{m,-m,m^d} \times (-1)^{m^d} \, G_{\ell^d \ell' \ell'}^{-m^d,m',-m'}\\
 &= C_\ell \, C_{\ell'} + \delta_{\ell \ell'} \, \frac{2 C_\ell \, C_{\ell'}}{2\ell+1} + \frac{1}{(2\ell+1)(2\ell'+1)} \sum_{m,m',\ell^d} \mathcal{P}^{(4)}(\ell,\ell,\ell',\ell',\ell^d)\\
 & \times \sum_{m^d} \frac{1}{\sqrt{4\pi}} \delta_{\ell^d,0} \delta_{m^d,0} \, \frac{1}{\sqrt{4\pi}} \delta_{\ell^d,0} \delta_{m^d,0}\\
 &= C_\ell \, C_{\ell'} + \delta_{\ell \ell'} \, \frac{2 C_\ell \, C_{\ell'}}{2\ell+1} + \frac{\mathcal{P}^{(4)}(\ell,\ell,\ell',\ell',\ell^d=0)}{4\pi}
\end{align*}
so that~:
\be
\mr{Cov}(\hat{C}_\ell , \hat{C}_{\ell'}) = \frac{2 C_\ell \, C_{\ell'}}{2\ell+1} \, \delta_{\ell \ell'} + \frac{\mathcal{P}^{(4)}(\ell,\ell,\ell',\ell',\ell^d=0)}{4\pi}
\ee
The first term is the Gaussian contribution and the second term the trispectrum contribution. Let us note that the $4\pi$ denominator allows the trispectrum contribution to have the same unit ($[\mathcal{P}^{(4)}] = K^4 \, \mr{sr}^3$) as the covariance ($[C_\ell^2] = K^4 \, \mr{sr}^2$). We can also note that the covariance matrix is diagonal in the Gaussian case, where the coefficients $a_{\ell m}$ and thus the power spectrum estimations are independent.

\subsubsection{Bispectrum}\label{Sect:bispcov}

The bispectrum estimator $\hat{b}_{123}$ is~:
\be
\hat{b}_{123} = \frac{1}{N_{123}} \sum_{m_{123}} G_{123} \, a_{123}
\ee
Hence we have~:
\be
\lbra \hat{b}_{123} \, \hat{b}_{456} \rbra = \frac{1}{N_{123} N_{456}} \sum_{m_{1\cdots 6}} G_{123} \, G_{456} \, \lbra a_{1\cdots 6} \rbra
\ee
Through the relations between correlation functions and Ursell functions, $\lbra a_{1\cdots 6} \rbra$ splits into four terms~:
\be
\lbra a_{1\cdots 6} \rbra = \lbra a_{1\cdots 6} \rbra_{2\times2\times2} + \lbra a_{1\cdots 6} \rbra_{3\times3} + \lbra a_{1\cdots 6} \rbra_{4\times2} + \lbra a_{1\cdots 6} \rbra_c
\ee
The first term corresponds to the case when $a_{\ell m}$ are grouped by pairs\,; it contains 15 terms of the form $\lbra a_{12} \rbra \lbra a_{34} \rbra \lbra a_{56} \rbra$. The second term corresponds to the case when $a_{\ell m}$ are grouped by triplets\,; it contains 10 terms of the form $\lbra a_{123} \rbra \lbra a_{456} \rbra $. The third term corresponds to the case when $a_{\ell m}$ are grouped into a 4-uplet (trispectrum) and a pair\,; it contains 15 terms of the form $\lbra a_{1234} \rbra_c \lbra a_{56} \rbra$. The last term is the sixth-order polyspectrum contribution.

The derivation is relatively long, so that I will proceed term by term and give the results without the derivation steps.
For the first term ($2\times2\times2$), three pairs have to be formed, which leads to two types of contributions~:
\begin{itemize}
\item 9 contributions where at last one of the pair belongs to one of the bispectrum triplet (e.g. (12)(34)(56) : (12) belongs to the triplet (123)). These contributions are null except if the monopole is considered (which is not the case in this thesis).\\
Example : $(12)(34)(56) \rightarrow \delta\lbra \hat{b}_{123} \, \hat{b}_{456} \rbra_{2\times2\times2} =0$ except if $\ell_3=\ell_4=0$.
\item 6 contributions where only multipoles from different triplets are paired (e.g. (14)(25)(36)). These contributions are non-zero when the multipoles paired are identical.\\
Example : $(14)(25)(36) \rightarrow \delta\lbra \hat{b}_{123} \, \hat{b}_{456} \rbra_{2\times2\times2} = \!\!\!\!\frac{C_{\ell_1}C_{\ell_2}C_{\ell_3}}{N_{123}} \delta_{\ell_1\ell_4}\;\!\delta_{\ell_2\ell_5}\;\!\delta_{\ell_3\ell_6}$
\end{itemize}
All this can be shortened into~:
\be\label{Eq:bispcovmat_term2}
\lbra \hat{b}_{123}^2 \rbra_{2\times2\times2} = \frac{C_{\ell_1}C_{\ell_2}C_{\ell_3}}{N_{123}} \times
\left\{ \begin{array}{ll} 6 & \mathrm{equilateral \ triangle}\\ 2 &
  \mathrm{isosceles \ triangle}\\ 1 & \mathrm{general
    \ triangle} \end{array}\right.
\ee
and $\lbra \hat{b}_{123} \, \hat{b}_{456} \rbra_{2\times2\times2} = 0$ if $\ell_{123}$ and $\ell_{456}$ are different triangles.
\newline

For the second term ($3\times3$), two triplets have to be formed, which leads to two types of contributions~:
\begin{itemize}
\item 1 contribution where the triplets corresponds to the original bispectrum triplets~:
$$(123)(456) \rightarrow \delta\lbra \hat{b}_{123} \, \hat{b}_{456} \rbra_{3\times3} = b_{123} \, b_{456}$$
This contribution cancels the term $\lbra\hat{b}_{123}\rbra\lbra\hat{b}_{456}\rbra$ in the definition of the covariance.
\item 9 other contributions where two multipoles of the original triplets have been permuted (e.g. (124)(356) : 3 and 4 have been permuted).\\
Example : $(124)(356) \rightarrow \delta\lbra \hat{b}_{123} \, \hat{b}_{456} \rbra_{3\times3} =\frac{b_{124} \, b_{356}}{2\ell_3+1} \delta_{\ell_3 \ell_4}$
\end{itemize}
Note that for a bispectrum of the form $b_{123} = \alpha_{\lu} \, \alpha_{\ld} \, \alpha_{\lt}$, the numerators of the 9 contributions (e.g. $b_{124} \, b_{356}$) are all equal to $b_{123} \, b_{456}$.\\
In a single expression, we have~:
\bea
\nonumber\mr{Cov}(\hat{b}_{123} , \hat{b}_{456})_{3\times3} &=& 
\frac{b_{156} \, b_{234}}{2\ell_1+1} \delta_{\ell_1 \ell_4}
+\frac{b_{146} \, b_{235}}{2\ell_1+1} \delta_{\ell_1 \ell_5}
+\frac{b_{145} \, b_{236}}{2\ell_1+1} \delta_{\ell_1 \ell_6}\\
\nonumber&&+\frac{b_{134} \, b_{256}}{2\ell_2+1} \delta_{\ell_2 \ell_4}
+\frac{b_{135} \, b_{246}}{2\ell_2+1} \delta_{\ell_2 \ell_5}
+\frac{b_{136} \, b_{245}}{2\ell_2+1} \delta_{\ell_2 \ell_6}\\
\label{Eq:bispcovmat_term3} &&+\frac{b_{124} \, b_{356}}{2\ell_3+1} \delta_{\ell_3 \ell_4}
+\frac{b_{125} \, b_{346}}{2\ell_3+1} \delta_{\ell_3 \ell_5}
+\frac{b_{126} \, b_{345}}{2\ell_3+1} \delta_{\ell_3 \ell_6}
\eea
This contribution to the covariance matrix has off-diagonal terms.
\newline

For the third term ($4\times2$), a pair and a quadruplet have to be formed which leads to two types of contributions~:
\begin{itemize}
\item 6 contributions where the pair belongs to on of the bispectrum triplet (e.g. (12)(3456)~: (12) belongs to (123)). These contributions are null except if the monopole is considered (which is not the case in this thesis).\\
Example : $(12)(3456) \rightarrow \delta\lbra \hat{b}_{123} \, \hat{b}_{456} \rbra_{4\times2} =0$ except if $\ell_3=0$\\
\item 9 contributions where multipoles paired are from different bispectrum triplets.\\
Example : $(14)(2356) \rightarrow \delta\lbra \hat{b}_{123} \, \hat{b}_{456} \rbra_{4\times2} = \frac{C_{\ell_1} \, \mathcal{P}^{(4)}(\ell_2,\ell_3,\ell_5,\ell_6,\ell^d=\ell_1)}{2\ell_1+1} \, \delta_{\ell_1 \ell_4}$
\end{itemize}

In a single formula, we have~:
\begin{align}
\nonumber\mr{Cov}(\hat{b}_{123} , \hat{b}_{456})_{4\times2} &= 
\frac{C_{\ell_1} \, \mathcal{P}^{(4)}(\ell_{2356},\ell_1)}{2\ell_1+1} \delta_{\ell_1 \ell_4}
+\frac{C_{\ell_1} \, \mathcal{P}^{(4)}(\ell_{2346},\ell_1)}{2\ell_1+1} \delta_{\ell_1 \ell_5}
+\frac{C_{\ell_1} \, \mathcal{P}^{(4)}(\ell_{2345},\ell_1)}{2\ell_1+1} \delta_{\ell_1 \ell_6}\\
\nonumber&+\frac{C_{\ell_2} \, \mathcal{P}^{(4)}(\ell_{1356},\ell_2)}{2\ell_2+1} \delta_{\ell_2 \ell_4}
+\frac{C_{\ell_2} \, \mathcal{P}^{(4)}(\ell_{1346},\ell_2)}{2\ell_2+1} \delta_{\ell_2 \ell_5}
+\frac{C_{\ell_2} \, \mathcal{P}^{(4)}(\ell_{1345},\ell_2)}{2\ell_2+1} \delta_{\ell_2 \ell_6}\\
\label{Eq:bispcovmat_term4} &+\frac{C_{\ell_3} \, \mathcal{P}^{(4)}(\ell_{1256},\ell_3)}{2\ell_3+1} \delta_{\ell_3 \ell_4}
+\frac{C_{\ell_3} \, \mathcal{P}^{(4)}(\ell_{1246},\ell_3)}{2\ell_3+1} \delta_{\ell_3 \ell_5}
+\frac{C_{\ell_3} \, \mathcal{P}^{(4)}(\ell_{1245},\ell_3)}{2\ell_3+1} \delta_{\ell_3 \ell_6}
\end{align}
This term of the covariance matrix contributes in the same off-diagonal entries as the $3\times3$ term.
\newline

The last term (from the 6-th order polyspectrum) has fortunately only one contribution~:
\be\label{Eq:bispcovmat_term6}
(123456) \rightarrow \delta\lbra \hat{b}_{123} \, \hat{b}_{456} \rbra_{c} = \frac{\mathcal{P}^{(6)}(\ell_{1\cdots 6}, \ell_1^d=\ell_3,\ell_2^d=0,\ell_3^d=\ell_4)}{4\pi}
\ee
\newline

I do not group all terms into a single equation, and I will only need to refer to Eq.\ref{Eq:bispcovmat_term2}-\ref{Eq:bispcovmat_term3}-\ref{Eq:bispcovmat_term4}-\ref{Eq:bispcovmat_term6} individually in the following. Note however that, similarly to the power spectrum, the covariance matrix is diagonal in the Gaussian case. However bispectrum measurements are then not independent but only linearly uncorrelated~: indeed they mix up different multipoles and e.g., $\hat{b}_{\lu \ld \lt}$ and $\hat{b}_{\lu \ld' \lt'}$ are obviously not independent. In the non-Gaussian case, the bispectrum covariance matrix is clearly non-diagonal.

\begin{figure}[htbp]
\begin{center}
\begin{tikzpicture}
%2*2*2
\draw [->, very thick,red] (-2,-1) -- node[above] {1} (-3.9,-1);
\draw [->, very thick,red] (-4.2,-1.2) -- node[below] {4} (-1.7,-1.2);
\draw [->, very thick,darkgreen] (-4,-1) -- node[right] {2} (-3.1,0.9);
\draw [->, very thick,darkgreen] (-3.1,1.3) -- node[left] {6} (-4.3,-1.15);
\draw [->, very thick,blue] (-3,1) -- node[left] {3} (-1.9,-0.9);
\draw [->, very thick,blue] (-1.6,-1.1) -- node[right] {5} (-3,1.3);
%3*3
\draw [->, very thick,red] (0.95,-1) -- node[below] {1} (-0.9,-1);
\draw [->, very thick,red] (-1,-1) -- node[left] {2} (-0.1,0.9);
\draw [->, very thick,red] (-0.05,0.95) -- node[left] {3} (1.1,-1);
\draw [->, very thick,blue]  (0.05,1.05) -- node[right] {4} (1.25,-0.95);
\draw [->, very thick,blue]  (1.3,-1) -- node[right] {5} (2,0.9);
\draw [->, very thick,blue] (2,1) -- node[above] {6} (0.2,1);
%4*2
\draw [->, very thick,blue] (4.95,-1) -- node[below] {1} (3.1,-1);
\draw [->, very thick,blue] (3,-1) -- node[left] {2} (3.9,0.9);
\draw [->, very thick,red] (3.95,0.95) -- node[left] {3} (5.1,-1);
\draw [->, very thick,red] (5.2,-0.9) -- node[right] {4} (4.05,1.05);
\draw [->, very thick,blue] (4.25,1) -- node[above] {5} (6,1);
\draw [->, very thick,blue]  (5.95,0.85) -- node[right] {6} (5.3,-1);
%6
\draw [->,very thick] (8.85,-0.1) -- node[below] {1} (7,-1);
\draw [->,very thick] (7,-0.85) -- node[right] {2} (7,1);
\draw [->,very thick] (7.15,0.95) -- node[above] {3} (8.96,0);
\draw [->,very thick] (9.1,0.1) -- node[above] {4} (11,1.05);
\draw [->,very thick] (11,0.9) -- node[right] {5} (11,-1);
\draw [->,very thick] (10.85,-0.9) -- node[below] {6} (9.04,0);
\end{tikzpicture}
\caption{Diagrams corresponding to the different terms of the bispectrum covariance.  From left to right $2\times2\times2$ (double-circulation triangle), $3\times3$ (parallel kite), $4\times2$ (opposite kite) and 6 (butterfly).}
\label{Fig:bispcov}
\end{center}
\end{figure}
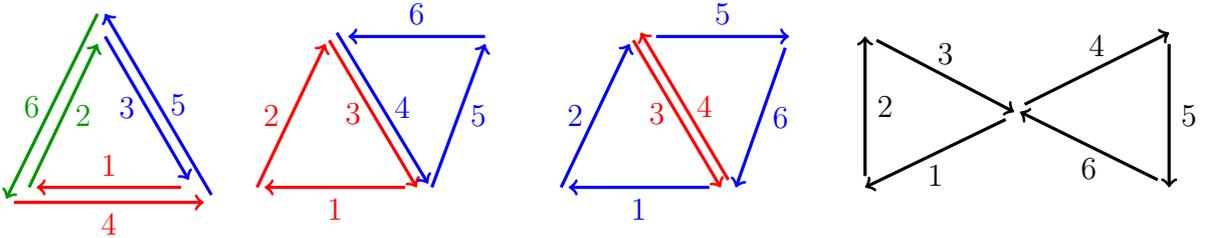

All these terms can be understood simply and graphically in the Fourier case where, as explained in Sect.\ref{Sect:2Dpolysp}, $n^\mr{th}$ order polyspectra correspond to a 2D polygon with $n$ sides. All terms of the bispectrum covariance correspond to possibilities of forming polygons with 6 vectors(1,2,3,4,5,6) given that (1,2,3) and (4,5,6) respectively form triangles. Fig.\ref{Fig:bispcov} shows diagrams representative of the different terms of the bispectrum covariance matrix.

In the same color are multipoles which are averaged together to form a polyspectrum. For example in the first diagram (1,4) (2,6) and (3,5) are averaged together respectively to form a power spectrum, yielding $C_{\lu}\delta_{1 4^*}$ $C_{\ld}\delta_{2 6^*}$ and $C_{\lt}\delta_{3 5^*}$ respectively, so that the total contribution of the diagram is $\frac{C_{\ell_1}C_{\ell_2}C_{\ell_3}}{N_{123}} \delta_{\ell_1\ell_4}\;\!\delta_{\ell_2\ell_6}\;\!\delta_{\ell_3\ell_5}$.

\subsubsection{Incomplete sky}\label{Sect:Cov-polysp_fsky}

In the incomplete sky case, we have seen that the estimation of a polyspectrum becomes complicated. The covariance matrix of this estimation is even more complicated\,; for example at the power spectrum level the covariance matrix may be computed with simulations or approximated analytically, but a full analytical computation is numerically virtually impossible \citep[see e.g.][]{Efstathiou2004}.

However, the situation becomes tractable if the polyspectrum varies slowly compared to the coupling matrix. Indeed, we have seen in the previous sections that the polyspectrum estimation is then biased by a factor $f_\mr{SKY}$. The covariance of this estimation is a particular case of the polyspectrum at double order, i.e., the covariance of the $n^\mr{th}$ order polyspectrum involves the 2$n^\mr{th}$ order polyspectrum. This covariance is hence also multiplied by a factor $f_\mr{SKY}$.\\
As we want an unbiased estimate of the polyspectrum, we need to divide the partial-sky measurement by $f_\mr{SKY}$. For this unbiased estimate, we then have~:
\be
\mr{Cov}_\mr{partial-sky} = \mr{Cov}_\mr{full-sky}/f_\mr{SKY}
\ee

This implies that the error bars scale as $1/\sqrt{n_\mr{pix}}$ (at fixed pixel size), i.e. going down with the number of observations.

This result has been derived in the case where the polyspectrum varies slowly compared to the mask coupling effect. In particular it is exact for flat polyspectra, e.g., that of a white noise. However it can be extended to larger masks with binned estimation, if the bin size renders the coupling between different bins negligible. The latter property is verified if the binned coupling matrix is mostly diagonal \citep{Hivon2002}. For example this is the case in Fig.\ref{Fig:mixmatbisp}. In the following, we will assume that the bin size is large enough to be in this situation.

%%%%%%%%%%%%%%%%%%%%%%%%%%%%%%%%%%%%%%%%%%%%%%%%
\section{Examples}
\subsection{Gaussian random field}
An isotropic Gaussian random field is entirely described by its mean $\lbra T \rbra$ and its power spectrum $C_\ell$. Furthermore all the statistical information is contained in these two correlation functions, i.e., the measurement of higher order correlation functions does not provide any improvement on the constraint of $\lbra T \rbra$ and $C_\ell$. \\
Ordering the harmonic coefficients in lexicographic order, the probability of a map reads~:
\be
P(\mathbf{a}=(a_{\ell m})) = \frac{1}{\sqrt{(2\pi)^N \det \mathbf{C}}} \exp\left[ ^T\bar{\mathbf{a}} \; \mathbf{C}^{-1} \, \mathbf{a} \right]
\ee
\clearpage
\be
\mr{with} \quad \mathbf{C} = 
\left(\begin{matrix}
C_1 & 0 & \cdots &  &  & \cdots & 0 \\
0 & C_1 & \ddots & & & & \vdots \\
\vdots & \ddots & \ddots & & & & \\
 & &  & C_\ell & & &\\
 & & & & \ddots & \ddots & \vdots \\
\vdots & & & & \ddots & C_\ell & 0 \\
0 & \cdots & & \tikzmark{first} &\cdots & \tikzmark{last} 0 & \ddots
\end{matrix}\right)
\ee
\begin{tikzpicture}[overlay, remember picture,decoration={brace,amplitude=5pt}]
\draw[decorate,thick] (last.south) -- (first.south)
      node [midway,below=5pt] {$\ell+1$ terms};
\end{tikzpicture}

\ \\  where $C_\ell$ is the power spectrum of the field and all higher order polyspectra are zero. Thus $a_{\ell m}$ are random complex numbers (real for $m=0$) with variance $C_\ell$, and importantly they are independent (except of course $a_{\ell m}$ and $a_{\ell,-m}$). The statistical information is contained in the moduli of the harmonic coefficients while their phases are independent and distributed randomly and uniformly in $[0,2\pi[$.

The important example in cosmology of a Gaussian random field on the sphere is the Cosmic Microwave Background (CMB), which will be described in Sect.\ref{Sect:CMB}. Another example is instrumental noise, which is often Gaussian to a good approximation. The noise may however not be homogeneous, as is the case for {\planck}, and as it is often uncorrelated between different pixels it is better characterised in pixel space than in harmonic space.

\subsection{White noise}\label{Sect:whitenoise}

For a white noise field all pixels are independent. If the white noise is homogeneous all pixels are drawn from the same underlying distribution and in particular the field is isotropic. This is for example the case of the emission of unclustered extragalactic point-sources. As explained in Sect.\ref{Sect:stat-whitenoise}, in this case the pixel space correlation function takes the form~:
\be
\lbra T(\hn_1) \cdots T(\hn_n) \rbra_c = \kappa_n(T) \, \delta_{\hn_1,\cdots,\hn_n}
\ee
with a Kronecker symbol being one iff $\hn_1 \cdots \hn_n$ are the same pixel. A harmonic transform then gives~:
\be
\lbra a_1 \cdots a_n \rbra_c = \kappa_n(T) \, \Omega_\mr{pix}^{n-1} \int \dd^2\hn \, Y_{1\cdots n}(\hn)
\ee
where $\Omega_\mr{pix}$ is the pixel area in steradian.\\
Thus a homogeneous white noise generates a diagonal-independent polyspectrum which is furthermore constant~:
\be
\mathcal{P}^{(n)}(\ell_1 \cdots \ell_n) = \kappa_n(T) \, \Omega_\mr{pix}^{n-1}
\ee
In particular, the power spectrum is $\sigma^2 \, \Omega_\mr{pix}$ (a well-known formula for instrumental noise), and the bispectrum is $\kappa_3 \, \Omega_\mr{pix}^2$. This also makes the polyspectrum unit obvious, e.g. if the field $T(\hn)$ is measured in Kelvins, the n-order polyspectrum has unit $\mr{K}^n \mr{sr}^{n-1}$.

The important example in cosmology of a white noise field on the sphere is a distribution of unclustered point-sources, such as extragalactic radio point-sources described in Sect.\ref{Sect:introps}. Instrumental noise may also be whiten in addition to its Gaussianity.

\subsubsection{Generating non-Gaussian fields}

Using white noises is a convenient way to generate highly non-Gaussian maps with a given bispectrum form. The first step is to generate a white noise such that its bispectrum has the highest possible signal-to-noise ratio. For this purpose, I have used a Poisson law (see Sect.\ref{Sect:statPoisson}) as the underlying distribution, since it is easily produced numerically by random number generators, and since the Poisson parameter $\lambda$ is easily tunable to obtain a high SNR. Indeed, compiling the results of Sect.\ref{Sect:statPoisson} and Sect.\ref{Sect:polyspcov}, for a Poisson white noise the SNR of a bispectrum coefficient is (up to some constants $A,B,C,D$)~:
\bea
\nonumber (\mr{SNR})^2 &\propto& \frac{\left(\lambda \Omega_\mr{pix}^2\right)^2}{A\left(\lambda \Omega_\mr{pix}\right)^3+B\left(\lambda \Omega_\mr{pix}^2\right)^2+C\lambda \Omega_\mr{pix}\,\lambda\Omega_\mr{pix}^3+D\lambda \Omega_\mr{pix}^5}\\
&=& \frac{1}{B+C+A\,\lambda/\Omega_\mr{pix} + D\,\Omega_\mr{pix}/\lambda}
\eea
The SNR is maximal for $\lambda\sim\Omega_\mr{pix}\propto1/n_\mr{pix}$. So a given bispectrum configuration has the highest signal-to-noise ratio for a Poisson distribution with one source per map on average. In reality I have runned simulations with an average thousand sources per map, so that the covariance matrix shall be closer to diagonal, and thus different bispectrum configurations shall be few correlated.

These highly non-Gaussian simulations can then be processed in the following way~:
\begin{itemize}
\item compute the harmonic transform of the white noise $\rightarrow a_{\ell m}^\mr{white}$
\item multiply the harmonic coefficient by an arbitrary function $\alpha_\ell$ : $a_{\ell m}^\mr{sim} = \alpha_\ell \, a_{\ell m}^\mr{white}$
\item compute the inverse harmonic transform : $T_\mr{sim}(\hn) = \sum_{\ell m} a_{\ell m}^\mr{sim} \, Y_{\ell m}(\hn)$
\end{itemize}
We then easily see that the final simulation has a bispectrum~:
\be
b_{123} = \alpha_{\lu} \, \alpha_{\ld} \, \alpha_{\lt} \, \lambda \Omega_\mr{pix}^3
\ee
with the same (high) SNR as the original white noises, and that this bispectrum fulfills the requirement of Eq.\ref{Eq:sepbisp4mixmat} for the coupling matrix estimation\footnote{The $\lambda \Omega_\mr{pix}^3$ can be incorporated in a redefinition of $\alpha_\ell$.}.

%%%%%%%%%%%%%%%%%%%%%%%%%%%%%%%%%%%%%%%%%%%%%%%%
\section{Relation with Fourier statistics}

This section describes the statistical tools introduced previously, in the case of a field $f(\xx)$ over $\mathbb{R}^d$ ($d=2,3$), instead of over the sphere. Indeed, sufficiently small patches of the sphere may be analysed as flat fields, with a correspondence between the Fourier modes and the spherical harmonics : the so-called `flat-sky' limit \citep[see e.g.][for a formalisation]{BPU2011}. Furthermore modelisations of cosmological field (e.g. the galaxy density, see Sect.\ref{Sect:HM_descrip}) are naturally set in 3D, hence I will describe 3D statistics and how they project onto the sphere, when observations cannot probe the radial distance.
In the following, the Fourier transform of the field will be noted $a_\kk~=~\int \dd^d\xx \, f(\xx) \, e^{i\kk\cdot\xx}$.

\subsection{2D polyspectra}\label{Sect:2Dpolysp}
\begin{figure}[htbp]
\begin{center}
\begin{tikzpicture}
%spectrum
\draw [->, very thick,red] (-6.1,-1) -- node[left] {$\vec{k}$} (-6.1,1);
\draw [->, very thick,red] (-5.9,1) -- node[right] {$-\vec{k}$} (-5.9,-1);
%bispectrum
\draw [->, very thick,darkgreen] (-4,-1) -- node[left] {2} (-3.1,0.9);
\draw [->, very thick,darkgreen] (-3,1) -- node[right] {3} (-1.9,-0.9);
\draw [->, very thick,darkgreen] (-2,-1) -- node[below] {1} (-3.9,-1);
%trispectrum
\draw [->, very thick,orange] (0,-1) -- node[left] {2} (0.9,0.9);
\draw [->, very thick, dash pattern=on 2pt off 3pt,orange] (1,1) -- node[left] {$1^d$} (2.1,-0.95); 
\draw [->, very thick,orange] (1.1,1) -- node[above] {3} (2.9,1);
\draw [->, very thick,orange]  (3,1) -- node[right] {4} (2.2,-0.85);
\draw [->, very thick,orange] (2,-1) -- node[below] {1} (0.1,-1);
%n-order
\draw [very thick,dash pattern=on 3pt off 4pt,blue] (8,0.8) --  (8,0);
\draw [->, very thick,blue] (8,-0.1) -- node[right] {n} (7.1,-0.9);
\draw [->, very thick,blue] (7,-1) -- node[below] {1} (6,-1);
\draw [->, very thick,blue] (5.9,-0.9) -- node[left,below] {2} (5,-0.1);
\draw [->, very thick,blue] (5,0) -- node[left] {3} (5.1,1.1);
\draw [very thick,dash pattern=on 3pt off 4pt,blue] (5.2,1.2) --  (5.8,1.8);
\draw [->, very thick, dash pattern=on 2pt off 3pt,blue] (7.05,-1) -- node[above] {$1^d$} (5.1,0);
\draw [->, very thick, dash pattern=on 2pt off 3pt,blue] (7.1,-0.95) -- node[right=5pt] {$2^d$} (5.2,1.1);
\end{tikzpicture}
\caption{Diagrams corresponding to the 2D polyspectrum. From left to right : orders \textcolor{red}{2},\textcolor{darkgreen}{3},\textcolor{orange}{4} and \textcolor{blue}{$n$}.}
\label{Fig:polygons2D}
\end{center}
\end{figure}
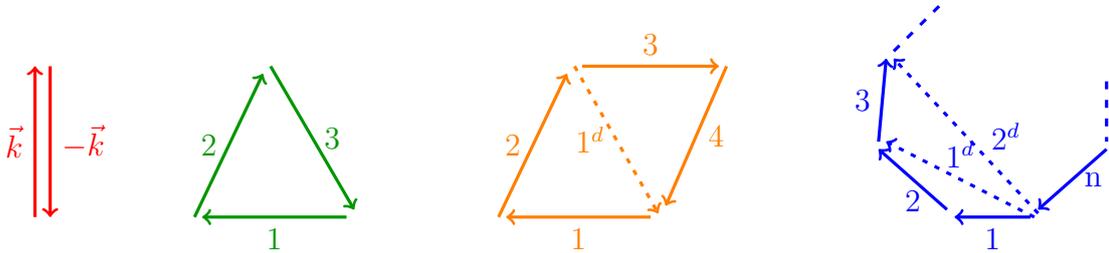

For a 2D field, the polyspectrum of order $n$ is defined through~:
\be\label{Eq:defpolysp2Draw}
\lbra a_{\kk_1} \cdots a_{\kk_n} \rbra_c = (2\pi)^2 \delta^{(2)}(\kk_1 + \cdots + \kk_n) \, \mathcal{P}^{(n)}(\kk_1,\cdots,\kk_n)\\
\ee
where $(\kk_1 , \cdots , \kk_n)$ form a polygon in 2D. Figure \ref{Fig:polygons2D} shows the corresponding polygon for the second (power spectrum), third (bispectrum), fourth (trispectrum) orders and the general case. At orders $\geq 4$, the polygon cannot be parametrised only by the length of its side, but some diagonals need to be fixed. This is why diagonal degrees of freedom appear in the polyspectrum. The choice of the fixed diagonals is somewhat arbitrary, for example for the trispectrum one can either fix $|\kk_1+\kk_2|=|\kk_3+\kk_4|$ or fix $|\kk_1+\kk_4|=|\kk_2+\kk_3|$. In analogy with my choice on the sphere, I here take a choice of diagonals inherited from the orders of the $\kk_i$ and shown in Fig.\ref{Fig:polygons2D} (i.e. $\kk_1^d= \kk_1+\kk_2$, $\kk_2^d=\kk_1^d-\kk_3$ etc).\\
Expanding the dirac $\delta^{(2)}(\kk_1 + \cdots + \kk_n)$ by decomposing the polygon in its sub-triangles, Eq.\ref{Eq:defpolysp2Draw} takes the following form~:
\begin{align}
\nonumber \lbra a_{\kk_1} \cdots a_{\kk_n} \rbra_c &= \int \frac{\dd^2\kk^d_1}{(2\pi)^2} \cdots \frac{\dd^2\kk^d_{n-3}}{(2\pi)^2} \; \mathcal{P}^{(n)}(k_1,\cdots,k_n,k^d_1,\cdots,k^d_{n-3}) \\
\nonumber & (2\pi)^2 \, \delta^{(2)}(\kk_1 + \kk_2 + \kk^d_1) \times (2\pi)^2 \, \delta^{(2)}(-\kk^d_1 + \kk_3 + \kk^d_2)\\
& \times \cdots \times (2\pi)^2 \, \delta^{(2)}(-\kk^d_{n-3} + \kk_{n-1} + \kk_n)
\end{align}
Let us note in particular that this is exactly analogous to the definition of the polyspectrum on the sphere (Eq.\ref{Eq:defpolyspectrum}-\ref{Eq:defG4polysp}) with the replacements~:
\begin{align*}
\sum_{\ell^d m^d} &\rightarrow \int \frac{\dd^2\kk^d}{(2\pi)^2}\\
G_{123} &\rightarrow (2\pi)^2 \, \delta^{(2)}(\kk_1 + \kk_2 + \kk_3)
\end{align*}
The last line reflects intuition as both terms are the integral of the basis functions~:
\be
G_{123} = \int \dd^2\hn \, Y_{123}(\hn) \quad \mr{and} \quad (2\pi)^2 \, \delta^{(2)}(\kk_1 + \kk_2 + \kk_3) = \int \dd^2\xx \, e^{i(\kk_1+\kk_2+\kk_3)\cdot\xx}
\ee

\subsection{3D polyspectra}\label{Sect:3Dpolysp}

For a 3D field, the polyspectrum of order $n$ is defined through~:
\be
\lbra a_{\kk_1} \cdots a_{\kk_n} \rbra_c = (2\pi)^3 \delta^{(3)}(\kk_1 + \cdots + \kk_n) \, \mathcal{P}^{(n)}(\kk_1,\cdots,\kk_n)\\
\ee
where $(\kk_1 , \cdots , \kk_n)$ forms a polygon in 3D so that the polyspectrum has $3n-6$ degrees of freedom\footnote{There are three d.o.f per vector, and six d.o.f. are discarded by isotropy, where six is the number of d.o.f. of a solid in 3D. Comparatively, a polyspectrum has $2n-3$ d.o.f. in 2D, where three is the number of d.o.f. of a solid in 2D.}.  A power spectrum still has one d.o.f. : $\mathcal{P}^{(2)}(\kk_1,\kk_2) = P(k=|\kk_1|)$ and can be visualised as two opposite vectors. A bispectrum still has three d.o.f. : $\mathcal{P}^{(3)}(\kk_1,\kk_2,\kk_3) =B(k_1,k_2,k_3)$ and can be visualised as a triangle.

To parametrise the polyspectrum d.o.f., we must choose a triangulation of the 3D polygon, then the polyspectrum can be parametrised by the length of the $n$ sides and $2n-6$ diagonals. We will index with $g=1\cdots 2n-5$ the triangles of the minimal triangulation chosen, and note $(\kk_1(g),\kk_2(g),\kk_3(g))$ the vectors of this triangle. This vectors $\kk_i(g)$ are either a $\pm k_i$ or a $\pm k^d_j$.\\
The Fourier $n$-point c.f. then takes the form~:
\bea
\nonumber \lbra a_{\kk_1} \cdots a_{\kk_n} \rbra_c &=& \int \frac{\dd^3\kk^d_1}{(2\pi)^3} \cdots \frac{\dd^3\kk^d_{2n-6}}{(2\pi)^3} \; \mathcal{P}^{(n)}(k_1,\cdots,k_n,k^d_1,\cdots,k^d_{2n-6}) \\
\label{Eq:polysp3Dwtriangul}&& \times \prod_g \, (2\pi)^3 \; \delta^{(3)}\left(\kk_1(g)+\kk_2(g)+\kk_3(g)\right)
\eea

\begin{figure}[htbp]
\begin{center}
\begin{tikzpicture}
%bispectrum
\draw [->, very thick,red] (-4,-1) -- node[left] {2} (-3.1,0.9);
\draw [->, very thick,red] (-3,1) -- node[right] {3} (-1.9,-0.9);
\draw [->, very thick,red] (-2,-1) -- node[below] {1} (-3.9,-1);
%trispectrum
\draw [->, very thick,darkgreen] (2.3,-1) -- node[below] {1} (0,-1);
\draw [->, very thick,darkgreen] (0.1,-0.85) -- (2,0.3);
\node[darkgreen] at (0.6,-0.3) {2};
\draw [->, very thick,dash pattern=on 2pt off 3pt,darkgreen] (2.1,0.3) -- node[right] {$1^d$} (2.6,-0.9);
\draw [->, very thick,darkgreen] (1.9,0.38) -- node[above] {3} (0.3,0.9);
\draw [->, very thick,darkgreen] (0.4,0.8) -- (2.5,-1);
\node[darkgreen] at (1.6,-0.5) {4};
\draw [->, very thick,dash pattern=on 2pt off 3pt,darkgreen] (0,-0.9) -- node[left] {$2^d$} (0.3,0.8);
\end{tikzpicture}
\caption{Diagrams corresponding to the 3D polyspectrum at \textcolor{red}{third} and \textcolor{darkgreen}{fourth} orders.}
\label{Fig:polygons3D}
\end{center}
\end{figure}
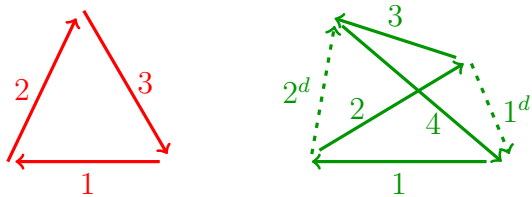

\noindent{\bf Examples :}\\
Fig.\ref{Fig:polygons3D} shows the polygons and their diagonals at third and fourth orders.

For the bispectrum, the polygon is a triangle\,; it is flat, so the number of d.o.f. coincides with the 2D bispectrum, and it has no diagonals. The triangulation is trivial : $g=1$ and $\kk_1(g)=\kk_1,\kk_2(g)=\kk_2, \kk_3(g)=\kk_3$. In this case Eq.\ref{Eq:polysp3Dwtriangul} takes the form~:
\be
\lbra a_{\kk_1} \cdots a_{\kk_3} \rbra_c = B(k_1,k_2,k_3) (2\pi)^3 \; \delta^{(3)}\left(\kk_1+\kk_2+\kk_3\right)
\ee

For the trispectrum, the polygon is a tetrahedron and there are two diagonals : $\kk_1^d=-\kk_1-\kk_2=\kk_3+\kk_4$ and $\kk_2^d=-\kk_1-\kk_4=\kk_2+\kk_3$.\\
There are hence 6 d.o.f. which already diverges from the 2D case (5 d.o.f.). The tetrahedron has four triangular faces but only three need to be fixed, so there is some arbitrary decision in the triangulation. If we choose as triangulation $(\kk_1,\kk_2,\kk^d_1),(-\kk^d_1,\kk_3,\kk_4),(\kk_1,\kk^d_2,\kk_4)$, Eq.\ref{Eq:polysp3Dwtriangul} takes the form~:
\bea
\nonumber \lbra a_{\kk_1} \cdots a_{\kk_4} \rbra_c &=& \int \frac{\dd^3\kk^d_1}{(2\pi)^3} \frac{\dd^3\kk^d_2}{(2\pi)^3} \, \mathcal{P}^{(4)}(k_{1\cdots4},k^d_1,k^d_2)
\; (2\pi)^3 \, \delta^{(3)}(\kk_1+\kk_2+\kk^d_1)\\
&& \times (2\pi)^3 \, \delta^{(3)}(-\kk^d_1+\kk_3+\kk_4) \; (2\pi)^3 \, \delta^{(3)}(\kk_1+\kk^d_2+\kk_4)
\eea

\subsection{Plane wave expansion}

The plane wave expansion, also called Rayleigh expansion, decomposes a 3D Fourier mode into the spherical and radial harmonic basis \citep[see e.g.][]{Brown2009}~:
\bea\label{Eq:rayleigh}
\nonumber e^{i\kk\cdot\xx} &=& (4\pi) \sum_{\ell m} i^\ell \; Y^*_{\ell m}(\hat{k}) \, j_\ell(kr) \, Y_{\ell m}(\hn)\\
&=& (4\pi) \sum_{\ell m} i^\ell \; Y_{\ell m}(\hat{k}) \, j_\ell(kr) \, Y^*_{\ell m}(\hn)
\eea
where 3D vectors are decomposed into their radial and tangential parts : $\kk = k \hat{k}$ and $\xx = r \hn$, and $j_\ell$ is the $\ell$-order spherical Bessel function \citep[see e.g.][]{Brown2009}.

This equation allows us to decompose a 3D field and its correlation functions, into its radial and tangential components.

\subsection{Projection of the past light cone}\label{Sect:pastlightproj}

A frequent situation in cosmology is that the observation in a direction $\hn$ is the cumulative emission along the line-of-sight. Because the signal travels at a finite speed, the contribution at a distance $r$ was emitted at a past time $t(r)$ \footnote{Because of the expansion history of the universe the time $t(r)$ is not simply $r/c$. We have in fact $r(t) = \int_t^{t_0} \frac{c\dd t}{a(t)}$ where $t_0$ is now a(t) is the scale factor and $a(t_0)=1$ by convention.}. The space-time locus of these emissions is called the past light-cone of the observer. Hence an emission integrated along the past light-cone yields~:
\be
T(\hn) = \int_0^\infty \dd r \, \frac{\dd T}{\dd r}(r\hn,t(r))
\ee
Noting for simplicity $f(\xx,t)\equiv\frac{\dd T}{\dd r}$, decomposing $f$ in Fourier, $T(\hn)$ in spherical harmonics and using the Rayleigh expansion, we find~:
\be
a_{\ell m} = i^\ell \int \dd r \, \frac{\dd^3\kk}{2\pi^2} \; f\left(\kk,t(r)\right) \, Y^*_{\ell m}(\hat{k}) \, j_\ell(kr)
\ee
Based on this equation, I have derived in \cite{Lacasa2013b} the relation between the observed 2D polyspectrum and the emission 3D polyspectrum, if the latter is diagonal-independent. Indeed, taking the $n$-point correlation function~:
\bea
\nonumber\lbra a_{\ell_1 m_1} \cdots a_{\ell_n m_n}\rbra_c &=& i^{\ell_1+\cdots+\ell_n} \int \dd r_{1\cdots n} \frac{\dd^3\kk_{1\cdots n}}{2\pi^2} \left[Y^*_i(\hat{k}_i) \, j_{\ell_i}(k_i r_i)\right]_{i=1\cdots n} \\
&&\qquad \times \, \mathcal{P}^{(n)}(\kk_{1\cdots n},t_{1\cdots n})\; (2\pi)^3 \, \delta^{(3)}(\kk_1+\cdots+\kk_n)
\eea
Involving the cross-bispectrum between different epochs $t_1 \cdots t_n$ (with $t_i=t(r_i)$). If the polyspectrum is diagonal-independent we have~:
\bea
\nonumber\lbra a_{\ell_1 m_1} \cdots a_{\ell_n m_n}\rbra_c &=& i^{\ell_1+\cdots+\ell_n} \int \left[\dd r_i \frac{k^2_i \dd k_i}{2\pi^2} \, j_{\ell_i}(k_i r_i) \right]_{i=1\cdots n} \mathcal{P}^{(n)}(k_{1\cdots n},t_{1\cdots n})\\
&&\times \underbrace{\int \dd^2\hat{k}_{1\cdots n} \left[Y^*_i(\hat{k}_i)\right]_{i=1\cdots n} (2\pi)^3 \, \delta^{(3)}(\kk_1+\cdots+\kk_n)}_{\equiv A^{(n)}(\ell_{1 \cdots n}, m_{1 \cdots n}, \kk_{1 \cdots n})}
\eea
where, after calculations detailed in \cite{Lacasa2013b}, I find~:
\be
A^{(n)}(\ell_{1 \cdots n}, m_{1 \cdots n}, \kk_{1 \cdots n}) = (4\pi)^n \, i^{\ell_1+\cdots+\ell_n} \int x^2 \dd x \, \left[ j_{\ell_i}(k_i x) \right]_{i=1\cdots n} \times \underbrace{\int \dd^2 \nn \left[ Y^*_{\ell_i m_i}(\nn) \right]_{i=1\cdots n}}_{\equiv \,\mathcal{G}(\mathbf{1}\cdots \mathbf{n})}
\ee
Hence, we find $\lbra a_{1\cdots n}\rbra_c \propto \mathcal{G}(\mathbf{1}\cdots \mathbf{n})$, which means that the 2D polyspectrum is also diagonal independent and we have~:
\be
\mathcal{P}^{(n)}(\ell_{1\cdots n}) = (-1)^{\ell_1+\cdots+\ell_n} \left(\frac{2}{\pi}\right)^n \int x^2\dd x\left[\dd r_i \, k^2_i \dd k_i \, j_{\ell_i}(k_i r_i)\, j_{\ell_i}(k_i x)\right]_{i=1\cdots n} \mathcal{P}^{(n)}(k_{1\cdots n},t_{1\cdots n})
\ee
This expression can be further simplified using the Limber approximation. Indeed, if we assume that, as a function of $k_i$, the 3D polyspectrum varies slowly compared to the quickly oscillating Bessel functions, it can be factorised out of the $k_i$ integral. Then, we can use the orthogonality relation of spherical Bessel functions, so that~:
\bean
\!\!\int\! \left[k^2_i \dd k_i \, j_{\ell_i}(k_i r_i)\, j_{\ell_i}(k_i x)\right]_{i} \mathcal{P}^{(n)}(k_{1\cdots n},t_{1\cdots n}) \!\!&\simeq&\!\! \mathcal{P}^{(n)}(k^*_{1\cdots n},t_{1\cdots n}) \int \left[k^2_i \dd k_i \, j_{\ell_i}(k_i r_i)\, j_{\ell_i}(k_i x)\right]_{i}\\
&\simeq& \mathcal{P}^{(n)}(k^*_{1\cdots n},t_{1\cdots n}) \left[\frac{\pi}{2x^2}\delta(r_i-x)\right]_{i}
\eean
with the 3D polyspectrum evaluated at the peak of the Bessel function : $k^*_i = \frac{\ell_i+1/2}{r_i}$.\\
Hence, the $r_i$ integrals are computed trivially and the 2D polyspectrum simplifies greatly~:
\be
\mathcal{P}^{(n)}(\ell_{1\cdots n}) = (-1)^{\ell_1+\cdots+\ell_n} \int \frac{x^2\dd x}{x^{2n}} \; \mathcal{P}^{(n)}(k^*_{1\cdots n},t(x))
\ee
which only involves the polyspectrum at a single epoch $t(x)$. Because of parity-invariance $\ell_1+\cdots+\ell_n$ is even so that $(-1)^{\ell_1+\cdots+\ell_n}=1$.\\
 Thus, the Limber approximation yields that the observed polyspectrum is an integral of the 3D polyspectrum over spherical shells. The weight $x^{-2n}$ of these shells increases strongly as the radius decreases.
% \newline
 
 {\bf Remarks} : Let us note that the Limber approximation is exact if the polyspectrum is constant, e.g., for a white noise.\\
For non-constant polyspectra, the Limber approximation may be expected to fail at low multipoles and most importantly at small radii x, as the spherical Bessel function are varying more slowly in these cases. The limit $x\rightarrow 0$ is particularly problematic as the weight $x^{-2n}$ blows up while the 3D polyspectrum is supposed to go to zero as $k_i\rightarrow\infty$. Hence, this limit is prone to numerical instabilities, in particular at high orders $n$.
\newline
\newline

%Conclusion of the chapter
In this chapter, I have shown that cosmological analysis, either theoretical or empirical, is often more naturally set in harmonic space. Indeed, the correlation function introduced in the first chapter have a simpler form in harmonic space where they are called polyspectra. I have shown how these polyspectra may be estimated at second and third order, dealing with the problem of partial sky coverage, and how error bars on these estimations may be computed. I have thus described a major part of the framework necessary to characterise statistically celestial observations. This characterisation would then need to be compared to theoretical predictions. In a first step towards this goal, I have described in this chapter the projection of 3D polyspectra onto the sphere. Indeed, theoretical models are most naturally set in 3D space, as will be described in the two following chapters.

\mychapter{Some cosmology}{Some cosmology}{chapt:cosmo}

%\lettrine{T}{}
This chapter introduces the physical background motivating the application to cosmology of the statistical tools introduced in the previous chapters. I first describe the current paradigm for the generation of the primeval cosmological perturbations and why it may yield non-Gaussian signatures. I then introduce the main observables of present cosmology, which stem from these primeval perturbations : the Cosmic Microwave Background and the Large Scale Structure. For both of them, I sketch what non-Gaussianity studies may bring to their understanding and to the understanding of the primordial perturbation generation.

%%%%%%%%%%%%%%%%%%%%%%%%%%%%%%%%%%%%%%%%%%%%%%%%%%%%%%%%%
\section{Friedman-Lema"tre space-time}

Observations of galaxy surveys and of the Cosmic Microwave Background show that the universe is isotropic (at least on very large scales). Combined with the Copernician principle stating that we do not live in a particular place in the universe, this leads to the conclusion that the universe is homogeneous \citep{Peebles1972,UzanPeter}.

These statements are meant to be statistical, that is \textit{on average} the universe is homogeneous and isotropic. Of course this is not true in the exact realisation we live in, as e.g. a human being represents an overdensity $\delta \rho/\rho \sim 10^{30}$ compared to the average universe density. This average, through an ergodic hypothesis, can be thought as the mean over a sufficiently large space area. That is, if we smooth the density field over larger and larger windows, the field becomes asymptotically homogeneous.

Although homogeneous, we know since Hubble that the universe is not static. This led Alexander Friedman and independently Georges Lema"tre to write down the most general metric (called FL metric in the following) describing a homogenous space-time~:
\be
\dd s^2 = -c^2 \dd t^2 + a^2(t) \dd \chi^2
\ee
where $a(t)$ is the so-called scale factor which governs the distance the distance between static observers. We normalise it to 1 at present time $a(t_0)=1$. $\dd \chi^2$ is the comoving distance on spatial hypersurfaces. As today's observations constrain the global spatial curvature to be negligible \citep{planck2013-cosmoparams}, we assume a euclidean metric $\dd \chi^2 = \dd x^2+\dd y^2+\dd z^2$.\\
From now on, unless explicitly specified, we work in natural units such that $c=1$.
\newline

The cosmic time is noted $t$, but we may introduce another time coordinate $\eta$ called conformal time, through $\dd \eta =\frac{\dd t}{a(t)}$. Photons travel along null geodesics : $0=\dd s^2 = a^2(\eta) (-\dd \eta^2+\dd\chi^2)$. Hence a photon travels a comoving distance $\chi=\eta$. This leads to the definition of the particle horizon \citep{Rindler1956} as the maximum distance that a photon may have travelled since the Big-Bang (defined at the end of this section)~:
\be
d_p(t) = a(t) \int_0^t \frac{\dd t'}{a(t')} 
\ee

Observationally, the time of emission is most often inferred from the redshift $1+z = a(t_0)/a(t)$ which is an observable.
In the following, a dot denotes derivative with respect to cosmic time (e.g. $\dot{a}$), while the occasional prime denotes derivative with respect to conformal time (e.g. $a'$).
\newline

The statement of homogeneity and isotropy requires that the energy-impulsion tensor describing the matter content of the universe has the form $T^{\mu}_{\phantom{\mu} \nu} = \mr{diag}(-\rho,P,P,P)$, where $\rho$ is the energy density and $P$ the pressure of the matter content. The Einstein equations of General Relativity (GR) then govern the dynamics through two coupled differential equations called Friedman equations~:
\bea
\label{Eq:FriedmannHubble} H^2 = \frac{8\pi G}{3} \rho + \frac{\Lambda}{3}\\
\label{Eq:FriedmannConservation} \dot{\rho} + 3 H (\rho+P) = 0
\eea
with $H=\dot{a}/a$ being the Hubble parameter, $G$ the gravitational constant and $\Lambda$ the cosmological constant.\\
The differential system is closed by specifying the equation of state linking $P$ and $\rho$. The matter content is typically composed of different fluids with equations of state $P = w \rho$, with $w=0$ for baryonic and dark matter, $1/3$ for radiation, and the cosmological constant can be considered as a fluid with $w=-1$.
If we index the different species by a label $i$, the first Friedman equation can be rewritten as an energy budget~:
\be
\sum_i \Omega_i = 1 \quad \mr{where} \quad \Omega_i = \frac{\rho_i}{\rho_c}
\ee
with the critical density $\rho_c(z)= \frac{3 H^2}{8\pi G}$.

Backward extrapolation of the Friedman equations with a standard matter content leads to an initial singularity : $a(t)$ hits zero in a finite cosmic time. This initial time, called the Big-Bang, is an essential singularity as geodesics cannot be continued through it. The current paradigm however is that GR ceases to be valid at early times when the temperature was of the order of the Planck temperature $T_P=\sqrt{\frac{\hbar c^5}{G k_B^2}}$ (with $\hbar$ the reduced Planck's constant and $k_B$ the Boltzmann constant). A quantum gravity theory (steming e.g. from Loop Quantum Gravity, string theory or non-commutative geometry) should in principle resolve the singularity. Thus the Big-Bang is an illegitimate extrapolation, which nevertheless serves as a referential origin.

%%%%%%%%%%%%%%%%%%%%%%%%%%%%%%%%%%%%%%%%%%%%%%%%%%%%%%%%%
\section{Inflation}

\subsection{General paradigm}\label{Sect:inflparadigm}
The Big-Bang model with standard energetic content (matter or radiation) yields a decelerated expansion. On the contrary, inflation is a proposed period of accelerated expansion in the very early universe, typically ending at an energy of the order $10^{16}$ GeV (Grand Unification scale). During this period the geometry is close to de Sitter meaning that the expansion is exponential $a(t) \propto e^{H t}$. The introduction of inflation in the 80s was meant to solve several conceptual problems that arose with the standard Big-Bang model \citep{Guth1981,Hawking1982}~:
\begin{itemize}
\item flatness problem : the spatial curvature is observationally constrained to be negligible ($\Omega_k \ll 1$) today \citep{planck2013-cosmoparams}. For a standard energetic content $\Omega_k$ grows with time, hence backward extrapolation would yield that spatial curvature becomes vanishingly small as time rolls back. This necessitates unnatural initial conditions (fine-tuning problem). Inflation solves this problem by washing out spatial curvature through accelerated expansion, so that $\Omega_k$ becomes arbitrarily small at the end of inflation if that period lasted long enough.
\item causal horizon : for a standard energetic content, the particle horizon at recombination (the time when the Cosmic microwave Background were last scattered, see Sect.\ref{Sect:CMB}) is small compared to the typical distance between two points of the last scattering surface. This means that these points did not have time to communicate, let alone relax to the same thermal equilibrium. Nevertheless, these points have a thermal equilibrium emission with extremely close temperature \citep[to $10^{-5}$ precision, see e.g.][]{Strukov1992,Smoot1992}. This again necessitates fine-tuned initial conditions which are unnatural. Inflation solves this problem, as the accelerated expansion means that these points were in fact much closer in a distant past, hence they had time to connect causally.
\item monopole problem : most extensions of the standard model of particle physics predict the creation in the early universe of magnetic monopoles (an elementary particle with the equivalent of an electric charge but sourcing the magnetic field $\vec{B}$ instead of $\vec{E}$). However no magnetic monopole has been observed to date in the universe. Inflation solves this apparent paradox by completely diluting the monopole density, so that there may be an arbitrarily small number of monopoles in the observable universe.
\end{itemize}

It has been realised rapidly that inflation also yields a process generating the perturbations of inhomogeneity \citep{Mukhanov1981,Linde1982,Guth1982,Starobinsky1982}. These are the very inhomogeneities that would later grow to form the CMB anisotropies, and much later on the large scale structure of the universe. In the next section, I illustrate briefly the implementation of inflation, and the generation of perturbations for the most standard inflationary model.

\subsection{Single field inflation}\label{Sect:singlefield}
The simplest implementation of inflation is mediated by a single scalar field $\varphi(\xx,t)$ minimally coupled to general relativity. Thus, the action takes the form~:
\be\label{Eq:actioninfl}
S=\frac{1}{16\pi G} \int \sqrt{-g} \, \dd^4x \left[R-\frac{1}{2} \partial_\mu \varphi \,\partial^\mu \varphi - V(\varphi)\right]
\ee
where $R$ is the Ricci scalar associated to the metric $g_{\mu \nu}$ and $V$ is the potential of the scalar field.

\subsubsection{Background evolution}
We first assume homogeneity to derive the background equations, so that $\varphi(\xx,t)=\varphi(t)$ and the metric is of the FL form.\\
The least-action principle gives the equations of motion. Namely by varying with respect to the metric $g_{\mu\nu}$ we get the Friedman equations, where the energy density and the pressure of the scalar field are~:
\bea
\rho(\varphi) &=& \frac{1}{2} \dot{\varphi}^2 + V(\varphi)\\
P(\varphi) &=& \frac{1}{2} \dot{\varphi}^2 - V(\varphi)
\eea
By varying the action with respect to $\varphi$, we get the Klein-Gordon equation~:
\be
\ddot{\varphi} + 3 H \dot{\varphi} + \frac{\dd V}{\dd \varphi} = 0
\ee
It is not the subject of this thesis to expose all the details of the inflationnary dynamics. Let us just state that if the potential $V(\varphi)$ is sufficiently flat, the dynamics naturally enters a so-called slow-roll regime where the kinetic energy is small compared to $V(\varphi)$. In this slow-roll regime, the inflaton equation of state is $P \approx - \rho$. Thus, through Friedman equations, the expansion is quasi-exponential and hence satisfies the requirements of Sect.\ref{Sect:inflparadigm}. The deviations to this exponential behaviour (where the Hubble parameter is constant and the potential is flat) are parametrised by the so-called slow-roll parameters~:
\be
\epsilon = -\frac{\dot{H}}{H^2} \quad \mr{and} \quad \delta = \frac{1}{3 H^2} \frac{\dd^2 V}{\dd \varphi^2}
\ee
We can note that these parameters also govern the magnitude of non-linearity in the Friedman and Klein-Gordon equations.

\subsubsection{Perturbation generation and primordial non-Gaussianity}

Inflation also generates the primordial density fluctuations around homogeneity. The basic idea is that quantum fluctuations of the inflaton are stretched out by the fast expansion and are frozen when their wavelength becomes larger than the Hubble length $c/H$. The details are complex and would necessitate a long description \citep[see e.g.][]{UzanPeter}. First the FL metric must be perturbed with gauge-invariant quantities. Then the action Eq.\ref{Eq:actioninfl} must be quantized around the background, and quantum initial conditions must be chosen (the standard choice being the Bunch-Davies vacuum, which is Gaussian). Nevertheless, one can show that the resulting density perturbations $\delta\rho/\rho$ (related to the gravitational potential $\Phi$ through the Poisson equation) have a nearly scale-invariant power spectrum~:
\be
P_\delta(k) \propto k^{n_\mr{S}} \quad \mr{with} \quad n_\mr{S} \approx 1
\ee
%where scale invariance corresponds to $n_\mr{S}=1$ and means that the gravitationnal %potential (related to the density through the Poisson equation $\Delta\Phi = 4\pi %G \rho$) has the same power per logarithmic k bin
More precisely it would be exactly scale invariant ($n_\mr{S}=1$, so that the dimensionless gravitational power spectrum $\frac{k^3 P_\Phi(k)}{2\pi^2} \propto k^{-1} P_\delta(k)$ is constant) if the background evolution was de Sitter, i.e. $a(t)\propto e^{Ht}$. The deviation to scale invariance is linked to this deviation from de Sitter space, i.e. to the slow-roll parameters~:
\be
n_\mr{S} - 1 = 2\delta - 4 \epsilon
\ee
This is a general prediction of inflation models \citep{Mukhanov1981,Hawking1982,Guth1982}, which yield a spectral index close to, but smaller than, unity\,; and indeed Planck has revealed a near scale invariant spectrum, detecting significantly the deviation to scale invariance (at 68\% C.L Planck CMB alone gives $n_\mr{S} = 0.9616 \pm 0.0094$ \citep{planck2013-cosmoparams}).\\
As for the background, the slow roll parameters govern the magnitude of non-linearity in the perturbation evolution equations\,; as the vacuum initial condition is Gaussian, it is not surprising that the resulting density perturbation are Gaussian to leading order and that non-Gaussianity scales proportionally to the slow-roll parameters. Indeed, \cite{Maldacena2003} has shown that the bispectrum amplitude in the squeezed limit ($k_1 \ll k_2\approx k_3$) is~:
\be\label{Eq:bispsinglefield}
B_\Phi(k_1,k_2,k_3) = \frac{5}{3} \frac{(1-n_\mr{S})}{2} \left[P_\Phi(k_1) P_\Phi(k_2) + P_\Phi(k_1) P_\Phi(k_3) + P_\Phi(k_2) P_\Phi(k_3) \right]
\ee
where the factor $\frac{5}{3}$ comes from the relation between the gravitational potential and the curvature perturbation.\\
Later on, \cite{Creminelli2004} have shown in fact that this equation holds in the squeezed limit for a whole class of inflation models, e.g., beyond slow-roll, as long as the inflaton is the only dynamical field (single clock). Hence, detecting primordial non-Gaussianity in the squeezed limit would rule out whole classes of inflationary models \citep[\textit{ibid.;}][]{Komatsu2002}.\\
In a general way, for inflation to generate non-Gaussian primordial perturbations (where the non-Gaussianity may peak in configurations other than squeezed, e.g. in equilateral), we need either the initial conditions to be non-Gaussian (non Bunch-Davies vacuum) or the evolution to be non-linear. The latter possibility is the case for exotic models with e.g., non-canonical kinetic term, multi-field, varying speed of sound etc \citep[see e.g.][and references therein]{Bartolo2004,Renaux-Petel2009,Lewis2011}. This further motivates the search for primordial non-Gaussianity as it may discriminate inflation models. Indeed these models are degenerate at the power spectrum level, as they are all built to reproduce near scale invariance.

%%%%%%%%%%%%%%%%%%%%%%%%%%%%%%%%%%%%%%%%%%%%%%%%%%%%%%%%%
\section{Cosmic Microwave Background anisotropies}\label{Sect:CMB}
The Cosmic Microwave Background (CMB), predicted by \cite{Gamow1948,Dicke1965} and then discovered by \cite{PenziasWilson1965}, is the relic radiation of the primordial plasma. Namely in the early universe (between $z\sim 10^8$ and $10^3$), baryons and photons were tightly coupled into a plasma through Compton scattering of the numerous photons ($\sim 10^9$ per baryon) off free electrons. Due to the high density of free electrons, the photon mean free path was small and the universe was opaque. Baryonic matter and radiation were in local thermal equilibrium and experienced acoustic oscillations through the competition of gravitational infall and radiation pressure, while dark matter was thermally decoupled as it interacts only gravitationnally.

With expansion the universe cooled down, at a redshift $z_\mr{rec}\sim 1000$ and a temperature $T\simeq 3000$ K, the photons were no more energetic enough to prevent the binding of electrons with available nuclei (in particular hydrogen and helium). This event is called recombination and leads to a dramatic decrease of the ionisation fraction so that photons could free stream in the universe (or nearly so, until reionisation took place and scattered some of them). These numerous photons still represent the vast majority of radiation energy in the universe today (see e.g. Fig.\ref{Fig:EBL}), and make up the CMB. The expansion has stretched their wavelength to the microwave band today, and as this stretching is achromatic they have kept a blackbody spectral distribution albeit with a current temperature of 2.725 K \citep{Fixsen2002}. 

The perturbations created by inflation lead to density, velocity and gravitational potential inhomogeneities in the baryon-photon plasma. These tiny inhomogeneities in the plasma at recombination lead to tiny temperature anisotropies (with relative amplitude $\sim 10^{-5}$) in the observed CMB today. There are several competing physical effects leading to the temperature anisotropies, and this can be summarised into an equation linking the initial gravitational potential fluctuations (where initial means end of inflation) to the harmonic coefficients of the observed relative temperature fluctuations~:
\be\label{Eq:Phitoalmcmb}
a_{\ell m}(\eta_0) = \int \frac{\dd^3\kk}{2\pi^2} \; i^\ell \Delta_\ell(k,\eta_i,\eta_0) \, \Phi(\kk,\eta_i) \, Y_{\ell m}(\hat{k})
\ee
where $\eta_0$ is today's conformal time, $\Phi(\kk,\eta_i)$ is the Fourier transform of the gravitational potential at the initial time (after inflation), and $\Delta_\ell(k)$ is the transfer function which encodes all the physics and can be computed by a Boltzmann code, e.g. CAMB \citep{Lewis2002} or CLASS \citep{Blas2011}.\\
For example, on large scales the Sachs-Wolfe effect \citep{SachsWolfe1967} is dominant. In the instantaneous recombination approximation it yields $\frac{\Delta T}{T}(\hn) = -\frac{\Phi(r_\mr{rec} \, \hn)}{3}$, and the transfer function takes the form~:
\be
\Delta_\ell(k) \simeq -\frac{1}{3} \, j_\ell (k r_\mr{rec})
\ee
up to growth factors (explained in Sect.\ref{Sect:lineargrowth}) giving the evolution of potentials between $\eta_i$ and $\eta_\mr{rec}$ for frozen super-Hubble modes.\\
From Eq.\ref{Eq:Phitoalmcmb}, we can then derive the CMB power spectrum~:
\be
C_\ell(\eta_0) = \frac{2}{\pi} \int k^2 \dd k \, \Delta^2_\ell(k,\eta_i,\eta_0) \, P_\Phi(k)
\ee

\begin{figure}[htbp]
\centering
\includegraphics[width=\linewidth]{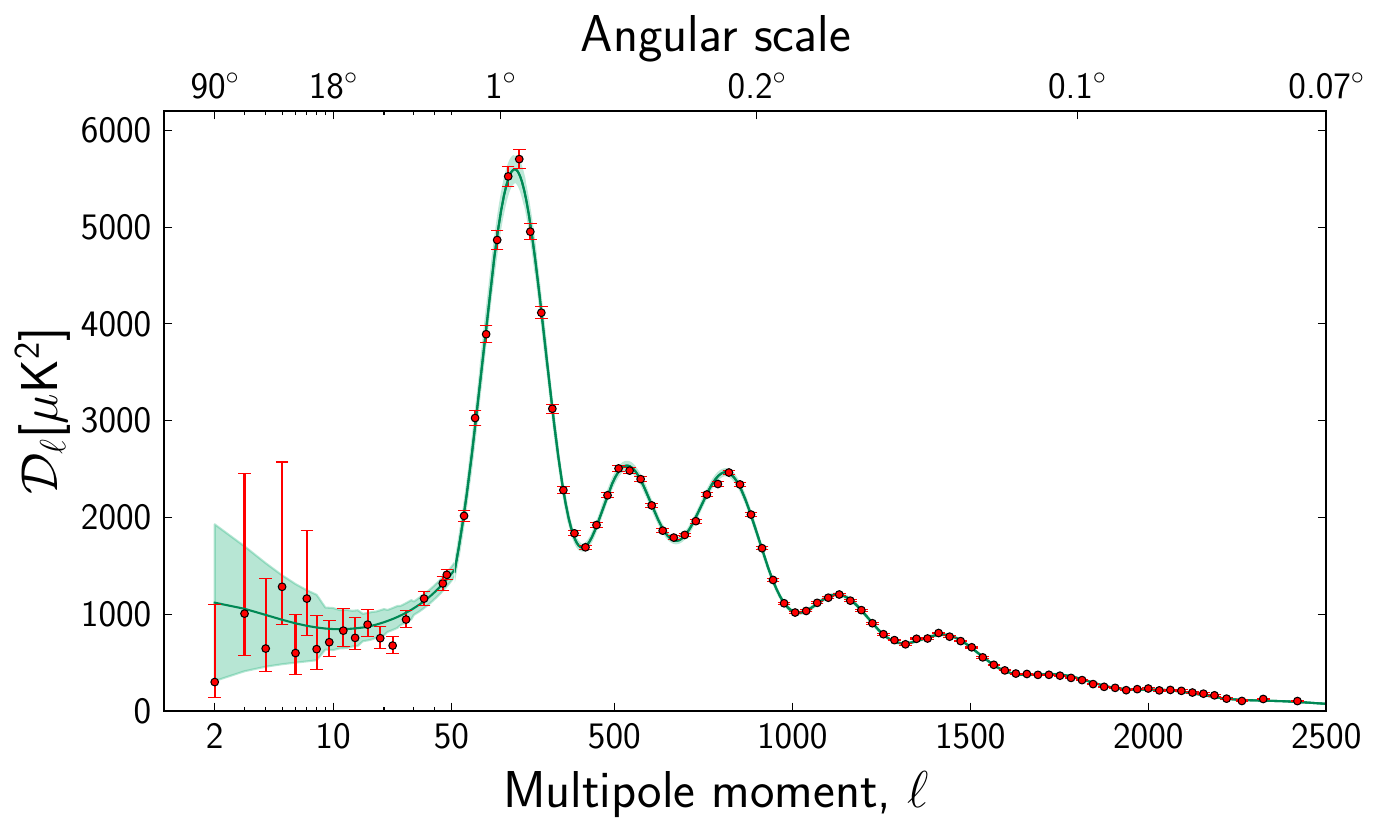}
\caption{ {\planck} CMB power spectrum (red dots, binned) along with the best fit cosmological model and its cosmic variance \citep{Planck2013-powerspectra}. Note the log-linear x-axis, and the y-axis is $\mathcal{D}_\ell= \frac{\ell(\ell+1)}{2\pi} C_\ell$.}
\label{Fig:Planck_cmb_Cl}
\end{figure}

Figure \ref{Fig:Planck_cmb_Cl} shows the CMB power spectrum measured by {\planck} \citep{Planck2013-powerspectra}. Specifically they plotted $\mathcal{D}_\ell= \frac{\ell(\ell+1)}{2\pi} C_\ell\,$, a now standard convention. Indeed for scale invariant fluctuations the Sachs-Wolfe effect produces $C_\ell \propto \frac{1}{\ell(\ell+1)}$ at low multipoles, and this convention reduces to the power per logarithmic multipoles bin $\frac{\ell(2\ell+1)}{4\pi} C_\ell$ when $\ell\gg 1$.

The power spectrum exhibits distinct characteristics which are the imprints of the primordial physics~:
\begin{itemize}
\item the Sachs-Wolfe plateau at low multipoles previously explained.
\item so-called acoustic peaks, which correspond to a dip in the real-space 2-point c.f. at an angle $\theta_s \simeq 1.2 \degree$. $\theta_s$ is the angle subtended by the sound horizon at last scattering, i.e. the distance sound waves traveled in the photon-baryon plasma between the generation of density perturbations and recombination.
\item so-called Silk damping at high multipoles \citep{Silk1968}, as the diffusion of photons during the period of recombination erases their inhomogeneity.
\end{itemize}

The transfer function $\Delta_\ell(k)$ encodes all these effects\,; as such it is highly sensitive to the primordial physics, hence to the cosmological parameters. The CMB is thus a powerful probe of cosmological parameters, and has been observed by several experiments in this prospect \citep[e.g.][]{Sievers2003,MacTavish2006,Komatsu2011}\,; the {\planck} results have put the most stringent constraints to date on them \citep{planck2013-cosmoparams}.
\newline

Higher order CMB polyspectra may also be interesting, to hunt for primordial non-Gaussianity. For example, from Eq.\ref{Eq:Phitoalmcmb} and assuming a primordial non-Gaussianity shape, one can derive the CMB bispectrum. For data analysis, the most popular non-Gaussianity shape is the local form, parametrised by a non-linearity parameter $\fnl$~:
\be\label{Eq:def-fnl-Phi}
\Phi(\xx) = \Phi_G(\xx) + \fnl \left(\Phi_G(\xx)^2 - \lbra \Phi_G(\xx)^2\rbra\right)
\ee
where $\Phi_G$ is a Gaussian field. To the first order in $\fnl$, this creates a gravitational bispectrum of the form~:
\be\label{Eq:BPhi-fnl}
B_\Phi(k_1,k_2,k_3) = \fnl \left[P_\Phi(k_1) P_\Phi(k_2) + \mr{5 \ perm.} \right]
\ee
i.e. the same form as Eq.\ref{Eq:bispsinglefield}\footnote{Although Eq.\ref{Eq:bispsinglefield} was only valid in the squeezed limit, the bispectrum peaks in these configurations so that this form captures the dominant behaviour.} so that for standard inflation $\fnl =\frac{5}{12} (1-n_\mr{S})$ (greater value of $\fnl$ may be generated e.g. by multifield models \citep{Bartolo2004}).\\
With calculations similar to that of Sect.\ref{Sect:pastlightproj}, one can show that the CMB bispectrum takes the form~:
\be\label{Eq:localCMBbisp}
b_{\ell_{123}} = \fnl \int r^2 \dd r \, \alpha_{\lu}(r) \, \beta_{\ld}(r) \, \beta_{\lt}(r) + \mr{5 \ perm.}
\ee
with~:
\bea
\alpha_{\ell}(r) &=& \frac{2}{\pi} \int k^2 \dd k \, \Delta_\ell(k) j_\ell(k r) \\
\beta_{\ell}(r) &=& \frac{2}{\pi} \int k^2 \dd k \, P_\Phi(k) \Delta_\ell(k) j_\ell(k r)
\eea
$r$ can be thought of as the comoving distance on the past light cone (although in the derivation it is just a bound variable), and indeed the filters $\alpha_\ell$ and $\beta_\ell$ peak at recombination $r\simeq r_\mr{rec}$. $\frac{\beta_\ell(r)}{C_\ell}$ is the optimal linear filter reconstructing the gravitational potential at comoving distance $r$ from the observed CMB $a_{\ell m}$ \citep{KSW2005}.\\
Other forms of primordial non-Gaussianity, as long as they produce a gravitational bispectrum of the form $B(k_{123}) = \sum_i f_i(k_1) g_i(k_2) h_i(k_3) + \mr{perm.}$, yield a CMB bispectrum with a functional form similar to Eq.\ref{Eq:localCMBbisp} albeit with different filters. For example, two other popular forms of non-Gaussianity are the so-called equilateral and orthogonal ones, which respectively arise for models with non-canonical kinetic terms or with non-standard vacuum initial conditions \citep[see e.g.,][]{Bartolo2004,Renaux-Petel2009,Lewis2011}. The CMB bispectrum that they yield involves two new filters traditionally noted $\gamma_\ell(r)$ and $\delta_\ell(r)\,$, see e.g. \cite{Curto2011} for details.\\
Non-Gaussianity in the observed CMB is also generated by non-linear evolution \citep{Pitrou2010,Huang2012} and by the secondary anisotropies, that is, anisotropies due to the distortion of the primordial CMB by the large scale structure along the line of sight. These secondary anisotropies will be reviewed in Sect.\ref{Sect:fgintro}. In particular, the iSW-lensing-lensing cross-bispectrum (shortened to ISW-lensing hereafter) has been shown to be important to consider for CMB NG studies \citep[e.g.][]{Smith2011,Hanson2009}, and has indeed been measured by \planck as will be explained in Sect.\ref{Sect:PNG_conta}.

\begin{figure}[htbp]
\centering
\includegraphics[width=.8\linewidth]{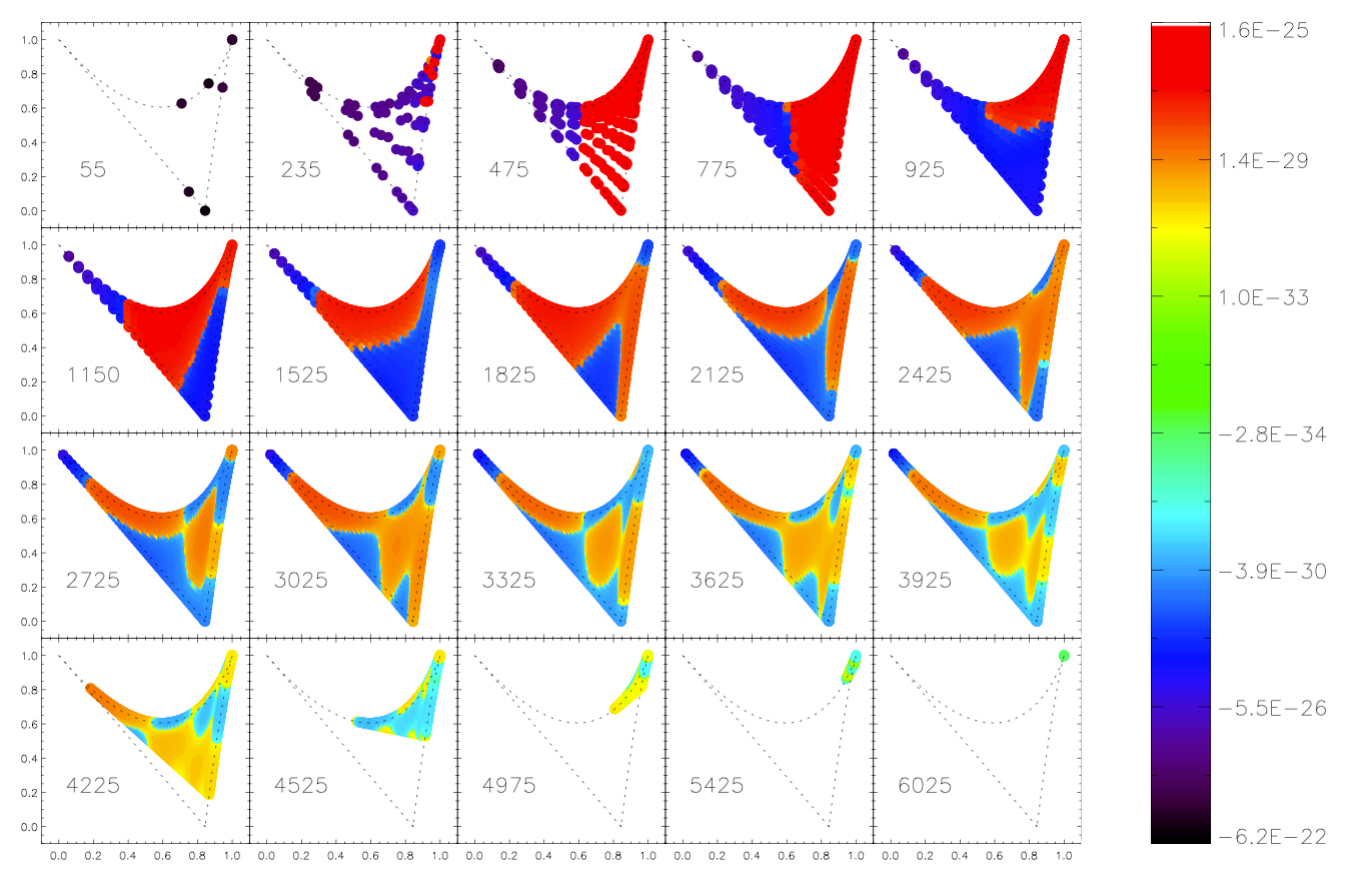}
\caption{Theoretical CMB bispectrum for local type primordial non-Gaussianity and $\fnl=1$. The perimeter value $\lu+\ld+\lt$ is indicated in each subplot. The color code is highly non-linear. The bispectrum peaks in the squeezed configurations with negative values (black-violet colors). Note that the maximum resolution condition $\ell \leq \ell_\mr{max}$ restrains the possible triangle shape at high perimeters, e.g. only near equilateral triangles are present in the last bin $P\sim 3\ell_\mr{max}$}
\label{Fig:local_cmb_bisp_inparam}
\end{figure}

Figure \ref{Fig:local_cmb_bisp_inparam} shows the theoretical CMB bispectrum for local type primordial non-Gaussianity and $\fnl=1$, plotted in the parametrisation proposed in \cite{Lacasa2012} and detailed in Sect.\ref{Sect:bisp_param}). The transfer function was computed using CAMB \citep{Lewis2002}.

Note that, for visualisation purpose, the color code is highly non-linear to highlight all features of the bispectrum. Each subplot corresponds to a perimeter bin\,; in that subplot the upper left corner corresponds to squeezed triangles, upper right to equilateral and lower right to folded triangles. The perimeter value, $\lu+\ld+\lt\,$,is indicated in the lower left corner of each subplot.\\
The bispectrum is dominated by the squeezed configurations where it is negative. Indeed one of the multipoles is small so that the transfer function is negative (Sachs-Wolfe regime $\frac{\Delta T}{T} = - \frac{\Phi}{3}$) while the two other multipoles are near equal so that their transfer function has same sign. The complex pattern which develops at increasing multipoles is the consequence of the acoustic peaks which change the sign of the transfer function. These sign changes could not be seen in the power spectrum as it is intrinsically positive. They occur at multipoles corresponding to local minima of $C_\ell$ and can be seen easily in the equilateral configurations of the bispectrum. In absolute value, the bispectrum is of order $\mathcal{O}(10^{-20})$ in the squeezed triangles, as can be expected from Eq.\ref{Eq:def-fnl-Phi} or \ref{Eq:BPhi-fnl}, given that the perturbations in $\Phi$ are of order $\mathcal{O}(10^{-5})$.

%%%%%%%%%%%%%%%%%%%%%%%%%%%%%%%%%%%%%%%%%%%%%%%%%%%%%%%%%
\section{Large Scale Structure}
The perturbations created by inflation and visible in the CMB grow through gravitational infall to form the structure of the universe that we see today \citep{UzanPeter,Bernardeau2007}. This section describes the growth of such structures and how to model the large scale structure of the universe.

\subsection{Linear growth}\label{Sect:lineargrowth}

A newtonian approach is valid on sub-Hubble scales ($\lambda < c/H$) where general relativity effects are negligible (see e.g. \cite{Weinberg1972} for a much more complete discussion). If a single pressureless fluid dominates the inhomogeneities, e.g., in matter- or $\Lambda$-dominated era\,; the Poisson equation, mass conservation and Euler equation take the form~:

%\begin{align}
%\Delta_\mr{phys} \Phi &= 4\pi G\; \delta\rho \\
%\left.\frac{\partial \rho}{\partial t}\right|_\mr{phys} &+ \mr{div}_\mr{phys}(\rho \, \vec{v}) = 0 \\
%\rho \left.\frac{\partial \vec{v}}{\partial t}\right|_\mr{phys} &+ \rho(\vec{v}\cdot\vec{\nabla}_\mr{phys})\,
%\vec{v} = -\rho \vec{\nabla}_\mr{phys} \Phi
%\end{align}
\begin{align}
\label{Eq:poisson}\Delta_\mr{p} \Phi &= 4\pi G\, \rho \\
\label{Eq:continuity}\left.\frac{\partial \rho}{\partial t}\right|_\mr{p} &+ \mr{div}_\mr{p}(\rho \, \vec{v}) = 0 \\
\label{Eq:Euler}\rho \left.\frac{\partial \vec{v}}{\partial t}\right|_\mr{p} &+ \rho(\vec{v}\cdot\vec{\nabla}_\mr{p})\,\vec{v} = -\rho \vec{\nabla}_\mr{p} \Phi
\end{align}
where $\vec{v}$ is the velocity field, and the subscript p denotes that derivatives are taken with respect to physical positions $\mathbf{x} = a(t) \mathbf{r}$, where $\mathbf{r}$ is the comoving position. The physical and comoving time-derivative are also different, as the time-derivative is meant to be taken at constant spatial coordinates. The conversion between physical and comoving derivatives, the latter being noted with a subscript c, are~:
\bea
\vec{\nabla}_\mr{p} &=& \frac{1}{a}\vec{\nabla}_\mr{c} \\
\left.\frac{\partial }{\partial t}\right|_\mr{p} &=& \left.\frac{\partial }{\partial t}\right|_\mr{c} - H \,\mathbf{r}\cdot \vec{\nabla}_\mr{c}
\eea

We will note $\delta_\mr{m} = \frac{\rho-\bar{\rho}}{\bar{\rho}}$ the relative perturbation of matter density, comprising baryonic matter and dark matter, and $\delta\vec{v}$ the velocity perturbation. If we linearise Eq.\ref{Eq:poisson}-\ref{Eq:continuity}-\ref{Eq:Euler} around the background solution\footnote{Except that the background variables are not solutions of the system, as they do not follow Newtonian dynamics but GR. Nevertheless, the Newtonian approach to perturbations produces the same perturbations equations as GR.} $(\rho=\bar{\rho}(t),\vec{v}=H\mathbf{x},\Phi=0)$, we get in comoving coordinates~:
\begin{align}
\label{Eq:poissonlin}\Delta_\mr{c} \Phi &= 4\pi G\, a^2 \bar{\rho} \, \delta_\mr{m} \\
\label{Eq:continuitylin}\left.\frac{\partial \delta_\mr{m}}{\partial t}\right|_\mr{c} &+ \frac{1}{a}\vec{\nabla}_\mr{c} \cdot \delta\vec{v} = 0\\
\label{Eq:Eulerlin}\left.\frac{\partial \delta\vec{v}}{\partial t}\right|_\mr{c} &+ H \, \delta\vec{v} = -\frac{1}{a}\vec{\nabla}_c\Phi
\end{align}
%Unlinearised equations :
%\begin{align}
%\Delta_\mr{c} \Phi &= 4\pi G\, a^2 \rho \\
%\left.\frac{\partial \rho}{\partial t}\right|_\mr{c} &+ 3H\rho+\frac{1}{a}\vec{\nabla}_\mr{c} \!\cdot\! \left(\rho\,\delta\vec{v}\right) = 0\\
%\left.\frac{\partial \delta\vec{v}}{\partial t}\right|_\mr{c} &+ H \, \delta\vec{v} + \frac{1}{a}\left(\delta\vec{v}\cdot\vec{\nabla}_c\right)\delta\vec{v} = -\frac{1}{a}\vec{\nabla}_c\Phi
%\end{align}
These equations can be combined to obtain a second-order differential equation on $\delta_\mr{m}$ only~:
\be\label{Eq:diffeq4lingrowth}
\ddot{\delta}_m + 2 H \,\dot{\delta}_\mr{m} = 4\pi G\, \bar{\rho} \, \delta_\mr{m}
\ee
where, for simplicity, the comoving time-derivative is noted with an overdot. As there are only time derivatives involved, this equation has separable solutions, i.e. of the form $\delta_\mr{m}(\xx,t) = D(t) f(\xx)$. Moreover, the differential equation is of second-order, so there are two independent solutions, with amplitude fixed by the initial conditions : 
\be
\delta_\mr{m}(\xx,t) = D^{+}(t)\,\delta^{+}_m(\xx,t_i)+D^{-}(t)\,\delta^{-}_m(\xx,t_i)
\ee
 Let us note that the right-hand side of  Eq.\ref{Eq:diffeq4lingrowth} can also be written $\frac{3}{2} H^2 \delta_\mr{m} $.\\
 Using the Friedmann equations \ref{Eq:FriedmannHubble} \& \ref{Eq:FriedmannConservation}, we see that the Hubble parameter $H$ is solution of Eq.\ref{Eq:diffeq4lingrowth}, hence $D^{-}(t)=H(t)/H(t_i)$. The other mode can be searched for in the form $D^{+}(t) = H(t) g(t)$, which leads to a first order differential equation on $\dot{g}$. Hence we find~:
\be
D^{+}(t) \propto H(t) \int_{t_i}^t \frac{\dd t}{\left[a(t) H(t)\right]^2} = H\int_{a_i}^a\frac{\dd a}{\left(a H\right)^3}
\ee
During matter domination $D^+ \propto a \propto t^{2/3}$ and $D^- \propto t^{-1}$, so that the corresponding solutions are respectively called the growing mode and the decaying mode.

As long as the matter perturbations are small,  they grow linearly without mixing Fourier modes. No extra non-Gaussianity is generated and the correlation functions of the density field are kept unchanged\,; this stands as well for the velocity and gravitational potential fields.\\
However, perturbations eventually grow out of the linear regime, and the non-linear dynamic is not solvable analytically. As the non-linear terms in Eq.\ref{Eq:poisson}-\ref{Eq:continuity}-\ref{Eq:Euler} all involve spatial derivatives, the non-linearity affects predominantly small scales. In the Fourier range, large scales are hence still in the linear regime while small scales are non-linear. As time goes by, the non-linear range extends to larger and larger scales.\\
The non-linearity can be tackled perturbatively, e.g. with Eulerian or Lagrangian perturbation theory \citep[see][for a review]{Bernardeau2002}. The linear approach described previously then gives the first order solution, and the system is solved iteratively order by order. Higher orders of computation, resummation of loops contributions etc, enable the domain of validity to extend to smaller scales. But perturbation theory eventually fails in the fully non-linear regime \citep{Buchert1993}. The purpose of the next two sections is to describe the current approach used to tackle the distribution of matter both at large and small scales, and in particular the distribution of galaxies.

\subsection{Local bias scheme}\label{Sect:localbias}
\begin{figure}[htbp]
\centering
\includegraphics[width=.5\linewidth]{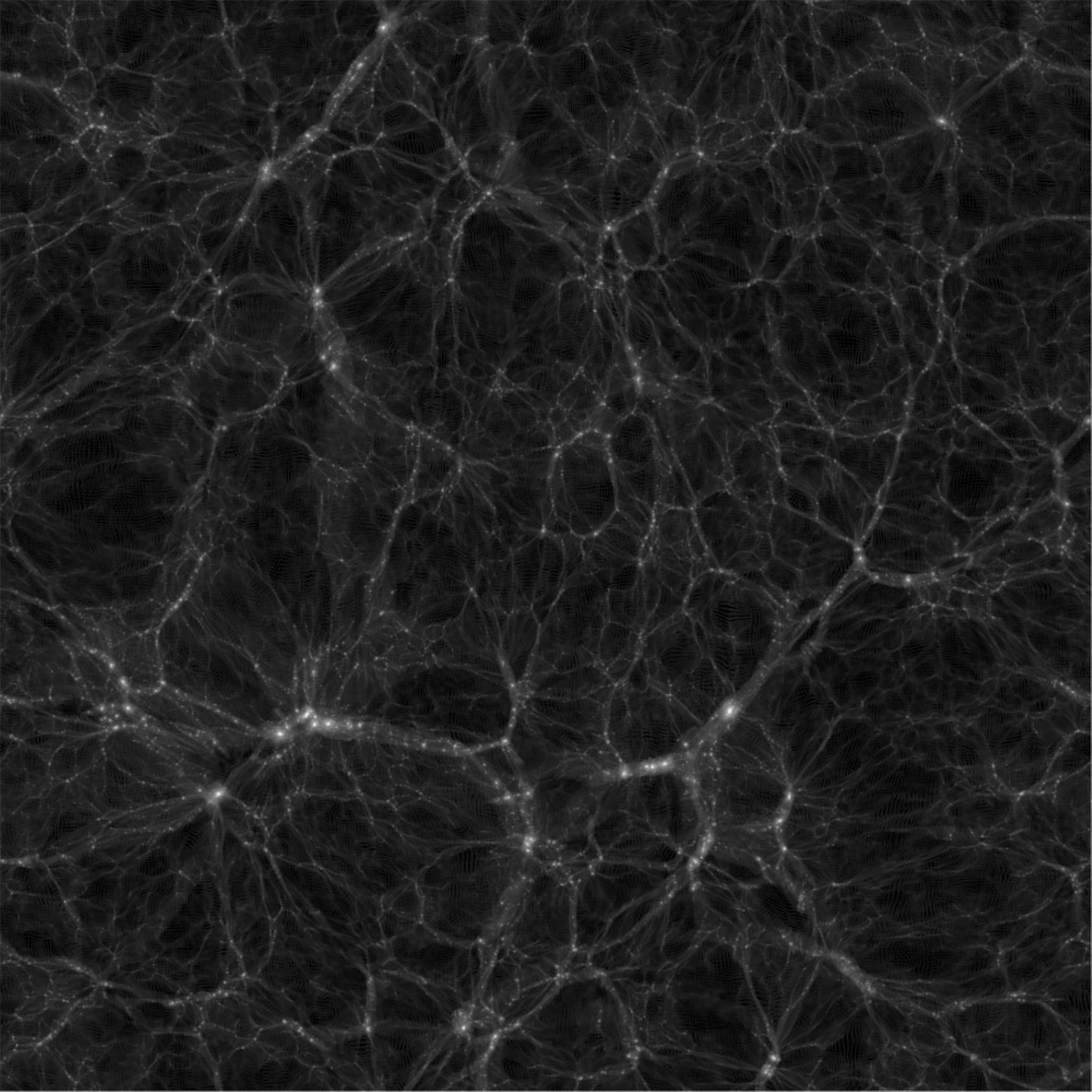}
\caption{Slice at $z$=0 of a dark matter simulation runned by the DEUS consortium \citep{Rasera2010}, showing the distribution of dark matter in filaments, voids and haloes.}
\label{Fig:DEUS}
\end{figure}

\begin{figure}[htbp]
\centering
\includegraphics[width=.5\linewidth]{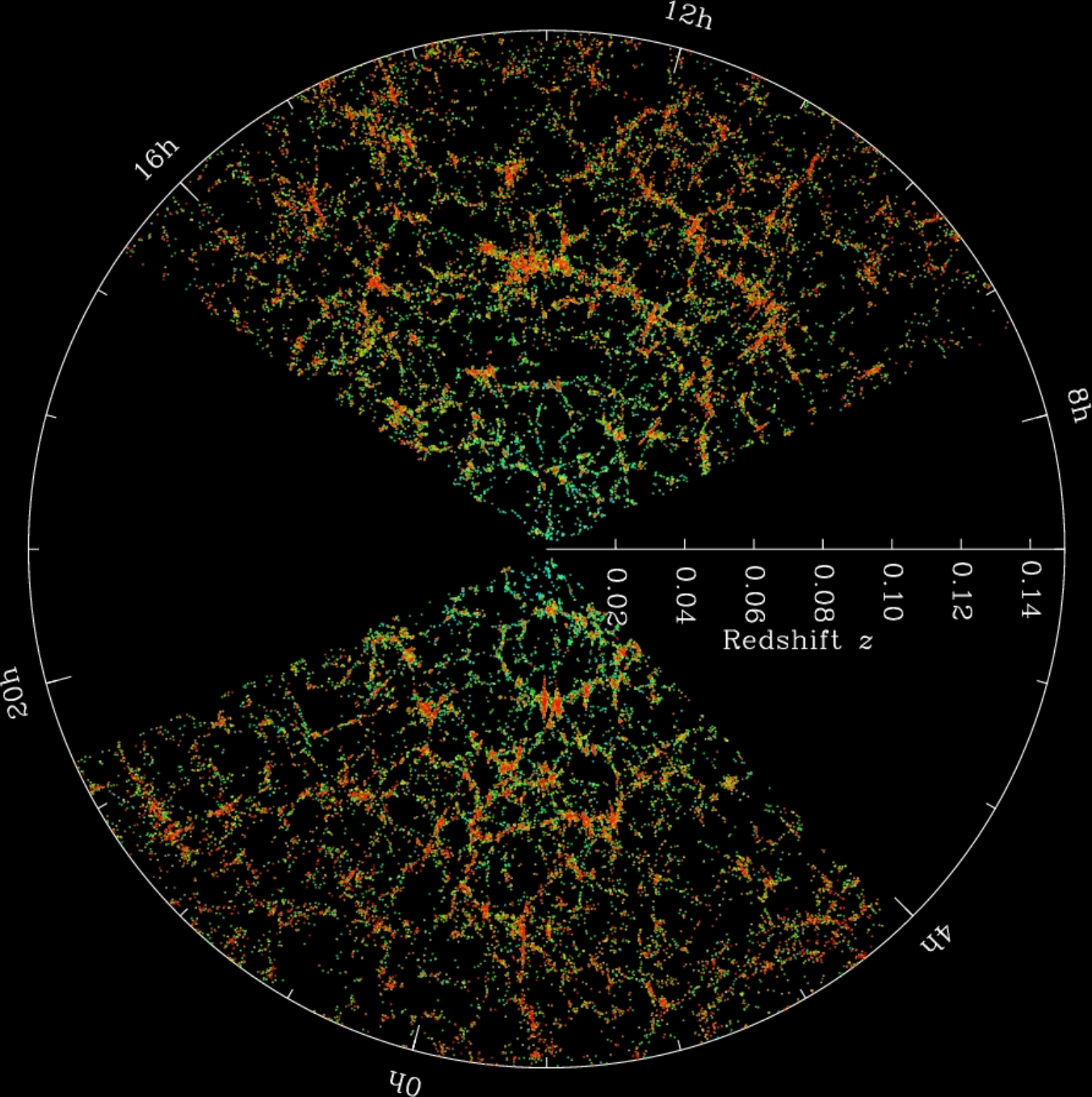}
\caption{Galaxy distribution observed by the Sloan Digital Sky Survey \citep{Tegmark2004}.}
\label{Fig:sdss_pie}
\end{figure}

At late times, the large scale structure of the universe forms a complex web : at large scales filaments of dark matter are separated by large void areas, haloes form in these filaments and host clusters of galaxies, the largest ones forming at the filaments intersections. For illustration, Fig.\ref{Fig:DEUS} shows a slice of a N-body dark matter simulation runned by the consortium DEUS \citep{Rasera2010}, and Fig.\ref{Fig:sdss_pie} shows the galaxy distribution observed by the Sloan Digital Sky Survey \citep{Tegmark2004}.

The study of the galaxy distribution in the universe is a long-standing field in cosmology \citep[e.g.][]{Peebles1972}. Their clustering has been particularly studied from the point of view of the real-space 2-point correlation function (\textit{ibid.}). Early observations showed that the galaxy correlation function was on large scale a biased version of the matter correlation function~: $\zeta_g(r) = b^2 \zeta_m(r)$ \footnote{This form is also expected for a thresholded Gaussian random field \citep{Kaiser1984}.}. This lead \cite{Fry1993} to propose that galaxies are a tracer of matter through a local non-linear transformation : $\delta_\mr{g}(\xx) = F(\delta_\mr{m}(\xx))$. This transformation may be Taylor-expanded to yield the so-called local biasing scheme :
\be\label{Eq:localbiasgal}
\delta_\mr{g}(\xx) = \sum_{k=0}^{+\infty} \frac{b_k}{k!} \delta^k_\mr{m}(\xx)
\ee
where $b_0=-\sum_{k=2}^{+\infty}\frac{b_k}{k!} \lbra \delta_\mr{m}^k\rbra$ so that $\lbra \delta_\mr{g}\rbra = 0$.
$b_i$ are the bias parameters, the most studied being $b_1$ also known as the linear bias. As the relation Eq.\ref{Eq:localbiasgal} is local, the bias parameters can be inferred from data, simulations or theoretical predictions solely from the 1-point function, i.e., the abundances.\\
This functional form holds on large scales for any selected objects, e.g., red galaxies, groups etc, and allows to describe their large scale distribution. However on smaller scales, the main driver of galaxy clustering is whether or not they live in the same halo. As haloes are the locus of galaxy formation, they are the key ingredient to model the galaxy distribution. The number and properties of the hosted galaxies are a function of the physical parameters of the halo, and in particular of the halo mass. It has been demonstrated by \cite{Manera2011} that, for haloes above a certain mass threshold, the halo distribution was well fitted by the local bias scheme with bias parameters derived from the halo abundances (also known as the halo mass function, described in the next section).\\
Thus the density field of haloes $\delta_\mr{h}$ at a given mass $M$ follows~:
\be
\delta_\mr{h}(\xx | M) = \sum_{k=0}^{+\infty} \frac{b_k(M)}{k!} \: \delta_\mathrm{DM}(\xx)^n
\ee
where $\delta_\mathrm{DM}$ is the dark  matter density field, the statistics of which are predicted through perturbation theory.\\
Note that this description breaks down on small scale when non-linearities become important.

\subsection{Halo model}\label{Sect:HM_descrip}
\cite{PressSchechter1974} pioneered the halo description of the Large scale Structure with a spherical collapse model. They gave an analytical formula for the abundance of haloes per mass bin and per comoving volume at a given redshift $\frac{\dd n_\mr{h}}{\dd M \dd V_\mr{c}}$, where they found the functional form~:
\be\label{Eq:PressSchechter}
\frac{\dd n_\mr{h}}{\dd M \dd V_\mr{c}}(M,z) \propto M^{-1-\alpha} \exp\left[-\mr{cste} (1+z)^2 M^{2-2\alpha}\right]
\ee
where $\alpha$ is a parameter between $\frac{1}{2}$ and $\frac{1}{3}$ \citep{PressSchechter1974}.\\
This functional form highlights the universal properties of the mass function : a relatively flat slope at low masses and a sharp cut-off at high masses, here with an exponential decay. Hence massive haloes are rare while low-mass haloes are numerous. The cut-off decreases with increasing redshift, so that massive haloes are even rarer at early times. Indeed they did not have enough time to form through gravitational infall.\\
Let us note that the exponential decay in Eq.\ref{Eq:PressSchechter} is the consequence of the assumption of Gaussian initial conditions, as massive haloes correspond to high extrema of the initial density field. Indeed, the exponential is written as $\exp\left[-\frac{\delta_\mr{c}^2(z)}{2\sigma^2(M,z)}\right]$ \footnote{$\delta_\mr{c}$ is the critical density, such that a spherical overdensity with that contrast collapses through gravitational infall instead of expanding with the Hubble flow. $\sigma(M,z)$ is the variance of the density field smoothed over a length R with $M = \frac{4}{3}\pi \bar{\rho} R^3$}. Hence, non-Gaussian initial conditions would impact the mass function\,; massive haloes are particularly sensitive to this possible effect and may be a probe of primordial NG \citep[e.g.,][]{Sefusatti2005}.
\newline

\begin{figure}[htbp]
\centering
\includegraphics[width=.8\linewidth]{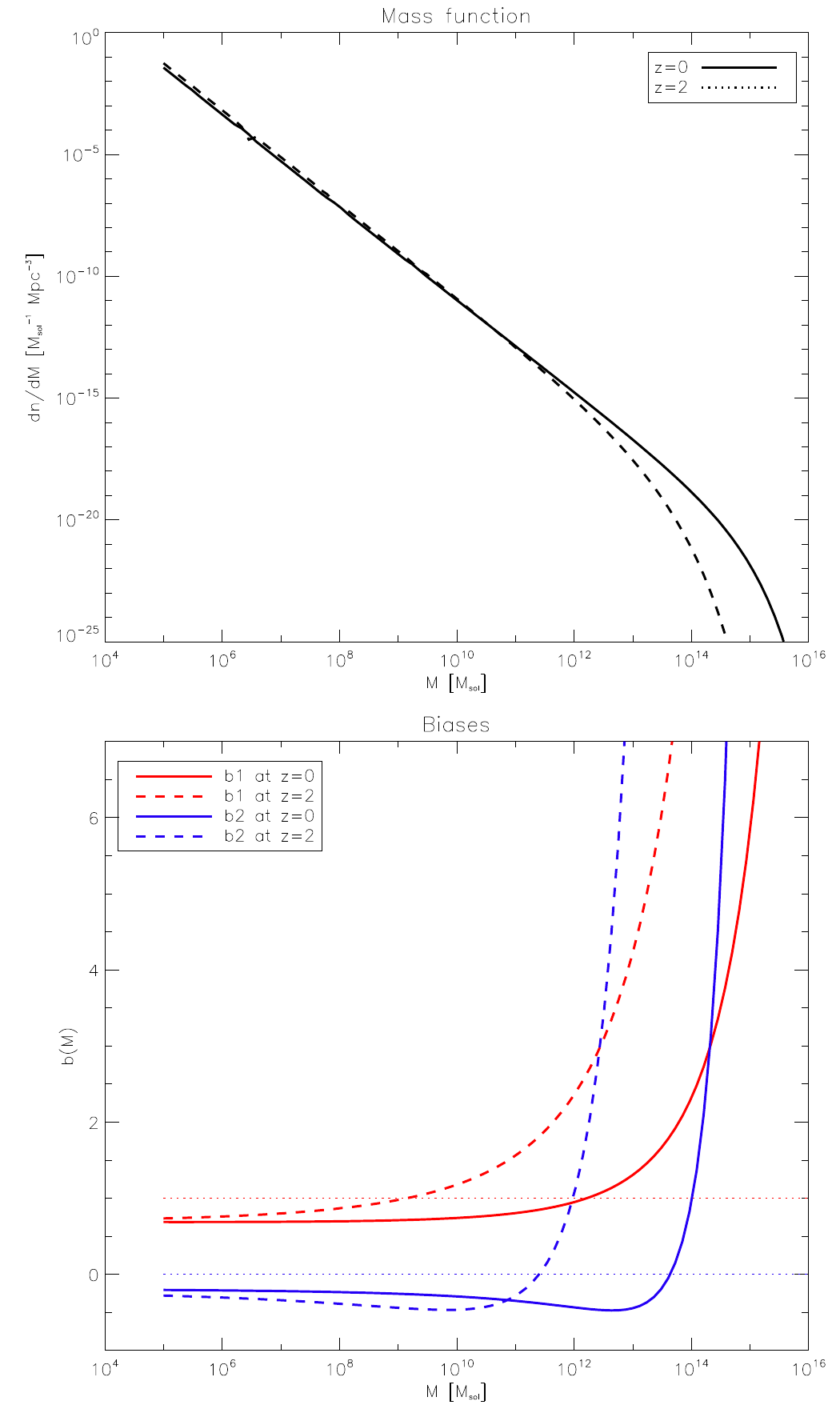}
\caption{\textit{Upper panel :} \cite{ShethTormen1999} mass function. \textit{Lower panel :} associated bias parameters at \textcolor{red}{first} and \textcolor{blue}{second} order, at z=0 (solid lines) and z=2 (dashed lines)\,; dotted lines show the mass averaged biases, see Eq.\ref{Eq:avb1}\&\ref{Eq:avbn}.}
\label{Fig:ShandT}
\end{figure}

Since \cite{PressSchechter1974}, more realistic mass functions have been proposed, beyond the spherical collapse model, with up to date cosmological parameters, and in agreement with N-body simulations. For this work I have been using the mass function proposed by \cite{ShethTormen1999}. Although it is not the most recent mass function, it is the only one for which the bias parameters beyond first order are available. In Fig.\ref{Fig:ShandT} the \cite{ShethTormen1999} mass function and associated bias functions are displayed at z=0 and 2 over a range of masses representative of galaxy clusters.

As noted by \cite{CooraySheth2002}, the average first order bias is one while higher order biases average to zero :
\bea
\label{Eq:avb1} \int \dd M \,\frac{\dd n_\mr{h}}{\dd M} \frac{M}{\bar{\rho}}\, b_1(M) &=& 1\\
\label{Eq:avbn} \int \dd M \,\frac{\dd n_\mr{h}}{\dd M} \frac{M}{\bar{\rho}}\, b_n(M) &=& 0 \quad \forall n>1
\eea
These averages are shown as dotted lines in Fig.\ref{Fig:ShandT}. We see that massive haloes are more biased than low-mass haloes. At higher redshifts, massive haloes are even more biased as they become extreme and very rare objects.
\newline

To describe the inner halo, we will assume it is spherical and is hence characterised by a radial profile. Numerical simulations show that dark matter haloes follow a universal profile named NFW for Navarro, Frenk and White \citep[see][]{NFW1997} :
\be
\rho(r|M) = \frac{\rho_\mr{s}}{(r/r_\mr{s})(1+r/r_\mr{s})^2}
\ee
where $\rho_\mr{s}(M)$ and $r_\mr{s}(M)$ (the scale radius) are the parameters of the model. These parameters are related to the virial radius $r_\mr{vir}$ \footnote{The virial radius is the radius such that inner particles can be considered gravitationally bound together so that the virial theorem applies\,; we consider $r_\mr{vir} = r_{200}$ the radius within which the average density is 200 times the critical density.} and virial mass $M_\mr{vir}$ of the halo, and in particular one defines the concentration parameter $c(M,z)$ such that $r_\mr{vir} = c r_\mr{s}$. In the following we adopt the mean concentration parameter given by \cite{Dolney2004}, and neglect scatter in this parameter.

We will also assume that the galaxy distribution follows the normalised dark matter profile $u(r|M)$ \footnote{$u(r|M) = \frac{\rho(r|M)}{\int \dd^3\mathbf{r} \,\rho(\mathbf{r}|M)}$}. While this may be controversial at first sight, we expect it to be a good approximation as galaxy formation is driven by initial overdensities in the dark matter density field.
\newline

\begin{figure}[htbp]
\centering
\includegraphics[width=.8\linewidth]{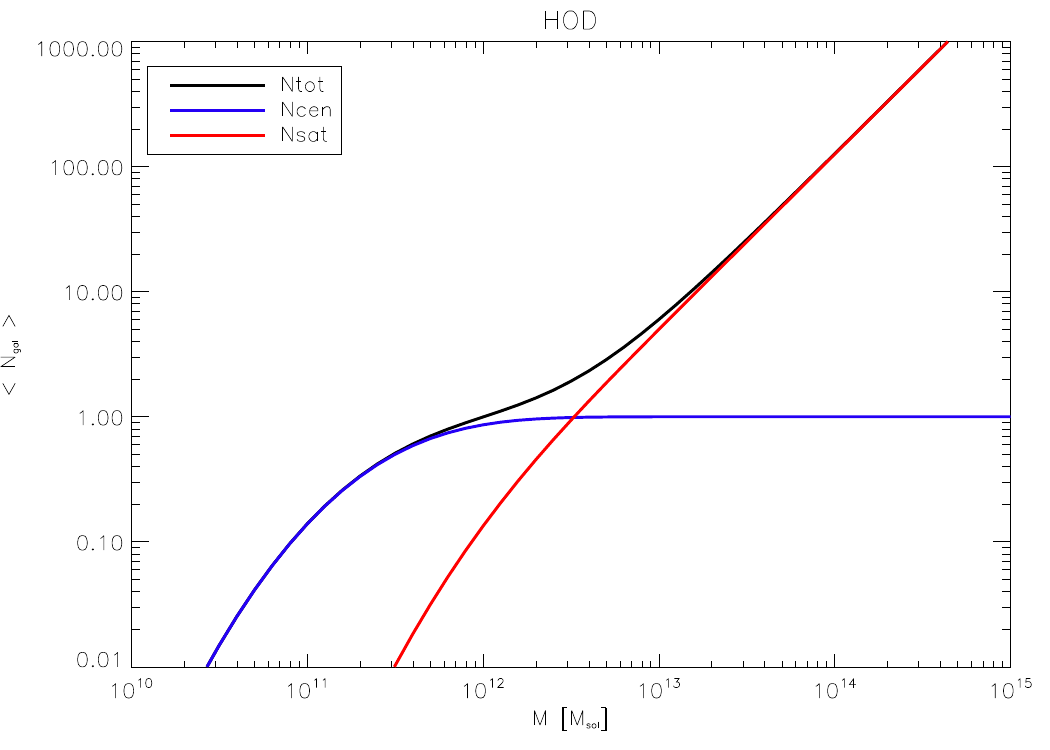}
\caption{Average number of galaxies in a halo depending on its mass, with the HOD described in Eq.\ref{Eq:NcenHOD}\&\ref{Eq:NsatHOD} and fiducial HOD parameters $M_\mr{min}= 10^{11.5} \Msun\,$, $M_\mr{sat}= 10 M_\mr{min}\,$, $\sigma_{\log M} = 0.65$ and $\alpha_\mr{sat}=1.4$.}
\label{Fig:HOD}
\end{figure}

The last element needed to describe the galaxy distribution is the number of galaxies inside a given halo. This is called the Halo Occupation Distribution (HOD hereafter). Following observations and numerical simulations \citep{Berlind2002,Zheng2005,VanDenBosch2007}, the number of galaxies is typically described as the sum of a central galaxy and possibly satellite galaxies : $N_\mr{gal} = N_\mr{cen} + N_\mr{sat}$ where both numbers are random variables drawn from a distribution depending on the halo mass. $N_\mr{cen}$ is either 0 or 1 with a mean value being a step-like function \citep{Kravtsov2004}. $N_\mr{sat}$ is drawn, conditionally to ${N_\mr{cen}=1}$, from a Poisson distribution \citep{Zheng2005} whose mean is a power law with the halo mass at high masses. Following \cite{Tinker2010,Penin2012}, we take~:
\bea
\label{Eq:NcenHOD} \langle N_\mathrm{cen} \rangle &=& \frac{1}{2} \left[ 1+\mathrm{erf}\left(\frac{\log M - \log M_\mathrm{min}}{\sigma_{\log M}}\right) \right]\\
\label{Eq:NsatHOD} \langle N_\mathrm{sat} \rangle &=& \frac{1}{2} \left[ 1+\mathrm{erf}\left(\frac{\log M - \log 2M_\mathrm{min}}{\sigma_{\log M}}\right) \right] \left( \frac{M}{M_\mathrm{sat}}\right)^{\alpha_\mathrm{sat}}
\eea
where $M_\mr{min}\,$, $M_\mr{sat}\,$, $\sigma_{\log M}$ and $\alpha_\mr{sat}$ are the four HOD parameters. $M_\mr{min}$ is the typical mass for which a halo starts to host a central galaxy, $\sigma_{\log M}$ governs the width of the step-like function for this central galaxy, $M_\mr{sat}$ is the typical mass above which a halo hosts a satellite galaxy, and $\alpha_\mr{sat}$ governs the slope of the number of satellite galaxies at high masses. We take as fiducial values $M_\mr{min}= 10^{11.5} \Msun\,$, $M_\mr{sat}= 10 M_\mr{min}\,$, $\sigma_{\log M} = 0.65$ and $\alpha_\mr{sat}=1.4$ as found by Cosmic Infrared Background studies \citep{Viero2009,Amblard2011,planck2011-CIB,Penin2012}. Heuristically, we would expect the number of galaxies to grow proportionally to the halo mass so that $\alpha_\mr{sat}=1$. However, CIB observations typically find best-fit values $\alpha_\mr{sat}>1$\,; this may be a modeling shortcoming. Figure \ref{Fig:HOD} shows the resulting average galaxy number depending on the halo mass.

\subsection{A diagrammatic method for galaxy correlation functions}\label{Sect:diagrammatique}
This section assembles the physical ingredients of the halo model described previously and proposes a new diagrammatic formalism allowing to compute the galaxy correlation functions/polyspectra.\\
The redshift dependence of all functions will be implicit in this whole section to lighten notations.
The galaxies number per comoving volume at a given point $\xx$ reads~:
\be\label{Eq:ngalhalobase}
n_\mathrm{gal}(\xx) = \sum_i n_\mathrm{gal}(\xx|i)
\ee
with $i$ being the halo index.\\
In the literature, $n_\mathrm{gal}(\xx|i)$ is assumed, implicitly or explicitly, to be a smooth distribution following the halo profile~:
\be
n_\mathrm{gal}(\xx|i) = N_\mr{gal} \, u(\xx |M_i)
\ee
However galaxies are discrete objects, so I write \citep{Lacasa2013b}~:
\be
n_\mathrm{gal}(\xx|i) = \sum_{j=1}^{N_\mathrm{gal}(i)} \theta(\xx-\xx_j)
\ee
where $j$ is the index of the random galaxies, $N_\mathrm{gal}(i)$ is the --random-- number of galaxies in the halo $i$, $x_j$ is the --random-- position of the $j$th galaxy, and $\theta(x)$ is the profile of a galaxy.\\
Accounting for this discreteness is necessary, as it allows to predict the shot-noise terms, and to do it self-consistently with the clustering terms. Eq.\ref{Eq:ngalhalobase} can then be rewritten as :
\be\label{Eq:ngalx}
n_\mathrm{gal}(\xx) = \int \dd M \,\dd^3\xx_\mr{h} \sum_i \delta(M-M_i) \,\delta^{(3)}(\xx_\mr{h}-\xx_i) \int \dd^3\xx_\mr{g} \sum_{j=1}^{N_\mathrm{gal}(i)} \delta^{(3)}(\xx_\mr{g}-\xx_j) \,\theta(\xx-\xx_j)
\ee
with $M_i$ and $\xx_i$ the mass and position of the halo $i$.\\
This equation serves as the basis for the computation of the galaxy clustering throughout this thesis.

Within this formalism, we have~:
\be
\lbra \sum_i \delta(M-M_i) \, \delta^{(3)}(\xx_\mr{h}-\xx_i)\rbra = \frac{\dd n_\mr{h}}{\dd M}(M)
\ee
which expresses the fact that the halo abundance is isotropic and given by the mass function.\\
Furthermore~:
\be
\lbra\delta^{(3)}(\xx_\mr{g}-\xx_j)\rbra_{\xx_j}=u(\xx_\mr{g}-\xx_i | M_i)
\ee
which expresses the fact that galaxy positions follow, on average, the halo profile.\\
I will further assume that galaxies are drawn independently in the halo.
\newline

Then galaxy correlation functions may be computed from Eq.\ref{Eq:ngalx}\,; for example the 2-point c.f. is~:
\be
\zeta_\mr{gal}(|\xx-\xx'|)=\frac{\lbra (n_\mr{gal}(\xx)-\overline{n}_\mr{gal})(n_\mr{gal}(\xx')-\overline{n}_\mr{gal}) \rbra}{\overline{n}_\mr{gal}^2}
\ee
These correlation functions may be Fourier transformed to yield the polyspectra.
\newline

I have proposed in \cite{Lacasa2013b} that these computations may be done through a diagrammatic method. It allows us to have a clear representation and an understanding of the different terms involved. It further allows us to avoid cumbersome calculations at high order, by replacing them with diagram drawings.\\
Note that, since Feynman diagrams \citep[see e.g.][]{Penco2006}, diagrammatic method can be found in other contexts in physics and cosmology when perturbation theory is involved. However the diagrams proposed here are of a different nature as they do not involve perturbation theory. For example there are a finite number of them, for a given correlation function, and they yield an theoretically exact result (compared to a perturbative expansion where the result is known to be approximate).

For the polyspectrum of order $n$, the first step is to draw in diagrams all the possibilities of putting $n$ galaxies in halo(s). Potentially, two or more galaxies can lie at the same point (``contracted") for the shot-noise terms. Then for each diagram, the galaxies should be labeled, e.g., from 1 to $n$, as well as the haloes e.g. with $\alpha_1$ to $\alpha_p$.

Each diagram produces a polyspectrum term. This term contains a prefactor $1/\overline{n}_\mathrm{gal}^{\,n}$ multiplied by an integral over the halo masses $\int \dd M_{\alpha_1 \cdots \alpha_p}$ of several factors. The following ``Feynman"-like rules prescribe these different factors~:
\begin{itemize}
\item for each halo $\alpha_j$ there is a corresponding~:
 \begin{itemize}
 \item halo mass function $\left.\frac{\dd n_\mr{h}}{\dd M}\right|_{M_{\alpha_j}}$
 \item average of the number of galaxy uplets in that halo.\\
 e.g. $\langle N_\mathrm{gal} \rangle$ for a single galaxy in that halo, $\langle N(N-1)\rangle$ for a pair etc.
 \item as many halo profile $u(k | M_{\alpha_j})$ as \emph{different} points, where $k=|\sum_{i\in\mr{point}} \kk_i|$.
\footnote{For example $k=k_i$ for a non-contracted galaxy i, while $k=|\kk_{i_1} + \cdots + \kk_{i_q}|$ for a galaxy contracted q times with labels $i_1 \cdots i_q$.}
 \end{itemize}
 \item the final factor is the halo polyspectrum of order $p$, conditioned to the masses of the corresponding haloes : 
 $$\mathcal{P}_\mathrm{halo}^{(p)}\left(\sum_{i \in \alpha_1} \kk_i \, , \cdots , \sum_{i \in \alpha_p} \kk_i \, | \, M_{\alpha_1} \, , \cdots , M_{\alpha_p} \right)$$
where the sum $\sum_{i \in \alpha_j} \kk_i$ runs over the indexes $i$ of the galaxies inside the halo $\alpha_j$.
\end{itemize}

Finally the possible permutations of the galaxy labels 1 to $n$ in the diagram should be taken into account : the contribution is the sum over permutations which produce different diagrams. For example we will see in Sect.\ref{Sect:CIB_HM} that some contributions to the bispectrum (namely 1-halo, 3-halo and shot1g) have a single term while others (namely 2-halo and shot2g) have 3 terms.
\newline

\begin{figure}[htbp]
\centering
\includegraphics[width=.6\linewidth]{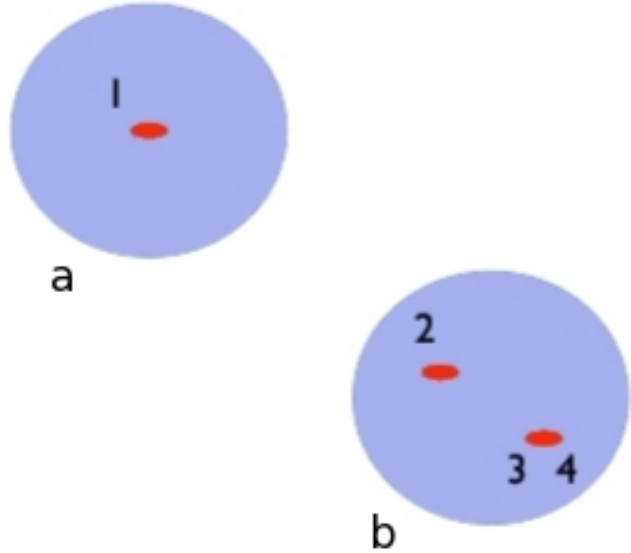}
\caption{A diagram involved in the computation of the galaxy trispectrum : 2-halo 3-galaxies with contraction in the second halo.}
\label{Fig:diag_trisp2h3g}
\end{figure}

For illustration, these rules are applied in the case of the diagram represented in Fig.\ref{Fig:diag_trisp2h3g}.

This is a diagram involved in the computation of the trispectrum, it is rather complicated which allows to illustrate all the rules. The diagram involves two haloes labeled \texttt{a} and \texttt{b} and four galaxies labeled 1,2,3,4 with galaxies 3 and 4 being contracted. The latter means that the 4-point correlation function hits twice the same point with its last arguments ($\xx_3=\xx_4$), so this is a shot-noise diagram. This diagram yields the following term for the trispectrum~:

\bea
\nonumber T(\vec{k}_1,\cdots,\vec{k}_4,z) &=& {\color[rgb]{0,0.5,1}\frac{1}{\overline{n}_\mathrm{gal}(z)^4}} \int dM_\texttt{ab} \, {\color{orange} \left.\frac{d n_\mr{h}}{d M}\right|_{M_\texttt{a}} \, \left.\frac{d n_\mr{h}}{d M}\right|_{M_\texttt{b}} }{\color{darkgreen} \langle N_\mathrm{gal}(M_\texttt{a}) \rangle \langle N_\mathrm{gal}(M_\texttt{b})(N_\mathrm{gal}(M_\texttt{b})-1)\rangle} \\
&&\quad\times {\color{red} u(k_1 | M_\texttt{a}) u(k_2 | M_\texttt{b}) u(|\vec{k}_3+\vec{k}_4| | M_\texttt{b})} {\color{blue} \, P_\mathrm{halo}(k_1 | M_\texttt{a} , M_\texttt{b})}
\eea
The \textcolor[rgb]{0,0.5,1}{prefactor} is dictated by the first ``Feynman"-like rule, then we have an integral over the mass of both haloes \texttt{a} and \texttt{b}, their \textcolor{orange}{mass functions}, their \textcolor{darkgreen}{galaxy occupation} (with an average number of duplets $N(N-1)$ in halo b as there are two different galaxies), their \textcolor{red}{profile}, and finally the polyspectrum linking the haloes, here the \textcolor{blue}{halo power spectrum} conditioned to the respective halo masses.
\newline

I have described in the previous sections the first elements (mass function, HOD, halo profile), the last elements needed are the halo polyspectra. They are biased versions of the dark matter polyspectra through deterministic biasing discussed in Sect.\ref{Sect:localbias} \footnote{As already pointed out in Sect.\ref{Sect:localbias}, this description of the halo polyspectra breaks down on small (non-linear) scale. However on these scales, the galaxy polyspectrum will be dominated by terms involving halo profiles and not halo correlations. For example at the power spectrum level, the description of the 2-halo term breaks down at high $k$, but this is, as it is meant to be, when the 1-halo and shot-noise terms become dominant.}.
At high order they have several possible origins : the first order biasing of the corresponding dark matter polyspectrum, either primordial (primordial non-Gaussianity) or from perturbation theory, or the higher order biasing of lower order dark matter polyspectra. This will be discussed in Sect.\ref{Sect:CIB_HM} when this formalism will be applied to the CIB \citep[see also][]{Lacasa2013b,Penin2013}.

A concluding remark for this section, is that we can resum shot-noise diagrams. For example, it is easily seen that a diagram for the n-th order polyspectrum with a contraction of order 2 corresponds to a diagram for the (n-1)th order polyspectrum without contraction. And the polyspectrum terms that these diagrams yield differ only by a factor $\frac{1}{\overline{n}_\mr{gal}(z)}$ and the different arguments $\kk_i$. Hence, when we sum all these diagrams with a contraction of order 2 we get the lower order polyspectrum~:
\be\label{Eq:resumcontrac2}
\mathcal{P}^{(n)}_\mr{gal,contrac2}(\kk_{1\cdots n},z) = \frac{1}{\overline{n}_\mr{gal}(z)} \, \mathcal{P}^{(n-1)}_\mr{gal,non-shot}(\kk_1+\kk_2,\kk_{3\cdots n},z) + \mr{perm.}
\ee 
where there are $\begin{pmatrix} 2 \\ n\end{pmatrix}$ permutation terms, and the (n-1)th order polyspectrum only contains non shot-noise terms, as the n-th order diagrams considered contain strictly a contraction of order 2 and no other contractions.\\
One can further develop resumation, which is of interest at high orders, but for the illustrations I will present in Sect.\ref{Sect:CIB_HM}, Eq.\ref{Eq:resumcontrac2} is sufficient.
\newline

Hence I have presented a diagrammatic method to compute galaxy polyspectra which brings us the power and simplicity of drawings to compute otherwise cumbersome equations at high order. Furthermore, this diagrammatic method can be easily adapted to model complexifications (different galaxy populations, additional galaxy and/or halo properties etc), as will be discussed in the perspectives.
\newline
\newline

%Conclusion of the chapter
In this chapter, I have described the very basis of the current cosmological description of the universe. I have described the inflation paradigm and how it provides a model producing the primeval cosmological perturbations. I have shown how these perturbations then evolve to produce the two main observables of modern cosmology : the Cosmic Microwave Background and the Large Scale Structure. Both observables can be described as random fields and are hence naturally fit for statistical characterisation with the tools developed in the previous chapters. In particular, I have shown how non-linearity of the physics generates non-Gaussian signatures in both these observables. Finally, I have presented a diagrammatic formalism allowing the prediction of high order polyspectra for a standard tracer of the Large Scale structure, the galaxy density field, and these results will be further used in the following chapter to describe the Cosmic Infrared Background.

\mychapter{Extragalactic foregrounds : tracers of the Large Scale Structure}{Extragalactic foregrounds : tracers of the Large Scale Structure}{chapt:LSS}

This chapter introduces the different signals that the Large Scale Structure produces at CMB frequencies. It focuses particularly on two of them, tracing respectively the ionised gas and the dust in the universe. I show how they can be described using tools and formalism introduced in the previous chapters, and particularly how their non-Gaussianity can be modeled in order to constrain their respective models.

\section{Foregrounds : introduction}\label{Sect:fgintro}
\begin{figure}[htbp]
\centering
\includegraphics[width=.9\linewidth]{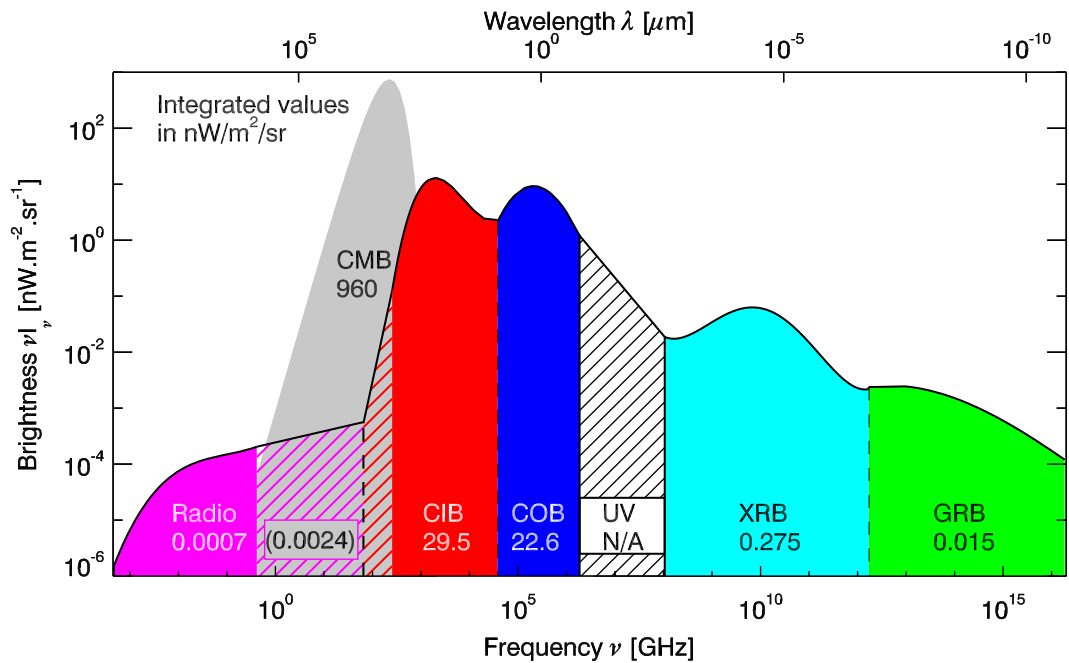}
\caption{Extragalactic background light energy density as a function of frequency. Courtesy of Herv\'e Dole. From left to right : radio background, Cosmic Microwave Background, Cosmic Infrared Background, Cosmic Optical Background, UltraViolet background, X-Ray Background and Gamma Ray Background.}
\label{Fig:EBL}
\end{figure}

The CMB follows a blackbody spectrum which peaks around 1 mm $\approx$ 300 GHz. However, the CMB is not the only signal in the sky. As shown in Fig.\ref{Fig:EBL}, the CMB clearly dominates the total power at millimeter wavelengths, while the radio/infrared background becomes dominant respectively at smaller/larger frequencies.

\begin{figure}[htbp]
\centering
\includegraphics[width=.9\linewidth]{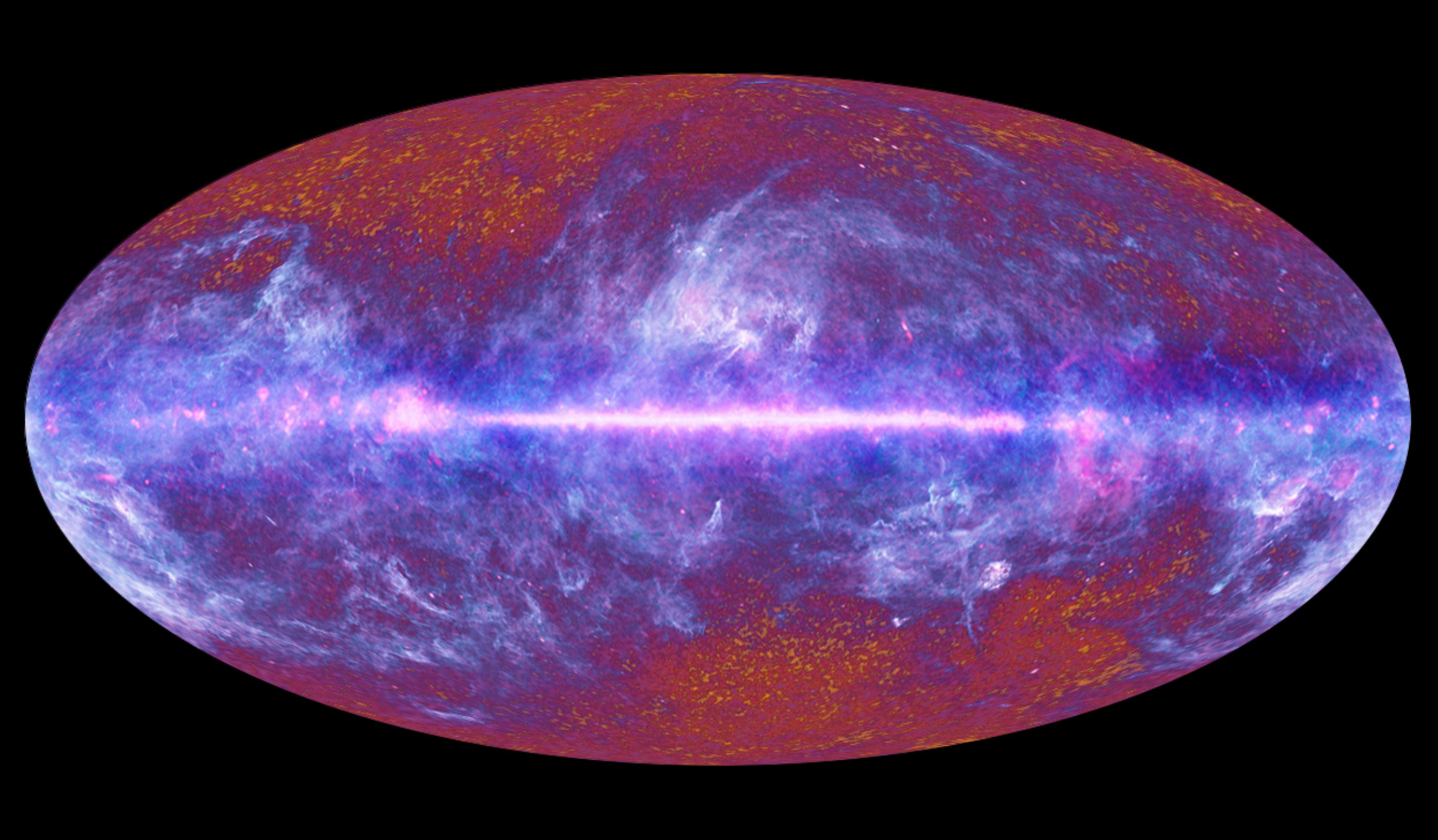}
\caption{Multifrequency image of the microwave sky by \planck (ESA/ LFI \& HFI Consortia, July 2010 press release). The galactic emission, mostly in blue, is clearly visible at the equator in this projection. It obscurs the CMB, which remains visible at high galactic latitudes.}
\label{Fig:multifreq-Planck}
\end{figure}

Although the CMB dominates in terms of total power, it is remarkably uniform over the sky. As stated before, its primary anisotropies only appear at the $10^{-5}$ level. Observationally the first detectable anisotropy, at the $10^{-3}$ level, is the so-called kinetic dipole due to the motion of the observer with respect to the surface where CMB photons were emitted (last scattering surface hereafter)\footnote{At first order in $\beta=\frac{v_\mr{obs}}{c}$ this effect creates a dipolar anisotropy, but also a whole aberration of the sky ; it furthermore yields a kinetic quadrupole and monopole at second order in $\beta$}. When this dipole is removed, the main feature visible on the sky is the Milky way, as can be seen for example on the first \planck released sky image Fig.\ref{Fig:multifreq-Planck}.

There are four main processes of galactic emission~:
\begin{itemize}
\item {\bf Synchrotron :} the  Bremsstrahlung emission of relativistic electrons spiraling in the galactic magnetic field. This emission peaks in the radio domain and decreases with frequency with a steep slope \citep[see][for a review]{Davies1998}.
\item {\bf Free-free :} emission generated when free electrons are scattered by ions, e.g. in an ionised hydrogen cloud. This emission peaks in the radio domain and decreases with frequency with a flatter slope \citep[see][for a review]{Bartlett1998}.
\item {\bf Thermal dust :} thermal emission from cool dust, typically at a temperature of 20 K. This emission follows a greybody law and peaks in the infrared domain \citep[see][for a review]{Draine2004}.
\item {\bf Anomalous microwave emission (AME) :} only detected in the last decade, and thought to be produced by spinning dust grains. It emits in the radio domain with a peak typically between 20 GHz and 40 GHz \citep[see][for a review]{Kogut1999}.
\end{itemize}

Other galaxies in the universe also emit through these processes, and may be visible outside of the galactic plane. Their emission, redshifted by the expansion, produces an extragalactic background light which traces the galaxy distribution. Galaxies are most often dominated by one emission type and are thus broadly classified into two populations~:
\begin{itemize}
\item {\bf radio sources} are dominated by synchrotron and/or free-free emission. They correspond to galaxies with a central Active Galactic Nucleus (AGN) \citep[see][for a review]{Toffolatti1999}.
\item {\bf infrared sources} are dominated by thermal dust emission. They correspond to dusty star-forming galaxies and form the Cosmic Infrared Background (CIB) \citep[see][for reviews]{Gispert2000,Lagache2005}.
\end{itemize}
Both populations will be described in more details in Sect.\ref{Sect:introps}.\\
The brightest of these sources are detected and characterised, the latest study being by \planck in \cite{planck2013-PCCS}. The study of the unresolved sources and their statistics has been an important part of my work.

Further anisotropies, called secondary, are present at CMB frequencies. They do not come from an additional emission process. Instead, they are distortions of the CMB emission, along its path from the last scattering surface to the observer. Three main distortions produce the most observable signals~:
\begin{itemize}
\item {\bf Lensing :} the trajectory of photons may deviate from a straight line, when they encounter gravitational potentials which produce a local spatial curvature. The angular scales of interest for the CMB do not probe magnification or multiple-imaging induced by strong lensing. Instead there is a distortion by the potentials on large scale, in the regime of weak lensing. The net effect is to remap the primordial anisotropies, i.e. $T_\mr{obs}[\hn] = T_\mr{prim}[\hn'(\hn)]$.\\
A review of the theory and observation of the CMB lensing is given by \cite{Hanson2010}, and the latest observations by \cite{planck2013-lensing} have produced the first estimated lensing map.
\item {\bf integrated Sachs-Wolfe effect (iSW) :} if the photons cross gravitational potentials that are evolving, the photons experience an energy shift. For example if the potential inhomogeneities flatten, a photon gains more energy falling down a well than it loses energy climbing up the exit. In a matter-dominated euclidean universe, gravitational potentials are however constant in the linear regime. Hence the iSW effect happens either when the dynamic differs from matter-dominated (e.g., when dark energy becomes important) or through non-linear gravitational infall (sometimes called Rees-Sciama effect \citep{Rees1968} or non-linear iSW). The iSW effect due to dark energy (sometimes called late iSW) peaks on large angular scales, and much more information about it can be found in the PhD thesis by \cite{Ilic2013}.\\
A review of the theory and observation of the iSW effect is given by \cite{Giannantonio2010}, and it has been detected and characterised with the latest observations by \cite{planck2013-ISW}.
\item {\bf Sunyaev-Zel'dovich effect (SZ) :} CMB photons may also be scattered by free electrons through Compton scattering. The thermal energy of these electrons creates $y$-type spectral distortions of the CMB blackbody named the thermal Sunyaev-Zel'dovich (tSZ) effect. Furthermore, the bulk velocity of these electrons creates a temperature anisotropy, named the kinetic Sunyaev-Zel'dovich (kSZ) effect, which depends on the orientation of the velocity with respect to the line of sight. \\
The tSZ effect will be described in more details in Sect.\ref{Sect:SZ}. 
Theory and observations reviews of the SZ effect are given by \cite{Cooray2005} and \cite{Bartlett2004}, and the latest observations \planck have produced the largest SZ-based cluster catalogue \citep{planck2013-SZcatalogue} and the first estimated full-sky SZ map \citep{planck2013-SZmap}.
\item Note that there are other processes leaving secondary imprint on the CMB, such as the Ostriker-Vishniac effect \citep{Ostriker1986} or inhomogeneous reionisation \citep{Aghanim1996}. They however have a much lower amplitude and peak at small angular scales. See \cite{Aghanim2008} for a thorough and complete review of the sources of secondary anisotropies.
\end{itemize}

Note that, as they trace the Large Scale Structure at low redshift ($z=0-1$), the secondary anisotropies are significantly correlated together and have a non-Gaussian distribution. As already mentioned, the iSW-lensing bispectrum is of particular importance for CMB NG studies. Indeed the large scale iSW signal modulates the small scale lensing power, and hence the bispectrum peaks in the squeezed limit and is a contaminant for local type primordial NG estimation (at the level $\Delta\fnl\sim10$, see \cite{Serra2008,Kim2013}).\\
Note also that the first three listed phenomena can potentially also affect photons that are produced by sources other than the CMB (as described previously : radio sources, CIB etc). However these latter signals were emitted much later on, so that along their way they did not cross appreciable gravitational potentials or free electron densities. Thus the distortion of foreground signals is mostly negligible, except for the occasional strong lensing of an extragalactic point-source, which magnifies its flux.
\newline

Simulations of these foregrounds have been produced, e.g. by CMB experiment teams to assess the foreground contamination to CMB analysis. In particular, I have been using two sets of simulations respectively aimed for the Atacama Cosmology Telescope \citep[ACT,][]{Kosowsky2003} and \planck experiments \citep{planck2011-mission}.\\
The first set is composed of all-sky simulated maps by \cite{Sehgal2010} at 30, 90, 148, 219 and 350 GHz, and publicly available\footnote{\url{http://lambda.gsfc.nasa.gov/toolbox/tb_cmbsim_ov.cfm}}. I have been using the simulated maps of the IR and radio point-sources and of the thermal Sunyaev-Zel'dovich effect. The maps are based on N-body simulations of the large scale structure, with a volume 1000 $h^{-1}$Mpc on a side, produced using a tree-particle mesh code. The simulated octant of the sky was replicated, which produces spurious effects on large scales, particularly up to the octopole $\ell=3$ ; however I have avoided these scales for my studies. Furthermore, the simulations reproduce the observables available at the time of their productions. Indeed for the CIB, they match the upper limit on the total intensity by \cite{Fixsen1998}, the dust parameters by  \cite{Knox2004,Dunne2000}, the source counts by \cite{Coppin2006,Austermann2010,Patanchon2009} and the constraint on the power spectrum by \cite{Lagache2007,Viero2009,Reichardt2009}. However they do not reproduce the scale dependence of the later measurements of the power spectrum \citep{planck2011-CIB}. For radio sources, the simulations are based on a radio model consistent with observations \citep{Lin2009,Lin2010} and reproduce the radio luminosity function \citep{Laing1983}, they however do not reproduce exactly the number counts found in the later results by \cite{planck2011-ERCSC,planck2013-PCCS}.\\
The second set of simulations is the \planck sky model \citep[PSM hereafter,][]{PSM2013} and consists of simulations of all the Galactic and extragalactic foregrounds to the CMB at all the \planck frequencies. That is, at 30-44-70 GHz for the Low Frequency Instrument (LFI) and at 100-143-217-353-545-857 GHz for the High Frequency Instrument (HFI). Note that the IR and radio point-sources simulations are not based on a large scale structure simulation but partly on observations and partly on random realisations. The PSM software is continuously evolving to reproduce observations, in particular \planck results. I have used the PSM simulations in particular for the \planck tSZ analysis, see Sect.\ref{Sect:meas_SZ_NG}.

\section{Sunyaev-Zel'dovich effect}\label{Sect:SZ}
\subsection{Introduction : ionisation and photon scattering}\label{Sect:SZintro}
%After recombination the universe became neutral, and remained so for a long period. As stars and quasars formed, the intergalactic gas got ionised in a process called reionisation, which is still little constrained today. As in the primordial plasma, the free electrons of the ionised gas scatter the CMB photons.\\
%When the universe is uniformly ionised, the CMB temperature anisotropies tend to be washed out. Indeed, absorption scales them with a factor $e^{-\tau}$ where $\tau$ is the optical depth, while Thomson scattering damps small-scale fluctuations in a similar manner to Silk damping (c.f. Sect.\ref{Sect:CMB}).

\begin{figure}[htbp]
\centering
\includegraphics[width=.7\linewidth]{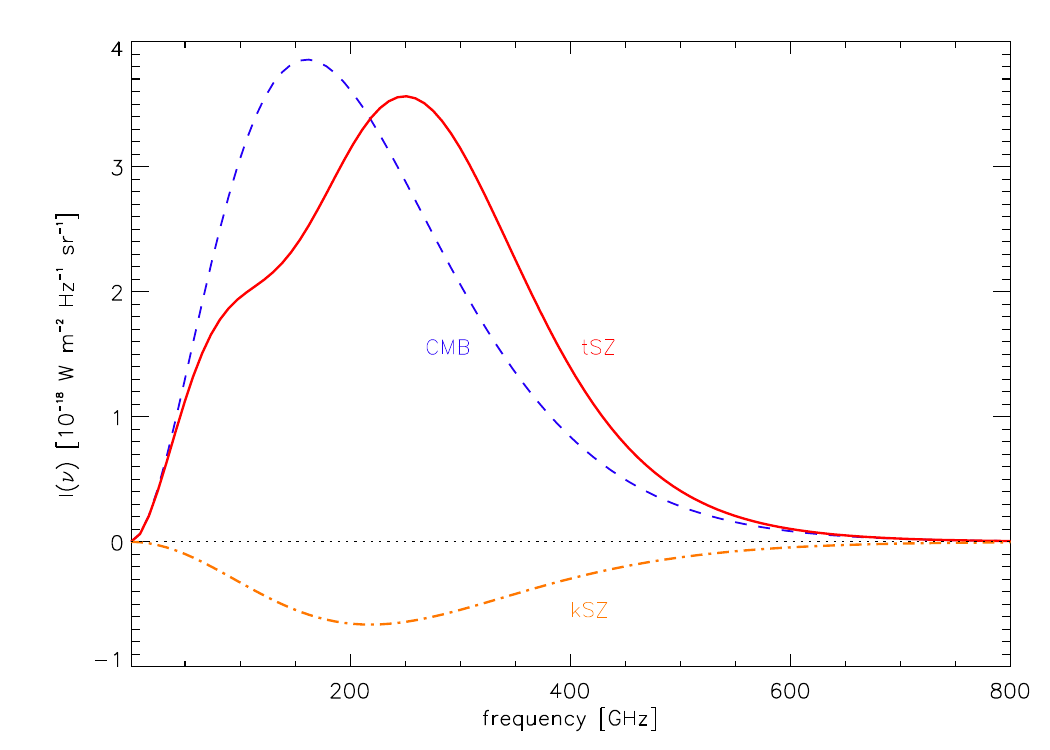}
\caption{Emission law for the primordial CMB (blue), for CMB photons distorted by a thermal SZ effect (red), and by the kinetic SZ effect (orange, $\Delta I(\nu)$) in the case of a recessing cluster.}
\label{Fig:SZ}
\end{figure}

Energetic free electrons are found in galaxy clusters, and as they are not distributed uniformly they leave secondary anisotropies in the CMB. Indeed \cite{SZ1970,SZ1972} have shown that thermal electrons inject energy in the CMB photons through inverse Compton scattering.
%\footnote{Sometimes named inverse Compton scattering to highlight that electrons inject energy to photons, while the term Compton scattering is reserved to the usual situation where photons inject energy to electrons.}
Through this scattering, the number of photons is conserved but they are transferred to higher frequencies.
 %Indeed, if they could sufficiently scatter the photons, the electrons would tend to thermalise the radiation at a higher temperature.
 The result is a distortion of the spectrum, with a depletion at low frequencies and an excess at high frequencies, compared to the CMB blackbody spectrum. In the case of optically thin clusters (on average photons experience a single scattering) and non-relativistic electrons, the spectral distortion has a universal form and is characterised by a single amplitude parameter, the Comptonisation parameter $y$~:
\be
y = \int n_\mr{e} \, \sigma_\mr{T} \,\frac{k_\mr{B} T_\mr{e}}{m_\mr{e} c^2}\, \dd l
\ee
where $n_\mr{e}$ and $T_\mr{e}$ are respectively the electron density and temperature in the cluster, $\sigma_\mr{T}$ is the Thomson cross-section, $m_\mr{e}$ is the electron rest-frame mass, and the integral runs over the line-of-sight.\\
A $y$-type distortion of the CMB spectrum is shown in Fig.\ref{Fig:SZ} for $y=0.1$, although the effect is much more subtle for real clusters ($y_\mr{rms} \sim 10^{-6}$).

The bulk velocity of the electrons in the clusters also leaves an imprint on the CMB photons. Indeed, the electrons tend to define a frame that the photons should be at rest with respect to. When this frame is moving with respect to the observer, the photons experience a Doppler redshift (outward velocity) or blueshift (inward velocity). This is the same process that leads to Doppler primary anisotropies of the CMB, due to the velocity of the plasma at the last scattering surface. The effect is achromatic and hence photons keep a blackbody spectrum albeit with a relative temperature change~:
\be
\frac{\Delta T}{T_\mr{CMB}} = \tau \, \frac{\vec{v}_\mr{e}\cdot \hn}{c}
\ee
where $\tau$ is the optical depth of the cluster. Fig.\ref{Fig:SZ} shows this kinetic SZ effect in the case of a recessing cluster.

Observationally, the tSZ effect has been detected for a long time \citep[e.g.,][]{Wilbanks1994,Lamarre1998}, and recently \planck has produced the first estimated map of the tSZ effect over the sky \citep{planck2013-SZmap}. But the detection of the kSZ effect is more challenging, due to its low amplitude and to its spectrum being indistinguishable from a legitimate CMB anisotropy. Its statistical detection over thousands of clusters has nonetheless been recently claimed by \cite{Hand2012}.
\newline

\subsection{Description with the halo model}\label{Sect:SZ_HM}
\begin{figure}[htbp]
\centering
\includegraphics[width=.7\linewidth]{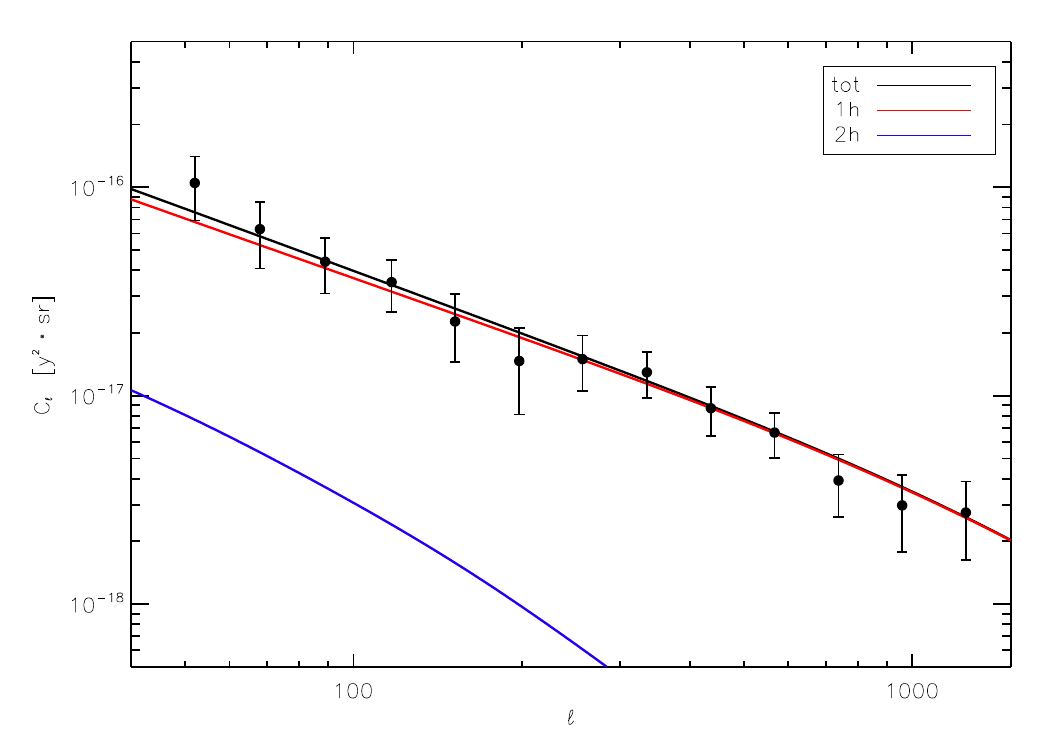}
\caption{tSZ angular power spectrum measured with \planck \citep[][black points with error bars]{planck2013-SZmap} and the best-fit power spectrum with the halo model (black curve). The red and blue curve show respectively the 1-halo and 2-halo term of the best-fit model.
}
\label{Fig:Cl_SZ_planck}
\end{figure}

If we are interested in the spatial variations of the thermal SZ signal, they can be described with the halo model previously introduced. Indeed the 3D Comptonisation parameter reads~:
\be
y(\xx,z) = \int \dd M \, \dd^3\xx_\mr{h} \; y(M,z) \, u(\xx-\xx_\mr{h}|M,z) \sum_i \delta(M-M_i) \, \delta^{(3)}(\xx_\mr{h}-\xx_i) 
\ee
where I splitted the y-profile into the total flux of the halo $y(M,z)$ and its normalised profile $u(\xx|M,z)$ in analogy with the CIB case (see Sect.\ref{Sect:CIB_HM}).\\
Thus the Fourier transform of $y(\xx,z)$ is~:
\be\label{Eq:ySZFourier}
y(\kk,z) = \int \dd M \, \dd^3\xx_\mr{h} \; y(M,z) \, u(k|M,z) e^{-i\kk\cdot\xx_\mr{h}} \sum_i \delta(M-M_i) \, \delta^{(3)}(\xx_\mr{h}-\xx_i)
\ee
where we assumed sphericity of the halos\footnote{Indeed the halo shape has been shown to have a small ($\leq 10\%$) effect on the dark matter bispectrum \citep{Smith2006}.} so that $u(\kk |M,z)=u(k|M,z)$.\\
The observed Comptonisation parameter $\tilde{y}$ is the line-of-sight integral of $y(\xx,z)$~:
\bea
\nonumber \tilde{y}(\hn) &=& \int \dd z \, \frac{\dd V_c}{\dd z \,\dd\Omega} \; y(r\hn,z)%\\
%\nonumber &=& \int \dd M \, \dd^3\xx_\mr{h} \, \frac{\dd^3\kk}{(2\pi)^3}\, \dd V_c \, y(M,z) \, u(k|M,z) e^{i\kk\cdot(r\hn-\xx_\mr{h})} \sum_i \delta(M-M_i) \, \delta^{(3)}(\xx_\mr{h}-\xx_i) \\
%\nonumber &=& \int \dd M\, \dd^3\xx_\mr{h} \, \frac{\dd^3\kk}{(2\pi)^3}\, \dd V_c \, y(M,z) \, u(k|M,z)  \, \left(\sum_i \delta(M-M_i) \, \delta^{(3)}(\xx_\mr{h}-\xx_i)\right) \\
%&& \qquad \times 4\pi \sum_{\ell m} i^\ell \, Y^*_{\ell m}(\hat{k}) \, j_\ell(kr) \, Y_{\ell m}(\hn) \ e^{-i \kk\cdot \xx_\mr{h}}
\eea
Hence, through Sect.\ref{Sect:pastlightproj}, if the 3D polyspectrum is diagonal-independent and with Limber's approximation, the angular polyspectrum reads~:
\be
\mathcal{P}^{(n)}_\mr{tSZ}(\ell_{1\cdots n}) = \int \frac{\dd z}{r^{2n-2}} \, \frac{\dd r}{\dd z} \, a^n(z) \;\mathcal{P}^{(n)}_\mr{tSZ,3D}(\kk^*_{1\cdots n},z)
\ee
with as usual $k^*_i = \ell_i/r(z)$.\\
The 3D polyspectra can be computed through the connected correlation functions of Eq.\ref{Eq:ySZFourier}. For illustration, the computation of the power spectrum is~:
\begin{align}
\nonumber (2\pi)^3 \delta^{(3)}(\kk_1+\kk_2) \, P_\mr{tSZ}(k_1,z) &= \lbra y(\kk_1,z) \,y(\kk_2,z) \rbra \\
\nonumber &= \int \left[\dd M_i \, \dd^3\xx^\mr{h}_i \; y(M_i,z) \, u(k_i|M_i,z) e^{-i\kk_i\cdot\xx^\mr{h}_i}\right]_{i=12}\\
&\times \lbra \sum_{i_1 i_2} \delta(M_1-M_{i_1}) \, \delta^{(3)}(\xx^\mr{h}_1-\xx_{i_1}) \delta(M_2-M_{i_2}) \, \delta^{(3)}(\xx^\mr{h}_2-\xx_{i_2})\rbra
\end{align}
It splits into two terms : $P(k) = P^\mr{1h}(k)+P^\mr{2h}(k)$, as the sum $\sum_{i_1 i_2}$ is splitted. The first term is the 1-halo term, it corresponds to $\sum_{i_1=i_2}\,$, while the second term is the 2-halo term and corresponds to $\sum_{i_1\neq i_2}\,$ \citep{CooraySheth2002,Ma2002}. We find~:
\be
P^\mr{1h}_\mr{tSZ}(k,z) = \int \dd M \, \frac{\dd n_\mr{h}}{\dd M} \, \left[y(M,z) \, u(k|M,z)\right]^2
\ee
and~:
\be
P^\mr{2h}_\mr{tSZ}(k,z) = \int \dd M_{12} \left[y(M_i,z) \, u(k|M_i,z) \, \left.\frac{\dd n_\mr{h}}{\dd M}\right|_{M_i}\right]_{i=12} P_\mr{halo}(k | M_1,M_2, z)
\ee
Correspondingly the angular power spectrum splits into 1-halo and 2-halo terms :
\be
C_\ell^\mr{tSZ} = C_\ell^\mr{1h} + C_\ell^\mr{2h}
\ee
Fig.\ref{Fig:Cl_SZ_planck} shows the first measurement of the tSZ angular power spectrum by \planck, and the best-fit halo model from \cite{planck2013-SZmap}, fitting for the cosmological parameters $\sigma_8$ and $\Omega_m$ which have the strongest influence. Because the tSZ signal is dominated by massive low-redshift clusters, the 1-halo term completely dominates the power spectrum, except marginally on the very first multipoles. (Note that this may change if applying a mass cut, i.e. masking detected clusters above a certain mass)

I also derive here the equations for the tSZ bispectrum. The 3-point correlation function of Eq.\ref{Eq:ySZFourier} splits into three terms, depending on the splitting of the sum over halo indices $\sum_{i_1 i_2 i_3}\,$. First, the 1-halo term corresponds to $\sum_{i_1=i_2=i_3}\,$, and after some computations it writes~:
\be
B_\mr{tSZ}^\mr{1h}(k_{123},z) = \int \dd M \frac{\dd n_\mr{h}}{\dd M} \, y(M,Z)^3 \, u(k_1|M,z) \, u(k_2|M,z) \, u(k_3|M,z)
\ee
The 2-halo term corresponds to $\sum_{i_1=i_2\neq i_3} + \mr{2 \; permutations}$ (perm. hereafter) and reads~:
\bea
\nonumber B_\mr{tSZ}^\mr{2h}(k_{123},z) &=& \int \dd M_\texttt{ab} \left.\frac{\dd n_\mr{h}}{\dd M}\right|_{M_\texttt{a}} \, \left.\frac{\dd n_\mr{h}}{\dd M}\right|_{M_\texttt{b}} \, y(M_\texttt{a},Z)^2 \, y(M_\texttt{b},z) \, u(k_1|M_\texttt{a},z) \\
&& \times \; u(k_2|M_\texttt{a},z) \, u(k_3|M_\texttt{b},z) \, P_\mr{halo}(k_3 | M_\texttt{a}, M_\texttt{b}, z) + \mr{2 \; perm.}
\eea
Finally, the 3-halo term corresponds to $\sum_{i_1\neq i_2\neq i_3}\,$, and writes~:
\be
B_\mr{tSZ}^\mr{3h}(k_{123},z) = \int \dd M_{123} \left[y(M_i,z) \, u(k_i|M_i,z) \, \left.\frac{\dd n_\mr{h}}{\dd M}\right|_{M_i}\right]_{i=123} B_\mr{halo}(k_{123} | M_{123},z)
\ee
I do not enter into further details in this section --in particular on the halo power spectrum and bispectrum--, as most of the issues will be treated in Sect.\ref{Sect:CIB_HM} for the CIB with similar equations.

%Let us note that the tSZ signal is dominated by massive low-redshift clusters, which explains why the 1-halo term dominates the power spectrum, except on the very first multipoles. As the same weighting of clusters applies at higher orders, I expect higher order polyspectra to be also dominated by the 1-halo term, except possibly when low multipoles are involved (e.g. in the squeezed configurations for the bispectrum).

Because the tSZ signal is dominated by massive low-redshift clusters, I expect high order polyspectra to be dominated by the 1-halo term as for the power spectrum, except possibly when low multipoles are involved (e.g. in the squeezed configurations for the bispectrum).
Measurement performed in Sect.\ref{Sect:meas_SZ_NG} will be shown to support this view.\\
Moreover, because it is dominated by massive low-redshift clusters, the tSZ signal is highly non-Gaussian \citep{Komatsu2002SZ}. This will also be shown observationally in Sect.\ref{Sect:meas_SZ_NG}, and motivates the use of non-Gaussianity studies of the tSZ observations to extract more statistical information, and hence more powerful model constraints.

\section{Extragalactic point-sources}
\subsection{Introduction}\label{Sect:introps}
As already discussed in Sect.\ref{Sect:fgintro}, the galaxies' emission at microwave frequencies yields a foreground to the CMB. They are classified as extragalactic point-sources as, with the resolution of microwave experiments, their angular size is much smaller than the beam (except for the nearest galaxies, such as Andromeda).
\newline

Radio sources include galaxies with a strong radio emission due to synchrotron and/or free-free processes. In these galaxies an Active Galactic Nucleus, most probably resulting from a super-massive black hole, produces a magnetised jet which ejects ionised matter out of the galactic core. The strong magnetic field in the core and the jet produces synchrotron emission, while the lobes of the jet (shock front) emit through free-free process. Depending on the orientation of the jet with respect to the observer, on the temperature of the lobes, on the velocity of electrons in the jet, on the absorption by the galactic disk etc, the observed flux may be dominated either by synchrotron or free-free, or be a mix, and be more or less powerful. The typical Spectral Energy Distribution (SED) of a radio source is steep \citep[$I(\nu)\propto \nu^{-\alpha}$ with $\alpha\sim 0.7$, see e.g.][]{Marscher1996,Tucci2011}, hence few of them are visible at high frequencies.

At microwave frequencies the detectable radio sources are the brightest ones, with a few hundreds being detected by \textit{WMAP} \citep{Wright2009} and about a thousand detected by \planck \citep{planck2013-PCCS}, and they dominate the extragalactic radio emission \citep{Toffolatti1998}. They reside at somewhat low redshifts \citep[peak at $z\sim1$, see][]{Tucci2011}, and can be considered as randomly distributed over the sky \citep{Toffolatti1998,Gonzalez-Nuevo2005}. Hence they form a white noise entirely characterised by its 1-point p.d.f., i.e., the number counts $\frac{\dd n}{\dd S}$. Thus, their polyspectra are constant and given by~:
\be\label{Eq:radpolysp}
\mathcal{P}^{(n)}_\mr{RAD} = \int S^n\, \frac{\dd n}{\dd S} \, \dd S
\ee
which stems from the properties of cumulants and the fact that the sources are independent. Let us note that the integral runs from $S=0$ to $S_\mr{cut}$. The flux cut, $S_\mr{cut}$, is the flux of the faintest detected source, i.e. all sources below that flux are present in the map and all brighter sources are assumed to be masked\footnote{The situation is often more complex than a flux cut consideration, and requires a selection function which expresses the estimated detection probability depending on the flux. This may happen for example when the noise and/or contaminants are anisotropic. I will not enter into these considerations here, let us just note that the flux cut case is the special case where the selection function is a Heaviside step function.}. Indeed, for the study of signal other than that of radio sources, one wants to minimise the radio contamination.\\
In particular, the bispectrum of unresolved radio-sources is traditionally noted $b_\mr{PS}$, with~:
\be\label{Eq:counts_to_bps}
b_\mr{PS} = \int_0^{S_\mr{cut}} S^3 \, \frac{\dd n}{\dd S} \, \dd S
\ee
This is a convention dating back to \textit{WMAP} \citep{KSW2005}, as radio sources are the only contributing sources at \textit{WMAP} frequencies.

\begin{figure}[htbp]
\centering
\includegraphics[width=.7\linewidth]{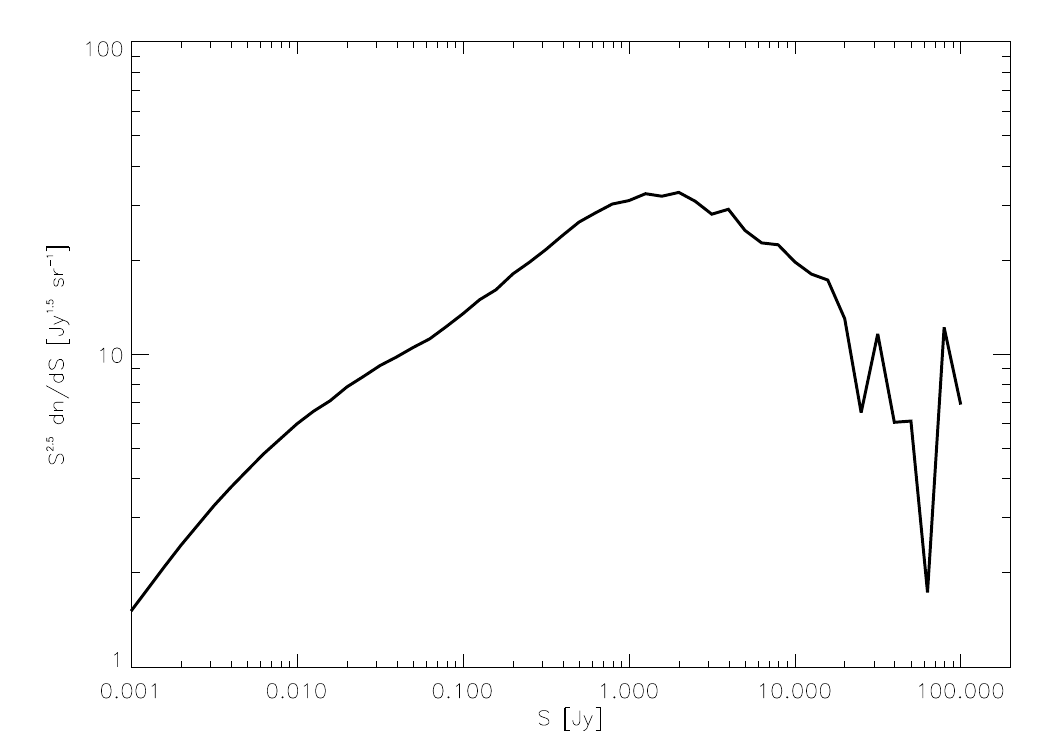}
\caption{Number counts for radio sources at 30 GHz with the model by \cite{Tucci2011}, in the Euclidean normalisation (see text for details).}
\label{Fig:counts_Tucci}
\end{figure}

During my work, I have been using number counts from the radio sources model by \cite{Tucci2011}, which is the most recent one and reproduces the observed number counts. Fig.\ref{Fig:counts_Tucci} shows these number counts at 30 GHz, the lowest \planck frequency.

In the figure I plot $S^{5/2} \, \frac{\dd n}{\dd S}\,$, this is the so-called Euclidean normalised convention. The reason for this convention is that in a flat static universe (Euclidean) the number counts go as $\frac{\dd n}{\dd S} \propto S^{-5/2}$. As a matter of fact, if we consider galaxies with a given luminosity $L$ uniformly distributed with density $\rho_L$, an observer sees $n(>S) = \rho_L \times \frac{4}{3}\pi R^3$ sources above flux S, with $S=\frac{L}{4\pi R^2}$. Hence $n(>S) \propto S^{-3/2}$ and derivation yields $\frac{\dd n}{\dd S} \propto S^{-5/2}$. This stands for all $L$, so we can integrate over $L$ which does not change the functional form but simply the proportionality constant.\\
Thus any departure from this behaviour comes from expansion and/or redshift evolution of sources. In particular the decrease in Fig.\ref{Fig:counts_Tucci} when $S\rightarrow 0$ comes from the fact that sources started emitting a finite time ago, so that there are no extremely distant and faint sources.
\newline

\begin{figure}[htbp]
\centering
\includegraphics[width=.6\linewidth]{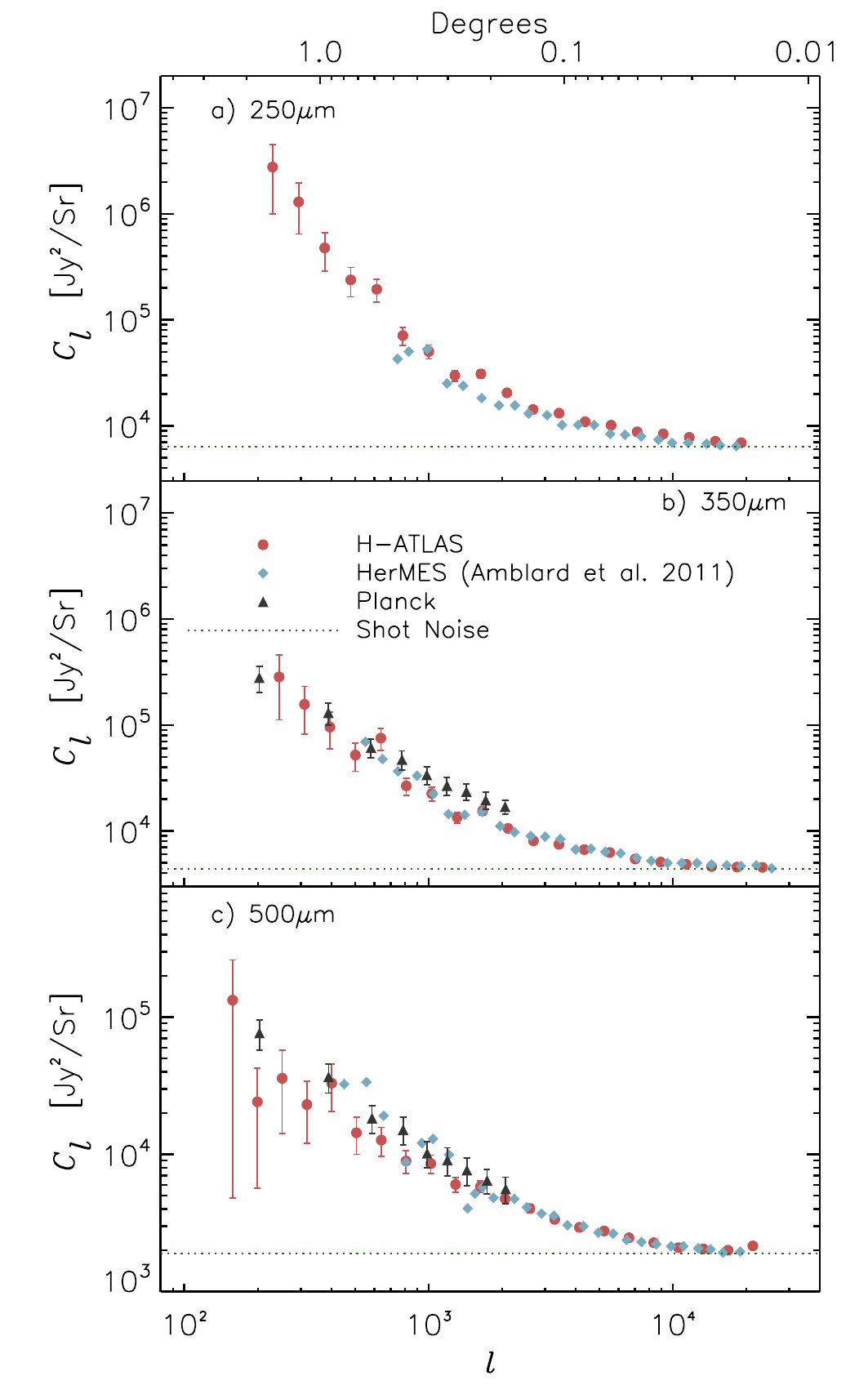}
\caption{Different measurements of the CIB power spectrum at 1200, 857 and 550 GHz (from top to bottom). Figure by \cite{Thacker2013}. The \planck data points are from the early results \citep{planck2011-CIB}.}
\label{Fig:CIB-Cl-Thacker}
\end{figure}

Infrared (IR) sources are galaxies with a strong emission from thermal dust. This dust emits with a greybody law at a temperature between 10 and 30 K, heated by the UV emission from young stars \citep{Gispert2000}. These sources are hence called dusty star-forming galaxies (DSFG), and, compared to radio sources, they reside at higher redshifts : the star formation peaks towards $z\sim 2$ \citep{Lagache2005}.\\ 
IR sources have been characterised at the number counts level, and statistical methods have been proposed to try to constrain the number counts below the detection limit, e.g. with so-called $P(D)$ analysis \citep{Patanchon2009} or with stacking \citep{Bethermin2011}. A few thousands of them have been detected individually with \planck higher frequencies channels \citep{planck2011-ERCSC,planck2013-PCCS}, but there are in fact countless unresolved sources in a \planck beam\footnote{For example, using the number counts by \cite{Bethermin2011}, at 545 GHz there are a thousand sources between 100 m Jy and the flux cut of the Early Release Compact Source Catalog \citep{planck2011-ERCSC} in a \planck beam of 4.72 arcmin.}. IR sources are strongly clustered in dark matter halos, so that their spatial distribution yields a non-trivial power spectrum. In the latest years, the CIB power spectrum has been more and more precisely measured by different experiments : BLAST \citep{Viero2009}, Spitzer \citep{Kashlinsky2012}, Herschel \citep{Amblard2011,Thacker2013} and \planck \citep{planck2011-CIB} for the most recent. Fig.\ref{Fig:CIB-Cl-Thacker} by \cite{Thacker2013} summarises the current state of the art. New measurements of this power spectrum by \planck will soon become public, I describe in Sect.\ref{Sect:meas_CIB_NG} my measurement of the CIB bispectrum.

The CIB power spectrum can be modeled with the halo model, and it splits into three terms : 2-halo, 1-halo and shot-noise. I will describe more precisely how these terms arise in Sect.\ref{Sect:CIB_HM}. Let us note however that the 2-halo term dominates on large angular scales, then the 1-halo term dominates and finally the shot-noise dominates on the smallest angular scales. For the \planck resolution, the shot-noise is subdominant over the whole range of multipoles of interest, as can be seen in Fig.\ref{Fig:CIB-Cl-Thacker}. The shot-noise is a term coming from the discreteness of galaxies and is independent of their clustering~; it corresponds to Eq.\ref{Eq:radpolysp} for $n=2$ and with IR number counts. In fact, in the case of radio sources, the shot-noise is the only contribution to the polyspectra as their clustering is negligible.\\
As the clustering of IR sources is dominated by faint unresolved sources, the flux cut has few effect on the CIB power spectrum. It mostly affects the shot-noise as the latter is sensitive to bright sources, but the effect on the clustering terms is small. More generally, as the CIB is mostly unresolved at CMB frequencies, it is the stealthiest contaminant for most CMB studies ; it is taken special care of e.g. for primary CMB and cosmological parameter constraints \citep[e.g.,][]{planck2013-cosmoparams} but also for the tSZ \citep{planck2013-SZmap}, lensing studies \citep{planck2013-CIBlensing} etc.

\subsection{Prescription for the non-Gaussianity of clustered and unclustered sources}\label{Sect:presp}
This section is based on results published in \cite{Lacasa2012}, using the previously described simulations by \cite{Sehgal2010}.
%The results are also based on simulations by \cite{Sehgal2010}. The latter are all-sky simulated maps of the IR and radio point sources at 30, 90, 148, 219 and 350 GHz, and publicly available\footnote{\url{http://lambda.gsfc.nasa.gov/toolbox/tb_cmbsim_ov.cfm}}. The maps are based on N-body simulations of the large scale structure, with a volume 1000 $h^{-1}$Mpc on a side, produced using a tree-particle mesh code. The simulated octant of the sky was replicated, which produces spurious effects on large scales, particularly up to the octopole $\ell=3$ ; however I have avoided these scales for my studies.

I have proposed a phenomenological prescription for the distribution of clustered sources, based on their number counts and power spectrum, following pioneering work by \cite{Argueso2003}. Namely, a map of a population of clustered sources is a white-noise drawn from the number counts, which has then been convolved by the power spectrum. By construction this prescription reproduces the 1- and 2- point correlation functions.

At the bispectrum level, the prescription yields~:
\be\label{Eq:presp}
b_{123}^\mr{clust} = \alpha \, \sqrt{C_{\lu}^\mr{clust} \, C_{\ld}^\mr{clust} \, C_{\lt}^\mr{clust}}
\ee
with~:
\be\label{Eq:def_alphapresp}
\alpha = \frac{b_{123}^\mr{white}}{\sqrt{C_{\lu}^\mr{white} \, C_{\ld}^\mr{white} \, C_{\lt}^\mr{white}}}  = \frac{\int S^3 \,\frac{\dd n}{\dd S}\, \dd S}{\left(\int S^2 \,\frac{\dd n}{\dd S}\, \dd S\right)^{3/2}}
\ee
Note that if the sources are not correlated, the prescription reduces to $b_{123}=\int S^3 \,\frac{\dd n}{\dd S}\, \dd S\,$, i.e. the shot-noise value.\\
If we have two populations of sources contributing to the same frequency, one clustered and one unclustered, as is the case with IR and radio sources, we assume that the populations are independent so that their bispectra add up~:
\be
b_{123} = b^\mr{RAD} + \alpha \sqrt{C_{\lu}^\mr{IR} \, C_{\ld}^\mr{IR} \, C_{\lt}^\mr{IR}}
\ee

\begin{figure}[htbp]
\centering
\includegraphics[width=.6\linewidth]{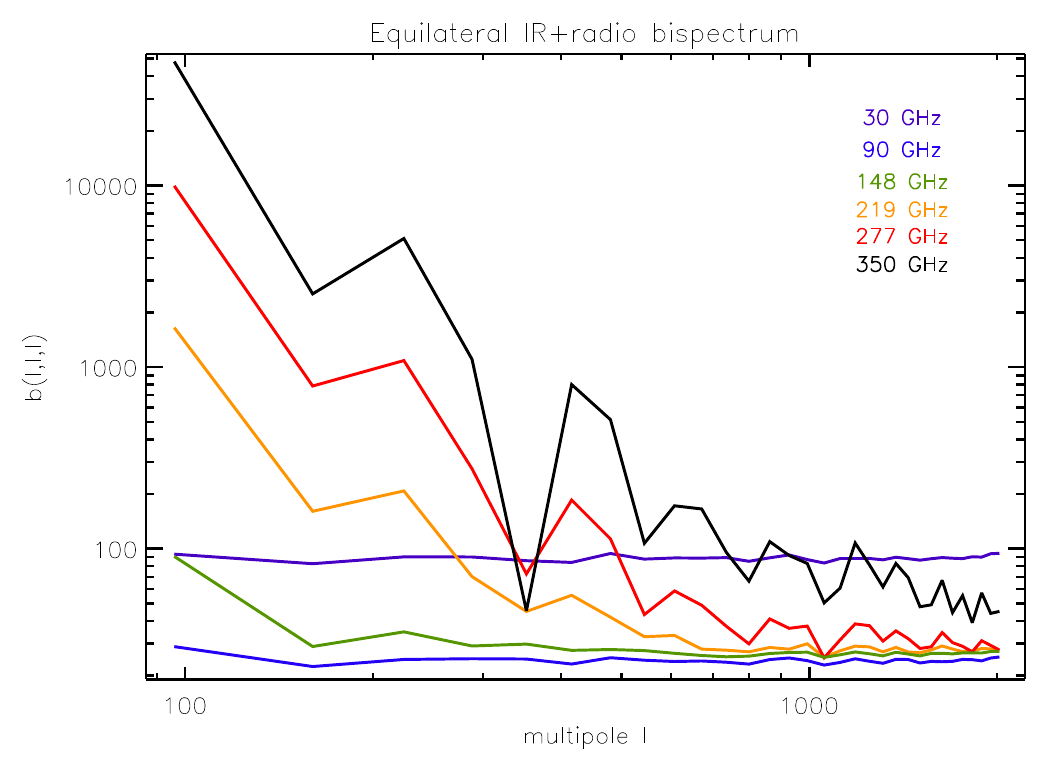}
\caption{IR + radio equilateral bispectrum from \cite{Lacasa2012} with simulations by \cite{Sehgal2010}, depending on frequency.}
\label{Fig:equibisp-IRAD}
\end{figure}

I have developed a tool to compute the bispectrum of full-sky maps in the HEALPix format \citep{Gorski2005}. I computed the bispectra of the simulated maps by \cite{Sehgal2010} at each frequency, for radio sources, infrared sources and their combination. We indeed see that the radio bispectrum is flat at all frequencies, whereas the IR bispectrum strongly depends on the multipoles, with orders of magnitude difference between large scales and small scales. For illustration I show in Fig.\ref{Fig:equibisp-IRAD} the bispectrum of the superposition of radio and IR maps in the equilateral configuration ($b_{\ell \ell \ell}$) depending on frequency.

At low frequencies, 30 \& 90 GHz, the bispectrum is flat, as it is dominated by the radio contribution. As frequency increases, the IR contribution becomes increasingly important, first at low multipoles but it finally dominates the whole multipole range at 350 GHz. The \planck central CMB channels correspond nearly to the 148 \& 219 GHz frequencies. At 148 GHz, we see that the IR contribution is subdominant at 148 GHz except at very low multipoles ($\ell \lesssim 100$). At 219 GHz the IR contribution is important at low multipoles but becomes subdominant beyond $\ell\sim500$.

\begin{figure}[htbp]
\centering
\includegraphics[width=.7\linewidth]{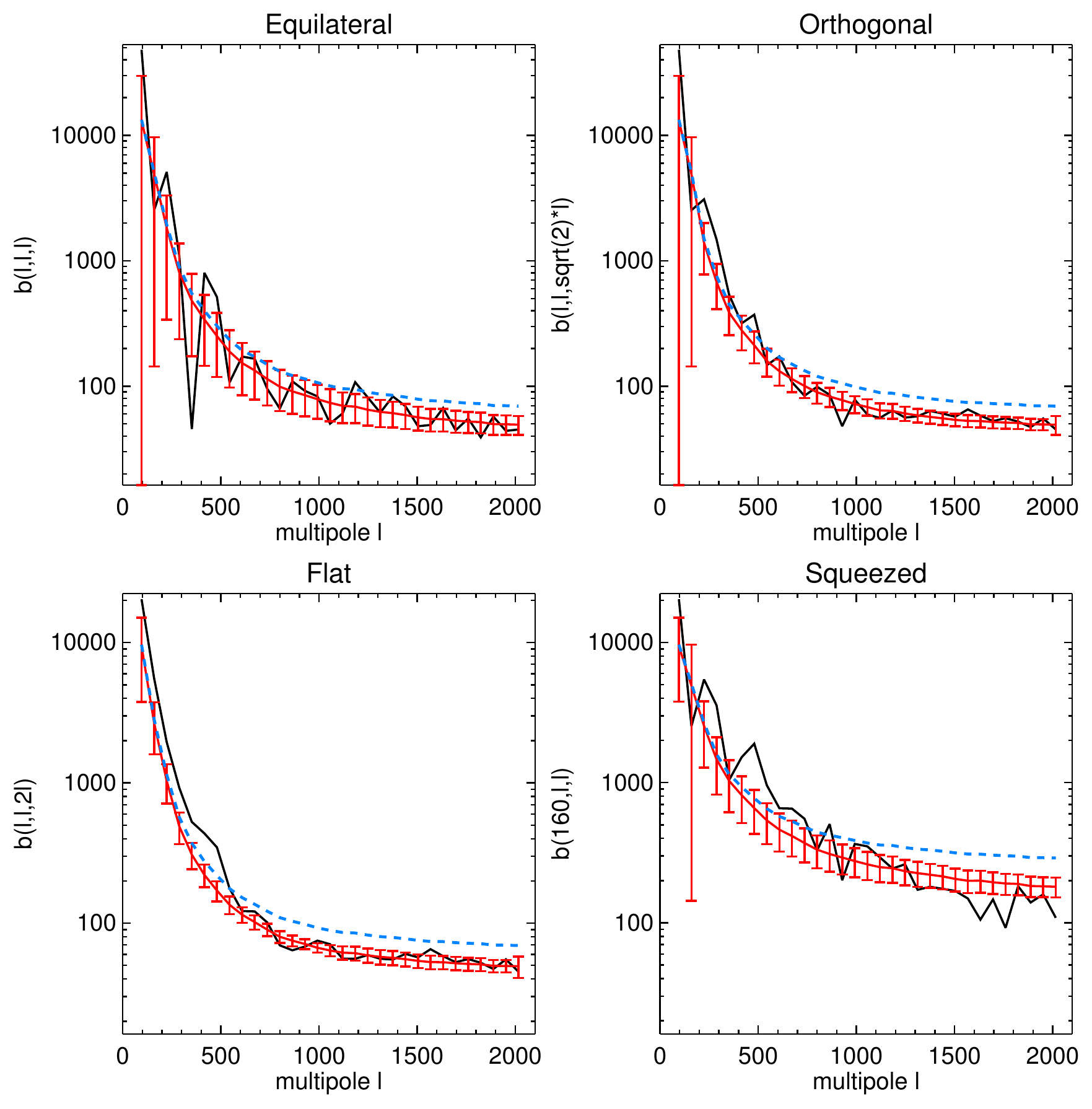}
\caption{From \cite{Lacasa2012}, IR + radio bispectrum at 350 GHz in some chosen configurations.}
\label{Fig:speconfirad350-erronmod}
\end{figure}

I have compared these measured bispectra with the prediction of the prescription (for radio, IR, IR+radio), where I used the power spectrum measured on the simulations and I fitted for the parameter $\alpha$. I also computed an expected value of $\alpha$, base on the sources catalog provided by \cite{Sehgal2010} and on Eq.\ref{Eq:def_alphapresp}. I found that the fitted $\alpha$ was consistent with the expected value, although somewhat lower : e.g. at 277 GHz I found $\alpha_\mr{fit} = 3.1\cdot 10^{-3}$ and $\alpha_\mr{exp}=4.7 \cdot 10^{-3}$. For the superposition of IR and radio signals, Fig.\ref{Fig:speconfirad350-erronmod} shows the measured bispectrum at 350 GHz in some configurations (black line), together with the prescription and its cosmic variance (red with error bars). The cosmic variance was computed with the $C_{2\times2\times2}$ and $C_{3\times3}$ terms (see Sect.\ref{Sect:bispcov}), and an $f_\mr{SKY}=1/8$ approximation to account for the octant replication in the \cite{Sehgal2010} simulations. Figure \ref{Fig:speconfirad350-erronmod} also shows in blue dashed line what the prescription would yield if the radio and IR populations were considered as $100\%$ correlated random fields.

We see that the prescription reproduces fairly well the measured bispectrum over the whole multipole range and for all configurations. The relative discrepancy between the prescription and the measurement is typically within $\pm$2\% at 30 \& 90 GHz and increases up to $\pm$30\% at 350 GHz. Figure \ref{Fig:speconfirad350-erronmod} shows in fact this latter worst case. We see that the difference between the prescription and the measurement is compatible with being caused by the cosmic variance of the measurement.\\
Clearly also, the blue dashed curve overestimates the IR+radio bispectrum at high multipoles, while the red curve does not underestimate it. Hence, the IR and radio populations have to be considered independent, as the prescription does. Indeed, the main contribution to the signals does not come from galaxies in the same redshift range, as the radio sources reside at $z\lesssim 1$ while the IR galaxies reside at $z\sim 1-5$.

\begin{figure}[htbp]
\centering
\includegraphics[width=\linewidth]{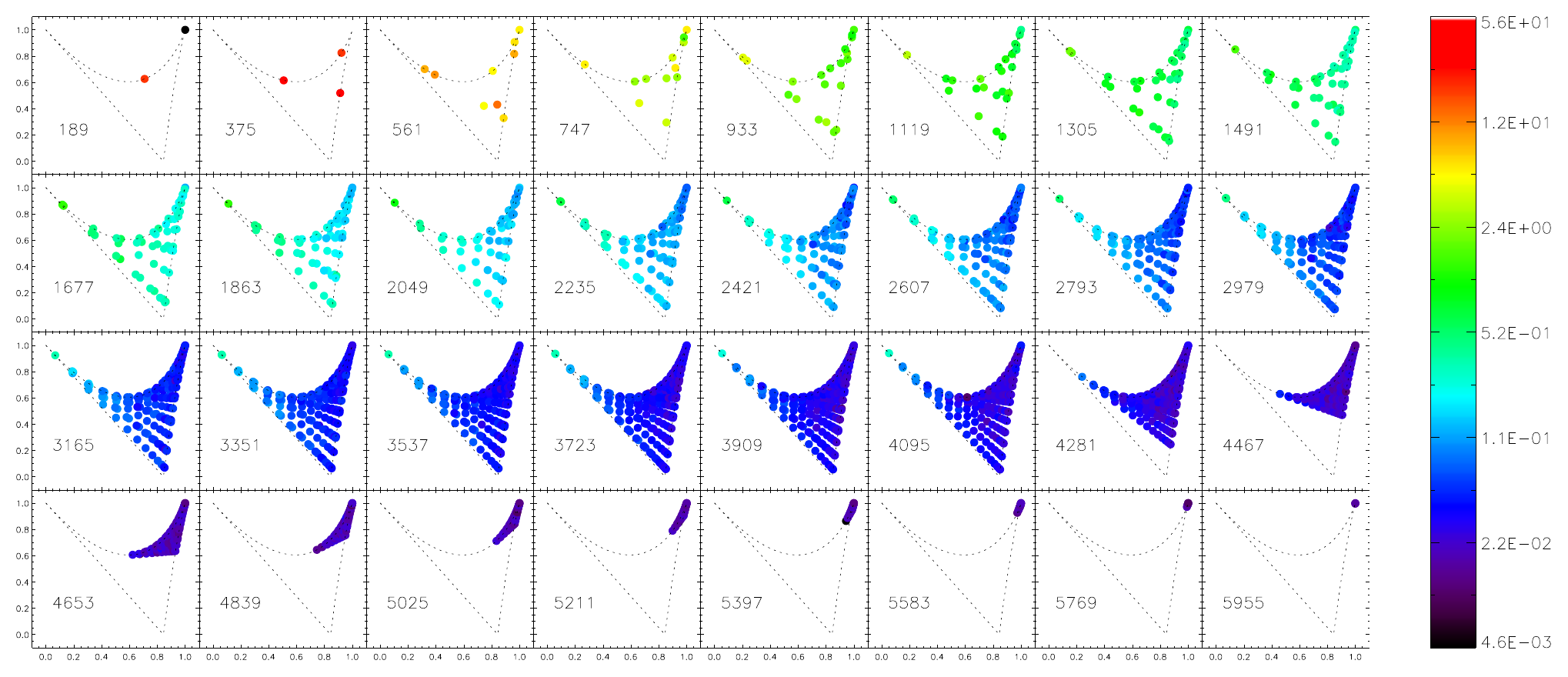}
\caption{From \cite{Lacasa2012}, IR bispectrum at 148 GHz in Jy$^3$/sr plotted with the parametrisation described in Sect.\ref{Sect:bisp_param}.}
\label{Fig:paramb_IR-148}
\end{figure}

While the radio bispectrum is constant, it is interesting to explore the configuration dependence of the IR bispectrum, which is not readily seen with the previous plots. For this, I have plotted in Fig.\ref{Fig:paramb_IR-148} the IR bispectrum at 148 GHz with the parametrisation described in Sect.\ref{Sect:bisp_param}.

We remind the reader that each subplot corresponds to a slice in a perimeter bin ($\lu+\ld+\lt \in [P,P+\Delta P]$), with perimeter increasing from left to right and top to bottom. The color code ranges between violet-blue for the lowest  values to red for the highest values. Figure \ref{Fig:paramb_IR-148} shows that the IR bispectrum generally decreases with scale, as already noted. Most interestingly in a given perimeter bin, we see that the bispectrum peaks in the squeezed configuration (upper left corner of each subplot). This is an important result, as the CMB bispectrum also peaks in squeezed for local type non-Gaussianity.

Hence, we expect the IR bispectrum to have a shape more correlated\footnote{Note that the correlation notion here is that of bispectrum shapes : whether the shapes have a significant overlap or not. This correlation will be quantified in Sect.\ref{Sect:AirandjointNG}. This correlation is independent of whether or not the signals are \emph{physically} correlated, and in the IR radio and CMB cases the signals can even be considered independent.} with the CMB one than the radio bispectrum does. This correlation, how it calls for a joint estimation of sources of non-Gaussianity and how it affects the estimation of CMB NG, will be presented in Sect.\ref{Sect:AirandjointNG} \& \ref{Sect:PNG_conta}.

\subsection{CIB description with the halo model}\label{Sect:CIB_HM}

The prescription is phenomenological and provides a quick and efficient prediction of the CIB bispectrum. However, a physical modelisation with the halo model would provide an actual model which consistently predicts the power spectrum and bispectrum (and possibly higher order), and hence can be constrained from data.\\
Several modelisations of the CIB clustering with the halo model have been proposed, all of which focus on the power spectrum only. This section, based on two articles \citep[][to be submitted]{Lacasa2013b,Penin2013}, describes the work I have done on modeling CIB polyspectra. We obtain a theoretical description at all orders, and implement it numerically for the bispectrum. This section also shows the improvement in parameter constraint that the bispectrum yield over a power spectrum analysis.

The CIB intensity in a given direction $\hn$ is the line-of-sight integral of the emissivity per comoving volume $\mathtt{j}$~:
\be
I(\hn,\nu) = \int \dd r \, a(z) \, \mathtt{j}(r(z)\hn,z,\nu)
\ee
with $\mathtt{j}$ in Jy/Mpc so that $I(\hn,\nu)$ has units of Jy/sr and may be converted to a temperature elevation at CMB frequencies through Planck's law. Through Sect.\ref{Sect:pastlightproj}, the angular CIB polyspectra can then be computed from the 3D $\mathtt{j}$ polyspectra, as long as the latter are diagonal~:
\be\label{Eq:jpolysptoCIBpolysp}
\mathcal{P}^{(n)}_\mr{CIB}(\ell_{1\cdots n}) = \int \frac{r^2 \, \dd r}{r^{2n}} \, a^{n}(z) \, \mathcal{P}^{(n)}_\mathtt{j}(k^*_{1\cdots n},z)
\ee
The IR emissivity traces the galaxy distribution but galaxies at the same redshift or in the same halo may have different IR fluxes. The simplest approach is to consider that the IR flux is stochastic with a distribution given by the number counts. Furthermore, the flux stochasticity is considered independent of the position stochasticity, e.g. the fluxes of two galaxies are not correlated whether they are close together or not. However, the flux distribution depends on the redshift, as the number counts do.

With these assumptions, all terms of the $n$-th order $\mathtt{j}$ polyspectrum which involve $n$ \emph{different} galaxies (i.e. non shot-noise) take the form~:
\be\label{Eq:polygaltopolyj}
\mathcal{P}^{(n)}_\mathtt{j}(k^*_{1\cdots n},z) = \overline{\mathtt{j}}(z)^n \, \mathcal{P}^{(n)}_\mr{gal}(k^*_{1\cdots n},z)
\ee
where $\overline{\mathtt{j}}(z)$ is the mean IR emissivity, given in terms of the number counts~:
\be
\overline{\mathtt{j}}(z) =\frac{1+z}{\frac{\dd r}{\dd z}} \int S \frac{\dd^2 n}{\dd S \, \dd z} \, \dd S
\ee
and where the galaxy polyspectrum terms are the non-shot noise ones and can be computed with the diagrammatic method described in Sect.\ref{Sect:diagrammatique}.

I have shown in \cite{Lacasa2013b} that the shot-noise contributions to $\mathcal{P}^{(n)}_\mathtt{j}$ can be tackled with equations to Eq.\ref{Eq:polygaltopolyj}. For example, if we consider a diagram with a contraction of second order, i.e. the same galaxy is `hit' twice by the correlation function, we need to consider the average flux squared $\lbra S^2 \rbra$ instead of the average flux $\lbra S \rbra$ as $\overline{\mathtt{j}}(z)$ does. This can be done by defining the p-th order emissivities~:
\be\label{Eq:porder_emm}
\mathtt{j}^{(p)}(z) = \frac{(1+z)^p \, r(z)^{2p-2}}{\frac{\dd r}{\dd z}} \int S^p \frac{\dd^2 n}{\dd S \, \dd z} \, \dd S
\ee
These are generalisations of the mean emissivity, as e.g., $\mathtt{j}^{(1)}(z) = \overline{\mathtt{j}}(z)$.\\
Then, the shot-noise 3D polyspectra should be multiplied by factors $\overline{n}_\mr{gal}(z)$ in accordance with the degree of p-th order emissivities introduced. For the above example of diagrams with a contraction of second order, we get~:
\begin{align}
\nonumber \mathcal{P}^{(n)}_{\mathtt{j},\mr{contrac2}}(k^*_{1\cdots n},z) &= \overline{\mathtt{j}}(z)^{n-2} \, \mathtt{j}^{(2)}(z) \ \overline{n}_\mr{gal}(z) \ \mathcal{P}^{(n)}_\mr{gal,contrac2}(k^*_{1\cdots n},z)\\
&= \overline{\mathtt{j}}(z)^{n-2} \, \mathtt{j}^{(2)}(z) \, \left(\mathcal{P}^{(n-1)}_\mr{gal,non-shot}(|\kk_1+\kk_2|^*,k^*_{3\cdots n},z) + \mr{perm.}\right)
\end{align}
where I used the resummation of shot-noise diagrams, described in Sect.\ref{Sect:diagrammatique}, for the second equality.\\
Let us note in passing, that this derivation of the shot-noise terms is consistent with the terms which were already known in the literature, i.e. the case where the correlation function 'hits' a single galaxy everytime (while this formalism also allows the derivation of the more complex shot-noise terms which arise at high orders). Indeed e.g. at the power spectrum level we have~:
\be
C_\ell^\mr{shot} = \int \frac{\dd r}{r^2} \, a^2(z) \, \mathtt{j}^{(2)}(z) \times 1 = \cdots = \int S^2 \frac{\dd n}{\dd S	} \dd S
\ee
i.e. the well-known shot-noise equation \citep[e.g.][]{Knox2001}.
\newline

\begin{figure}[htbp]
\centering
\includegraphics[width=.7\linewidth]{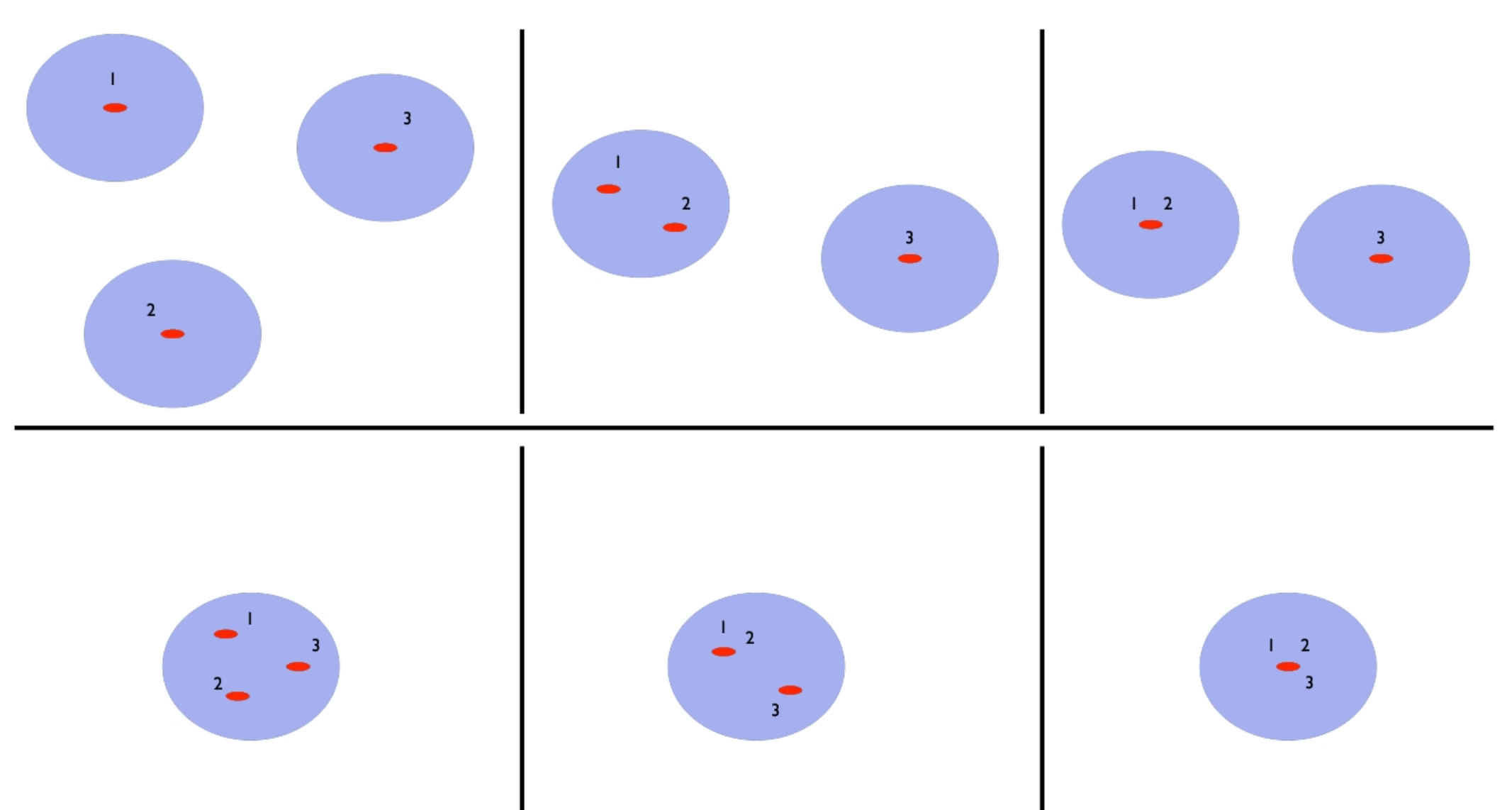}
\caption{Bispectrum diagrams. From left to right and top to bottom : 3-halo (3h), 2-halo 3-galaxies (2h3g), 2-halo 2-galaxies (2h2g), 1-halo 3-galaxies (1h3g), 1-halo 2-galaxies (1h2g) and 1-halo 1-galaxy (1h1g).}
\label{Fig:diagram_bisp}
\end{figure}

Now, we can tackle the computation of the CIB bispectrum. All the bispectrum diagrams are shown in Fig.\ref{Fig:diagram_bisp} with, from left to right and top to bottom : 3-halo (3h), 2-halo 3-galaxies (2h3g), 2-halo 2-galaxies (2h2g), 1-halo 3-galaxies (1h3g), 1-halo 2-galaxies (1h2g) and 1-halo 1-galaxy (1h1g).

The third and fifth diagrams (2h2g \& 1h2g) are shot-noise diagrams which can be resummed together, and in the following I will call shot2g (for shot-noise with two galaxies) this combination. I will also call simply 2-halo the 2h3g term, 1-halo the 1h3g term, and shot1g (shot-noise with one galaxy) the 1h1g term. The equations for these terms are respectively :
\begin{align}
\nonumber B_\mathtt{j}^\mr{1h}(k_{123},z) &= \mathtt{j}^{(1)}(z)^3 \int \dd M \,\frac{\lbra N_\mr{gal} (N_\mr{gal}-1) (N_\mr{gal}-2)\rbra}{\overline{n}^3_\mr{gal}(z)} \, \frac{\dd n_\mr{h}}{\dd M} \\
& \qquad \qquad \qquad \times u(k_1|M,z) \, u(k_2|M,z) \, u(k_3|M,z) \\
\nonumber B_\mathtt{j}^\mr{2h}(k_{123},z) &= \mathtt{j}^{(1)}(z)^3 \int \dd M_\texttt{ab} \frac{\lbra N (N-1)\rbra(M_\texttt{a})}{\overline{n}^2_\mr{gal}(z)} \, \frac{\lbra N \rbra(M_\texttt{b})}{\overline{n}_\mr{gal}(z)}\, \left.\frac{\dd n_\mr{h}}{\dd M}\right|_{M_\texttt{a}} \, \left.\frac{\dd n_\mr{h}}{\dd M}\right|_{M_\texttt{b}} \\
& \qquad \times u(k_1|M_\texttt{a},z) \, u(k_2|M_\texttt{a},z) \, u(k_3|M_\texttt{b},z) \ P_\mr{halo}(k_3,z) \ + \mr{perm.}\\
B_\mathtt{j}^\mr{3h}(k_{123},z) &= \mathtt{j}^{(1)}(z)^3 \int \dd M_{123} \left[\frac{\lbra N_\mr{gal} \rbra(M_i)}{\overline{n}_\mr{gal}(z)}\, \left.\frac{\dd n_\mr{h}}{\dd M}\right|_{M_i} \!\!\! u(k_i|M_i,z) \right]_{i=123} \!\!\!\! B_\mr{halo}(k_{123},z)\\
B_\mathtt{j}^\mr{shot2g}(k_{123},z) &= \mathtt{j}^{(1)}(z)^3 \, \mathtt{j}^{(2)}(z) \left(P_\mr{gal}^\mr{non-shot}(k_1) + P_\mr{gal}^\mr{non-shot}(k_2) + P_\mr{gal}^\mr{non-shot}(k_3)\right) \\
B_\mathtt{j}^\mr{shot1g}(k_{123},z) &= \mathtt{j}^{(3)}(z)
\end{align}

The halo polyspectra are biased versions of the dark matter polyspectra with the local bias scheme described in Sect.\ref{Sect:localbias}, while the dark matter polyspectra are computed from perturbation theory. I have been working at tree-level, that is to take the lowest order in the biasing scheme and perturbation theory which gives a non-zero result. For example, for the halo power spectrum we just need to go at first order in local biasing and in perturbation theory (i.e., linear theory described in Sect.\ref{Sect:lineargrowth}), we get~:
\be
P_\mr{halo}(k|M_1,M_2,z) = b_1(M_1) \, b_1(M_2) \, P_\mr{lin}(k,z)
\ee
For the halo bispectrum, linear theory predicts a vanishing dark matter bispectrum for Gaussian initial conditions. Hence, we need to go to second order in local biasing and/or perturbation theory, we get~:
\begin{align}
\nonumber B_\mr{halo}(k_{123}|M_{123},z) &= b_1(M_1) b_1(M_2) b_1(M_3) \, B_\mr{2PT}(k_{123},z) \\
& \quad + b_1(M_1) b_1(M_2) b_2(M_3) P_\mr{lin}(k_1,z) \, P_\mr{lin}(k_2,z) + \mr{perm.}
\end{align}
where the dark matter bispectrum at second order in perturbation theory (2PT) is \citep{Fry1984}~:
\be\label{Eq:Bhalo_2PT}
B_\mr{2PT}(k_{123},z) = F^s(\kk_1,\kk_2) \, P_\mr{lin}(k_1) \, P_\mr{lin}(k_2) + \mr{2\ perm.}
\ee
with the 2PT kernel~:
\be\label{Eq:Fskernel}
F^s(\kk_i,\kk_j) = \frac{5}{7} + \frac{1}{2} \cos \left(\theta_{ij}\right) \left(\frac{k_i}{k_j} + \frac{k_j}{k_i}\right) + \frac{2}{7} \cos^2\left(\theta_{ij}\right)
\ee
Note that although Eq.\ref{Eq:Fskernel} is singular in the squeezed limit, Eq.\ref{Eq:Bhalo_2PT} is not, as divergences cancel themselves in the sum over permutations.\\
Hence, there are two sources for the halo non-Gaussianity : the non-linearity of halo biasing with respect to dark matter, and the non-linearity of gravitational infall producing dark matter NG.\\
Note that considering primordial NG (PNG) would be more complex as it introduces a momentum dependence of the bias factors \citep{Dalal2008}, we however checked that the corresponding 3-halo term was negligible for $\fnl$ up to 50. Furthermore \planck results constrained PNG to be consistent with zero.

With these equations, the 2-halo and 3-halo term can be rewriten~:
\begin{align}
\nonumber B_\mr{gal}^\mr{2h}(k_{123},z) &=  \mathcal{G}_1(k_1,k_2,z) \; P_\mathrm{lin}(k_3 , z) \, \mathcal{F}_1(k_3,z) +\ \mathcal{G}_1(k_1,k_3,z)\; P_\mathrm{lin}(k_2 , z) \,\mathcal{F}_1(k_2,z) \\
& \qquad +\ \mathcal{G}_1(k_2,k_3,z) \; P_\mathrm{lin}(k_1 ,z ) \, \mathcal{F}_1(k_1,z)\\
\nonumber B_\mr{gal}^\mr{3h}(k_{123},z) &= \mathcal{F}_1(k_1,z)\, \mathcal{F}_1(k_2,z)\, \mathcal{F}_1(k_3,z) \times \left[F^s(\mathbf{k_1},\mathbf{k_2})\, P_\mathrm{lin}(k_1,z)\, P_\mathrm{lin}(k_2,z) +\mathrm{perm.}\right] \\
\nonumber & \qquad +\ \mathcal{F}_1(k_1,z)\, \mathcal{F}_1(k_2,z)\, \mathcal{F}_2(k_3,z) \times P_\mathrm{lin}(k_1 , z)\, P_\mathrm{lin}(k_2 , z)\\
\nonumber & \qquad +\ \mathcal{F}_1(k_1,z)\, \mathcal{F}_2(k_2,z)\, \mathcal{F}_1(k_3,z) \times P_\mathrm{lin}(k_1 , z)\, P_\mathrm{lin}(k_3 , z)\\
& \qquad +\ \mathcal{F}_2(k_1,z)\, \mathcal{F}_1(k_2,z)\, \mathcal{F}_1(k_3,z) \times P_\mathrm{lin}(k_2 , z)\, P_\mathrm{lin}(k_3 , z)
\end{align}
where I introduced the notations~:
\be
\mathcal{F}_i(k,z) = \int \dd M \frac{\langle N_\mathrm{gal}(M)\rangle}{\overline{n}_\mathrm{gal}(z)} \,\frac{dN_h}{dM}(M,z) \,b_i(M,z) \, |u(k|M,z)|
\ee
and~:
\be
\mathcal{G}_1(k_1,k_2,z) = \int \dd M \frac{\langle N_\mathrm{gal}(N_\mathrm{gal}-1)\rangle}{\overline{n}_\mathrm{gal}(z)^2} \,\frac{dN_h}{dM}(M,z) \, b_1(M,z) \, |u(k_1|M,z) \, u(k_2|M,z)|
\ee
In the following, I will call ``3hcos'' the part of the 3h term containing the 2PT kernel $F^s$, and simply ``3h'' the part with $\mathcal{F}_2$ factors coming from second order halo biasing.
\newline

Based on the formalism described above, I have developed a fast and complete a code to compute the galaxy 3D power spectrum and bispectrum, and the CIB angular power spectrum and bispectrum. As stated previously, I use the \cite{ShethTormen1999} mass function and associated bias parameters\footnote{Indeed, this is the most recent mass function for which the second order bias is available.}, the \cite{NFW1997} halo profile. I use the WMAP7 cosmological parameters \citep{Komatsu2011}. The IR emissivities are computed with number counts by \cite{Bethermin2011}, and the HOD is described in Sect.\ref{Sect:HM_descrip}. I use fiducial HOD parameters fitted to reproduce the \planck CIB angular power spectrum \citep{planck2011-CIB} : $M_\mr{min}= 10^{11.5} \Msun\,$, $M_\mr{sat}= 10 M_\mr{min}\,$, $\sigma_{\log M} = 0.65$ and $\alpha_\mr{sat}=1.4$.

\begin{figure}[htbp]
\centering
\includegraphics[width=.9\linewidth]{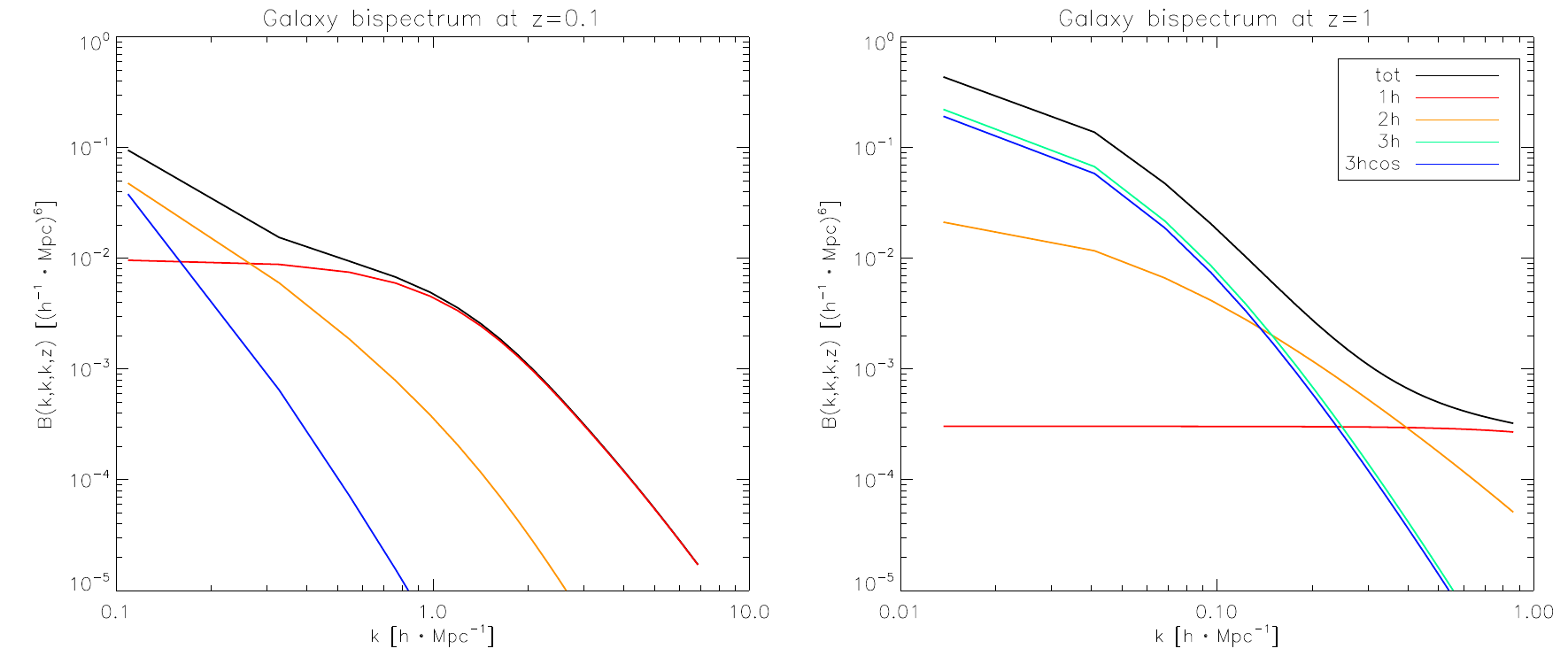}
\caption{From \cite{Lacasa2013b}, contributions to the galaxy equilateral bispectrum at $z$=0.1 (left panel) and $z$=1 (right panel). Note that the $k$-range depends on redshift, as these are the modes that will project onto the multipoles of interest (see text for details). The 3h term is negative at $z$=0.1 and thus absent of the left panel.}
\label{Fig:Bk2z}
\end{figure}

Figure \ref{Fig:Bk2z} shows the resulting galaxy bispectrum and its different components in the equilateral configuration at $z$=0.1 and $z$=1. Note that the $k$-range depends on redshift. Indeed, in view of the computation of the angular bispectrum and to speed up an otherwise time expensive code, we only compute in the code the 3D power spectra and bispectra for modes which project on the multipoles grid (through $k^*=(\ell+1/2)/r(z)$).

We see that the 1-halo term is flat on large scales and decreases after $\sim 1$ h$\cdot$Mpc$^{-1}$, a scale corresponding to the typical size of a massive halo ($M=10^{12-13}\Msun$). The 2-halo and 3-halo terms decrease monotonically with scale, with the 3h and 3hcos terms having the steeper slopes while the 2h term has an intermediate slope. Note that the 3h term is negative at low redshifts as, as shown in Fig.\ref{Fig:ShandT}, the second order bias is negative on a broader mass range at lower redshifts.

\begin{figure}[htbp]
\centering
\includegraphics[width=.6\linewidth]{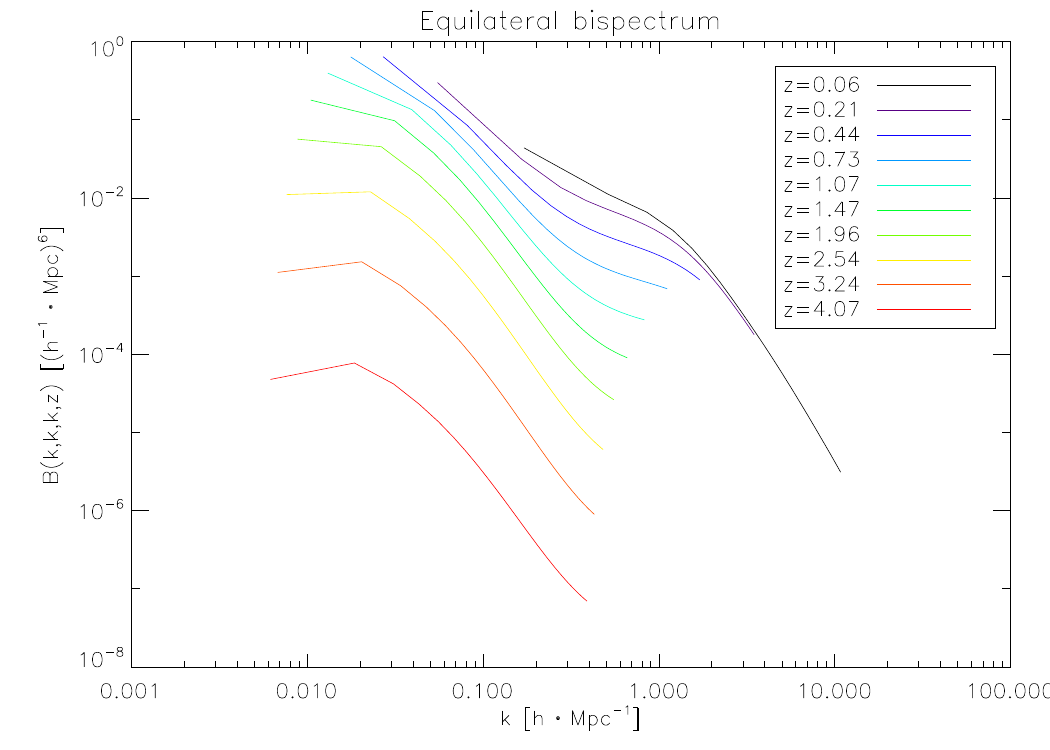}
\caption{From \cite{Lacasa2013b}, total galaxy equilateral bispectrum as a function of redshift.}
\label{Fig:Bequi_z}
\end{figure}

Figure \ref{Fig:Bequi_z} shows the evolution of the total galaxy bispectrum in the equilateral configuration, on the range of redshift where there is significant IR emission.

We see that the bispectrum increases considerably with time (whereas the power spectrum increases much more slowly), as non-Gaussianity builds up with non-linear physics entering in action. The figure also shows clearly the projection effect through which the $k$-range evolves with redshift (particularly at low redshift, afterwards the comoving distance increases more slowly). Note that the ``bump'' visible on large scales corresponds to the wavevector $k_\mr{eq}$ \footnote{$k_\mr{eq}$ is the wavevector whose physical wavelength becomes equal to the Hubble distance at the matter-radiation equality, i.e. $\lambda_\mr{eq} = c / H(a_\mr{eq})$.} where the linear power spectrum slope breaks from $k$ to $k^{-3}$ \cite[see][whose fit I use]{Eisenstein1999} \footnote{The \cite{Eisenstein1999} fit I use does not incorporate the effect of Baryonic Acoustic Oscillations, however these oscillations would be washed out by the redshift integration.}.
%Interestingly the ``bump'' visible on large scales at high redshift is a consequence of the bump in the linear power spectrum when its slope changes from $k$ to $k^{-3}$ \cite[see][whose fit we use]{Eisenstein1999}.
\newline

Using the IR emissivities computed from number counts by \cite{Bethermin2011}, we can integrate over redshift Eq.\ref{Eq:jpolysptoCIBpolysp} to compute the CIB angular bispectrum.\\
However, we are interested in unresolved galaxies hence we should account for the flux cut effect. Indeed, the flux cut discards the brightest sources, hence it has an effect on the average emissivity and higher order  emmisivities through Eq.\ref{Eq:porder_emm}, where the integral runs from $S=0$ to $S_\mr{cut}$. As it discards sources, the flux cut also has an effect on the galaxy clustering~: there are less galaxies to consider, particularly at low redshift, as they are resolved (and hence masked as we focus on unresolved galaxies). The flux cut effect is modeled with an effective redshift cut in Eq.\ref{Eq:jpolysptoCIBpolysp}, considering that galaxies below $z_\mr{cut}$ are resolved, where we have~:
\be
S_\mr{cut} = \frac{L_*}{4\pi \,  d^2_L(z_\mr{cut})}
\ee
where $d_L$ is the luminosity distance and $L_*$ is the typical galaxy IR luminosity (technically~: the knee of the luminosity function).\\
Thus in Eq.\ref{Eq:jpolysptoCIBpolysp} the redshift integral runs from $z_\mr{cut}$ to $z_\mr{max}$ ($z_\mr{max}=7$ with the number counts we use, although the contribution of the last redshift bins is negligible).

\begin{figure}[htbp]
\centering
\includegraphics[width=.7\linewidth]{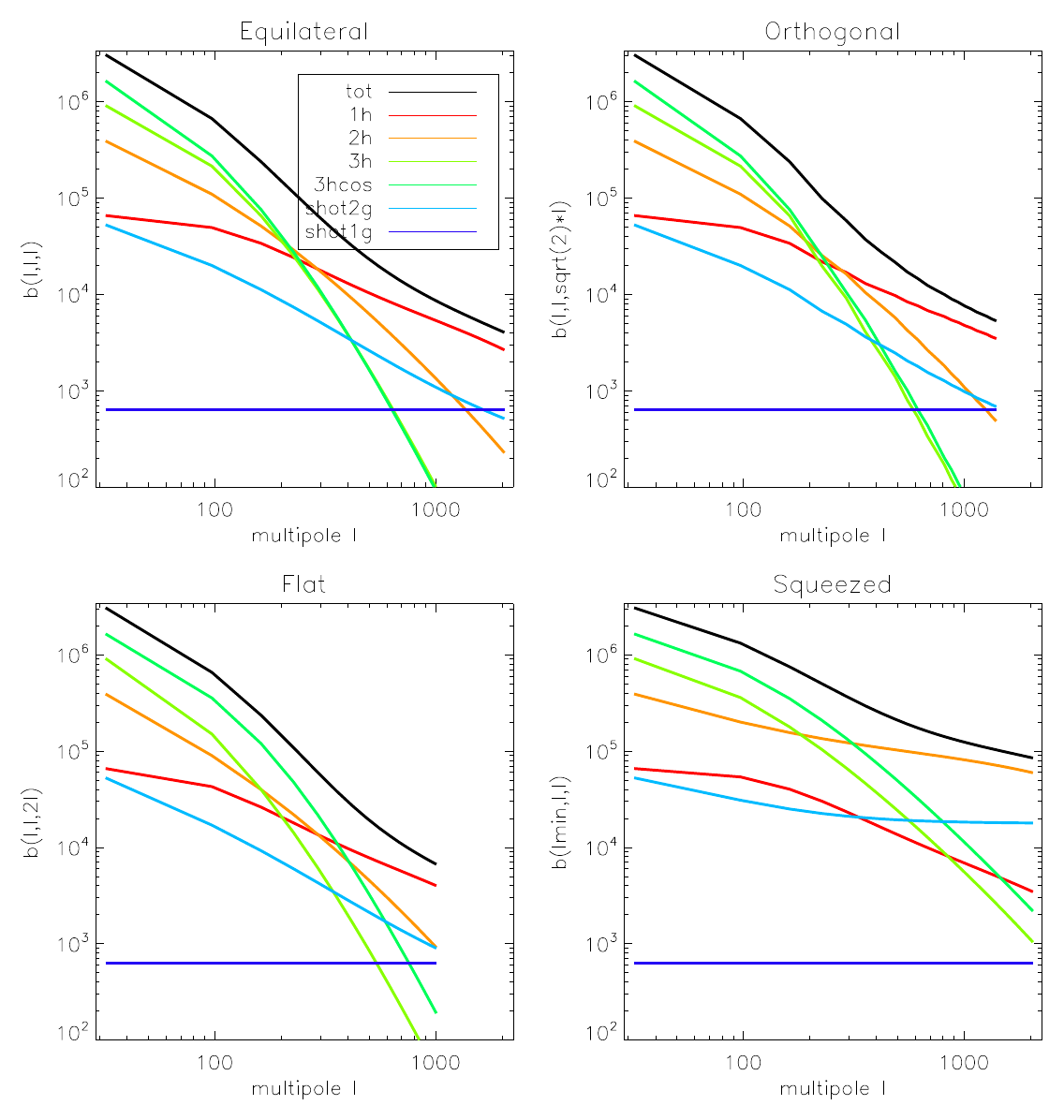}
\caption{From \cite{Lacasa2013b}, CIB angular bispectrum at 857 GHz in some chosen configurations, and the different terms contributing to it. The black line is the total bispectrum while the color lines give the different terms : shot-noise with 1 galaxy in violet, shot-noise with 2 galaxies in blue, 1-halo in red, 2-halo in orange, 3-halo from non-linear halo biasing in light green, and 3-halo from perturbation theory in darker green.}
\label{Fig:speconf_IR350}
\end{figure}

The CIB total angular bispectrum and its different terms, computed over a range of multipoles of interest for \planck ($\ell = 32 \cdots 2048$ with a step $\Delta\ell=64$), are shown in Fig.\ref{Fig:speconf_IR350} in some chosen configurations at 857 GHz.

We see that the 3h and 3hcos terms (green curves) dominate on large angular scales, the 2-halo term (orange) is important on intermediate scales (towards $\ell\sim 300$ but on a limited multipole range), and the 1-halo term (red) dominates on smaller scales. Finally, the shot-noise term shot1g (dark blue) dominates on the smallest scales, which are not accessible to the \planck resolution but are of interest for the \textit{Herschel} telescope.\\
It is worth emphasizing that these trends do depend on the HOD parameters used. For example, by increasing $\alpha_\mr{sat}\,$\footnote{to values which are nevertheless not consistent with N-body simulations.} we give more weight to massive halos at low redshift. It is quite possible for the 1-halo term to dominate the whole multipole range (as it happens for the tSZ signal at the power spectrum level).\\
Interestingly, in the present case we see that the 3hcos term dominates the 3h term in the flat isosceles and squeezed configurations. This is due to the $F^s$ kernel which is more important in flat triangles than in general ones. We also observe that the 2-halo term is often dominated by other terms in most configurations, but it takes its revenge in the squeezed limit, where it dominates at small angular scales ($\ell > 400$). At even smaller angular scales (not shown on the figure), the 2-halo term eventually crosses the shot2g term so that the later dominates the bispectrum in this case.
\newline

\begin{figure}[htbp]
\centering
\includegraphics[width=1\linewidth]{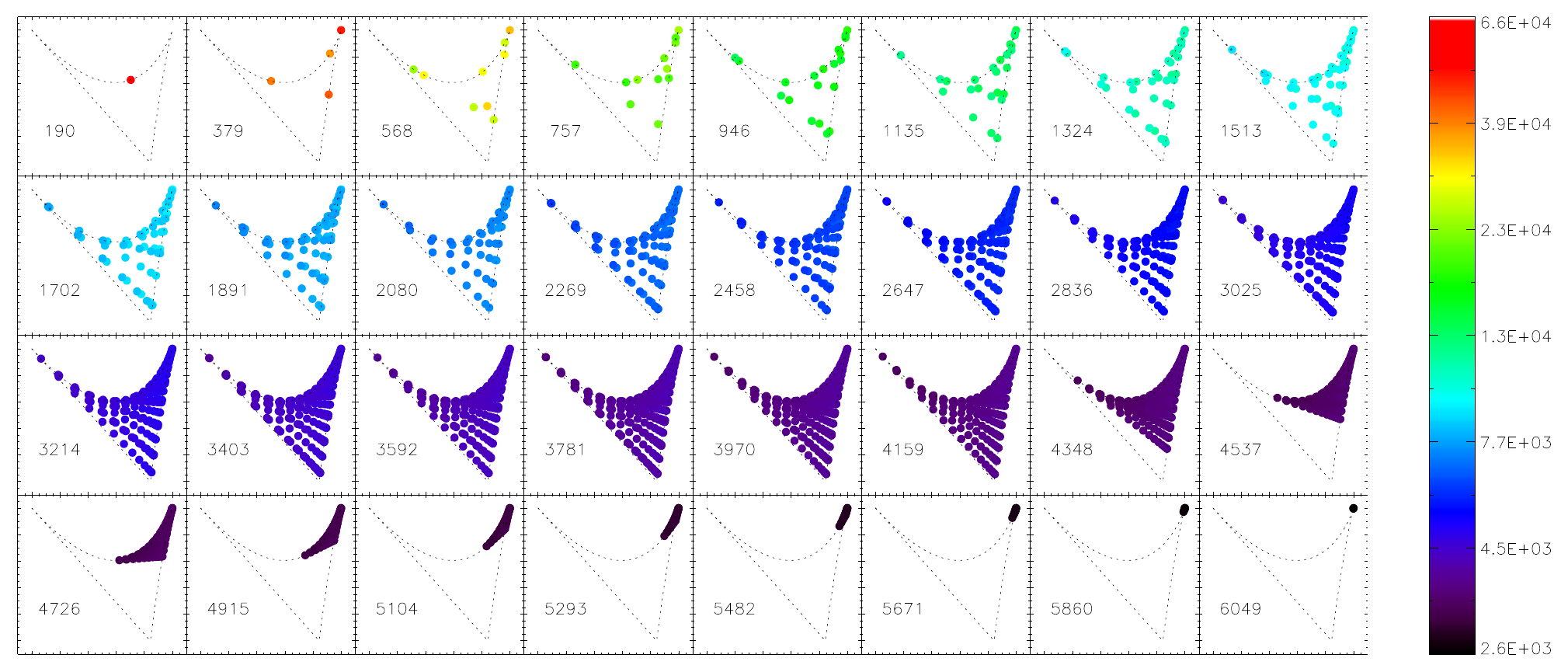}\\\
\\
\includegraphics[width=1\linewidth]{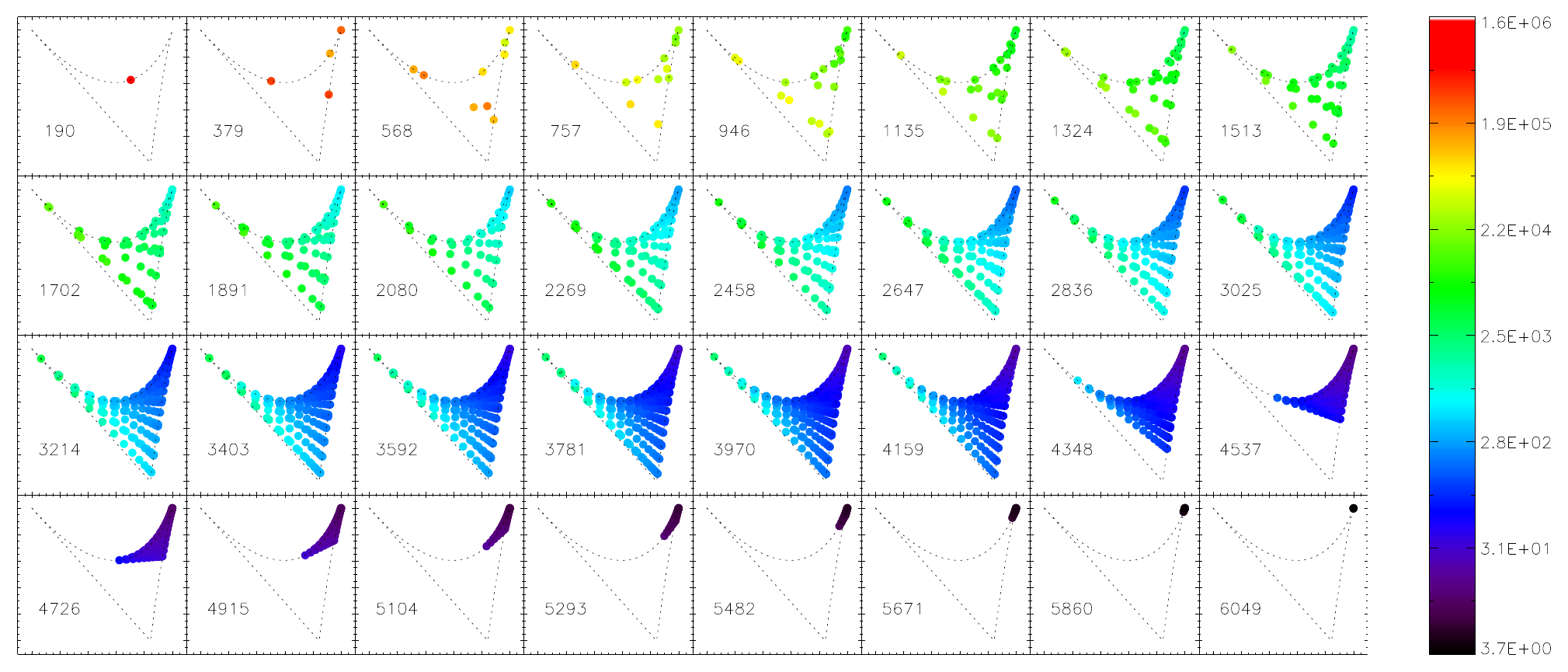}\\\
\\
\includegraphics[width=1\linewidth]{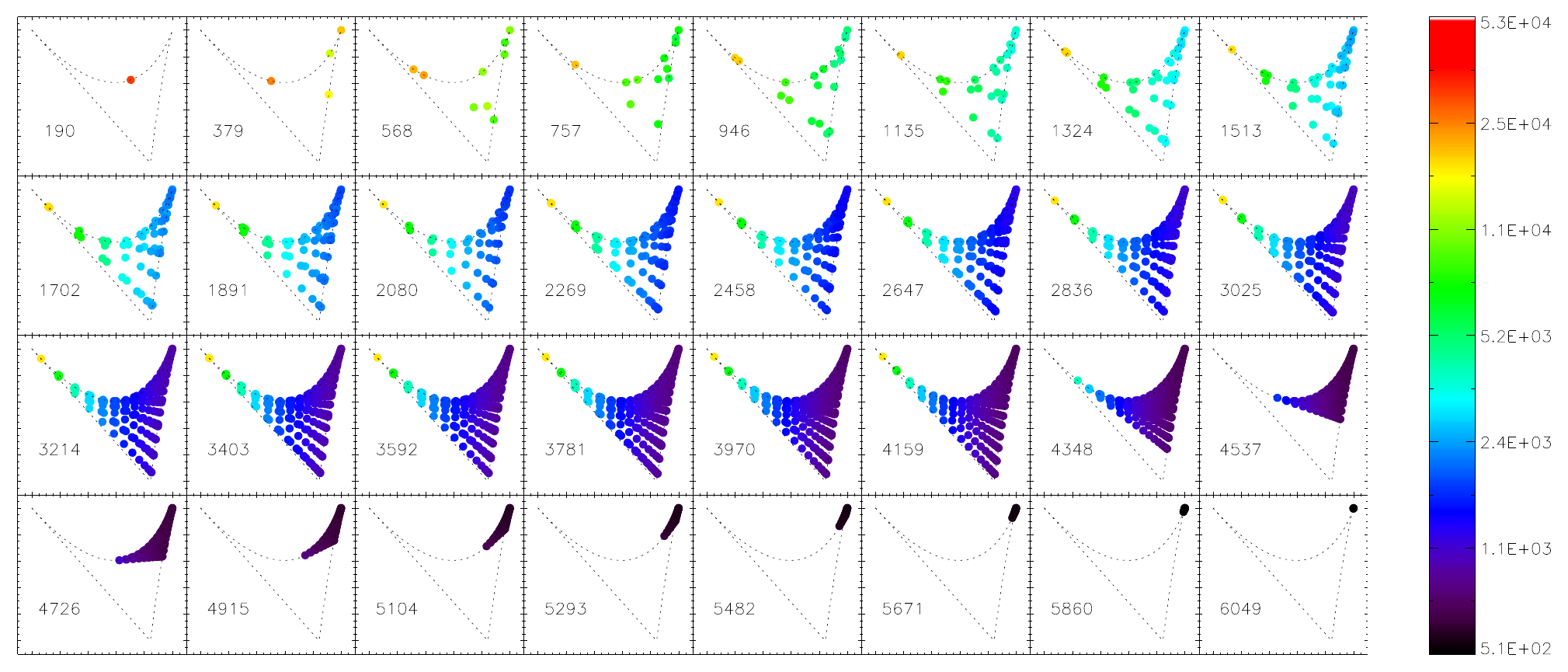}
\caption{From \cite{Lacasa2013b}, some terms of the CIB angular bispectrum plotted in the parametrisation. From top to bottom : 1h, 3hcos, shot2g. The color code is logarithmic and adapted to each term. Each of the 32 subplots corresponds to a perimeter bin, with the center of the bin indicated in it lower left corner ; the squeezed triangles are present in the upper left corner, the equilateral triangles in the upper right and the flat isosceles in the lower right.}
\label{Fig:bl_CIB_inparam}
\end{figure}

We can also examine the configuration dependence of the CIB bispectrum and its different terms. In agreement with Sect.\ref{Sect:presp}, we found that the CIB bispectrum peaks in the squeezed configurations. Figure \ref{Fig:bl_CIB_inparam} shows some of the terms plotted in the parametrisation, note that the color code is logarithmic and adapted to each term.

The upper panel shows the 1-halo term. We see that it is strongly scale dependent but has little dependence on the configuration, as it is nearly constant in a perimeter bin. The middle panel shows the 3hcos term. It also has a strong scale dependence and shows a peculiar configuration dependence : it peaks in the flat triangles (from squeezed to flat isosceles), because of the $F^s$ kernel, but the squeezed configuration takes over the flat isosceles in high perimeter bins, because of the halo profile which decreases with $\ell$. The lower panel shows the shot2g term. It peaks strongly in the squeezed configuration. The 2-halo term shows a similar behaviour, although with a different dynamic range and less difference between the squeezed and the other configurations. The 3-halo term also peaks in squeezed but less strongly than the 2-halo and the shot2g terms. The last term is the shot1g one but its configuration dependence is trivial as it is constant. Hence, we see that the scale and configuration dependence of each of the CIB bispectrum terms are different ; thus the different terms may be discriminated in measurements using this information.
\newline

In \cite{Penin2013}, we have looked at the influence of the IR emissivity on the bispectrum, using three different IR emissivity models. We have shown that the IR emissivity has few influence on the configuration dependence of the bispectrum terms, however it strongly affects the scale dependence and the amplitude of the bispectrum. In particular we have shown that models which may be degenerate at the power spectrum level (i.e. producing indistinguishable $C_\ell$ once the HOD is fitted) may be distinguished at the bispectrum level, as the bispectra they predict may differ by up to a factor of 10.
\newline

\begin{figure}[htbp]
\centering
\includegraphics[width=1\linewidth]{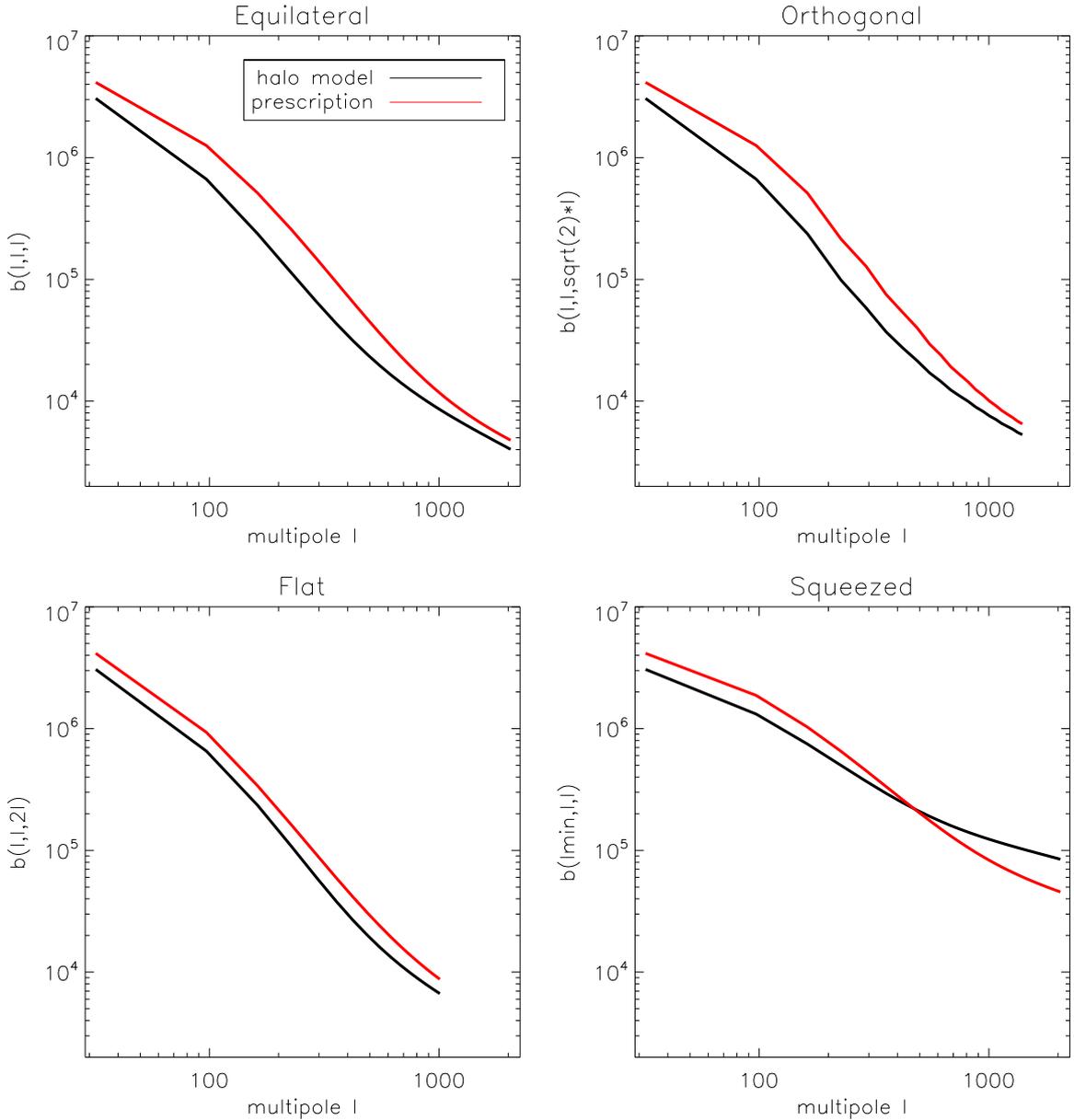}
\caption{From \cite{Lacasa2013b}, comparison between the halo-model bispectrum and the empirical prescription at 857 GHz in some chosen configurations.}
\label{Fig:speconf_comp_wpresp}
\end{figure}

I have also compared in \cite{Lacasa2013b} the CIB bispectrum computed from the halo model with the one derived from the empirical prescription. For this purpose, I used the same amplitude parameter $\alpha$ as the one deduced from \cite{Sehgal2010} simulations (see Sect.\ref{Sect:presp}). Furthermore I used the power spectrum computed with the halo model. In doing so, I am testing the functional form of the prescription. The comparison is displayed in Fig.\ref{Fig:speconf_comp_wpresp}.

The prescription reproduces the overall shape and the order of magnitude of the halo-model bispectrum. The prescription has generally a higher amplitude than the halo-model bispectrum, which can be compensated by rescaling the amplitude parameter $\alpha$ by a factor $\sim3/4$. Interestingly, the halo-model bispectrum departs from the prescription at multipoles $\ell>300$ in the squeezed limit, that is when the 2-halo term starts to dominate. Hence, the prescription is a good phenomenological first order approach, but as expected a comprehensive modeling of the bispectrum based on the halo model is necessary for detailed studies and to analyse actual measurements.

\begin{figure}[htbp]
\centering
\includegraphics[width=.325\linewidth]{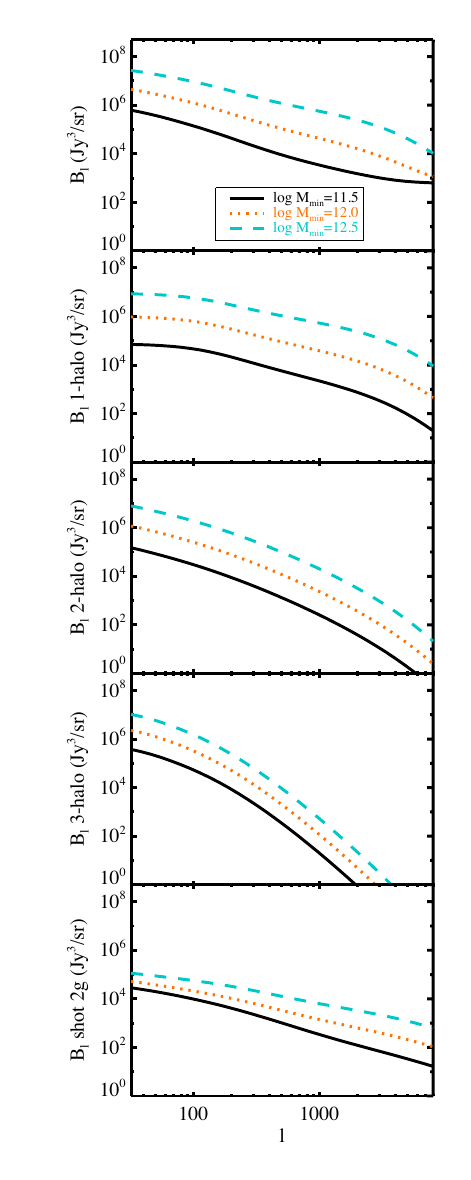}
\includegraphics[width=.325\linewidth]{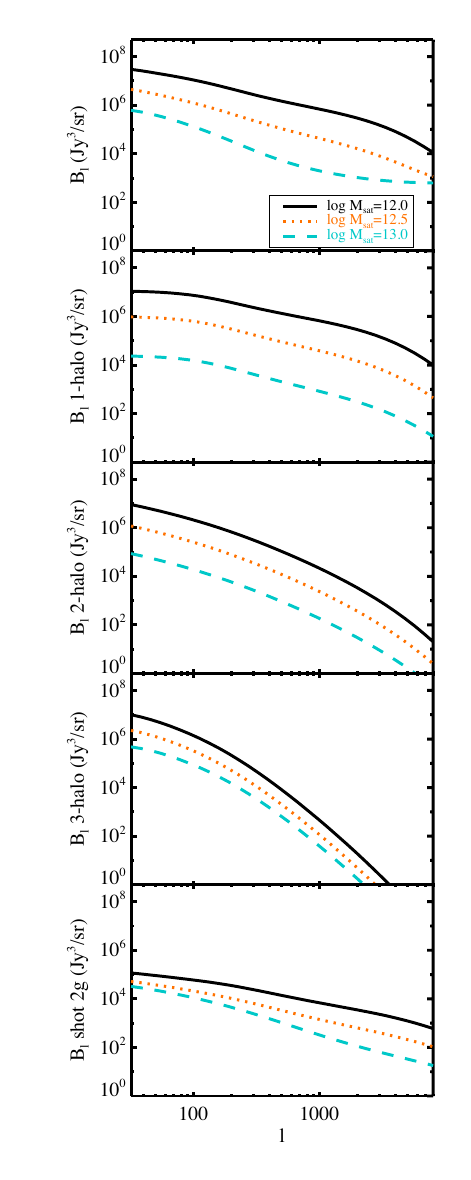}
\includegraphics[width=.325\linewidth]{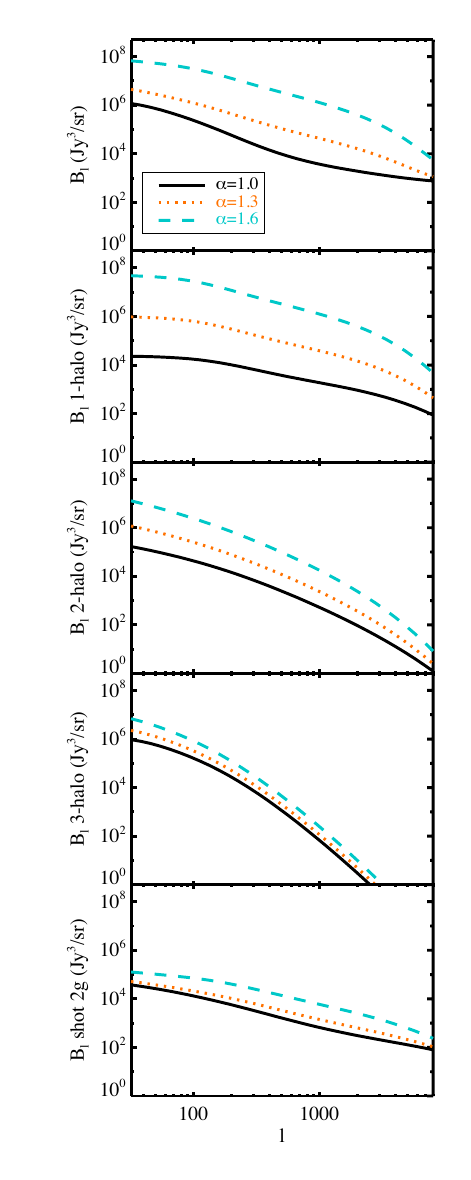}
\caption{From \cite{Penin2013}, variations of the CIB equilateral angular bispectrum and its different terms with the HOD parameters. From left to right : varying $M_\mr{min}$, $M_\mr{sat}$ and $\alpha_\mr{sat}$. The solid line shows the bispectrum for the lowest value of the parameter of interest, the dotted line for an intermediate value, and the dashed line for the highest value of the parameter.}
\label{Fig:bl_vs_HOD}
\end{figure}

Indeed, the goal of using a physically-base model of the bispectrum is to constrain it with observations. To this end it is interesting to concentrate on the variation of the CIB bispectrum with model parameters. For illustration, Fig.\ref{Fig:bl_vs_HOD} shows the variation of the CIB bispectrum an its different terms, in the equilateral configuration, when one HOD parameter is varied while the others are fixed.

From left to right we vary respectively : $M_\mr{min}$, $M_\mr{sat}$ and $\alpha_\mr{sat}$. These parameters govern respectively the minimal mass for a halo to contain a (central) galaxy, the minimal mass for a halo to contain satellite galaxies, and the number of satellite at high mass ($M \gg M_\mr{sat}$). For example, we see that varying $\alpha_\mr{sat}$ has a strong effect on the 1-halo term, and slightly less effect on the other terms. The 2-halo term is the most affected, and the 3-halo term (which here regroups the plain 3h and the 3hcos) is the least affected. Indeed increasing $\alpha_\mr{sat}$ affects mostly the massive jalos where the number of galaxies increases strongly ($N_\mr{sat} \propto M^{\alpha_\mr{sat}}$). As the 1-halo term is cubic with the number of galaxies at a given mass ($\lbra N(N-1)(N-2)\rbra$ factor), increasing $\alpha_\mr{sat}$ blows up $b^\mr{1h}$.
\newline

\begin{figure}[htbp]
\centering
\includegraphics[width=.49\linewidth]{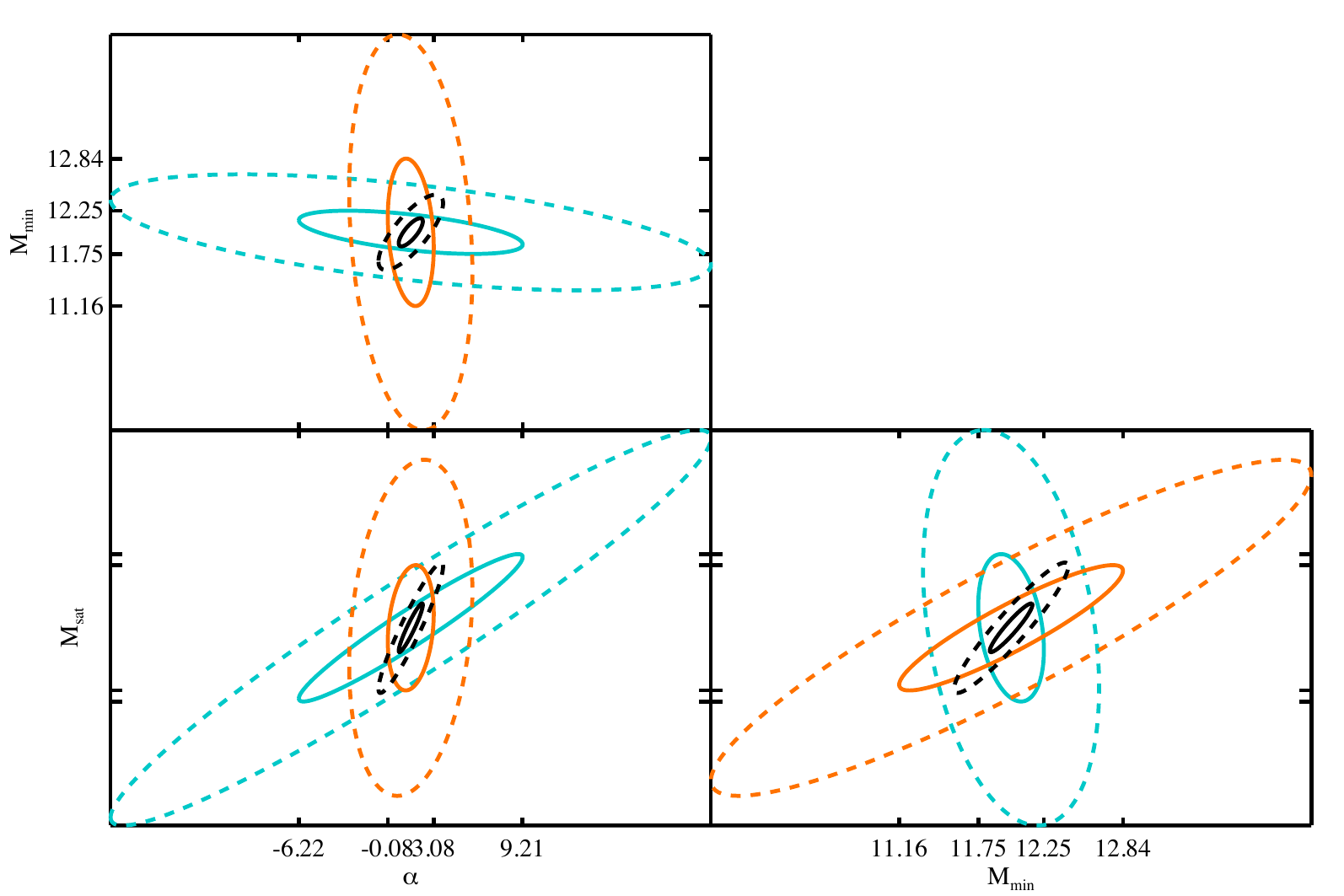}
\includegraphics[width=.49\linewidth]{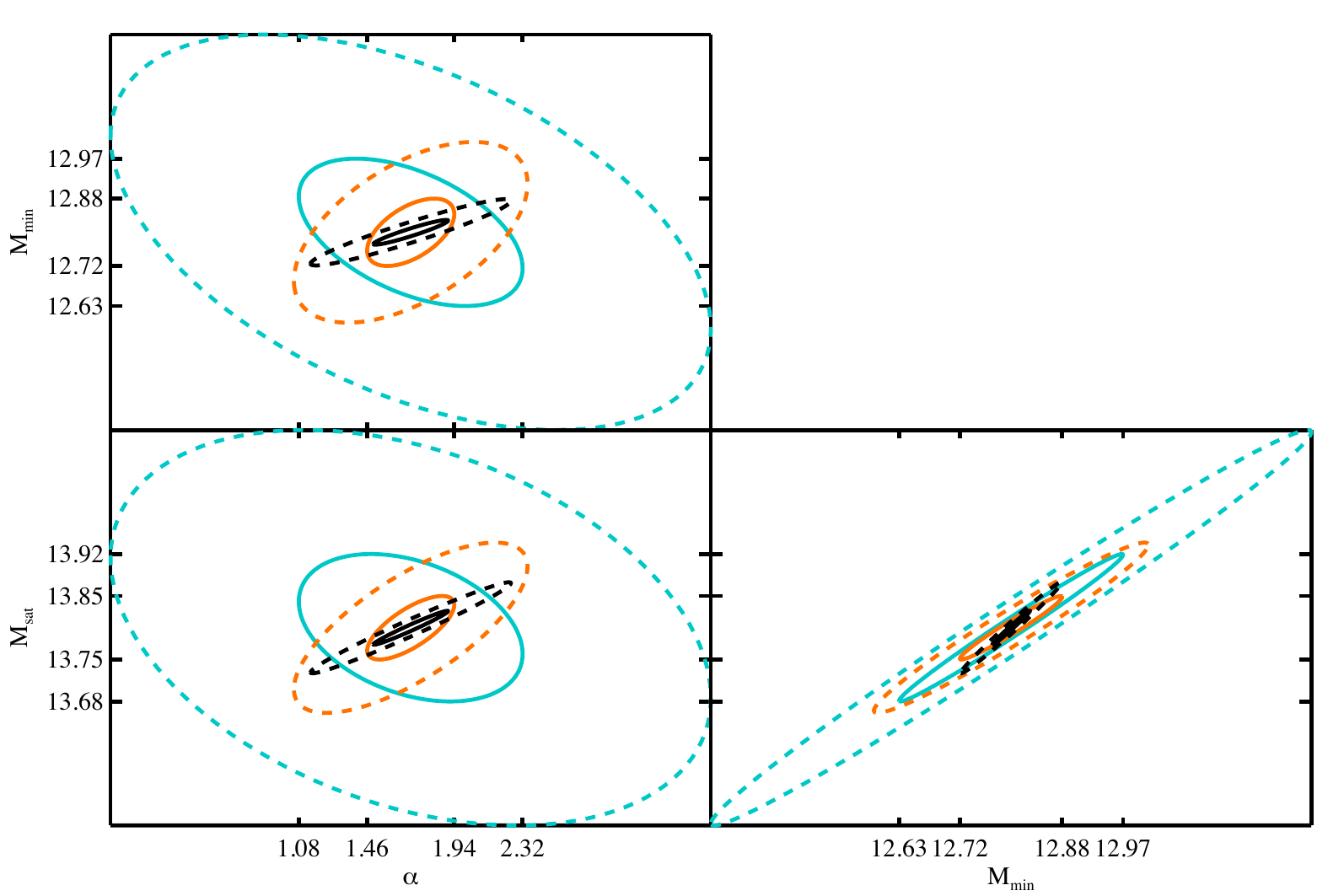}
\caption{Fisher matrix ellipses of HOD parameter constraints, with the power spectrum (blue), the bispectrum (orange) and combined (black). Solid lines are $1\sigma$ contours and dashed are $2\sigma$ contours. Left figures are at 217 GHz and right at 857 GHz.}
\label{Fig:ellipses_hod}
\end{figure}

We have observed that the variation of HOD parameters has more effect on the bispectrum than on the power spectrum. Hence, the bispectrum seems to be a powerful observable to constrain them. This is to be balanced however by the fact that, from the measurement point of view, bispectrum coefficients have much lower SNR than power spectrum ones, and the bispectrum is more difficult to measure. To quantify the constraint that the power spectrum and bispectrum may respectively yield on the HOD parameters, we have conducted in \cite{Penin2013} a Fisher analysis. The Fisher matrix of a set of parameters $(\theta_\alpha)_\alpha$ has entries~:
\be
F_{\alpha \beta} = \leftexp{T}{\frac{\partial \mathcal{O}}{\partial \theta_\alpha}} \cdot C^{-1} \cdot \frac{\partial \mathcal{O}}{\partial \theta_\beta}
\ee
where $\mathcal{O}$ is the observable (in our case the power spectrum or the bispectrum) ordered as a vector, and $C$ is its covariance matrix. If the observable can be considered to vary linearly with parameters within the error bars (e.g., if the error bars are sufficiently small), then the likelihood of the parameters $(\theta_\alpha)_\alpha$ is Gaussian and $F$ is its covariance matrix. In other words $F^{-1}_{\alpha \alpha}$ is the error bar on the parameter $\theta_\alpha$ when fitting to data. The off-diagonal coefficients give the correlation of error bars on different parameters.\\
To compute the Fisher matrices, we considered binned estimations of the power spectrum and bispectrum with $\ell=32\cdots2048$ and $\Delta\ell=64$, we considered \planck Bluebook's beam and noise values \citep{PlanckBluebook} and $f_\mr{sky}=50\%$\footnote{In the $f_\mr{SKY}$ approximation, lower sky fractions increase the error bars by a factor $\sqrt{f_\mr{SKY}}$ but do not change the relative power of the power spectrum and bispectrum constraints.}. For the bispectrum we considered the $C_{2\times2\times2}$ and $C_{3\times3}$ terms of the covariance matrix (see Sect.\ref{Sect:polyspcov}).\\
Then, one can compute from the Fisher matrix the 1, 2 or 3 $\sigma$ contours whose interior correspond to the 68, 95 or 99 \% confidence level (C.L.) respectively. In a 2-dimensional slice of the parameter space, these contours are ellipses. Fig.\ref{Fig:ellipses_hod} shows this ellipses for the three HOD parameters $\alpha_\mr{sat}$, $M_\mr{min}$ and $M_\mr{sat}$, at respectively 217 GHz (left panel) and 857 GHz (right panel).

%For example the upper left plot shows the contours in the $(\alpha_\mr{sat},M_\mr{min})$ plane. 
We see that the bispectrum brings very interesting constraint on the HOD parameters. At 217 GHz, its degeneracies are complementary to those of the power spectrum, so that the combination of both improves greatly the error bars, in an ideal case without Galactic contamination in particular. Indeed, the error bar on $M_\mr{min}$ is decreased by 30\% at 217 GHz and by a factor $\sim5.5$ at 857 GHz, by using a combined constraint compared to a power spectrum only constraint. For $M_\mr{sat}$ the corresponding improvement is a factor $\sim3$ at 217 GHz and a factor $\sim4$ at 857 GHz ; for $\alpha_\mr{sat}$ the improvements are a factor $\sim7.5$ at 217 GHz and $\sim3$ at 857 GHz. Note also that at 857 GHz, the bispectrum constraint on the parameters are better than those of the power spectrum. However at this frequency the measurement is severely impaired by the contamination by Galactic dust, as will be described in Sect.\ref{Sect:meas_CIB_NG}.\\
The fact that the bispectrum brings more stringent constraint at 857 GHz than at 217 GHz can be understood easily. For the CIB, the signal at higher frequencies has more contribution from low redshift sources \citep[see e.g.][]{Penin2012}. As the bispectrum is highly sensitive to low redshifts, it is particularly suited for the CIB at 857 GHz.
\newline

Thus, we have shown that the CIB bispectrum is potentially a well-suited observable to constrain CIB models. I have built a powerful theoretical framework to predict the CIB clustering at high orders with a consistent and physically motivated model. Furthermore, I have developed a fast and efficient code to compute the CIB angular power spectrum and bispectrum, which can hence be used to constrain the model with measurement of both quantities.

\mychapter{Measuring non-Gaussianity}{Measuring non-Gaussianity}{chapt:NGmeas}

In the previous chapter, I have described extragalactic foregrounds to the CMB, and e.g., I have shown how high order moments contain valuable information to constrain the CIB model. In this chapter, I describe the statistical methods I have proposed and the data analysis I have performed, for the measurement of the non-Gaussianity of extragalactic point-sources and of the thermal SZ signal with \planck data.
%I have been working on the \planck data, as such experimental characteristics of \planck and how to account for them are presented in Sect.\ref{Sect:planck_obs_consid}.
%Each section/subsection is in fact related to an article that I have either written or contributed to.

%%%%%%%%%%%%%%%%%%%%%%%%%%%%%%%%%%%%%%%%%%%%%
\section{Observational considerations for \planck}\label{Sect:planck_obs_consid}
This chapter concerns data analysis, in particular with \planck data. We thus need to account for observational limitations which appear in this context. The first limitation is the necessity of a mask due to the astrophysical contamination of areas of the sky, by the Galactic emission or by point-sources. The problem of the inherent incomplete sky coverage was tackled in Sect.\ref{Sect:estim_specandbisp}. There are then two limitations due to the instrumental characteristics that need to be considered~:
 \begin{itemize}
 \item Resolution : any instrument has a finite angular resolution, due in particular to the diffraction limit associated to the mirror width. Thus the observation of a point-like signal is not a point-like image but a smoother spot. This image is called the point-spread function or beam. If the beam is independent of the pointing direction, the observed sky is a convolution of the ``true sky'' by the beam. Furthermore if the beam is rotationally invariant, the observed harmonic coefficients read~:
 \be\label{Eq:almbeamconvol}
 a_{\ell m}^\mr{obs} = a_{\ell m}^\mr{sky} \times B_\ell
 \ee
 A common and legitimate assumption is that the beam is Gaussian, in which case it is characterised by its full width to half maximum (FWHM) and we have~:
 \be
 B_\ell =\exp\left(-\ell(\ell+1)\frac{\mr{FWHM}^2}{16\ln(2)}\right)
 \ee
 with FWHM in rad. FWHM measured for the \planck beams are given by \cite{planck2013-HFIbeams} (they range between 4 and 10 arcmin for the High Frequency Instrument and between 13 and 33 arcmin for the Low Frequency Instrument).
 \item Noise : even with a perfect acquisition chain, there is a fundamental limit to the detector precision which is photon noise. Photon noise arises from the fact that the number of photons hitting the detector in a given time interval is random (following a Poisson law, see Sect.\ref{Sect:statPoisson}). This produces variations of detected power and hence noise in the inferred sky intensity. There are many other possible sources of noise along the acquisition chain, but \planck has been designed so that the fundamental photon noise is its main instrumental limitation. If we consider the noise contribution, Eq.\ref{Eq:almbeamconvol} is modified into~:
 \be\label{Eq:almwbeamnandnoise}
 a_{\ell m}^\mr{obs} = a_{\ell m}^\mr{sky} \times B_\ell + a_{\ell m}^\mr{noise}
 \ee
 Because the CMB monopole is dominant, we can consider that the average number of photons is isotropic, and hence $a_{\ell m}^\mr{noise}$ is a random variable independent of $a_{\ell m}^\mr{sky}$. Furthermore, this number of photons is large so $a_{\ell m}^\mr{noise}$ can be considered Gaussian through the central limit theorem. Finally, if all areas of the sky are surveyed during the same time, the noise is isotropic and is hence solely characterised by its angular power spectrum $C_\ell^\mathrm{noise}$. This last point is not the case for \planck due to its scanning strategy, but it is a common assumption, that we will consider at first order.
 \end{itemize}

In the following, we note with calligraphical sign the power spectrum accounting for the noise and the beam effects~:
\be\label{Eq:Clwbeamandnoise}
{\cal C}_\ell = C_\ell\; B^2_\ell + C_\ell^\mathrm{noise}
\ee
We also note the corresponding quantity in bold for the bispectrum~:
\be\label{Eq:blwbeamandnoise}
\mbl = \bl \; B_{\ell_1} B_{\ell_2} B_{\ell_3}
\ee
Let us note that this last equation considers that the noise is Gaussian so that its bispectrum vanishes. I have indeed computed non-Gaussianity estimates on realist noise simulations by the \planck collaboration and found them consistent with zero.\\
Although the bispectrum (and higher orders) estimation is not biased by the noise, contrary to the power spectrum, its error bars are increased due to the noise. For example the $C_{2\times2\times2}$ term of the bispectrum covariance contains a product of power spectra, and these power spectra include a noise contribution as in Eq.\ref{Eq:Clwbeamandnoise}.
\newline

Observationally, the noise level can be estimated from the data itself by the so-called jacknife procedure. In the case of \planck, the scanning strategy produces data in the form of rings of the sky. Each pixel of the sky is observed several times (on average $\sim$1,000 times), so maps may be produced from the first half of these rings and compared to the map produced with the other half. These maps are called respectively first and last ring, and their difference (the jacknife) represents instrumental noise as the sky signal is cancelled. I will also make use of these maps for the CIB NG measurement in Sect.\ref{Sect:meas_CIB_NG}.

%%%%%%%%%%%%%%%%%%%%%%%%%%%%%%%%%%%%%%%%%%%%%
\section{Non-Gaussianity from extragalactic point-sources}

In Sect.\ref{Sect:presp} I presented the characterisation of the non-Gaussianity from extragalactic point-sources (radio and IR) theoretically and on simulations. The following sections describe the statistical analysis I have performed to measure this non-Gaussianity in the \planck context.

%%%%%%%%%%%%%%%
\subsection{IR amplitude and joint NG constraints}\label{Sect:AirandjointNG}

%This section describes the work which lead to my second article \citep{Lacasa2013a}.\\
As shown in Sect.\ref{Sect:presp}, the CIB bispectrum can be an important contribution to the total bispectrum of the extragalactic signals, especially above 217 GHz. Furthermore, it peaks in the squeezed configurations, i.e., the same as the CMB bispectrum for local type PNG. Hence, the estimation of primordial, CIB and radio non-Gaussianity may interfere with each other, i.e. each separate estimation may be contaminated by the others. For this reason, we need a tool to estimate the amplitude of the CIB bispectrum in a map, and a tool which can be combined consistently with those used to quantify the primordial and radio NG. This tool also needs to be computationally fast, to enable numerous tests on simulations as needed for \planck analysis.

Fast estimators for the parameters $\fnl$ and $\bps$, characterising respectively the amplitude of the primordial and radio bispectra, have been proposed by \cite{KSW2005}. In \cite{Lacasa2013a}, I have proposed an estimator for the CIB, following the same principles.

To this end, the parameter $\air$ is defined as the amplitude of the IR bispectrum compared to a template derived from the empirical prescription Eq.\ref{Eq:presp}. In practice, for a full-sky map where the CIB is assumed to be the sole NG source, the amplitude of the bispectrum is measured by minimising the $\chi^2$ of the measured bispectrum to the template. This yields~:
\be
\chi^2(\air) = \sum_{\lu \leq \ld \leq \lt} \frac{\left(\mbl - \air\,\mbl^\mathrm{IR}\right)^2}{\sigma^2(\lu,\ld,\lt)}
\ee
with~:
\be \label{Eq:defsig2l123}
\sigma^2(\lu,\ld,\lt) = \frac{{\cal C}_{\lu}^\mathrm{tot} {\cal C}_{\ld}^\mathrm{tot} {\cal C}_{\lt}^\mathrm{tot}}{N_{\lu \ld \lt}}\times \Delta_{\lu \ld \lt}
\ee
being the bispectrum variance in the weak non-Gaussianity limit (the term $C_{2\times2\times2}$ of the covariance, see Sect.\ref{Sect:bispcov}).\\
In this expression, ${\cal C}_{\ell}^\mathrm{tot}$ is the measured power spectrum, and we have~:
\bea
\nonumber\Delta_{\lu \ld \lt} &=& 1+ \delta_{\lu\ld}+ \delta_{\lu\lt}+ \delta_{\ld\lt}+ 2\,\delta_{\lu\ld\lt}\\
&=& \left\{ \begin{array}{ll} 6 & \mathrm{for \; equilateral \; triangle}\\ 2 &
  \mathrm{for \; isosceles \; triangle}\\ 1 & \mathrm{for \; general
    \; triangle} \end{array}\right.
\eea
By minimising the $\chi^2$, we get the maximum likelihood estimator~:
\be\label{Eq:hatAir}
\hat{A}_\mr{IR} = \sum_{\lu \leq \ld \leq \lt} \frac{\mbl \;\mbl^\mathrm{IR}}{\sigma^2(\lu,\ld,\lt)} \times \sigma^{2}(\hat{A}_\mr{IR})
\ee
with~:
\be\label{Eq:sig2_Air}
\sigma^2(\hat{A}_\mr{IR}) = \left(\sum_{\lu \leq \ld \leq \lt} \frac{\left(\mbl^\mathrm{IR}\right)^2}{\sigma^2(\lu,\ld,\lt)}\right)^{-1}
\ee
being a normalisation factor which gives the error-bar of the $\air$ estimator.\\
The bispectrum derived from the prescription, Eq.\ref{Eq:presp}, is separable. That is, it can be written 
in the form $b(\ell_1,\ell_2,\ell_3) = \sum_i f_i(\ell_1) \, g_i(\ell_2) \, h_i(\ell_3) + \mathrm{perm.}$ This is the case for the prescription bispectrum with the sum reduced to one element and $f_i(\ell) = g_i(\ell) = h_i(\ell) = \alpha^{1/3} \sqrt{C_\ell^\mathrm{IR}}$. Hence, a simpler computation of the numerator of Eq.\ref{Eq:hatAir} can be devised, inspired by \cite{KSW2005}. Indeed, defining the filtered map~:
\be\label{Eq:filteredF}
F(\nn) = \sum_{\ell m} \frac{\alpha^{1/3} \, \sqrt{C_\ell^{\mathrm{IR}}} \, b_\ell }{{\cal C}_\ell^\mathrm{tot}}\, a_{\ell m} \, Y_{\ell m}(\nn)
\ee
and~:
\be
\mathcal{S}_\mathrm{IR}= \int \dd^2\nn \; F(\nn)^3
\ee
some algebra leads to~:
\be \label{Eq:defSir}
\mathcal{S}_\mathrm{IR}= \sum_{\lu \ld \lt} \frac{\hat{\mathbf{b}}_{123} \; \mbl^\mathrm{IR}}{{\cal C}_{\lu}^\mathrm{tot} {\cal C}_{\ld}^\mathrm{tot} {\cal C}_{\lt}^\mathrm{tot}} \; N_{123} = 6 \times \sum_{\lu \leq \ld \leq \lt} \frac{\hat{\mathbf{b}}_{123} \; \mbl^\mathrm{IR}}{\sigma^2(\lu,\ld,\lt)}
\ee
The computations of the filtered map and of $\mathcal{S}_\mathrm{IR}$ take respectively $\mathcal{O}(N_\mathrm{pix}^{3/2})$ and $\mathcal{O}(N_\mathrm{pix})$ operations, while the denominator of Eq.\ref{Eq:hatAir} is independent of the map and can be precomputed. Hence, this approach is much faster than a full bispectrum analysis which scales as $\mathcal{O}(N_\mathrm{pix}^{5/2})$ operations.
\newline

In realistic cases of CMB analysis, the statistical isotropy of the signal is broken by inhomogeneous noise --e.g., due to the scanning strategy-- or by masking large areas of the sky --e.g., the galactic plane. In these cases, the estimator of the IR-bispectrum amplitude is no longer optimal and it is biased in a
non-trivial way. Nevertheless and similarly to the case of the $f_{\mathrm{NL}}$, the bias and the lack of optimality can both be tackled through adapted modifications to the estimator, similar to the bispectrum case (see Sect.\ref{Sect:linearterm_bisp} \& \ref{Sect:analyt_bispcouplmat}).
\newline

\noindent {\bf Bias}\\
When a covariance matrix of the map(s) can be estimated, 
\cite{Creminelli2006} have shown that Wiener filtering of the 
map(s) will debias the non-Gaussian estimator from anisotropic 
contaminants or noise. Specifically, if $\mathbf{C}$ is the estimated 
covariance matrix in harmonic space, the estimator Eq. \ref{Eq:hatAir} can 
be debiased by applying the modification~:
\be
\frac{a_{\ell m}}{{\cal C}_\ell^\mathrm{tot}} \rightarrow \left(\mathbf{C}^{-1}\cdot \mathbf{a}\right)_{\ell m} = \sum_{\ell' m'} C^{-1}_{\ell m, \ell' m'} \; a_{\ell' m'}
\ee
in the filtered map, Eq. \ref{Eq:filteredF}. The denominator of the estimator Eq.\ref{Eq:sig2_Air} must undergo a consistent modification. This denominator is then most easily computed through simulations. Note that $\left(\mathbf{C}^{-1}\cdot \mathbf{a}\right)_{\ell m}$ reduces to $\frac{a_{\ell m}}{{\cal C}_\ell^\mathrm{tot}}$ in the isotropic case.
\newline

\noindent {\bf Variance}\\
When isotropy is broken, the 3-point correlation function used to define the bispectrum no longer has minimal variance. It must be replaced by the Wick product of the three harmonic coefficients \citep{Donzelli2012} :
\be\label{Eq:wickprod_a123}
a_1 \, a_2 \, a_3 \rightarrow a_1 \, a_2 \, a_3 - \langle a_1 \, a_2 \rangle \, a_3 - \langle a_1 \, a_3 \rangle \, a_2 - \langle a_2 \, a_3 \rangle \, a_1
\ee
These additional terms are the ones which also demand a linear term for the bispectrum estimation, as described in Sect.\ref{Sect:linearterm_bisp}. These terms vanish in the isotropic case.\\
Only the 2-point correlation function is to be considered for the expectation values $\lbra\rbra$ in Eq.\ref{Eq:wickprod_a123}. Therefore in practice, the latter are obtained from a sufficiently large number of Gaussian realisations with the same power spectrum as that of the signal considered.\\
Inputting the linear corrections given by Eq. \ref{Eq:wickprod_a123} into the expression of $S_\mathrm{IR}$ in Eq. \ref{Eq:defSir} yields :
\be
S_\mathrm{IR} \rightarrow \tilde{S}_\mathrm{IR} = \int \dd^2\nn \; F(\nn)^3 -3\times \!\!\int \dd^2\nn \; F(\nn) \, G(\nn)
\ee
with~:
\be
G(\nn) = \langle F(\nn)^2 \rangle_\mathrm{MC}
\ee
where the brackets $\langle \rangle_\mathrm{MC}$ stand for the average over Gaussian simulations.
\newline

I now present the expected detection significance of the CIB NG in terms of the signal-to-noise ratio (SNR) for the $\air$ estimator, through $\mr{SNR} = 1 / \sigma(\hat{A}_\mr{IR})$. To this end, I use the values of $\alpha$ and $C_\ell^\mathrm{IR}$ measured on the simulations of \cite{Sehgal2010}, see Sect.\ref{Sect:presp}, with a maximum multipole $\ell_\mr{max}=2048$.

First, I present an ideal case : a full-sky CIB map without noise and with perfect angular resolution. In this case, one sees that the signal-to-noise ratio increases with the number of available bispectrum configurations~:
\be
\mathrm{SNR} \propto \sqrt{N_\mathrm{tot}(\ell_\mathrm{max})}
\qquad \mr{with} \qquad
N_\mathrm{tot}(\ell_\mathrm{max}) = \!\!\!\!\!\!\!\!\!\sum_{\ell_\mathrm{min}\leq\lu \leq \ld \leq \lt\leq\ell_\mathrm{max}}\!\!\! N_{\lu \ld \lt}
\ee
The resulting SNR at the frequencies of the simulations by \cite{Sehgal2010} are shown in the first line of Table \ref{Table:SNR}. The obtained SNR are of order $\sim$1000 and do not vary much with frequency.

\begin{table}[htbp!]
\centering
\begin{tabular}{|c|c|c|c|c|}
\hline
frequency (GHz) & 150 & 220 & 280 & 350 \\
\hline
Ideal case & 1218 & 1157 & 1161 & 1159 \\
Full-sky IR & 15.5 & 98 & 336 & 833 \\
Full-sky with CMB & 0.39 & 6.7 & 55 & 387 \\
50\% sky with CMB & 0.28 & 4.7 & 39 & 274 \\
50\% sky with 10\% CMB & 3.67 & 45.9 & 210 & 577 \\
\hline
\end{tabular}
\caption{Expected SNR at $\ell_\mathrm{max}=2048$. Ideal case stands for a full-sky cosmic-variance limited CIB map without noise nor contaminations. All following rows adopt the instrumental characteristics specified in Table \ref{Table:instruspec}. Full-sky IR represents the case of the CIB map as seen by the instrumental set-up. Full-sky with CMB is when CMB is present, contaminating the CIB map. 50\% sky with CMB stands for the case of masking half of the sky. 50\% sky with 10\% CMB stands for a half sky CIB map where 90\% of the CMB (in amplitude) has been removed.}
\label{Table:SNR}
\end{table}

In a more realistic case, the CIB map is convolved by the instrumental beam and the signal is contaminated by noise, but we do not include astrophysical contamination. To model these effects, I used a Gaussian beam and white noise whose values are specified in Table \ref{Table:instruspec}. Both are representative of a \planck-like experiment.\\
I also present partial-sky coverage cases, assuming that the optimisation and the debiasing described previously have been applied, the SNR scales as $f_\mathrm{SKY}^{-1/2}$.

\begin{table}[htbp!]
\centering
\begin{tabular}{|c|c|c|c|c|}
\hline
Beam & Noise & $C_\ell^\mathrm{noise}$ \\
\hline
5 arcmin & $10^{-8}$ $\Delta T/T \cdot \mathrm{sr}^{1/2}$ & $7.4 \cdot10^{-4} \mu {\mathrm{K}}^2\cdot\mathrm{sr}$\\
\hline
\end{tabular}
\caption{Instrumental specifications used throughout this section.}
\label{Table:instruspec}
\end{table}

The Table \ref{Table:SNR} shows the case of a convolved but ``perfect" full-sky CIB map, a full-sky case but including CMB contamination, a case with a 50\% sky-fraction mimicking a galactic mask, and finally the case where 90\% of the CMB is removed. As a matter of fact, component separation methods can efficiently estimate the CMB signal from multifrequency observation with an error in amplitude below 10\% \citep[see][and references therein]{Delabrouille2007,Remazeilles2011,Bobin2013}. The estimated CMB map can then be substracted from frequency maps to reveal the weaker IR signal from unresolved sources as performed for example by \citet{planck2011-CIB}. \newline

Until now, I have considered that the CIB was the only source of non-Gaussianity. However when several non-Gaussian signals are present, a joint estimation of their amplitudes taking their covariances into account is necessary. In this section I focus on the main extragalactic non-Gaussian signals : unclustered sources (radio but more generally any population exhibiting flat moments), IR point-sources and the primordial NG of the local-type. A joint estimation then requires minimisation of the $\chi^2$~:
\be\label{Eq:chi2fnlbpsAIR}
\chi^2(\fnl, \bps, \air) = \sum_{\lu \leq \ld \leq \lt} \frac{\left(\mbl^\mathrm{obs} - \mbl^\mathrm{model}(\fnl, \bps, \air)\right)^2}{\sigma^2(\lu,\ld,\lt)}
\ee
with~:
\be
\mbl^\mathrm{model}(\fnl, \bps, \air) = 
\fnl \,\mbl^\mathrm{CMB} + \bps \,\mbl^\mathrm{RAD} + \mathrm{A}_\mathrm{IR}\,\mbl^\mathrm{IR}
\ee
and (involved in the definition of $\sigma^2(\lu,\ld,\lt)$)~:
\be
{\cal C}_\ell^\mathrm{tot} = \left(C_\ell^\mathrm{CMB}+C_\ell^\mathrm{RAD}+C_\ell^\mathrm{IR}\right) B_\ell^2 + C_\ell^\mathrm{noise}.
\ee
For backward-compatibility with the literature, I use the standard parameter $\bps$ instead of noting  $A_\mathrm{RAD}$. Note that the radio NG amplitude is heavily dependent on the flux cut through Eq.\ref{Eq:counts_to_bps}. The flux cut thus needs to be specified to predict $\bps$ theoretically.

Let us define the scalar product between two bispectra $\mathbf{b}^\alpha$ and $\mathbf{b}^\beta$ :
\be\label{Eq:defscalprod}
< \mathbf{b}^\alpha , \mathbf{b}^\beta > = \sum_{\lu \leq \ld \leq \lt} \frac{\mbl^\alpha \; \mbl^\beta}{\sigma^2(\lu,\ld,\lt)}
\ee
Minimising Eq.\ref{Eq:chi2fnlbpsAIR} corresponds to solving the linear system :
\bea \label{Eq:JointNG_scalprod}
\nonumber &\left(\begin{array}{ccc}
<\mathbf{b}_\mathrm{CMB} , \mathbf{b}_\mathrm{CMB}> & <\mathbf{b}_\mathrm{RAD} , \mathbf{b}_\mathrm{CMB}> & <\mathbf{b}_\mathrm{IR} , \mathbf{b}_\mathrm{CMB}> \\
<\mathbf{b}_\mathrm{CMB} , \mathbf{b}_\mathrm{RAD}> & <\mathbf{b}_\mathrm{RAD} , \mathbf{b}_\mathrm{RAD}> & <\mathbf{b}_\mathrm{IR} , \mathbf{b}_\mathrm{RAD}> \\
<\mathbf{b}_\mathrm{CMB} , \mathbf{b}_\mathrm{IR}> & <\mathbf{b}_\mathrm{RAD} , \mathbf{b}_\mathrm{IR}> & <\mathbf{b}_\mathrm{IR} , \mathbf{b}_\mathrm{IR}>
\end{array}\right)
\cdot
\left(\begin{array}{c}
\fnl \\
\bps \\
\air
\end{array}\right)
\\ &\quad =
\left(\begin{array}{c}
<\mathbf{b}_\mathrm{obs} , \mathbf{b}_\mathrm{CMB}> \\
<\mathbf{b}_\mathrm{obs} , \mathbf{b}_\mathrm{RAD}> \\
<\mathbf{b}_\mathrm{obs} , \mathbf{b}_\mathrm{IR}>
\end{array}\right)
\eea
If I introduce the estimators which consider only one source of non-Gaussianity, noting them with upper tilde,  e.g.~:
\be
\tilde{A}_\mathrm{IR} = \frac{<\mathbf{b}_\mathrm{obs} , \mathbf{b}_\mathrm{IR}>}{<\mathbf{b}_\mathrm{IR} , \mathbf{b}_\mathrm{IR}>}
\ee
then Eq.\ref{Eq:JointNG_scalprod} can be rewritten to define the joint NG estimators --noted with upper hat-- as :
\be
\left(\begin{array}{c}
\hat{f}_\mathrm{NL} \\
\hat{b}_\mathrm{PS} \\
\hat{A}_\mathrm{IR}
\end{array}\right)
=
\mathcal{M}^{-1} \cdot
\left(\begin{array}{c}
\tilde{f}_\mathrm{NL} \\
\tilde{b}_\mathrm{PS} \\
\tilde{A}_\mathrm{IR}
\end{array}\right)
\ee
with~:
\be
\mathcal{M} = \left(\begin{array}{ccc}
1 & \Delta \fnl^\mathrm{RAD} &  \Delta \fnl^\mathrm{IR}\\
\Delta \bps^\mathrm{CMB} & 1 & \Delta \bps^\mathrm{IR} \\
\Delta \air^\mathrm{CMB} & \Delta \air^\mathrm{RAD} & 1
\end{array}\right)
\ee
and~:
\bea\label{Eq:mixmat_NG_elts}
\nonumber \Delta \fnl^\mathrm{RAD} = \frac{<\mathbf{b}_\mathrm{RAD} , \mathbf{b}_\mathrm{CMB}>}{<\mathbf{b}_\mathrm{CMB} , \mathbf{b}_\mathrm{CMB}>} \qquad
\Delta \fnl^\mathrm{IR} =\frac{<\mathbf{b}_\mathrm{IR} , \mathbf{b}_\mathrm{CMB}>}{<\mathbf{b}_\mathrm{CMB} , \mathbf{b}_\mathrm{CMB}>}\\
\nonumber \Delta \bps^\mathrm{CMB} = \frac{<\mathbf{b}_\mathrm{CMB} , \mathbf{b}_\mathrm{RAD}>}{<\mathbf{b}_\mathrm{RAD} , \mathbf{b}_\mathrm{RAD}>} \qquad
\Delta \bps^\mathrm{IR} = \frac{<\mathbf{b}_\mathrm{IR} , \mathbf{b}_\mathrm{RAD}>}{<\mathbf{b}_\mathrm{RAD} , \mathbf{b}_\mathrm{RAD}>}\\
\Delta \air^\mathrm{CMB} = \frac{<\mathbf{b}_\mathrm{CMB} , \mathbf{b}_\mathrm{IR}>}{<\mathbf{b}_\mathrm{IR} , \mathbf{b}_\mathrm{IR}>} \qquad
\Delta \air^\mathrm{RAD} = \frac{<\mathbf{b}_\mathrm{RAD} , \mathbf{b}_\mathrm{IR}>}{<\mathbf{b}_\mathrm{IR} , \mathbf{b}_\mathrm{IR}>}
\eea
Note in particular that the estimators $\tilde{f}_\mathrm{NL}$ and $\tilde{b}_\mathrm{PS}$ are the ones defined in \cite{KSW2005}, while $\tilde{A}_\mathrm{IR}$ is the newly proposed estimator.

For illustration, assuming ERCSC flux cuts \citep{planck2011-ERCSC} and including the instrumental beam and noise specified Table \ref{Table:instruspec}, the mixing matrix at 220 GHz with $\ell_\mathrm{max}=2048$ is~:
\be\label{Eq:mixingmatrix220}
\mathcal{M} = \left(\begin{array}{ccc}
1 & 0.0461 &  0.245 \\
1.21\!\cdot\! 10^{-4} & 1 & 2.12 \\
1.31\!\cdot\! 10^{-4} & 0.435 & 1
\end{array}\right)
\ee
where $\bps$ is normalised to its expected value so that all amplitude parameters are of order $\mathcal{O}(1)$. That is, the matrix is understood to be applied to $\bps/\bps^\mr{exp}$, where the expected $\bps^\mr{exp}$ was computed following formula \ref{Eq:counts_to_bps} and using number counts for radio sources by \cite{Tucci2011}.

The matrix $\mathcal{M}$, Eq.\ref{Eq:mixingmatrix220}, shows, at this frequency, that the contamination by primordial NG to point-sources NG estimators is negligible (see first column). Furthermore and unless $\air$ and $\bps/\bps^\mr{exp}$ are much bigger than one, the contamination by point-sources to the primordial NG estimator is also negligible compared to its error bars ($\sigma(\fnl)\sim 5$). Besides, the RAD-IR submatrix, with a condition number $\sim$100, shows much more coupling between the $\bps$ and $\air$ estimations. They are relatively degenerate. At this frequency, we see that the CIB NG is slightly dominant, by a factor $\sim$2. These conclusions however are only valid for this frequency, for local type primordial NG, and for this flux cut.\\
I have examined the cases at other frequencies and other flux cuts. Lower flux cuts upweights the CIB NG compared to the radio NG. That is, $\mathcal{M}_{23}$ is increased while $\mathcal{M}_{32}$ is decreased. Indeed IR sources are faint, so that the CIB bispectrum is mostly unaffected by the flux cut. On the contrary, the radio bispectrum is more affected, as it is dominated by sources just below the flux cut. However, this effect is less important than  the frequency dependence. Indeed, the frequency behaviour of both signals, and the fact that the bispectra have a cubic dependence in the emission law, well-defined domination depending on frequency. That is, the radio NG dominates below 220 GHz while the IR dominates above, and 220 GHz is the only frequency at which their bispectra are of same order. Thus, 220 GHz is the only frequency at which $\mathcal{M}_{23}$ and $\mathcal{M}_{32}$ are of order $\mathcal{O}(1)$, at other frequencies this entries are either $\ll 1$ or $\gg 1$.

\begin{figure}[htbp]
\begin{center}
\includegraphics[width=.7\linewidth]{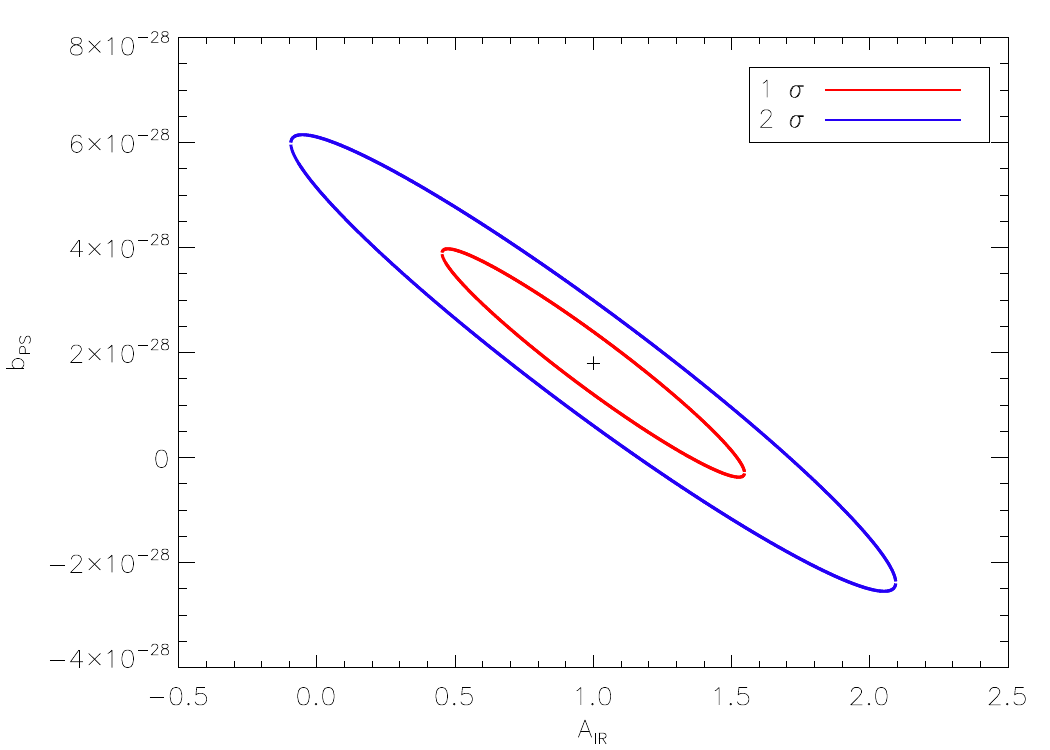}
\caption{1$\sigma$ and 2$\sigma$ confidence level for ($\air$ ,
  $\bps$) in a joint estimation at 220 GHz $\ell_\mathrm{max} = 2048$
  with 5 arcmin Gaussian beam and $10^{-8}\, \Delta T/T$ noise}
\label{Fig:ellipse_bpsAir}
\end{center}
\end{figure}

At 220 GHz, the primordial NG estimation decouples from the point-source estimation, however $\air$ and $\bps$ need to be estimated jointly. For illustration, Fig.~\ref{Fig:ellipse_bpsAir} shows the likelihood contours in the ($\air$ , $\bps$) plane, showing the expected combined constraints on these parameters at 220 GHz.  $\air$ and $\bps$ estimations are quite degenerate, as expected from the high correlations of their templates. This significantly degrades the constrain that can be put on each of them independently.\\
Indeed, if the radio and CIB bispectrum templates were not correlated, $\bps$ would be detected with a $\sim3\sigma$ significance and $\air$ would be detected with a $\sim6.7\sigma$ significance. The combined constraint decreases these significances to a SNR$\sim0.8$ for $\bps$ and a SNR$\sim1.8$ for $\air$. There are nevertheless some possibilities to improve the constraints~: including the physically-motivated prior $\bps \geq 0$, and also to some extent $\air \geq 0$, and CMB removal may also help by decreasing the variance of both estimators (although not their correlation). Experimentally, the latter point is realistically achievable, and the maps I will analyse in Sect.\ref{Sect:meas_CIB_NG} \& \ref{Sect:meas_SZ_NG} will indeed be cleaned from CMB contamination.
%discuss CMB removal

%%%%%%%%%%%%%%%
\subsection{Contamination of primordial NG estimation}\label{Sect:PNG_conta}

As described in the previous section, the non-Gaussianity of extragalactic point-sources biases the estimation of primordial NG. This has been one of the concerns for the \planck NG study, whose main scientific goal is the primordial NG estimation, and is generally a concern for any NG study with CMB data. From the \planck point of view, the main interest was to quantify this contamination and, mostly for radio sources, to assess whether the masking of detected sources --either with the ERCSC \citep{planck2011-ERCSC} or with the PCCS \citep{planck2013-PCCS}-- was sufficient to control it.

In this section, I will describe the different estimates of primordial NG contaminations that I have computed through different methods.% I will present these estimates in chronological order. Note that, for the first estimates, I had not yet developed the concept and tool for joint NG estimation with the mixing matrix as described in the previous section.
\newline

In a first step,  I applied the KSW estimator to the radio and IR simulated maps by \cite{Sehgal2010}. These estimates do not consider any instrumental beam nor noise. Furthermore, for these computations $\sigma^2(\lu,\ld,\lt)$ in Eq.\ref{Eq:defscalprod} only contains $C_\ell^\mr{CMB}$. That is, these computations give the bias $\Delta\fnl$ that would arise \emph{if the map was thought to be pure CMB}. Moreover, I performed different estimations depending on whether or not a flux cut is applied and the maximum multipole of analysis. The results are shown in Table \ref{Table:Dfnl_2012} for radio and IR sources, in the case where the flux cut is implemented.

\begin{table}
\centering
\begin{tabular}{l|cccccc}
\hline
\multicolumn{7}{c}{Radio sources}\\
\hline
\hline
$\nu$ (GHz) & $\ $30 & 90 & 148 & 219$\ $  & 277$\ $ & 350\\
\hline
$\ell_\mathrm{max}=700$ & 108 & 0.17 & 0.0071 & 0.0031 & 0.0035 & 0.0064\\
$\ell_\mathrm{max}=2048\ $ & 4930 & 7.5 & 0.31 & 0.14 & 0.16 & 0.29 \\
\hline
\hline
\multicolumn{7}{c}{CIB}\\
\hline
\hline
$\nu$ (GHz) & $\ $30 & 90 & 148 & 219$\ $  & 277$\ $ & 350\\
\hline
$\ell_\mathrm{max}=700$ & $\,$-$1.3 \!\cdot\! 10^{-6}$ & -0.00019 & -0.0033 & -0.078 & -0.74 & -11\\
$\ell_\mathrm{max}=2048\,$ &  $\,$-$1.8 \!\cdot\! 10^{-5}$& -0.0026 & -0.039 & -0.67 & -6.3 & -66 \\
\hline
\end{tabular}
\caption{From \cite{Lacasa2012}, bias of the $f_\mathrm{NL}$ estimator due to radio and IR sources, measured on \cite{Sehgal2010} simulations.}
\label{Table:Dfnl_2012}
\end{table}

\noindent For the flux cut, I used the ERCSC values, or the nearest value when the frequency did not match, and masked the maps at pixels above the threshold. Note that the flux cut did not have any effect for IR sources below 219 GHz, as all sources were fainter than the flux cut at these frequencies.\\
I also considered different maximum multipoles. 
%At $\ell_\mr{max}=50$ the CMB bispectrum is dominated by the Sachs-Wolfe effect and is negative, as noted in Sect.\ref{Sect:CMB}. This case serves as a consistency check : indeed we find negative $\Delta\fnl$ in this case, either for radio or IR sources, which is expected as both bispectra are positive.
The case $\ell_\mr{max}=700$ is representative of \textit{WMAP}, in this case we see that the bias is always negligible compared to the error bar $\sigma(\fnl)\approx 20$ \citep{Komatsu2011} except at the frequencies 30 \& 350 GHz. The case \mbox{$\ell_\mr{max}=2048$} is representative of \planck, in this case we see that the bias is often comparable to, or greater than, the error bar $\sigma(\fnl)\approx 5$ \citep{planck2013-NG}. In details, the bias is important for radio sources at 30 to 90 GHz, and at 277 to 350 GHz for IR sources. I have produced similar estimations without implementing the flux cut, for comparison. We then see that the flux cut greatly reduces the radio bias (I.e. the bias is much more important without flux cut), while it mostly does not affect the IR bias. This means that at high frequencies, it is important to consider the CIB NG even with a conservative masking of sources.\\
One caveat with these computations, are the limitations of the \cite{Sehgal2010} simulations. For example the CIB power spectrum in these simulations is too steep ($C_\ell^\mr{IR} \sim \ell^{-2}$ while \cite{planck2011-CIB} has shown that $C_\ell^\mr{IR} \sim \ell^{-1}$). Furthermore, the octant replication yields an artificial break of power at low multipoles, to which the KSW estimator is sensitive as local type primordial NG peaks in the squeezed configurations (where one multipole is small). Hence my estimations may be artificially biased by this simulation's shortcoming.
\newline

%Dfnlrad report These/planck/WG-NG/report-DfnlRAD-Tucci/report-DfnlRAD-Tucci.pdf
%\url{http://hfilfi.planck.fr/index.php/WG/NonGaussianityLog?action=download&upname=Dfnl-RAD.pdf}
% Attention, pas de IR ici, ce qui peut changer les biais car pas de cl_IR dans les barres d'erreurs
%ConcatŽnation du tableau unres bluebook pour LFI et du tableau unres Tristram Cl pour HFI
We now turn to analytic computation of the primordial NG bias using Eq.\ref{Eq:mixmat_NG_elts}. First, I computed the bias on $\fnl$ estimation, but also its error bar, due to radio sources. This error bar is given by~:
\be
\sigma^2(\fnl) = \left(\sum_{\lu\leq\ld\leq\lt} \frac{\left(b_{123}^\mr{CMB}\right)^2}{\sigma^2(\lu,\ld,\lt)}\right)^{-1}
\ee
analogously to Eq.\ref{Eq:sig2_Air}. The results are shown in Table \ref{Table:DfnlunresTristram}, for Gaussian beams with FWHM from \cite{planck2013-HFIbeams}, noise power spectra measured on \planck data, $\ell_\mr{max}=2048$, ERCSC flux cuts, and radio number counts by \cite{Tucci2011}.

\begin{table}[ht]
\centering
\begin{tabular}{c||c|c|c|c|c|c|c|c|c|c|c|c|}
\hline
 $\nu$ (GHz) & 30 & 44 & 70 & 100  & 143 & 217 & 353 & 545 & 857\\
 \hline
$\Delta f_\mathrm{NL}^\mathrm{RAD}$ & 74.5 & 31.9 & 2.93 & 0.82 & 0.28 & 0.096 & 0.074 & 0.63 & 195 \\
\hline
$\sigma(f_\mathrm{NL})$ & 23.5 &  19.4 & 11.4 & 7.70 & 5.63 & 5.48 & 8.62 & 42.2 & 40827 \\
\hline
\end{tabular}
\caption{$\Delta f_\mathrm{NL}^\mathrm{RAD}$ and $\sigma(f_\mathrm{NL})$ at \planck frequencies, considering only unresolved radio sources. Using Gaussian beams with FWHM from \cite{planck2013-HFIbeams}, noise power spectra measured on \planck data, $\ell_\mr{max}=2048$, ERCSC flux cuts, and radio number counts by \cite{Tucci2011}.}
\label{Table:DfnlunresTristram}
\end{table}

We see that the bias is under control except at the extreme \planck frequencies 30-44-857 GHz, in particular the bias is negligible at the central CMB frequencies 100-143-217 GHz, where the error bars on $\fnl$ are the smallest. Note that the numbers in Table \ref{Table:DfnlunresTristram} would change with a more aggressive masking, e.g. with PCCS flux cuts \citep{planck2013-PCCS}, although the main conclusions would be unchanged.
\newline

%wiki post \url{http://hfilfi.planck.fr/index.php/WG/JointNGCMB-radio-CIB}
Finally, I have computed mixing matrices, described in Sect.\ref{Sect:AirandjointNG}, for \planck HFI frequencies. For the computation of these matrices, I used the same instrumental and radio characteristics as for Table \ref{Table:DfnlunresTristram}, and for the CIB I used power spectra (needed for the prescription for $b^\mr{CIB}_{123}$ and for the error bars $\sigma^2(\lu,\ld,\lt)$) provided by the CIB model described in \cite{planck2011-CIB} and reproducing the \planck measurement (\textit{ibid.}). I do not present all matrix elements here but just those associated with the bias on $\fnl$ estimation. They are shown in Table \ref{Table:Dfnl_wikipost}, along with the correlation coefficient $r$(IR,RAD), between the radio and IR bispectra, defined as~:
\be
r(\mr{IR,RAD}) = \frac{<\mathbf{b}_\mr{IR},\mathbf{b}_\mr{RAD}>}{\sqrt{<\mathbf{b}_\mr{IR},\mathbf{b}_\mr{IR}> \, <\mathbf{b}_\mr{RAD},\mathbf{b}_\mr{RAD}>}}
\ee

\begin{table}
\centering
\begin{tabular}{|c||c|c|c|c|c|c|c|c|c|}
\hline
 $\nu$ (GHz) & 100  & 143 & 217 & 353 & 545 & 857\\
 \hline
 $\Delta\fnl^\mr{RAD}$ & 0.82 & 0.27 & 0.089 & 0.046 & 0.016 & -391 \\
 \hline
 $\Delta\fnl^\mr{IR}$ & $8.2\cdot 10^{-4}$ & $5.6\cdot 10^{-3}$ & 0.11 & 19.3 & -1749 & $-7.43\cdot 10^7$ \\
 \hline
 $r$(IR,RAD) & 94.3\% & 94.8\% & 94.8\% & 95.2\% & 92.8\% & 91.9\% \\
\hline
 \end{tabular}
\caption{Bias of the $f_\mathrm{NL}$ estimator for local type primordial NG due to radio and IR sources, and correlation between the radio and IR bispectrum templates, computed with the mixing matrices described in Sect.\ref{Sect:AirandjointNG}. Using Gaussian beams with FWHM from \cite{planck2013-HFIbeams}, noise power spectra measured on \planck data, $\ell_\mr{max}=2048$, ERCSC flux cuts, and radio number counts by \cite{Tucci2011}.}
\label{Table:Dfnl_wikipost}
\end{table}

Note that Table \ref{Table:Dfnl_wikipost} only considers \planck HFI frequencies, as the CIB signal is completely negligible at LFI frequencies.\\
One can in particular compare the obtained $\Delta\fnl^\mr{RAD}$ with that shown in Table \ref{Table:DfnlunresTristram}. We see that they are in agreement at 100 \& 143 GHz and start gradually disagreeing afterwards, until having opposite sign at 857 GHz. The reason for this is that in Table \ref{Table:DfnlunresTristram} only radio sources were considered, so that $\sigma^2(\lu,\ld,\lt)$ contained $C_\ell^\mr{CMB}$ and $C_\ell^\mr{RAD}$, while for Table \ref{Table:Dfnl_wikipost} we also have a contribution of $C_\ell^\mr{IR}$ to $\sigma^2(\lu,\ld,\lt)$. Therefore the estimates are in agreement at low frequencies, where $C_\ell^\mr{IR}$ is negligible, and start to disagree as frequency increases.\\
The sign reversals of $\Delta\fnl^\mr{RAD}$ at 857 GHz between Table \ref{Table:DfnlunresTristram} and \ref{Table:Dfnl_wikipost}, and of $\Delta\fnl^\mr{IR}$ between 353 and 545 GHz show the sensitivity of the biases to the conditions at high multipoles. Indeed, the radio and IR power spectrum become dominant over the CMB at high multipoles, hence the $\fnl$ estimator downweights these configurations. When this domination extends to lower multipoles, e.g. when $C_\ell^\mr{IR}$ is added to $C_\ell^\mr{RAD}$ or at high frequencies, the configurations with low multipoles are favoured, and the CMB is negative in these configurations.\\
The same reasoning also shows the importance of the noise model. Indeed, the noise power spectrum used for Table \ref{Table:DfnlunresTristram} and \ref{Table:Dfnl_wikipost} is not flat but decreasing with multipoles. In particular, its value in the first 100 multipoles is much larger than at higher multipoles where it flattens. This changes the weighting of bispectrum configurations in the estimators and in the matrix elements Eq.\ref{Eq:mixmat_NG_elts}, particularly in the squeezed limit. Indeed, assuming a white-noise changes the values of $\Delta f_\mathrm{NL}^\mathrm{RAD}$ and $\sigma(f_\mathrm{NL})$, and may induce for example a sign reversal of $\Delta f_\mathrm{NL}^\mathrm{RAD}$ at 857 GHz.\\
Because of the oscillations and sign-changes of $b_{123}^\mr{CMB}$, the radio and CMB bispectra are nearly orthogonal, and the IR and CMB bispectra also. Hence, some changes of the weighting of bispectrum configurations in the estimators may easily change the value and sign of the scalar product. This also explains how, at 545 GHz in Table \ref{Table:Dfnl_wikipost}, $\Delta f_\mathrm{NL}^\mathrm{RAD}$ can be positive while $\Delta\fnl^\mr{IR}$ is negative, even if the bispectra are $\sim$93\% correlated, as illustrated schematically in Fig.\ref{Fig:scheme_bispRADIRCMB}

\begin{figure}[htbp]
\begin{center}
\begin{tikzpicture}
%2*2*2
\draw [->, very thick,darkgreen] (0,0) -- node[left] {$b_\mr{CMB}$} (0,2);
\draw [->, very thick,blue] (0,0) -- node[above] {$b_\mr{RAD}$} (2.1,0.2);
\draw [->, very thick,red] (0,0) -- node[below] {$b_\mr{IR}$} (2.1,-0.2);
\end{tikzpicture}
\caption{Schematic illustration of how $\Delta f_\mathrm{NL}^\mathrm{RAD}$ can be positive while $\Delta\fnl^\mr{IR}$ is negative. Indeed, the bispectrum vectors $b_\mr{IR}$ and $b_\mr{RAD}$ are nearly colinear, but the scalar product $\lbra b_\mr{RAD}, b_\mr{CMB}\rbra$ is positive while $\lbra b_\mr{IR}, b_\mr{CMB}\rbra$ is negative.}
\label{Fig:scheme_bispRADIRCMB}
\end{center}
\end{figure}

%In this thesis I only focused on primordial NG of the local type. However inflationary models can produce other NG templates, e.g., the equilateral and orthogonal ones as introduced in Sect.\ref{sect:CMB}. Furthermore, the CMB secondary anisotropies can produce observable non-Gaussianity as they are sourced by the Large Scale Structure at late times which is non-Gaussian. Out of all the possible secondaries' bispectra, the cross-bispectrum ISW-lensing-lensing (shortened to ISW-lensing hereafter) has been shown to be the most important to consider \citep{Smith2011,Hanson2009}. Indeed, there is a high correlation between the ISW and lensing effect as both are produced by Large Scale Structure mostly in the range $z=0-1$, and the ISW produces large scale power while lensing peaks on small scales. So the ISW-lensing peaks in the squeezed configuration and is a contaminant of local-type $\fnl$ estimation (at the level $\Delta\fnl\sim10$, see \cite{Serra2008,Kim2013}).

The \planck NG article \citep{planck2013-NG} has used the \planck data to put constraint on the different sources of non-Gaussianity in the \planck CMB maps. To this end, different CMB maps produced by several component separation methods have been considered, as well as the raw 143 GHz channel. Several estimators based on the 3-point function (except for Minkowsky functionals which test all orders) have been applied to the data and extensively tested on realistic Gaussian and non-Gaussian simulations. Furthermore, a joint NG estimation has been used considering the following sources of non-Gaussianity : primordial NG of the local, equilateral and orthogonal types (see Sect.\ref{Sect:CMB})\,; $\bps$ template\,; and ISW-lensing (see Sect.\ref{Sect:CMB} and \ref{Sect:fgintro}).\\
It can be noted that the CIB template was not considered. This choice was made because of the following considerations : time constraints, the high correlation ($>90\%$) between the radio and CIB templates, and the fact that the maps used were expected by the working group to be more contaminated by radio than by IR sources (in particular the 143 GHz channel).
%In light of the results of this section and of Fig.\ref{Fig:scheme_bispRADIRCMB}, the second point may be debatable. However the maps used for the NG study had low noise, in which case configurations with high multipoles are the most important for $\Delta f_\mathrm{NL}^\mathrm{RAD}$ and $\Delta\fnl^\mr{IR}$, so that both bias have the same sign. Hence we expect $\bps$ estimation to essentially capture the amplitude of IR NG if any is present, and we expect that the joint estimation will correctly debias PNG estimators at first order.
Furthermore, the robustness of PNG estimations to point-sources is also ensured by the fact that the obtained $\fnl$ values do not change much if $\bps$ is not considered in the joint estimation (at most by 0.3 $\sigma$).

However, the obtained $\bps$ estimation can difficultly be used for point-source constraints for the following reasons. First, it is potentially a mix of $\bps$ and $\air$ as already outlined. Then it was obtained with a conservative masking (union of sources detected at any frequency) which does not translate into a flux cut or a single selection function. Finally, except for the 143 GHz value, it was obtained on component separated maps which are designed to decrease the foregrounds signals and whose exact effect on a point-source bispectrum is unknown.

For the Cosmic Infrared Background, a dedicated NG study is thus necessary to use the information present in the \planck maps, particularly above 217 GHz.

%%%%%%%%%%%%%%%
\subsection{Measuring the CIB NG with \planck data}\label{Sect:meas_CIB_NG}

This section describes the \planck data analysis performed to detect the CIB bispectrum, as part of the results published in \cite{planck2013-CIB}.\\
%Description of the CIB paper
Let us first describe the \planck CIB article : the study is based on the \planck maps as well as the IRAS maps at 3000 GHz \citep{Neugebauer1984} reprocessed by \cite{Miville2005}. For the \planck channels, two maps are used per frequency, constructed respectively from the first and last rings (see Sect.\ref{Sect:planck_obs_consid}). The goal of the article is to measure the CIB anisotropies and model them so as to constrain the star formation history of the universe. To this end, the CIB anisotropies are unveiled through a cleaning of the maps detailed below. The power spectrum analysis proceeds to estimate the maps spectra and cross-spectra through cross-correlation of the first and last maps. Some contaminations are further accounted for at the power spectrum level (tSZ and CIB leakage, see sect.\ref{Sect:GASSpost}), before a halo model is fitted to the data and the results are interpreted in term of galaxy bias and star formation history.

%This section describes specifically the measurement of the CIB bispectrum on the \planck data.

To extract the CIB from the \planck maps, two components need to be removed : the CMB and the Galactic dust. The CMB was  removed by substracting an estimated CMB map obtained through Wiener filtering of the \planck 100 GHz map~:
\be
a^\mr{corr}_{\ell m}(\nu) = a_{\ell m}(\nu) - w_\ell \; a_{\ell m}(100\ \mr{GHz})
\ee
with $w_\ell$ the 100 GHz Wiener filter.\\
This simple technique allows propagation of the errors and of contaminations (by the CIB and tSZ present in the 100 GHz map in particular). On the contrary, for a CMB map estimated by multi-frequency component separation, the contamination is unknown\footnote{In particular the contamination of the estimated CMB by the CIB itself is problematic.}.\\
The Galactic dust component was removed by constructing a dust model with atomic hydrogen ({\sc Hi})observations, as \cite{Boulanger1996} reported a tight dust-{\sc Hi} correlation except for high {\sc Hi} column densities. Two different sets of {\sc Hi} observations were used for the studies I will present. First, {\sc Hi} observations by the Green Bank Telescope \citep{Boothroyd2011} have a $\sim9.5$ arcmin resolution and define nine fields with a total area of $\sim$0.6\% of the sky. Second, Parkes telescope observations define the GASS {\sc Hi} survey \citep{GASS2009}, with much larger sky coverage ($\sim$10\% of the sky) albeit with a lower (16.2 arcmin) resolution.

\subsubsection{Small fields}

\begin{figure}[htbp]
\begin{center}
\includegraphics[width=.7\linewidth]{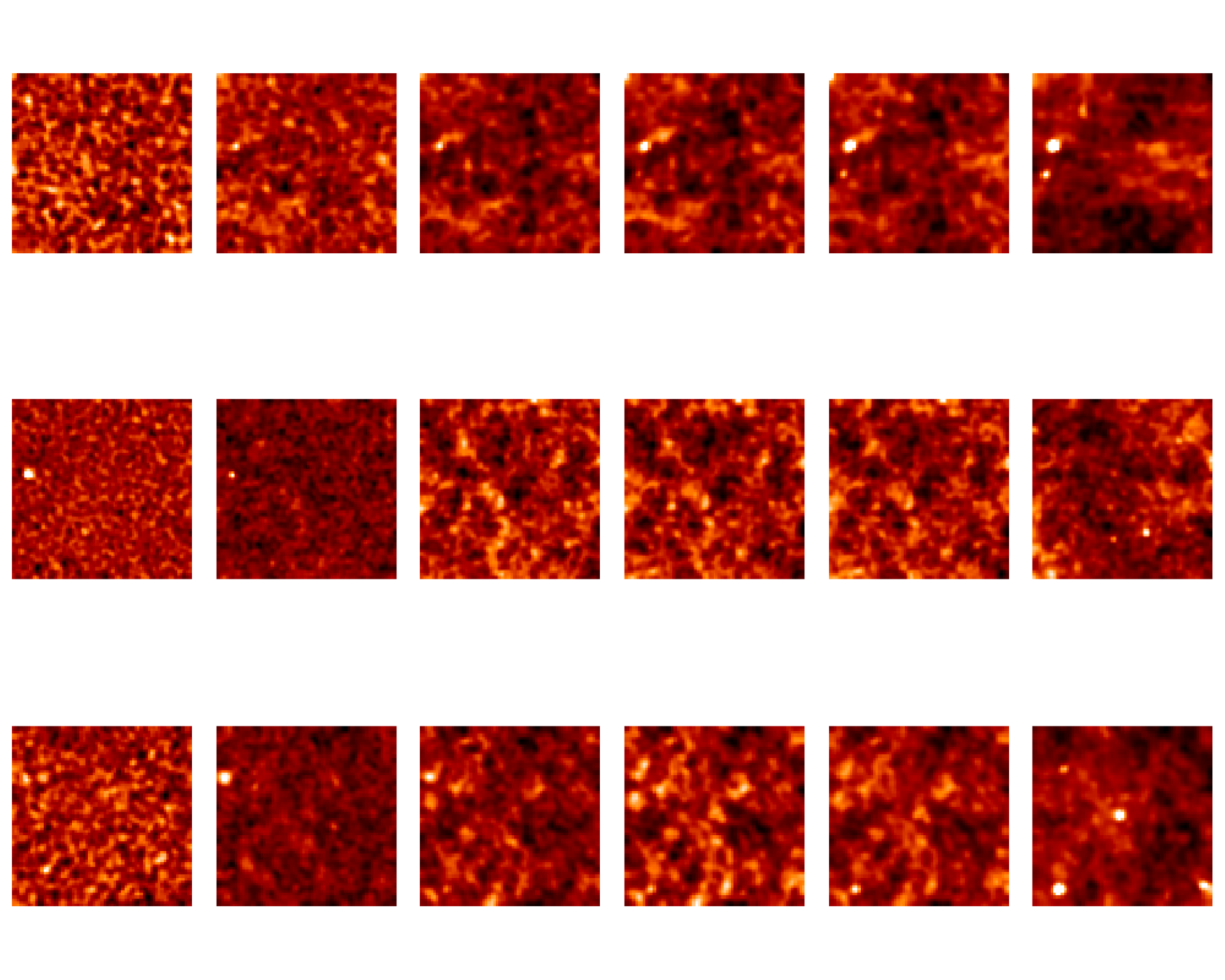}
\caption{Three of the cleaned small CIB fields. \textit{From top to bottom :} SPC4, N1 and LH2. The maps at the different frequencies are shown, \textit{from left to right} : \planck 143-217-353-545-857 GHz and IRAS 3000 GHz.}
\label{Fig:petitschamps}
\end{center}
\end{figure}

The first fields I used for the bispectrum analysis were small patches (hence hereafter called small fields) cleaned with the high-resolution Green Bank Telescope {\sc Hi} observations. %The high resolution of these observations allow us not to degrade the \planck maps, however the sky coverage is small as the total area of all fields is $\sim$0.6\% of the sky.
There are nine fields, with historical names : N1, AG, SP, NEP4, SPC4, SPC5, LH2, MC and Bootes. Their area ranges from $0.04$ to $0.1$\% of the sky, with a total area $\sim$0.6\% of the sky. For illustration, Fig.\ref{Fig:petitschamps} shows three of the fields at the frequencies used for the power spectrum analysis (i.e., \planck 143-217-353-545-857 GHz and IRAS 3000 GHz).

\begin{figure}[htbp]
\begin{center}
\includegraphics[width=.6\linewidth]{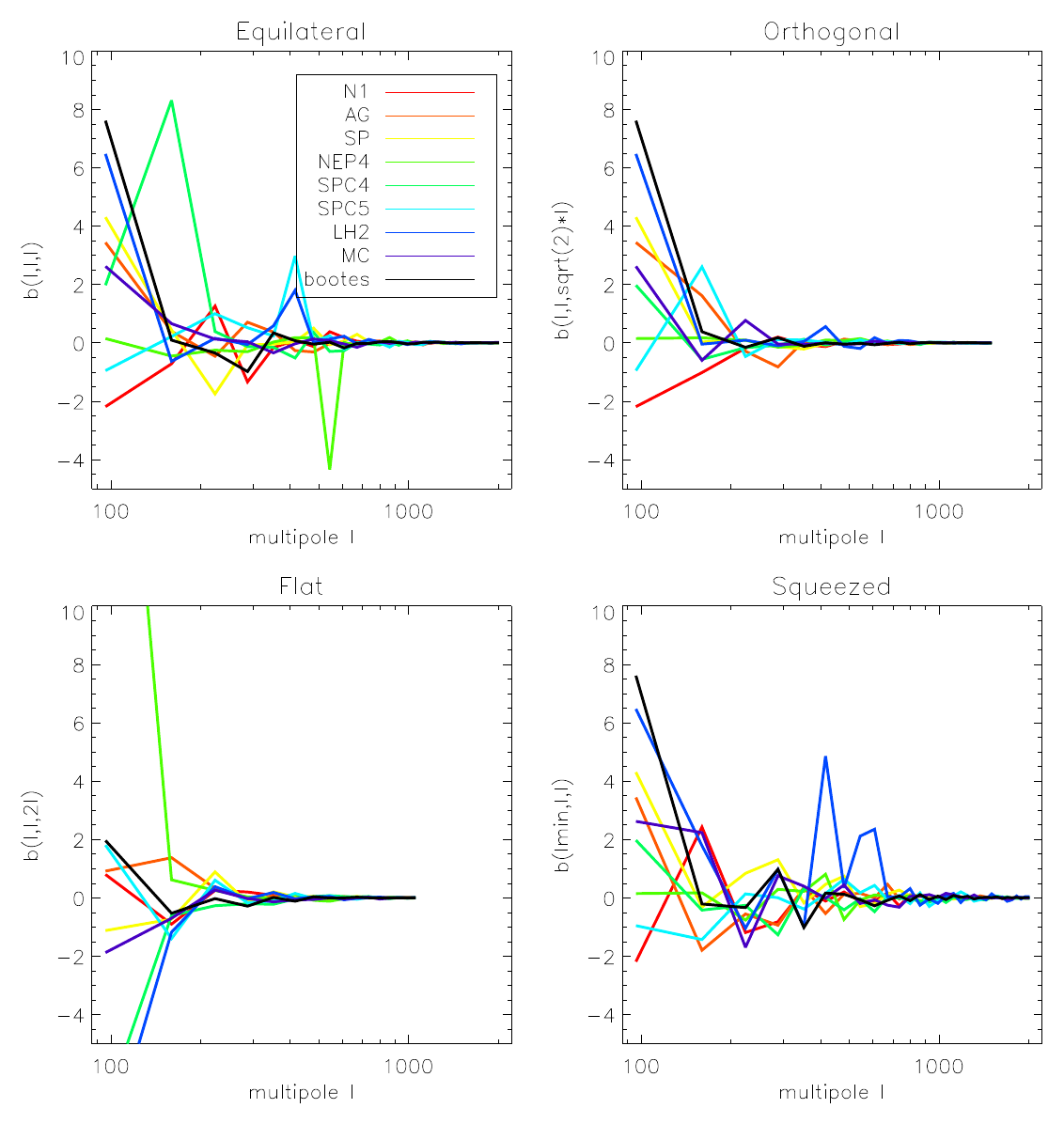}
\caption{Bispectra of the nine different small CIB fields at 545 GHz.}
\label{Fig:bisp_petitschamps_545GHz}
\end{center}
\end{figure}

To use these patches, I developed a flat sky bispectrum estimator. Fig.\ref{Fig:bisp_petitschamps_545GHz} shows the resulting bispectra obtained for each field, at 545 GHz.

\begin{figure}[htbp]
\begin{center}
\includegraphics[width=.7\linewidth]{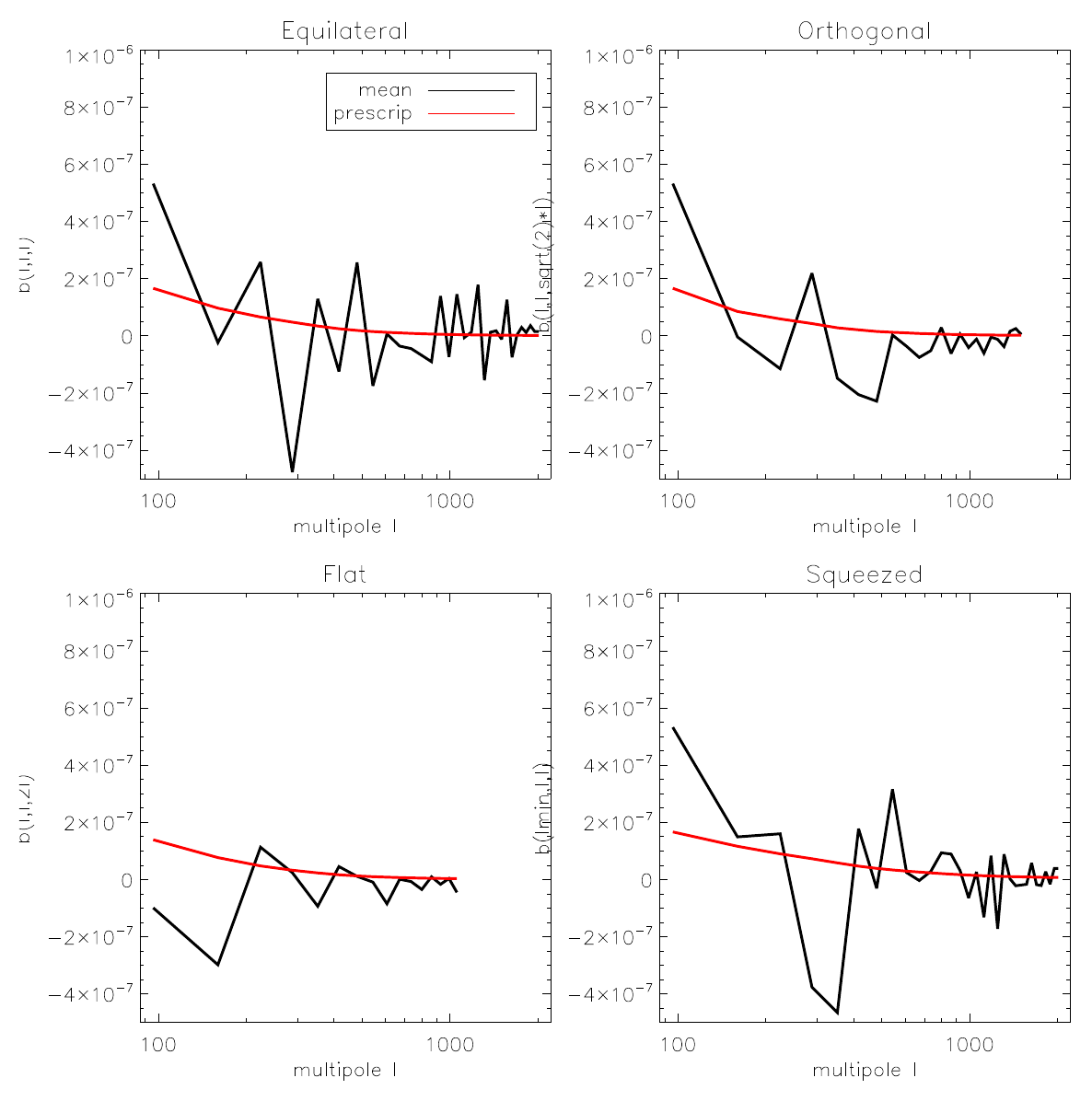}
\caption{$1/f_\mr{SKY}$ weighted mean of the small CIB fields bispectra at 217 GHz (black), with the empirical prescription for the bispectrum overplotted in red for comparison.}
\label{Fig:bisp_moypetitschamps_217}
\end{center}
\end{figure}

We see a large dispersion between the bispectrum estimates, as they are variance-dominated due to the small sky coverage. The estimates however tend to be positive and larger at low multipoles than at high multipoles. To obtain a single estimate with a lower variance, I have computed an inverse-variance weighted average of the bispectra. More precisely, we have seen in Sect.\ref{Sect:bispcov} that, in the weak NG case, the bispectrum variance is~:
\be
\mr{Var}(b_{123})= \frac{C_{\lu}\,C_{\ld}\,C_{\lt}}{N_{123}} \, \Delta_{\lu \ld \lt}
\ee
In the flat sky case, the expression is analog but $N_{123}$ is replaced by $N(k_1,k_2,k_3)$ the number of triplets of Fourier modes such that $\vec{k}_1+\vec{k}_2+\vec{k}_3=\vec{0}$. This latter number depends on the size and geometry of the field considered : the smaller the field, the fewer Fourier modes. Hence the inverse-variance weighting of the bispectra corresponds to a $N(k_1,k_2,k_3)$ weighting. The resulting average bispectrum at 217 GHz is shown in Fig.\ref{Fig:bisp_moypetitschamps_217} along with the expected CIB bispectrum (computed with the prescription) for comparison.

We see that the inverse-variance weighted average has importantly reduced the variance of the bispectrum estimate which seems to follow the prescription at the order of magnitude level. However, the estimate is still too noisy because of the limited sky coverage.% for this reason I have turned to work on a larger field.
In the next section, I describe the analysis of a larger field.

\subsubsection{GASS field : bispectrum estimation pipeline}

\begin{figure}[htbp]
\begin{center}
\includegraphics[width=.5\linewidth]{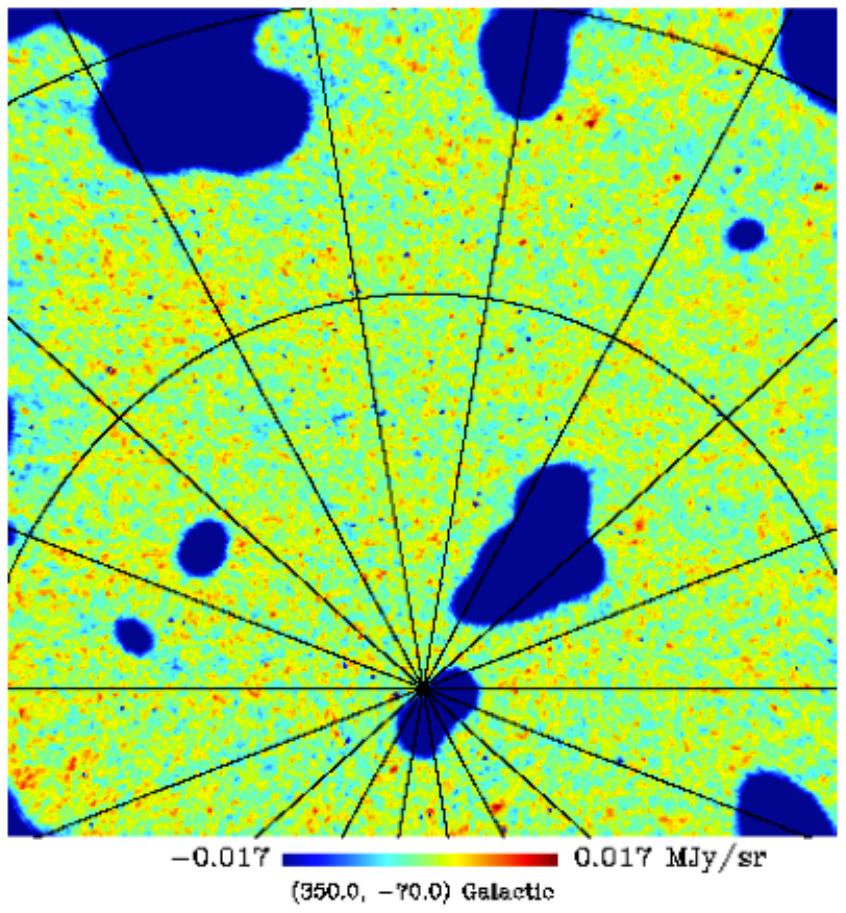}
\caption{From \cite{planck2013-CIB}, cleaned CIB map at 217 GHz in the GASS field.}
\label{Fig:GASS_field_217}
\end{center}
\end{figure}

The other field I have analysed was cleaned with the GASS {\sc Hi} survey \citep{GASS2009}. An estimated map of the Galactic dust emission was constructed with this survey, as detailed in \cite{dustmodel,planck2013-CIB}. This estimated map will be refered to as the dust model hereafter. Compared to the small fields described in the previous section, the GASS field provides a much larger sky fraction, $\sim$10\% of the sky, but with lower resolution (16.2 arcmin). For illustration, figure \ref{Fig:GASS_field_217} shows the cleaned CIB map obtained by \cite{planck2013-CIB} at 217 GHz.

The map is centered around the Galactic South pole and at $N_\mr{side}=512$ HEALPix resolution, so that the maximum multipole of analysis is $\ell_\mr{max}=1024$. With this resolution, the computation of a bispectrum is relatively fast which allows us to easily run tests and simulations.

\begin{figure}[htbp]
\begin{center}
\includegraphics[width=.7\linewidth]{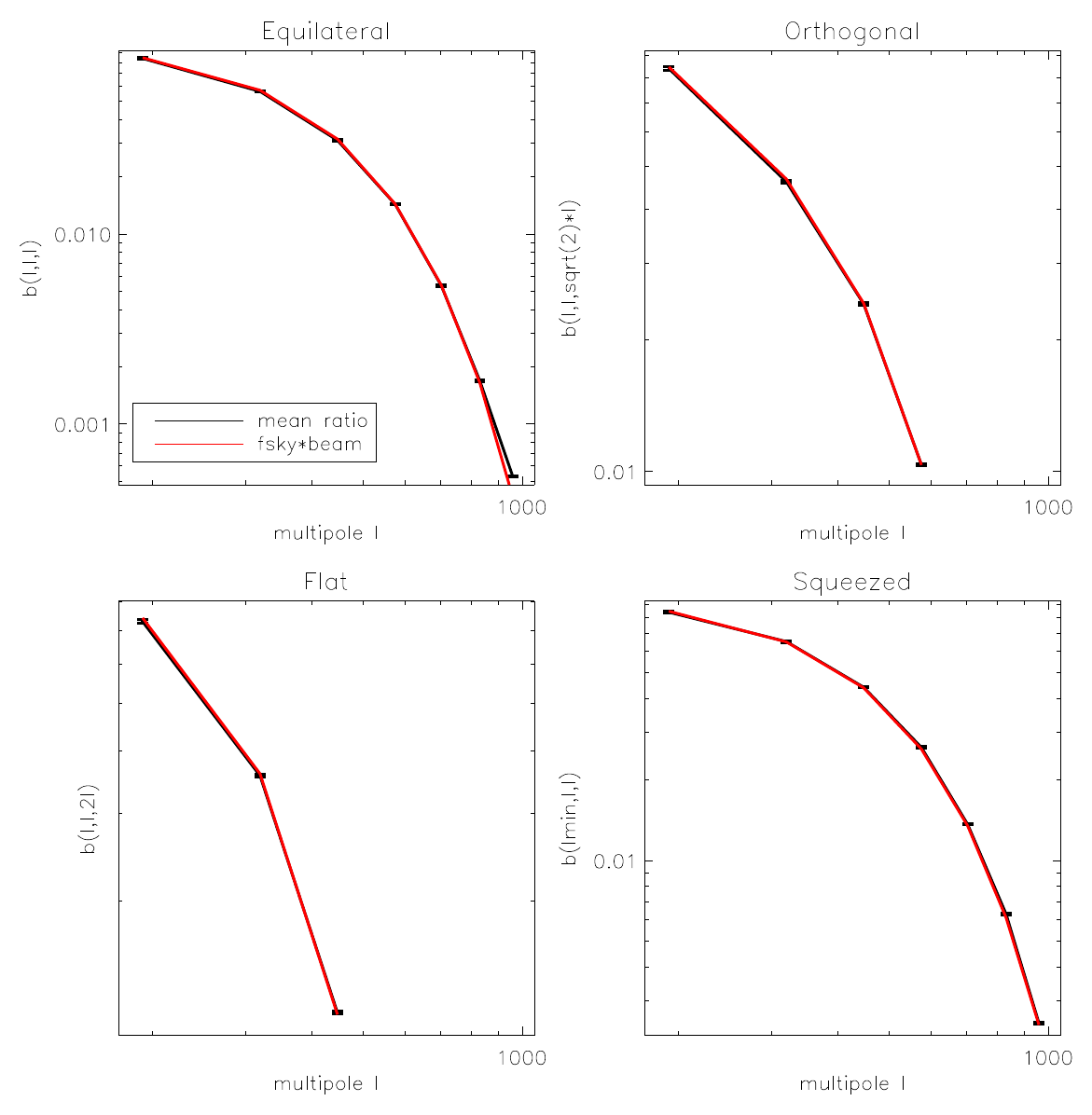}
\caption{Ratio of the masked bispectrum to the full-sky bispectrum (black), compared with the $f_\mr{SKY}$ approximation (red).}
\label{Fig:ratio_invmask}
\end{center}
\end{figure}

The GASS field is large, but a complex mask is needed as seen in Fig.\ref{Fig:GASS_field_217}. As shown in Sect.\ref{Sect:analyt_bispcouplmat} and \ref{Sect:bispcouplmat_sim}, this may induce a bias to the bispectrum estimation, and the coupling matrix cannot be computed at this resolution. To tackle this issue, I first chose a binning size large enough to reduce the coupling between bins, but not too large so as not to lose information on the variations of the bispectrum. I computed the power spectrum coupling matrix to estimate the necessary minimal bin size and found two choices : $\Delta\ell=64$ and $\Delta\ell=128$. Since the estimates with $\Delta\ell=64$ were still noisy (SNR per configuration lower than one), I chose $\Delta\ell=128$. Indeed I wanted to be able to visualise the bispectra, to detect possible contaminations and problematic configurations. I checked that this bin size did not significantly lose bispectrum features\footnote{With the power-law model found in Sect.\ref{Sect:GASSanalysis}, I computed that there is a $>96\%$ correlation between the binned and the full bispectra}. Then I built an average debiasing of the bispectra, assuming a template for it. To do so, I have generated maps with a bispectrum given by the empirical prescription, smoothed them at the GASS resolution (16.2 arcmin), masked them, measured the masked bispectrum and compute the ratio masked/full-sky in each configuration, and average this ratio over simulations. Figure \ref{Fig:ratio_invmask} shows this ratio averaged over 50 simulations, which were enough to reach convergence. It also shows the small error bars on the estimation of this ratio and the analytical ratio in the case of the $f_\mr{SKY}$ approximation.

We see that the measured ratio follows very closely the $f_\mr{SKY}$ approximation indicating that the bin-coupling is small and that we have no complex mask effect. However, it is worth emphasising that this result relies on the assumption that the underlying bispectrum is properly described by the empirical prescription. In the following, I will debias CIB bispectrum estimates with this ratio, and this will provide unbiased estimates as long as the empirical prescription describes correctly the CIB NG.\\
As shown in Sect.\ref{Sect:linearterm_bisp}, a linear correction to the bispectrum is normally also needed to reduce the variance of the bispectrum estimate. I have computed this linear term with 50 simulations and found it completely negligible compared to the cubic term. Indeed, the linear term is meant to correct for the coupling between different bins, but the large bin size decreases this bin coupling.
%I neglected this linear correction hereafter, as it would need a much larger number of simulations for convergence, hence impractical CPU time, and it would remain completely negligible.

\begin{figure}[htbp]
\begin{center}
\includegraphics[width=.7\linewidth]{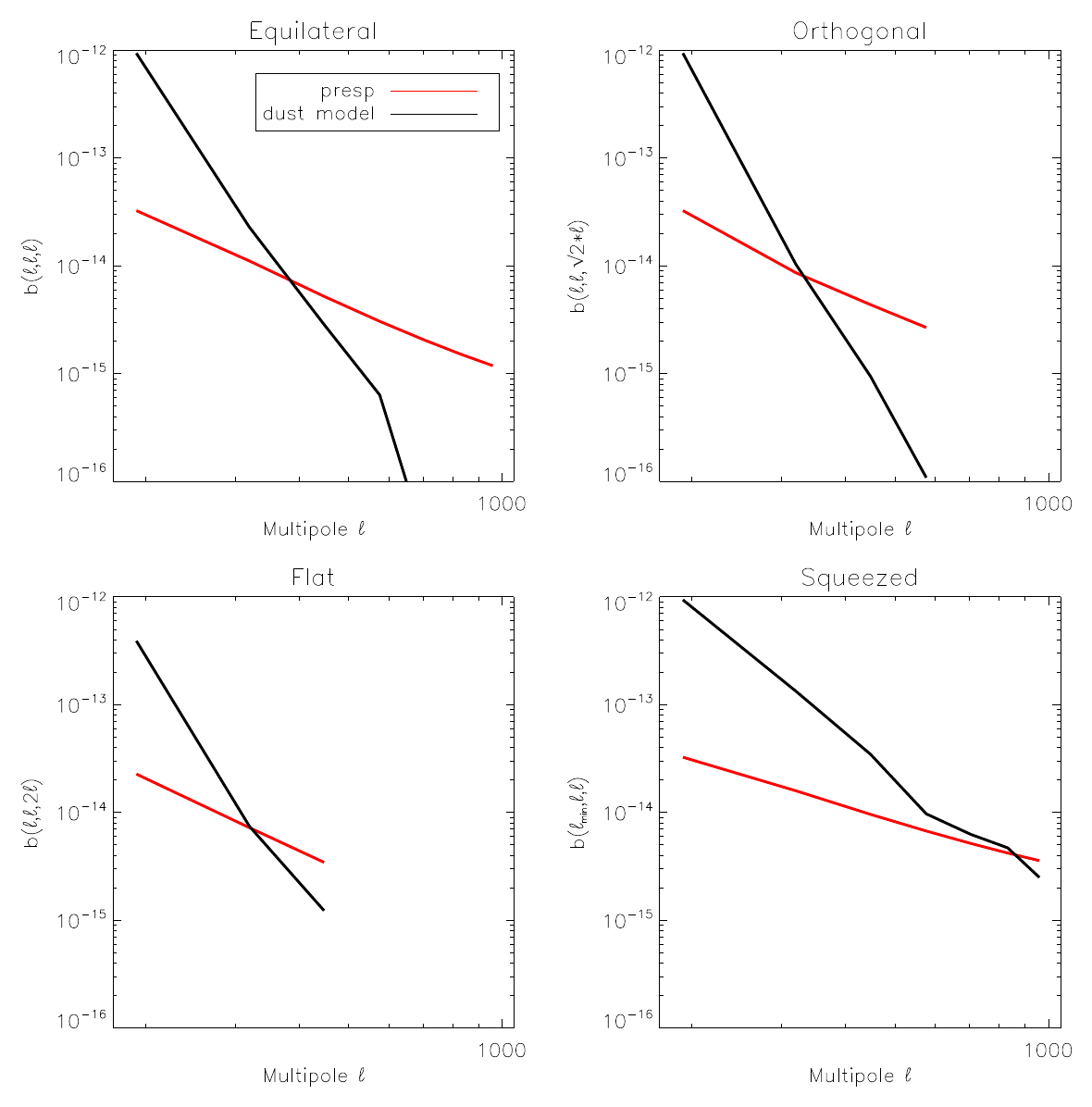}
\caption{Bispectrum of the Galactic dust model at 857 GHz, compared with the empirical prescription for the CIB bispectrum.}
\label{Fig:dustmodel_857}
\end{center}
\end{figure}

Another problem to address in the CIB analysis is the presence of dust residuals. Indeed, the dust model used to clean the \planck maps is not perfect, and there are residuals of Galactic dust in the maps, due in particular to ``dark gas'' not traced by {\sc Hi} \citep{planck2011-darkgas}. Galactic dust has a highly non-Gaussian, and non-isotropic, distribution \citep{planck2011-galdust}. For example, I have computed the bispectrum of the dust model map introduced previously, it is shown in Fig.\ref{Fig:dustmodel_857} at 857 GHz. The figure also shows the empirical prescription for the CIB bispectrum for comparison.

We see that the dust bispectrum dominates the expected CIB bispectrum on large angular scales. In particular in this plot, the first bin ($\ell=2\cdots128$) where the dust bispectrum is too important was discarded. In all my CIB bispectrum estimations, due to the high dust signal as compared to the CIB, this first bin has been discarded. In the second bin, we see that the dust bispectrum must be reduced by a factor $\sim100$ to uncover the CIB signal, which means that the dust model must account for $\sim$80\% of the dust. \cite{planck2013-CIB} did estimate that there are $\sim$20\% dust residuals in the maps compared to the dust model, by comparing the obtained power spectrum with different dust masks. I obtained the same estimate by comparing the raw map bispectrum (without dust cleaning) and the dust model bispectrum at 857 GHz. Hence, the bispectrum estimates involving the second bin may be contaminated by dust, and I include this effect in the error bars of the CIB bispectrum estimation. We see however that CIB bispectrum estimates are safe from dust contamination at higher multipoles, as the dust bispectrum has a steep decrease.

A possibility to minimise the contamination by dust residuals was to use a mask leaving a smaller sky area ($f_\mr{SKY}$=4.6\%) that was designed for the power spectrum analysis. After extensive testing I found that the results were not better than with $f_\mr{SKY}$=10\% and did not justify losing sky area and increasing cosmic variance.

%first-last explanation
%A detail I did not mention until now is that there are in fact two CIB maps per frequency. Indeed a common practice for \planck maps is to build so-called `first ring' and `last ring' maps. These maps are produced with half of the \planck time-ordered data to be precise, i.e. the raw power received by the bolometers as a function of time. They are useful in particular for jacknife tests (the difference of these maps should be pure noise) and for cross-correlation. 
As already mentioned, there are two \planck maps per frequency for the CIB analysis, respectively built from the first and last rings. The CIB power spectrum was estimated with the cross power spectrum of those maps, as this debiases the measurement from noise contamination (as the noise in the first rings is not correlated with that in the last rings, note however that the noise still increases the variance of the measurement). Since the noise is Gaussian, we do not need such considerations for the bispectrum. So I have computed auto-bispectra for the first and last rings respectively, and this allowed me to perform consistency checks by looking at their difference.

\subsubsection{GASS field : results and post-treatment}\label{Sect:GASSpost}

\begin{figure}[htbp]
\begin{center}
\includegraphics[width=.7\linewidth]{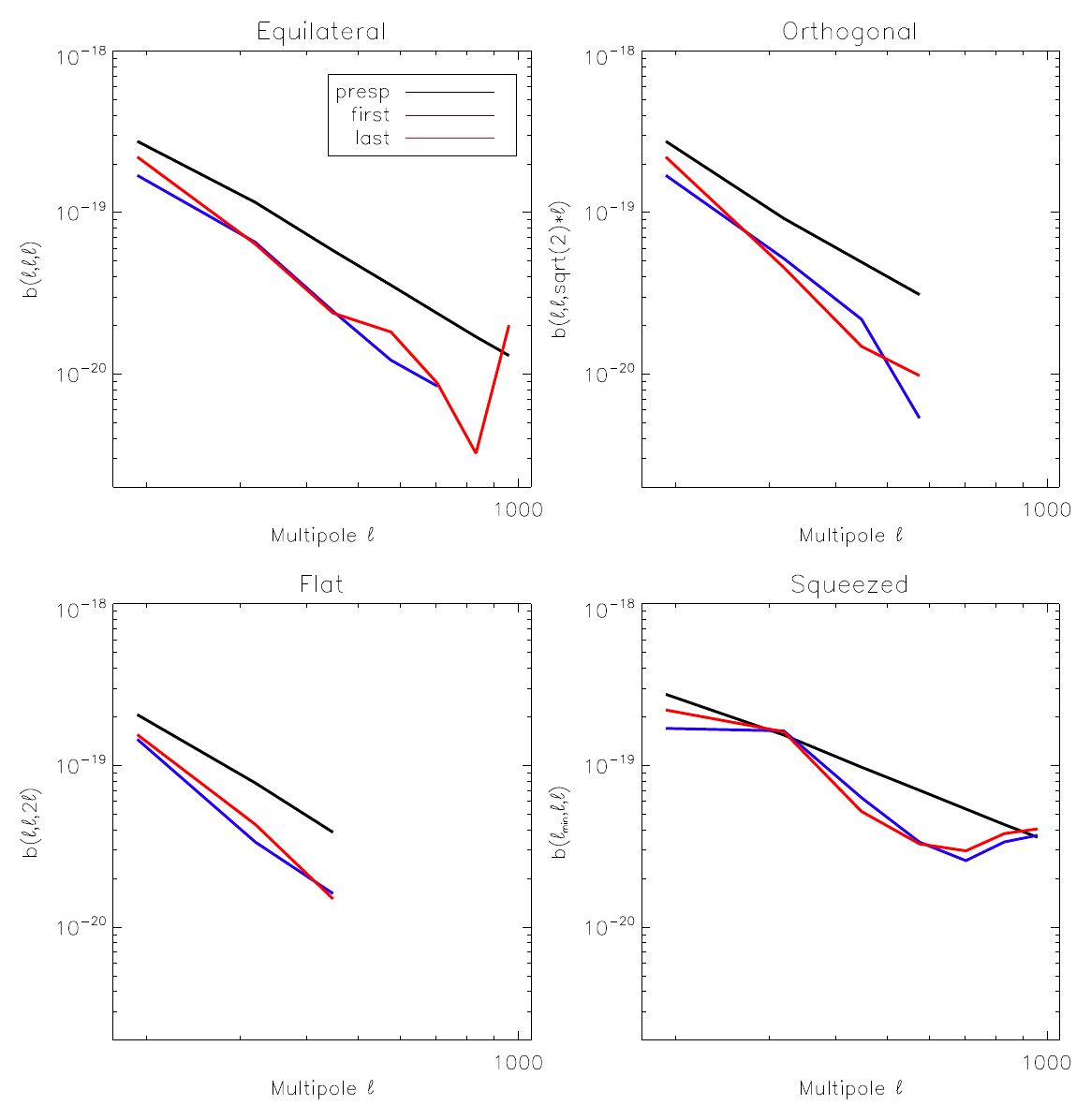}
\caption{Estimated CIB bispectrum at 545 GHz with the first and last rings, along with the prescription for comparison.}
\label{Fig:bisp_GASS_545}
\end{center}
\end{figure}

Considering all the aforementioned details, Fig.\ref{Fig:bisp_GASS_545} shows the estimates of the CIB bispectrum at 545 GHz, along with the empirical prescription for comparison.

We see that the first and last-ring estimates are consistent, except possibly at high multipoles in the equilateral configuration. Moreover, they have the same order of magnitude as the prescription and a similar slope, but they are nevertheless lower.\\
I have then combined these estimates and computed error bars, in fact a covariance, for the measurement. As described in \cite{planck2013-CIB}, the covariance contains the following terms~:
\begin{itemize}
\item cosmic variance : $C_{2\times2\times2}$ term with the $f_\mr{SKY}$ approximation. For the power spectra involved, I estimated the first and last-ring auto-spectra, containing the noise power spectrum, as necessary.
\item cosmic variance : $C_{3\times3}$ term with the $f_\mr{SKY}$ approximation. The bispectrum involved is the estimated CIB bispectrum.
\item $C_{4\times2}$ and $C_6$ terms of the cosmic variance : I have neglected these terms since I did not have estimates of the CIB trispectrum nor 6-point function.
\item dust residual : I used the estimated dust residual fraction and the bispectrum of the dust model. Conservatively, I estimated the bispectrum on the whole GASS field and with the two masks previously described, and took the maximum estimate per configuration.
\end{itemize}

The 143 GHz measurement is dominated by noise : the bispectrum has positive and negative values, with absolute value up to ten times bigger than the prescription prediction. The prescription values are within the error bars, but the noise is too large for the measurement to be of any interest. This is not surprising as the 143 GHz analysis was already difficult at the power spectrum level \citep{planck2013-CIB}. Therefore, I did not consider this frequency any further.\\
I also discarded the 857 GHz frequency for CIB analysis. In this case the bispectrum coefficients involving the second bin are negative. In the following bins, the bispectrum follows the prescription, but exceeds it and rises at high multipoles. I attribute this behaviour to dust contamination. It is indeed found by computing the bispectrum of the masked dust model and of simulations with an input bispectrum as steep as the dust one. Although the $f_\mr{SKY}$ approximation works well for a ``shallow'' bispectrum as shown in Fig.\ref{Fig:ratio_invmask}, for a steep bispectrum the mask can yield a leakage of large-scale power into the small-scales. That is, even if the bin coupling is small, the large-scale power is so important that it contaminates significantly smaller scales. The measurements in the 857 GHz channel have nevertheless allowed me to pinpoint the configurations which are the most prone to dust contamination, and to have an estimation of the contamination pattern.

\begin{figure}[htbp]
\begin{center}
\includegraphics[width=.7\linewidth]{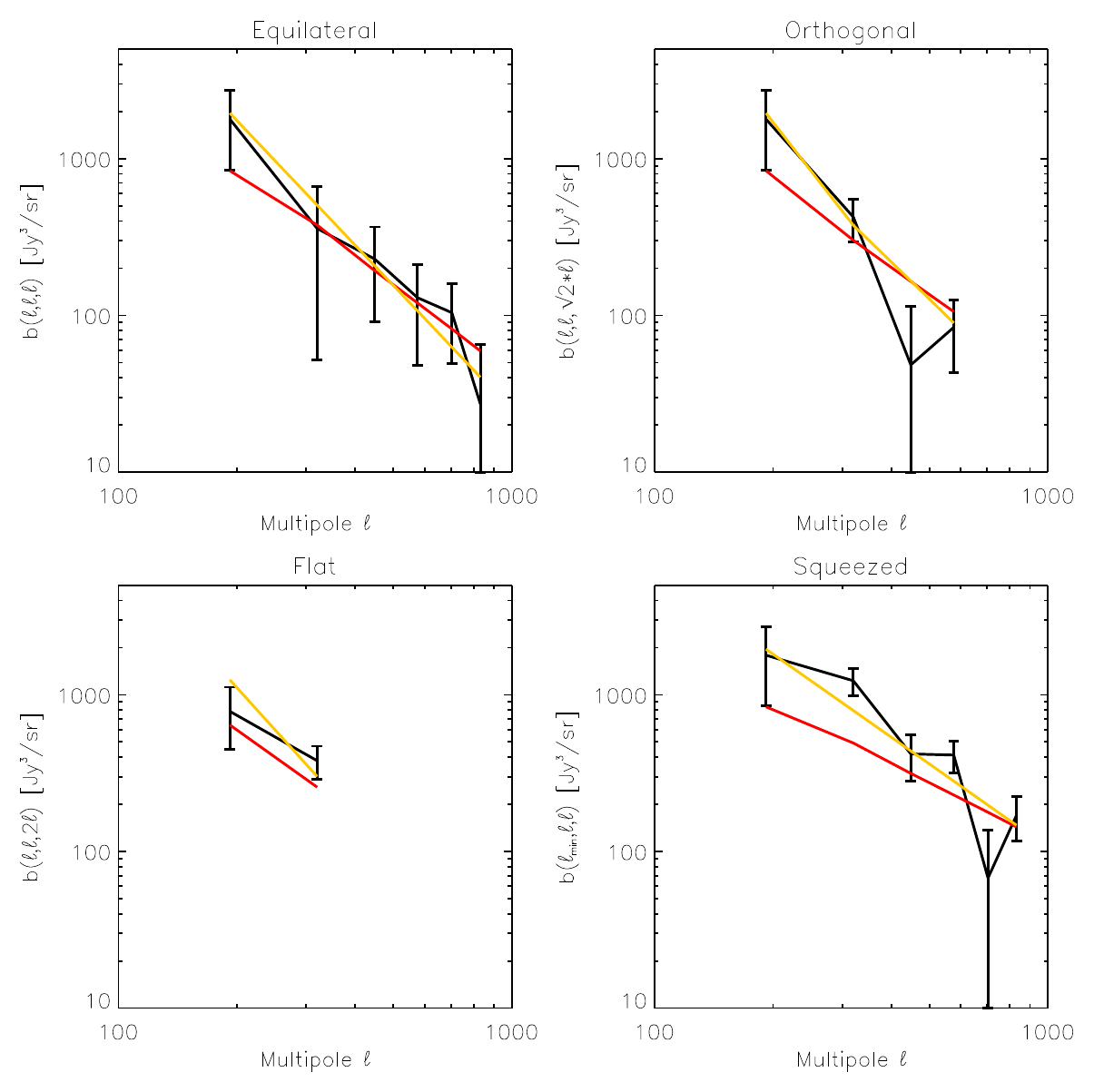}
\caption{From \cite{planck2013-CIB}, measured CIB bispectrum at 353 GHz, along with the best-fit prescription (red) and power law (orange).}
\label{Fig:prespNplfit_Jy_353}
\end{center}
\end{figure}

The CIB bispectrum measurement at 353 GHz with its error bars is shown in Fig.\ref{Fig:prespNplfit_Jy_353}, along with some fits described hereafter.

%discard frequencies and configs
The measurements at 217 and 545 GHz show a similar behaviour, although with larger error bars at 217 GHz and smaller ones at 545 GHz.\\
I then proceeded to check whether the dust contamination pattern visible at 857 GHz was present in the measurements at 217 353 and 545 GHz, and to discard the configurations that seemed affected, depending on the frequencies. I found that some configurations were indeed affected, mostly at high multipoles. There were a bit more of them at 545 GHz than at 217 and 353 GHz. This result is expected as the dust emission falls more steeply than the CIB one when frequency decreases, so that the dust contamination is more important at high frequencies. I also discarded configurations where the bispectrum was lower than a minimum acceptable value\footnote{The minimum acceptable value was 1\% of the minimum value of the empirical prescription bispectrum.}, and where the ratio shown in Fig.\ref{Fig:ratio_invmask} was too different from the $f_\mr{SKY}$ approximation. However, this latter condition did not result in any exclusion, as the $f_\mr{SKY}$ approximation performs well in all configurations of this analysis.

%correction SZ
Nevertheless, the measured bispectra need an additional correction. Even if they are free from CMB and Galactic dust contamination, the bispectra measure the non-Gaussianity of all the other foregrounds. Namely, there is contamination from radio and tSZ, at the observed frequency but also leaking from 100 GHz. Indeed, as stated in the introduction of this section, the CMB component is cleaned with a Wiener-filtering of the 100 GHz map. This 100 GHz map contains contributions from the radio tSZ and CIB signals which will leak into the frequency CIB maps. Hence the observed spherical harmonic coefficients take the form~:
\be
a_{\ell m}(\nu) = a^\mr{CIB}_{\ell m}(\nu) + a^\mr{RAD}_{\ell m}(\nu) + a^\mr{tSZ}_{\ell m}(\nu) - w_\ell \left[a^\mr{CIB}_{\ell m}(100) + a^\mr{RAD}_{\ell m}(100) + a^\mr{tSZ}_{\ell m}(100)\right]
\ee
where $w_\ell$ is the Wiener filter.\\
First, the radio contamination is negligible. The $\bps$ value for the CIB fields computed with the \cite{Tucci2011} number counts is much lower than the CIB measurements. Indeed, as the GASS field is at high galactic latitudes, the point-sources are detected down to lower fluxes than on the rest of the sphere and are thus more efficiently masked.\\
Then, I neglected the leakage from CIB at 100 GHz. Indeed, the IR emission decreases quickly with wavelength so that the intensity at 100 GHz is much smaller than at 217 GHz, and even more so at higher frequencies. Including the CIB leakage would be a small correction, as for the power spectrum \citep{planck2013-CIB}, which dominant term would be the $100\times217\times217$ cross-bispectrum.\\
Finally, the tSZ leakage leads to a contamination by the tSZ bispectrum and by the tSZ-CIB cross bispectra. At the power spectrum level, the cross tSZ-CIB correction is found negligible compared to the tSZ power spectrum correction \citep[3\% vs 15\%, see][]{planck2013-CIB}. For the bispectrum analysis, I neglected the cross-bispectra and focused on the correction of the tSZ bispectrum. For this purpose, I used the bispectrum measured on the projected tSZ catalogue \citep[see Sect.\ref{Sect:meas_SZ_NG} and][]{planck2013-SZcatalogue} as a template so that~:
\be
b_{123}^\mr{corr-tSZ}(\nu) = (g_{\nu}-g_{100}\, w_{\lu}) (g_{\nu}-g_{100}\, w_{\ld}) (g_{\nu}-g_{100}\, w_{\lt}) \, b_{123}^\mr{template}
\ee
where $g_\nu$ is the factor converting Compton parameter $y$ to $\mu K_\mr{CMB}$ at that frequency.\\
The tSZ correction is negligible at 353 and 545 GHz, but important at 217 GHz. It is worth noting that the tSZ signal deos not exactly vanish at 217 GHz. $g_{217}$ is non-zero when we account for the \planck bandpass \footnote{The signal integrated over the frequency response does not vanish exactly.}. Accounting for this effect produces a $\sim15\%$ correction to $b_{123}^\mr{corr-tSZ}$.\\
The last element of the measurement is the assessment of the error bar on the tSZ substraction. For this, I considered a 80\% error bar on the overall amplitude of the bispectrum template, motivated by the difference between the bispectrum produced by the tSZ catalogue, the measured tSZ bispectrum and that of the \planck tSZ simulation (see Sect.\ref{Sect:meas_SZ_NG} for details). The error bars corresponding to this tSZ correction are totally correlated (the corresponding covariance matrix has rank 1). I added this covariance matrix to the previously mentioned one. The total covariance matrix thus becomes highly non-diagonal at 217 GHz.
\newline

%SNR
Finally, we can now assess the significance of the measurement at the three  frequencies 217, 353 and 545 GHz. To this end, I have computed the mean SNR per configuration and the total SNR, which show the significance of the detection. The results are reported in Table \ref{Table:bispSNR}. Note that $\mr{SNR}_\mr{tot}^2 \neq n_\mr{config} \times \mr{SNR}^2_\mr{config}\,$, as the covariance matrix is non-diagonal, and as it is the mean $\mr{SNR}_\mr{config}$ not the rms which is reported. Note also that there were originally 43 configurations before discarding/flagging the contaminated ones.

\begin{table}[htdp]
\begin{center}
\begin{tabular}{|c|c|c|c|}
frequency & number of config. & mean SNR per config. & total SNR \\
\hline
217 GHz & 38 & 1.24 & 5.83 \\
353 GHz & 40 & 2.85 & 19.27 \\
545 GHz & 36 & 4.59 & 28.72
\end{tabular}
\end{center}
\caption{From \cite{planck2013-CIB}, detection significance of the CIB bispectra at each frequency.}
\label{Table:bispSNR}
\end{table}

We find a very significant detection of the CIB bispectrum, even per configuration. Note that the significance increases with frequency as the noise becomes lower relative to the signal. 
%The exploitation of these measurements is tackled in Sect.\ref{Sect:GASSanalysis}
These measurements can be compared with a recently anounced detection of the CIB non-Gaussianity by \cite{Crawford2013} with South Pole Telescope (SPT) data. They computed a template fitting of the bispectrum measured on SPT data at 95-150-220 GHz, with three components :  a $\bps$ template, a template for the CIB derived from the empirical prescription of \cite{Lacasa2012} and a template for the tSZ bispectrum derived from \cite{Bhattacharya2012}. They reported a $\sim5\sigma$ detection of the CIB bispectrum with the multi-band analysis.\\
In comparison with the SPT analysis, the measurements with \planck data provide a detection at each frequency instead of a multi-band analysis. The measurements have a higher significance, the least-significant being a $5.8\sigma$ detection, while the most significant is a $28.7\sigma$ detection. Also, I considered the non-Gaussian $C_{3\times3}$ contribution to the covariance, while the SPT analysis only considers the Gaussian contribution, i.e., a variance but no off-diagonal terms. As will be argued in Sect.\ref{Sect:meas_SZ_NG}, this probably overestimates their detection significance, in particular because of the tSZ contribution in their multi-band analysis, as the tSZ is highly non-Gaussian.

\subsubsection{GASS field : exploitation of the measurements}\label{Sect:GASSanalysis}

I compared in Fig.\ref{Fig:bisp_GASS_545} the bispectrum measured on the CIB maps with the empirical prescription, described in Sect.\ref{Sect:presp}. With the amplitude parameter $\alpha\approx3\!\cdot\!10^{-3}$, the value found on \cite{Sehgal2010} simulations, the prescription overestimates the measurement at all frequencies. So I fitted for $\alpha$ through a $\chi^2$ minimisation with the covariance matrix described previously. The results are found in Table \ref{Table:fitalphabisp}, along with the error bar on $\alpha$, the best-fit $\chi^2$ and the number of degrees of freedom.

\begin{table}[htdp]
\begin{center}
\begin{tabular}{|c|c|c|c|}
frequency & $\alpha$ & $\chi^2$ & d.o.f. \\
\hline
217 GHz & $(1.89\pm0.53)\!\cdot\! 10^{-3}$ & 21.4 & 37 \\
353 GHz & $(1.21\pm0.07)\!\cdot\! 10^{-3}$  & 45.5 & 39 \\
545 GHz & $(1.56\pm0.06)\!\cdot\! 10^{-3}$ & 95.9 & 35
\end{tabular}
\end{center}
\caption{From \cite{planck2013-CIB}, best-fit amplitude parameter for the empirical prescription, its error bar and the chi-square value of the fit.}
\label{Table:fitalphabisp}
\end{table}

Note that the obtained values of $\alpha$ are consistent with each other. This is a strong indication that the CIB measurements have little to no residual contamination. Indeed, the intrinsic level of non-Gaussianity is found consistent between frequencies, as expected. Note also that the $\chi^2$ degrades as frequency increases because the noise decreases with frequency. At 217 GHz, the error bars may be overestimated, in particular due to the tSZ correction, so that the $\chi^2$ is underestimated.

The prescription fit does not provide exceptional $\chi^2$ values, furthermore the measurement seem to have a constant slope in each configuration and to be steeper than the prescription, see Fig.\ref{Fig:bisp_GASS_545} and \ref{Fig:prespNplfit_Jy_353}. I have thus also fitted for a separable power law~:
\be
b^\mr{pl}_{123} = A \times \left(\frac{\lu \, \ld \, \lt}{\ell_0^3}\right)^{-n}
\ee
where the amplitude A and the index $n$ are the two parameters. I chose a pivot scale $\ell_0=320$, which is the center of my third multipole bin\footnote{Thus, A is the best-fit value for the CIB bispectrum in the equilateral configuration (320,320,320). A change of pivot scale amounts to a rescaling of A.}. I have fitted for A and $n$ through $\chi^2$ minimisation with the aforementioned covariance matrix, and I computed their error bars and their correlation with a Fisher matrix. I checked that the Fisher matrix reproduces adequately the 1-2-3$\sigma$ contours of the likelihood. The results are shown in Table \ref{Table:fitpowerlawbisp}~: best-fit amplitude and index, their errors and correlation, and the best-fit $\chi^2$ and numbers of d.o.f..

%remarque : ce sont les bonnes dernires valeurs (au 10/7/2013)
\begin{table}[htdp]
\begin{center}
\begin{tabular}{|c|c|c|c|c|c|}
frequency & A $[\mathrm{Jy}^3/\mathrm{sr}]$ & $n$ & correlation & $\chi^2$ & d.o.f. \\
\hline
217 GHz & $(1.46\pm0.68) \!\cdot\! 10^{1}$ & 0.822 $\pm$ 0.145 & 81.4\% & 20.6 & 36 \\
353 GHz & $(5.06\pm0.49) \!\cdot\! 10^{2}$ & 0.882 $\pm$ 0.070 & 82.3\% & 34.3 & 38 \\
545 GHz & $(1.26\pm0.09) \!\cdot\! 10^{4}$ & 0.814 $\pm$ 0.050 & 85.4\% & 82.8 & 34
\end{tabular}
\end{center}
\caption{From \cite{planck2013-CIB}, best fit amplitude A and index $n$ of a power law fit to the bispectra, their error bars and associated correlation, and the chi-square value of the fit. The pivot scale is $\ell_0=320$}
\label{Table:fitpowerlawbisp}
\end{table}

We see that the indices are consistent with each other : $n\sim0.8-0.9$, indicating again that there is little or no residual contamination at these frequencies. Note that the power law provides a significantly better fit to the data at 353 and 545 GHz than the prescription with lower $\chi^2$ values. Interestingly, we find indication that the CIB bispectrum is steeper than the prescription. The obtained indices are significantly larger than those of the prescription (which predicts $n\sim0.6$, as $C_\ell \propto \ell^{-1.2}$), which is an interesting result.\\ %This is an interesting result, which will need to be confirmed.\\
Indeed, we have seen that the modelisation with the halo model, Sect.\ref{Sect:CIB_HM}, produces a bispectrum in general agreement with the prescription for the emissivity models considered. A possible solution to the disagreement between the halo model and the measurement would be to change the IR emissivities. For example, \cite{Bethermin2013} recently described a model for the CIB power spectrum different from what is presented in Sect.\ref{Sect:CIB_HM}, considering IR emissivities depending on the halo mass (and on the galaxy type : central/satellite). They showed that the IR emissivity decreases in massive haloes due to the quenching of star formation. This leads to a decrease of the power spectrum 1-halo term, so that the 2-halo term dominates over the multipole range of interest for \planck. This was confirmed by the \planck CIB power spectrum results compared to a similar modeling \citep{planck2013-CIB}. For the bispectrum, this effect would also decrease the 1-halo term for the bispectrum, and most probably also the 2-halo term, compared to the 3-halo term. These changes would be consistent with a steeper-than-prescription bispectrum~: e.g., the 3-halo term goes as $P(k)^2$ in equilateral which falls more steeply than the prescription going as $P(k)^{3/2}$.
\newline

To conclude, these CIB NG measurements open an interesting window for CIB science. They provide the opportunity to constrain CIB models with both 2-point and 3-point correlations, and they may indicate that the IR emissivity models need to depend not only on the redshift but also on other explicative parameters.

%%%%%%%%%%%%%%%%%%%%%%%%%%%%%%%%%%%%%%%%%%%%%
\section{Measuring the thermal Sunyaev-Zel'dovich NG}\label{Sect:meas_SZ_NG}

As mentioned in Sect.\ref{Sect:SZ}, the thermal Sunyaev-Zel'dovich (tSZ) effect is an important secondary anisotropy of the CMB, leaving a negative signal below 217 GHz and positive above\footnote{for relativistic electrons the zero frequency increases slightly with temperature}. As it probes preferentially the most massive galaxy clusters at low redshift ($z$=0-1), the tSZ signal is highly non-Gaussian. This section describes the tSZ bispectrum measurement that I have performed, in particular in the \planck context.

%%%%%%%%%%%%%%%
\subsection{On simulations}
Although the tSZ signal is highly non-Gaussian, its contribution to the NG in CMB maps is in general relatively lower than the other foregrounds (radio sources, CIB). For illustration, I have computed the bias that the tSZ introduces to local-type primordial NG estimation with \cite{Sehgal2010} simulations, and found that $|\Delta\fnl|<1.5$ at all frequencies (see Table \ref{Table:Dfnl_tSZ}).

\begin{table}[htbp!]
\begin{center}
\begin{tabular}{c|c|c|c|c|c|c|}
$\nu$ (GHz) & 30 & 90 & 148 & 219 & 277 & 350 \\
\hline
$\ell_\mr{max}=700$ & -0.34 & -0.18 & -0.04 & $5\cdot10^{-7}$ & 0.04 & 0.47 \\
$\ell_\mr{max}=2048$ & -0.92 & -0.51& -0.11 & $1\cdot10^{-6}$ & 0.11 & 1.30
\end{tabular}
\end{center}
\caption{Bias on the primordial NG parameter $\fnl$ due to the tSZ signal simulated by \cite{Sehgal2010}.}
\label{Table:Dfnl_tSZ}
\end{table}

I have also computed the tSZ bispectrum on a simulation by the \planck Sky model \citep[PSM,][]{PSM2013}, with a binning $\Delta\ell=64$ and $\ell_\mr{max}=2048$. The \cite{Sehgal2010} simulations yield the same behaviour.

\begin{figure}[htbp]
\begin{center}
\includegraphics[width=\linewidth]{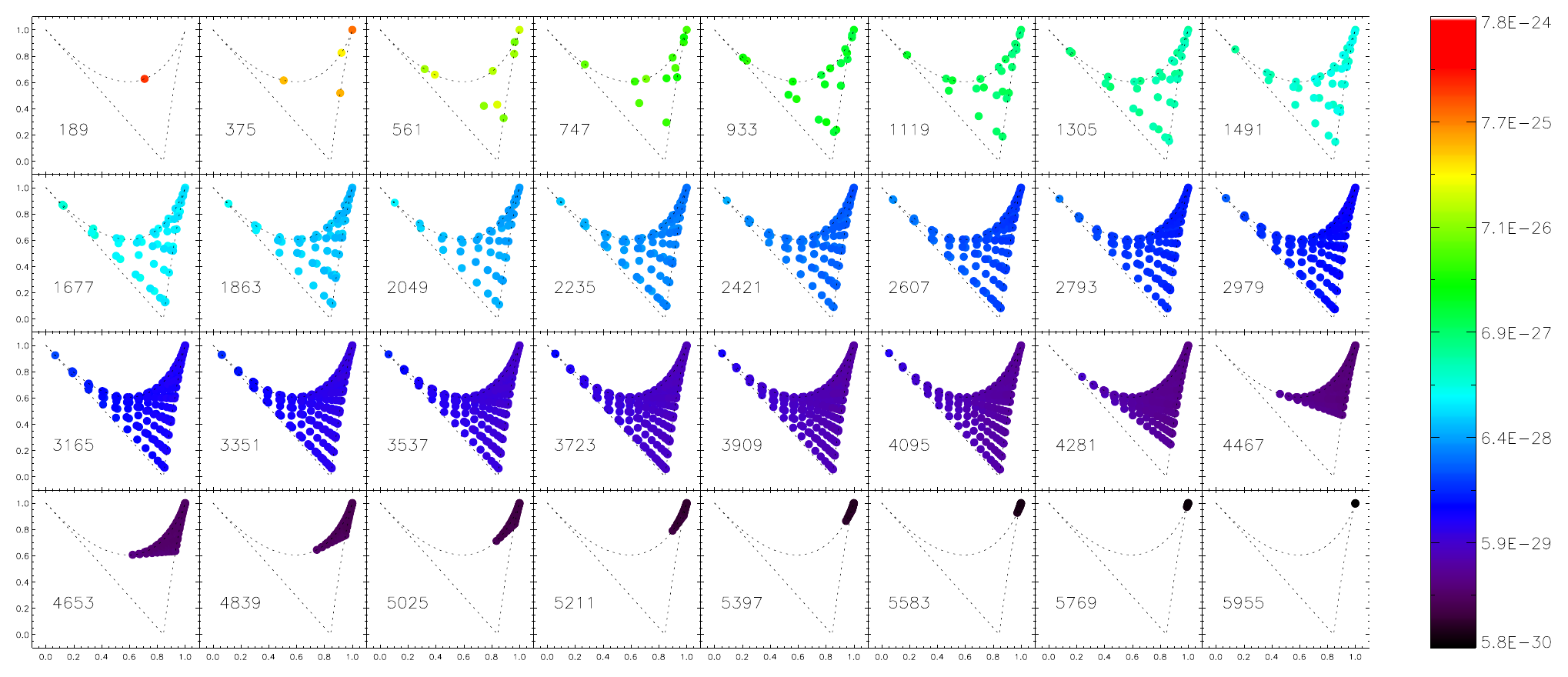}
\caption{tSZ bispectrum computed from a simulation by the \planck Sky Model, plotted in the geometrical parametrisation.}
\label{Fig:param_TSZ-psm}
\end{center}
\end{figure}

We see from Fig.\ref{Fig:param_TSZ-psm} that the tSZ bispectrum is scale dependent but has little dependence on the shape of the triangle, as its value is nearly constant in each subplot. By analogy with the CIB, this means that the 1-halo term the bispectrum (see e.g. Fig.\ref{Fig:bl_CIB_inparam}). %, as it does at the power spectrum level \citep[see e.g.][]{planck2013-SZmap}.

We see that the tSZ signal has a particular NG signature in comparison with the other signals studied previously. Indeed, the tSZ bispectrum depends on scale contrary to the radio bispectrum, and contrary to the CIB bispectrum it does not peak in squeezed and has little/no dependence on configuration. Non-Gaussianity is thus potentially a powerful tool to discriminate these foregrounds, in particular the tSZ and CIB whose power spectra have similar slopes on the range of multipoles of interest for \planck \citep{planck2013-SZmap}.

%%%%%%%%%%%%%%%
\subsection{\planck tSZ data : characteristics and processing pipeline}

\begin{figure}[htbp]
\begin{center}
\includegraphics[width=.7\linewidth]{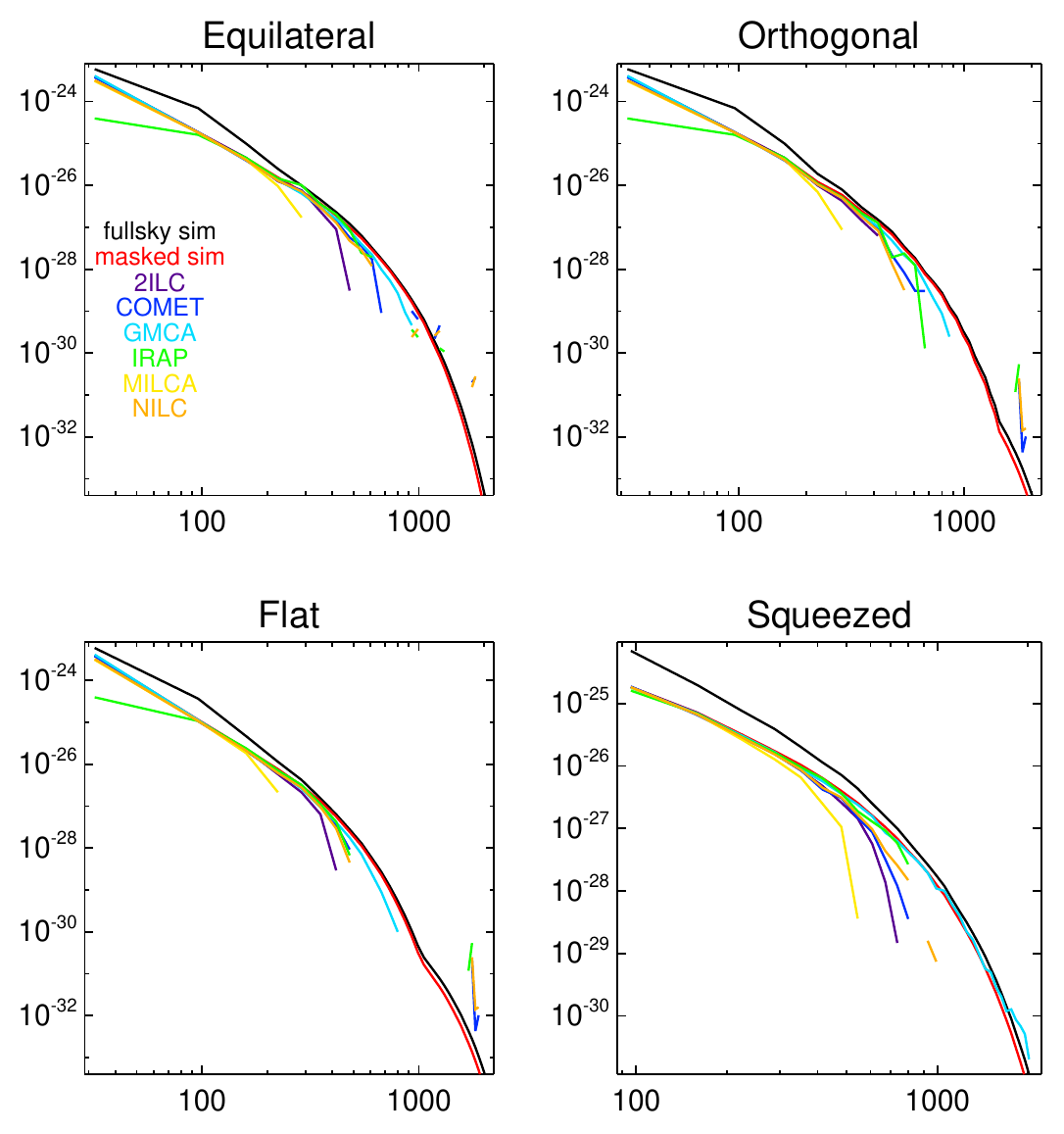}
\caption{tSZ bispectrum from the input tSZ map (black full-sky, red masked) and its measurement on tSZ reconstructions by several component separation methods with the PSM simulations \citep{PSM2013}, plotted in some chosen configurations.}
\label{Fig:comp_compsep_4SZ}
\end{center}
\end{figure}

Due to its unique spectral signature, the tSZ signal can be detected with adapted filters. The large frequency range of \planck  yields a data set of high quality for this effort. For example, the tSZ effect of about a thousand clusters has been revealed by \planck through adapted detection algorithms \citep{Herranz2002,Melin2006,Carvalho2012}. This sample and its characterisation form the \planck SZ catalogue \citep{planck2013-SZcatalogue}.\\
Another project of the \planck Collaboration is the estimation of an all-sky tSZ map, mapping the Compton parameter $y$. To this end, numerous component separation methods have been proposed. I have participated in the assessment of their relative quality, the quality criterion being the conservation of the tSZ NG. Namely, the PSM produced maps of the different sky components, which were combined to yield \planck simulated frequency maps. Then the different component separation methods took these frequency maps as input and estimated a tSZ map. Different tests were performed~: recovery of the power spectrum, recovery of the fluxes and profile of the input clusters, and preservation of the NG character. I took charge of comparing the non-Gaussianity performance of the different component separation methods. I measured the bispectrum of the map produced by each method, with a galactic mask and a binning $\Delta\ell=64$ and $\ell_\mr{max}=2048$, and compared it with the bispectrum of the original tSZ map. Figure \ref{Fig:comp_compsep_4SZ} shows the bispectrum of the full-sky original tSZ map (black), of the masked tSZ map debiased by $f_\mr{SKY}$ (red), and of the different component-separated maps debiased by $f_\mr{SKY}$ (colors).

We see that the methods have different performances. Most of them allow a good recovery of the bispectrum at low multipoles, although the IRAP method underestimates the first bin. At higher multipoles the methods tend to underestimate the bispectrum, then the line breaks when the bispectrum becomes negative. It afterwards becomes noise-dominated, alternating between positive and negative values. This break happens between $\ell=250$ and $\ell=900$ depending on the method, with NILC \citep[Needlet Internal Linear Combination,][]{Remazeilles2011} and GMCA \citep[General Morphological Component Analysis,][]{Bobin2013} performing the best, as they extend tohigher multipoles.\\
%One caveat is that I did not have the time to perform an additional test masking the detectable point-sources (in addition to the galactic mask)\,; this is expected to improve the performance of the methods. In particular, 
It is worth noting that radio-sources are known to leave negative residuals in the MILCA maps \citep[Modified Internal Linear Combination Algorithm,][]{Hurier2013}, and hence a negative bispectrum which should explain the break at high multipoles in Fig.\ref{Fig:comp_compsep_4SZ}.

Finally, based on the different tests, the component separation methods that were selected for the analysis of \planck data were NILC and MILCA.
\newline

%y-map paper description
The \planck y-map article \citep{planck2013-SZmap} presents the first estimated full-sky maps of the tSZ signal. It uses the \planck data as well as the FFP6 (Full Focal Plane and version 6) simulations. These simulations use the PSM as signal model, and take it through the \planck data processing pipeline, from time-ordered input to the mapmaking stage, adding noise and various instrumental artefacts along the way, finally producing \planck-like frequency maps. \cite{planck2013-SZmap} further quantifies the contamination of the recovered tSZ maps by foregrounds and assesses their consistency with the PCCS tSZ catalogue by comparing the detection of sources in the maps and the corresponding tSZ flux. Then the maps are analysed with the following statistics : angular power spectrum and bispectrum, and 1D p.d.f.. The power spectrum is estimated through cross-correlation of the first and last rings maps, and fitted with a halo model description with 1-halo and 2-halo terms (see Sect.\ref{Sect:SZ_HM}). As the tSZ power spectrum depends on the cosmology, the measurement allows to infer constraints on the cosmological parameters. Namely, the main cosmological parameters having an effect on $C_\ell^\mr{tSZ}$ are $\sigma_8$ and $\Omega_\mr{m}$. The tSZ power spectrum analysis found the following constraint on these parameters : $\sigma_8 = 0.74 \pm 0.06$ and $\Omega_\mr{m} = 0.33 \pm 0.06$, with a strong degeneracy along $\sigma_8 \cdot \Omega_\mr{m}^{0.39} = \mr{cste}$. The 1D p.d.f. analysis compared the recovered tSZ maps skewness to that of the FFP6 simulation and assumed that this skewness scaled as $\sigma_8^{10.7-11.1}$, it found the constraint $\sigma_8 = 0.779 \pm 0.015$. In the following, I describe the bispectrum analysis of the recovered tSZ maps.  
\newline

\begin{figure}[htbp]
\begin{center}
\includegraphics[width=.7\linewidth]{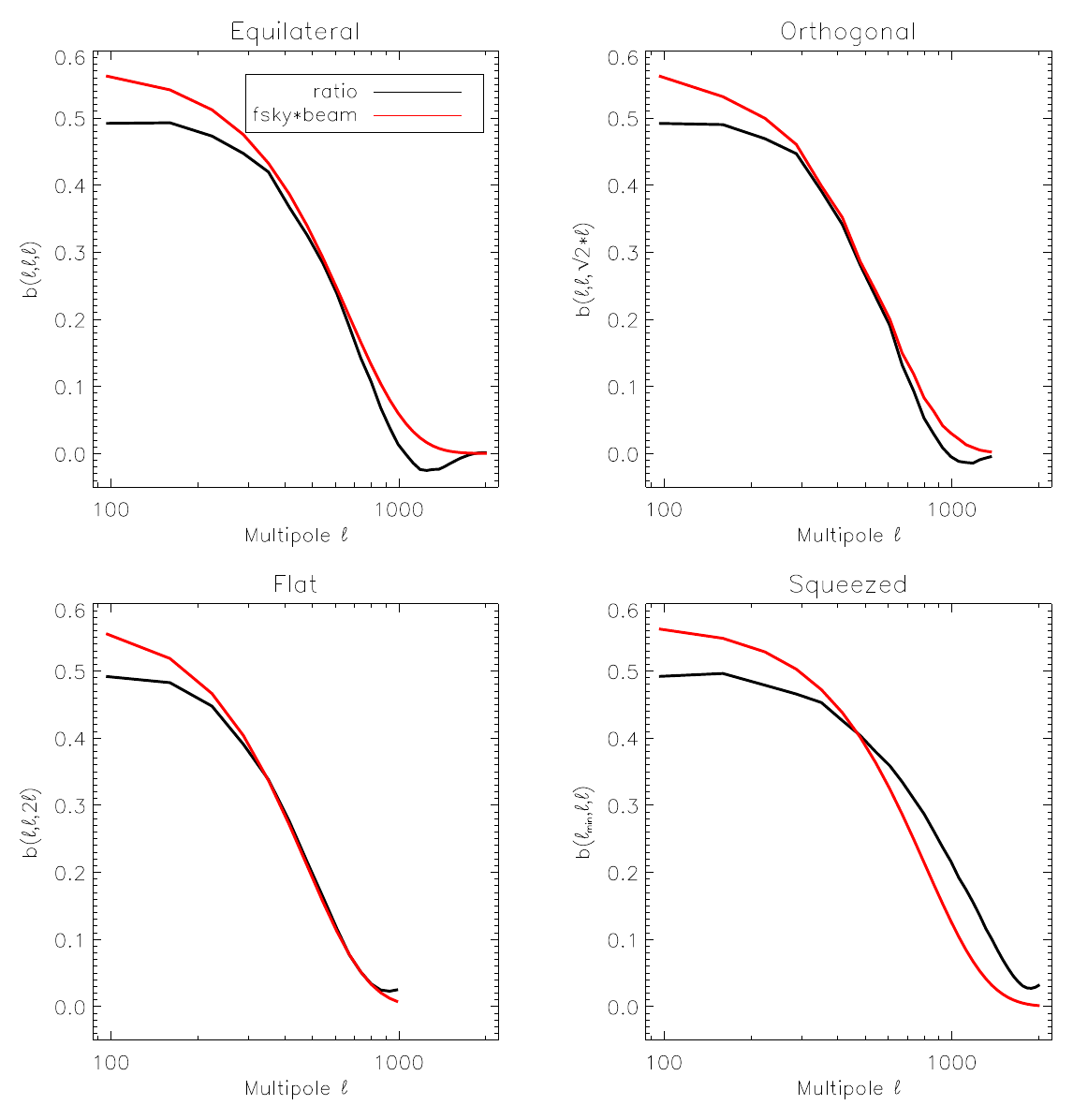}
\caption{Masked sky to full-sky bispectrum ratio with simulations of a tSZ-like bispectrum, compared with the $f_\mr{SKY}$ approximation.}
\label{Fig:ratio_tofull}
\end{center}
\end{figure}

%Mask
As already shown in Fig.\ref{Fig:mixmatcl}, a Galactic and point-source mask was used for the tSZ analysis, leaving $\sim 60\%$ of the sky available. In order to reduce the variance of the bispectrum estimation, I measured the linear correction term discussed in Sect.\ref{Sect:linearterm_bisp}. This correction is negligible in most configurations except in squeezed, where it reaches 3.5\% \footnote{The smallness of the linear correction is due to the fact that the multipole bins are large, while the correction peaks in squeezed because the mask induces mostly a leakage of large scales power to small scales}.\\
The other problem induced by the mask is the bias, see Sect.\ref{Sect:analyt_bispcouplmat}. To correct for this, I have runned simulations using the method described in Sect.\ref{Sect:whitenoise}. These simulations had a tSZ-like bispectrum with a high SNR, and I measured the ratio of the masked bispectrum to the full-sky bispectrum. The results are shown in Fig.\ref{Fig:ratio_tofull} and compared to the $f_\mr{SKY}$ approximation.

We see that contrary to the CIB (Fig.\ref{Fig:ratio_invmask}), the ratio does not follow the $f_\mr{SKY}$ approximation. Instead, compared to the $f_\mr{SKY}$ approximation, the ratio shows a deficit at low multipoles and an excess in the squeezed configurations at high multipoles ($\ell\sim1000$). The ratio becomes negative in equilateral and orthogonal for some high multipoles configurations. This situation is due to the fact that the tSZ bispectrum is much steeper than the CIB one, so that the large scale power leaks onto the small scales. The negative values of the ratio can be interpreted as resulting from the point-source mask. Indeed, the latter yields numerous small areas with zero value (while the $y$-map is positive) yielding negative skewness on small angular scales.\\
When the measured ratio is too different from the $f_\mr{SKY}$ approximation, the mask-induced leakage is too important and we cannot uncover the tSZ bispectrum. This happens in particular at high multipoles, where the tSZ bispectrum estimates indeed diverge. In the analysis, I have discarded affected configurations, for which the measured ratio differs from the $f_\mr{SKY}$ approximation by more than 50\%. This excluded most configurations with a multipole higher than 1024, and I could run new simulations at $\ell_\mr{max}=1024$ and use a resolution of $N_\mr{side}=512$. I ran 100 simulations at $N_\mr{side}=512$. They confirmed Fig.\ref{Fig:ratio_tofull}, and showed that the ratio converged.
\newline

\begin{figure}[htbp]
\begin{center}
\includegraphics[width=.7\linewidth]{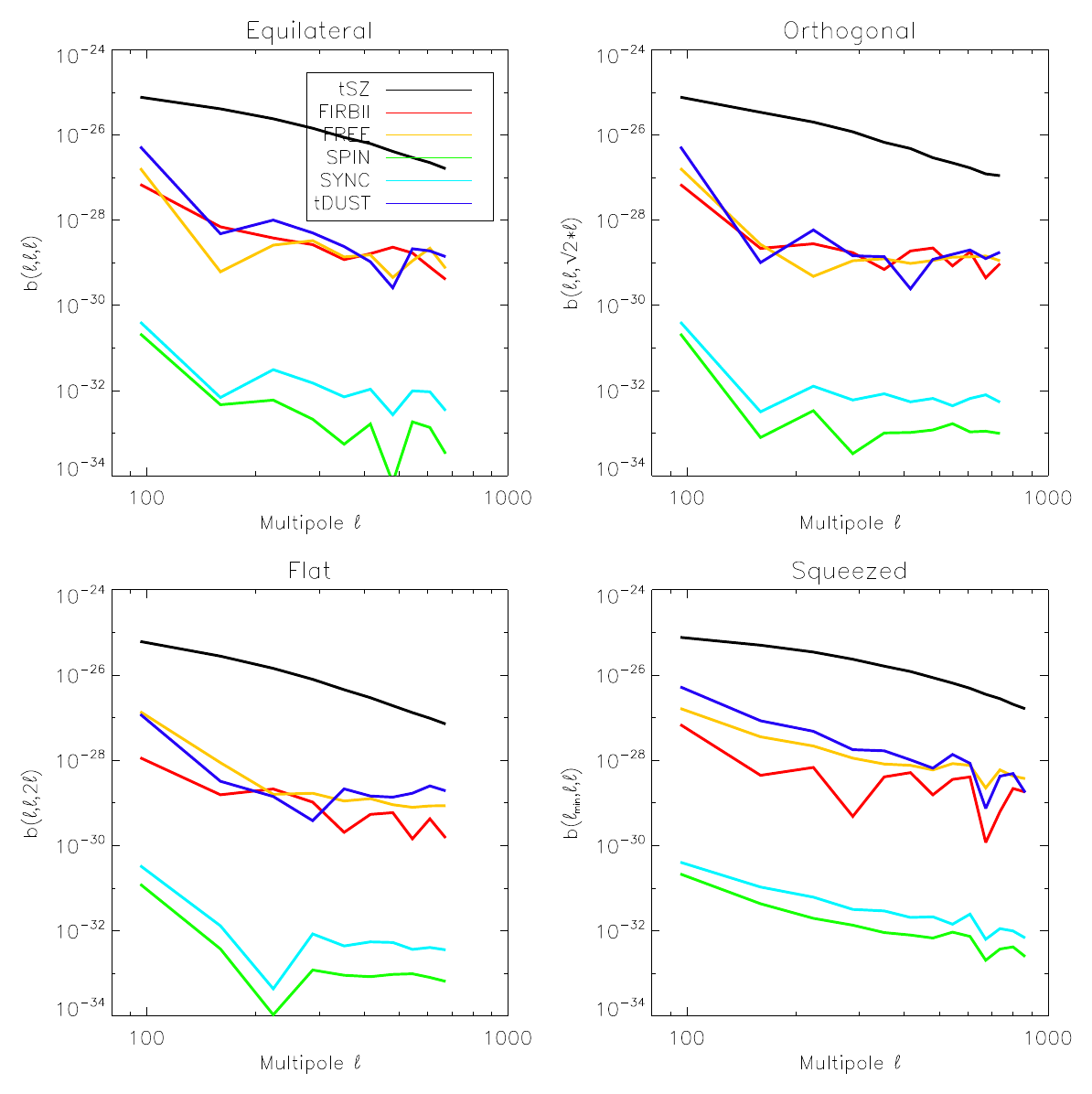}
\caption{From \cite{planck2013-SZmap}, comparison of the tSZ bispectrum and the foregrounds residuals (in absolute value) for the MILCA \citep{Hurier2013} component separation method on FFP6 simulations.}
\label{Fig:FFP6_fg_MILCA}
\end{center}
\end{figure}

Having taken care of the mask effect, we also need an in-depth assessment of the foreground effects, i.e., of their possible contamination to the bispectra. To this end, the FFP6 (Full Focal Plane and version 6) simulations were used. These simulations use the PSM as signal model, and take it through the \planck data processing pipeline, from time-ordered input to the mapmaking stage, adding noise and various instrumental artefacts along the way, finally producing \planck-like frequency maps. NILC and MILCA methods use the frequency maps as input and estimate a $y$-map. To test the foreground contamination, NILC/MILCA weights were applied to the original foregrounds maps to build residuals maps (since the methods are linear). I have computed the bispectrum of the residuals maps and of the tSZ map, and Fig.\ref{Fig:FFP6_fg_MILCA} shows these results in the case of the MILCA method (the absolute value of the bispectrum is shown for the residuals).

%Note that most high multipoles configurations are discarded due to the mask leakage, as previously described previously. 
The configurations involving the first multipole bin are too contaminated by foregrounds\,; these configurations are already discarded in Fig.\ref{Fig:FFP6_fg_MILCA} as a result of this observation.
For the other configurations, Fig.\ref{Fig:FFP6_fg_MILCA} shows that the bispectrum of foregrounds residuals is much smaller than the tSZ bispectrum. Hence the bispectrum estimates of the recovered $y$-maps are safe from foreground contamination in the considered configurations.\\

Then, I have discarded the configuration which fulfilled any of the following conditions~:
\begin{itemize}
\item masked ratio differing from the $f_\mr{SKY}$ approximation by more than 50\%, or negative
\item one bin is the first one, and is too contaminated by foregrounds
\item one of the multipole is greater than 1000
\item the NILC or MILCA estimate is lower than the minimum acceptable value $10^{-29}$ \footnote{This minimum acceptable value is $\sim2$ orders of magnitude below the minimum measured value in Fig.\ref{Fig:bispSZ_finalestim}.}
\end{itemize}
where the third and fourth conditions were put in place to discard some remaining configurations which had escaped the first two tests but were nonetheless unreliable.

%%%%%%%%%%%%%%%
\subsection{\planck tSZ data : results and characterisation}

\begin{figure}[htbp!]
\begin{center}
\includegraphics[width=.7\linewidth]{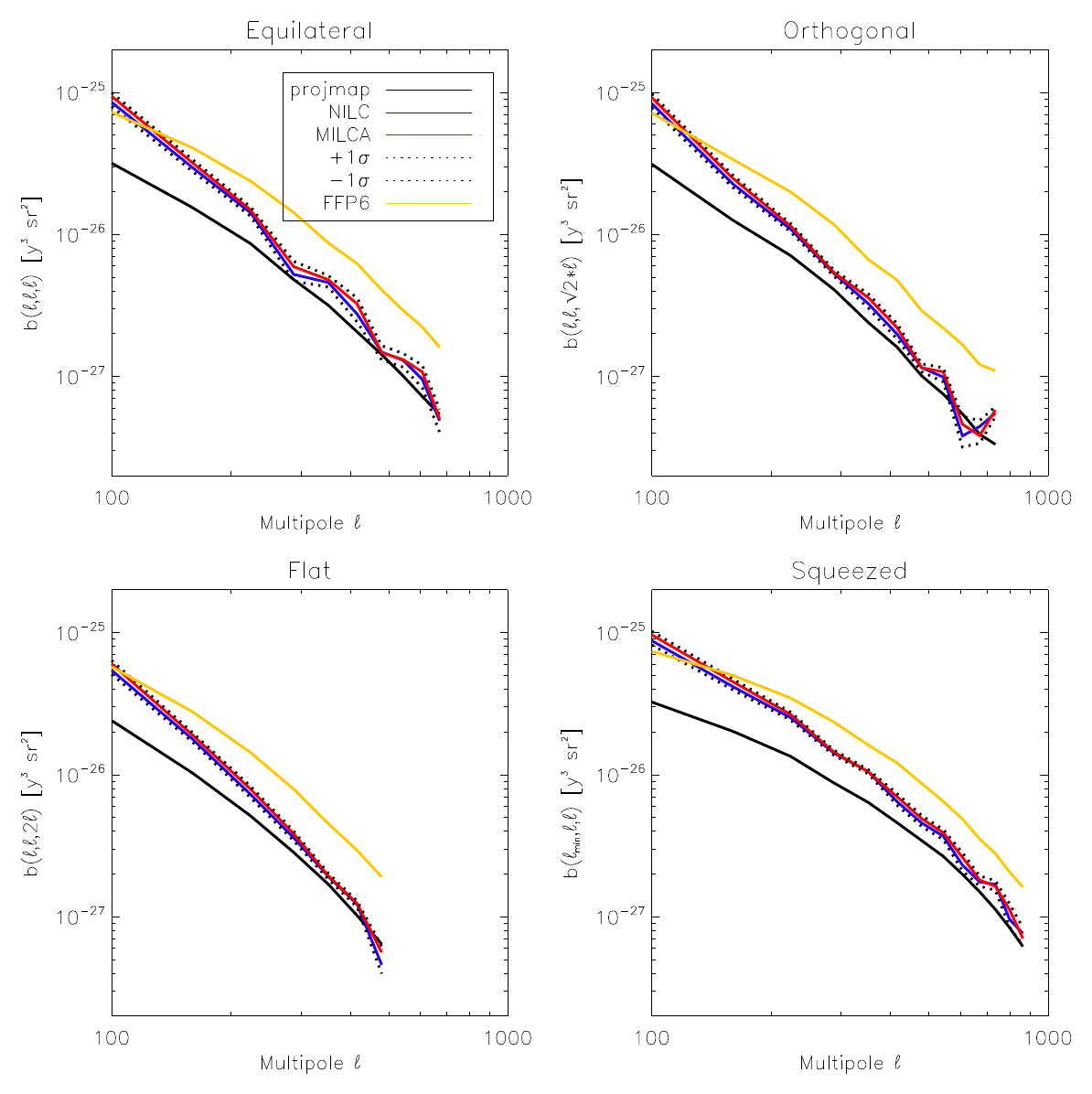}
\caption{From \cite{planck2013-SZmap}, tSZ bispectrum estimates with NILC and MILCA recovered $y$-maps, with their error bars in dotted lines. Overplotted are the bispectrum of the projected catalogue from \cite{planck2013-SZcatalogue} and of the FFP6 simulation.}
\label{Fig:bispSZ_finalestim}
\end{center}
\end{figure}

Considering all the corrections and flagging described in the previous section, Fig.\ref{Fig:bispSZ_finalestim} shows my final estimates of the tSZ bispectrum for the NILC and MILCA recovered $y$-maps with their error bars.

We see that the bispectrum estimates for the NILC and MILCA $y$-maps are consistent with each other. The estimated bispectrum for NILC is lower on average, and they have the same order of magnitude as the FFP6 tSZ simulation, although being steeper.\\
I have also overplotted (in black) an estimation of the tSZ bispectrum from the detected clusters present in the \planck SZ catalogue \citep{planck2013-SZcatalogue}. This estimate was computed with a $y$-map provided the detected (and confirmed) clusters were projected with the tSZ flux and radius from the catalogue. We see that the bispectrum estimates from NILC and MILCA have the same order of magnitude as the projected map, but are significantly larger. This indicates a contribution from undetected clusters or diffuse tSZ emission, or possibly foreground contamination at low multipoles.\\
The $\pm1\sigma$ error bars around the bispectrum estimates from NILCA and MILCA $y$-maps are shown with dotted lines, they were computed with the following terms~:
\begin{itemize}
\item cosmic variance : $C_{2\times2\times2}$ term with the $f_\mr{SKY}$ approximation. For the power spectrum involved, I estimated the NILC and MILCA $y$-map auto-spectra (containing the noise power spectrum, as necessary), and averaged them.
\item cosmic variance : $C_{3\times3}$ term with the $f_\mr{SKY}$ approximation. The bispectrum involved is the estimated average of the bispectra from NILC and MILCA $y$-maps.
\item cosmic variance : the $C_{4\times2}$ and $C_6$ terms were neglected, as I did not have estimates of the tSZ trispectrum nor 6-point function.
\item systematic effects : to quantify the systematic uncertainty due to the component separation method, I used the difference between the bispectra from NILC and MILCA $y$-maps.
\end{itemize}
Note that the $C_{3\times3}$ term produces a non-diagonal covariance matrix, hence the error bars in Fig.\ref{Fig:bispSZ_finalestim} are correlated.

The significance of the tSZ NG detection is quantified by computing the SNR in different cases summarised in Table \ref{Table:SNR_tSZ}.

\begin{table}[htbp!]
\begin{center}
\begin{tabular}{c|c|c|}
SNR & NILC & MILCA \\
\hline
full non-Gaussian & 210.61 & 223.29 \\
Gaussian & 332.49 & 351.62 \\
non-Gaussian diagonal & 319.77 & 338.47 \\
\end{tabular}
\end{center}
\caption{Detection SNR for the tSZ NILC and MILCA bispectra, in different cases. The `Gaussian' case neglects the $C_{3\times3}$ term, while the `non-Gaussian diagonal' only considers its contribution to the variance, so that the covariance matrix is diagonal.}
\label{Table:SNR_tSZ}
\end{table}

The first line (`full non-Gaussian') shows the SNR with the full covariance matrix described above. We see that the result are highly significant. There are 416 non-flagged configurations, each being on average a 14.2$\sigma$ (15.3$\sigma$) detection for the NILC (MILCA) bispectrum. This can be compared favorably with the $\sim 10\sigma$ announced detection of the tSZ bispectrum by \cite{Crawford2013}, who consider all configurations and all SPT frequencies.\\
Furthermore, the \cite{Crawford2013} analysis only accounts for Gaussian contributions to the error bars. For comparison, I computed the SNR that would be obtained by dismissing the $C_{3\times3}$ term of the covariance matrix, it is shown in the first line of Table \ref{Table:SNR_tSZ}. I also computed the SNR that would be obtained considering $C_{3\times3}$ contribution to the bispectrum variance but not the off-diagonal contributions, it is shown in the second line of the same Table. We see that considering the full non-Gaussian covariance matrix is important, since we otherwise overestimate the detection significance by $\sim60\%$. We also see that the decrease of SNR mostly comes from the off-diagonal terms of the covariance matrix. The non-Gaussian term may thus seem small when considering the variance, but it is important to consider them for the whole significance computation.
\newline

\begin{figure}[htbp!]
\begin{center}
\includegraphics[width=\linewidth]{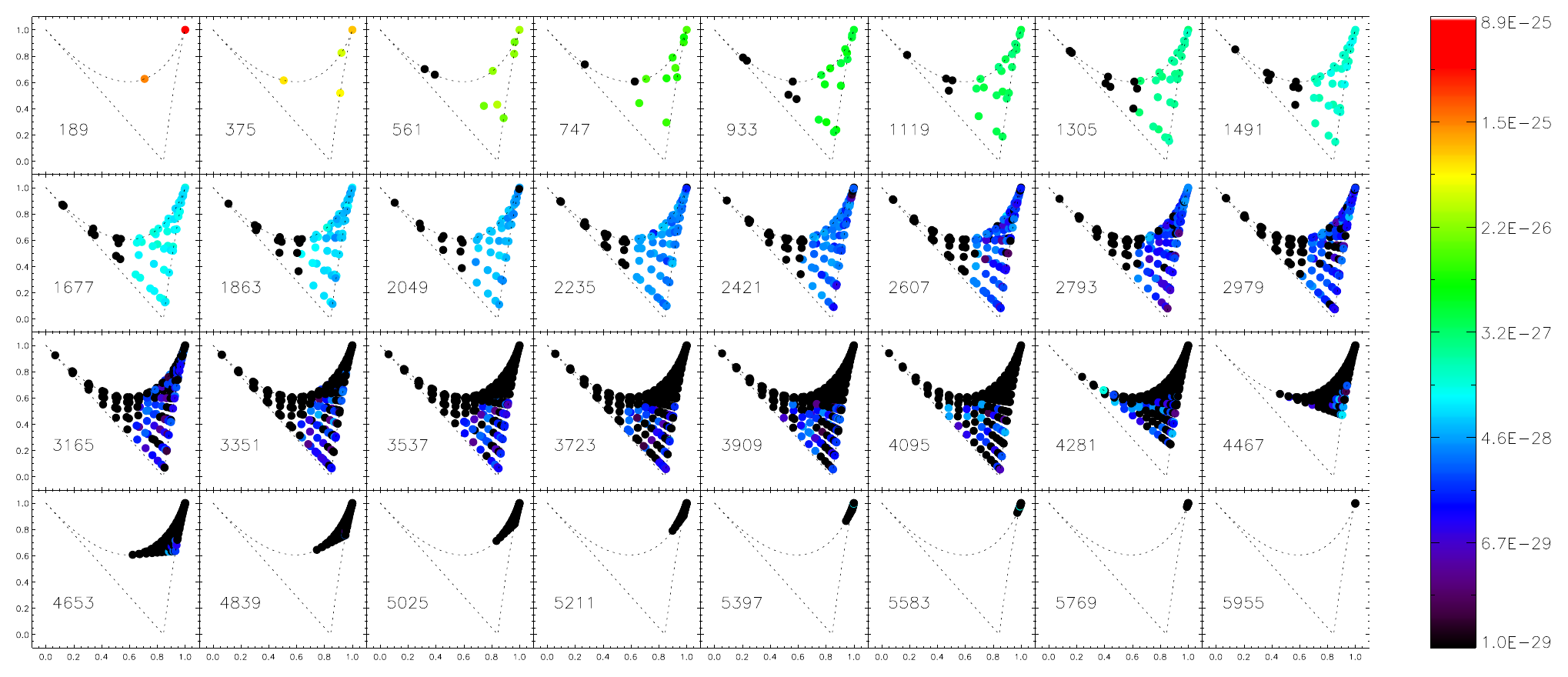}
\caption{tSZ bispectrum estimate for MILCA $y$-map, plotted in the geometrical parametrisation. Black dots denote flagged configurations.}
\label{Fig:MILCA_inparam}
\end{center}
\end{figure}

We can now consider the configuration dependence of the tSZ bispectrum measurements. To this end, I have plotted the bispectra for NILC and MILCA $y$-maps with the geometrical parametrisation. For illustration, the MILCA bispectrum is shown in Fig.\ref{Fig:MILCA_inparam} \footnote{The NILC bispectrum shows the same behaviour.}.

This figure shows a behaviour remarkably consistent with Fig.\ref{Fig:param_TSZ-psm}. The bispectrum varies with scale but does not depend on the shape of the triangle. This gives further confidence in the robustness of the measurement, since the bispectra of other foregrounds do not exhibit this behaviour. We also note the heavy flagging of configurations, especially for squeezed triangles and at high multipoles.
\newline

Finally, I used these non-Gaussianity measurements for cosmological inference. Indeed, \cite{Bhattacharya2012} published a theoretical study of the tSZ bispectrum and showed its dependence with cosmological parameters. Specifically, considering the most influential parameters, they showed that the amplitude of the bispectrum scales as $\sigma_8^{11.6} \, \Omega_b^{4.1}$. I compared the amplitude of the measured bispectrum with that of the FFP6 simulation (whose cosmological parameters are known) in the equilateral configurations. I could thus put a constrain on these parameters $(\sigma_8,\Omega_b)$. Considering the measurement error bars, the dispersion between NILC and MILCA, and an uncertainty on the $\sigma_8$ exponent\footnote{\cite{Bhattacharya2012} give an uncertainty on the $\sigma_8$ exponent, being between 11 and 12. I was conservative and took a range [10,13] for this exponent.}, I have the conservative constrain \citep{planck2013-SZmap}~:
\be
\sigma_8 \, \left(\frac{\Omega_b}{0.049}\right)^{0.35} = 0.74 \pm 0.04
\ee
It is remarkably consistent with the other constraints on $\sigma_8$ from tSZ measurements, either from the power spectrum or the skewness \citep{planck2013-SZmap}, or from the catalogue of detected clusters \citep{planck2013-SZcounts}.
\newline

I have thus provided the first measurement of the tSZ bispectrum by configuration. The measurement over two different $y$-map estimation methods are found consistent together, and the shape of the bispectrum is found consistent with expectations from simulations. Furthermore, the excellent signal-to-noise ratio of the detection opens up the possibility of using tSZ non-Gaussianity to constrain tSZ physics and cosmology, and I indeed realised first constraints on the cosmological parameters $\sigma_8$ and $\Omega_b$.

A full inference analysis with a tSZ model would however be necessary to fully exploit these bispectrum measurements, and it is indeed one of my next goals to build such a model and the related inference analysis.

\chapter*{Conclusion}
\addstarredchapter{Conclusion}

In this thesis, I have shown how random fields can be characterised by their (connected) correlation functions. Especially, in harmonic space these correlation functions are called polyspectra, and I have shown their properties on the sphere. I have shown how they can be estimated at second and third order (power spectrum and bispectrum), even in the presence of incomplete sky coverage. For the bispectrum case, the incomplete sky coverage is computationally problematic as the coupling matrix cannot be computed numerically from its analytical definition. I have designed a method to compute this matrix through simulations, but it was still unmanageable numerically for a \planck-like resolution. Thus, when analysing \planck data I have introduced an approximate method to debias the bispectrum estimate and locate cases where the approximation fails.

For cosmology, the study of high order correlation functions is of interest for the CMB, in particular as a possibility to discriminate the primordial process generating the cosmological perturbation (e.g. inflation models). As the foregrounds to the CMB have a non-Gaussian distribution, they may bias/contaminate the study of the primordial NG. My study has focused on the extragalactic foregrounds, and first on radio point-sources and the Cosmic Infrared Background. I have proposed in \cite{Lacasa2012} an empirical and analytical prescription for the non-Gaussianity of clustered point-sources accounting for possible multiple populations. Based on this prescription, I built in \cite{Lacasa2013a} a fast estimator for the amplitude of the CIB bispectrum. I further showed how to combine this estimator with corresponding estimates of the radio and primordial NG, in order to build a joint and robust estimation of non-Gaussianity. It has allowed me in particular to place upper limits on the contamination of point-sources to the primordial NG estimation \citep{planck2013-NG}.

I have also developed tools necessary for physical modeling of high orders for LSS tracers. I have indeed shown how 3D polyspectra can be projected onto the sphere, and developed a diagrammatic method to compute the polyspectra of the galaxy density field. I have shown the resulting equations for the bispectrum of the tSZ signal, and more thouroughly for the Cosmic Infrared Background. For the latter, I have implemented a numerical computation of the bispectrum, which allowed the first predictions of the CIB bispectrum and its variation with model parameters \citep{Lacasa2013b}. I have shown the interest of this quantity, as the constraints on CIB model parameters were of similar quality compared to the constraint coming from the power spectrum \citep{Penin2013}.

Finally, I have made the first measurement of the foregrounds NG on \planck data. I have computed the first measurement of the tSZ bispectrum, using tSZ maps estimated by \planck. This measurement proves consistent with expectations from the \planck tSZ catalogue and with available tSZ simulations\,; it is furthermore of high significance. Although I did not have a theoretical model, and its numerical implementation, to compare to the measurement, I was able to use the measurement to put a constraint on cosmological parameters, specifically on a combination of $\sigma_8$ and $\Omega_b$ \citep{planck2013-SZmap}.\\
I have also realised the first measurements of the CIB bispectrum at several frequencies, accounting for the problem of partial sky coverage and contamination by Galactic dust in particular. The detection is significant at each frequency, with significance increasing from 217 to 545 GHz, and the estimates are consistent together. Indeed the intrinsic level of non-Gaussianity as well as the slope of the bispectrum are found consistent between frequencies \citep{planck2013-CIB}. The measured bispectrum is however found significantly steeper than the empirical prescription or the halo model prediction presented in this thesis, which hence calls for a complexification of the IR emissivity model.

\chapter*{Perspectives}
\addstarredchapter{Perspectives}

There are several possible developments and improvements of the work that I have presented in this thesis. I will list below some possibilities, grouped by thematics.

First, the effects on non-Gaussianity studies of partial sky coverage need to be better accounted for. Indeed, the problem of the leakage of large scale power onto small scales has been limiting my estimates of the tSZ bispectrum with \planck data. It is  also limiting the CIB bispectrum estimates because of the Galactic dust contamination. Coupling matrix computations are needed to quantify this effects, and the numerical method I have presented to compute this matrix becomes too CPU intensive at high resolution. However this method may be manageable with a better choice of bispectrum basis and/or new methods to generate non-Gaussian maps. We may look for a method to optimise the $\alpha_\ell$ in my simulated bispectra, e.g. by orthogonalising the bispectra with the Gram-Schmidt process, or we may use cross-bispectra between different simulations. Alternatively other methods may be searched for, the separable modal methodology \citep{Fergusson2007,Fergusson2009} may provide useful tools for example.\\ 
For the \planck data analysis presented in this thesis, a first possibility to mitigate the mask effect would be a high-pass filtering by estimating low multipoles through maximum likelihood and substracting them. The robustness of the results with respect to this filtering should then be tested. This filtering may in particular mitigate the dust contamination to the CIB bispectrum estimates.\\
Using other bases, e.g. wavelets/needlets \citep[e.g.][]{Pietrobon2009}, or other quantities, e.g. Minkowsky functionals \citep[e.g.][]{Ducout2013}, is another possibility, as they may account more easily for the mask effect. However, theoretical predictions would need to be propagated to the new basis, which may not be straightforward. For example the wavelet-based $\fnl$ estimator for \planck relied on simulations to compute the theoretical third order wavelet statistics produced by primordial NG. The modal methodology has also emerged as a powerful possibility, allowing a fast near-complete and model-independent description of the NG at third order. It may be even more powerful at higher orders compared to current methods, although physical understanding of the results and control of the systematics may be more difficult in this case.

Then, I would like to develop a model for the tSZ bispectrum with its numerical implementation, as I did for the CIB bispectrum. This model is needed to exploit the information present in the \planck tSZ bispectrum measurement. In particular we could combine the information coming from the power spectrum measurement with that of the bispectrum, increasing the constraint on model parameters. More generally, I am interested in building a more optimal analysis of the tSZ signal, which is highly non-Gaussian. This may either be done through non-linear data transformation \citep{Leclercq2013,Carron2013,Simpson2013}, or Bayesian analysis may be possible if we neglect halo correlations.\\
The tSZ bispectrum model would also serve to explore physically motivated additions to the $\Lambda$CDM cosmology. For example, massive neutrinos \citep{Mak2013} and primordial NG \citep{Mak2012,Trindade2013} have been shown to have a significant impact on the tSZ signal, while the halo triaxiality \citep{Limousin2013} has an impact on the galaxy bispectrum \citep{Smith2006} and thus most probably also on the tSZ bispectrum.

Furthermore, the model for the CIB bispectrum I have presented, based on the halo model, need to be complexified to account for the \planck measurements. As already discussed in Sect.\ref{Sect:meas_CIB_NG} a promising track to reconcile model and data is to consider new IR emissivities, e.g. depending on the halo mass and/or on the galaxy type (central/satellite) such as in the model by \cite{Bethermin2013} and \cite{planck2013-CIB}. This more complex modeling can be incorporated in the formalism described in Sect.\ref{Sect:CIB_HM}, and in particular the diagrammatic method can be adapted straightforwardly. We may also want to assess the effect of halo triaxiality which has a noticeable effect on the galaxy bispectrum as already mentioned.\\
Besides, the cross-correlation between the CIB and tSZ effect would be interesting to explore. Indeed, their cross-spectrum has been predicted by \cite{Addison2012}, and was needed to debias the \planck power spectrum estimates both for the tSZ study and for the CIB study. The \cite{Addison2012} model revealed unpractical to use within \planck, and for a consistent treatment the modeling of the tSZ and CIB signals must be the same as in respectively \cite{planck2013-SZmap} and \cite{planck2013-CIB}. Moreover a cross-spectrum is not the only quantity of interest scientifically. For example, one would want to assess the CIB contamination to the estimated tSZ flux in detected clusters. Also, as both signals are non-Gaussian, a cross-spectrum does not allow for an optimal analysis of the cross-information between the signals.

Moreover, CMB and galaxy lensing are research areas where the tools I have developed could be adapted. Indeed, the study of these signals involves the modeling of higher orders of the distribution of the gravitational potential sourced by the large-scale structure of the universe. Thus non-Gaussianity tools as well as the halo model at high orders are of particular interest in this prospect.

Finally, there are several foregrounds or secondary anisotropies tracing the LSS : CMB lensing, iSW, tSZ, CIB, and surveys in other wavelengths (radio, galaxy surveys in the optical and near-IR, cluster in X-rays...). Most of their cross-spectra have been or are currently been studied, however the question remains open as of how to best combine these probes. In this prospect, I am thus also interested in looking for statistical methods to extract more information from a joint analysis of these signals.

\chapter*{Glossary}
\addstarredchapter{Glossary}

%{\bf Symbols}
\begin{table}[htdp]
\begin{center}
\begin{tabular}{|c|c|}
\hline
{\bf Symbol} & {\bf Definition} \\
\hline
$\hat{A}$ & estimator of a quantity $A$\\
$A_\mr{IR}$ & CIB bispectrum amplitude\\
$a_i$, $a_{\ell_i m_i}$ & harmonic coefficient \\ 
$B_\ell$ & beam profile in harmonic space \\
$b_{123}$, $b_{\ell_{123}}$, $b_{\lu \ld \lt}$ & angular bispectrum \\
$\mathbf{b}_{123}$ & bispectrum accounting for the beam effect \\ 
$\bps$ & radio sources bispectrum amplitude \\
$C_{2\times2\times2}$, $C_{3\times3}$, $C_{4\times2}$, $C_6$ & terms of the bispectrum cosmic covariance \\
$\mathcal{C}_\ell$ & power spectrum accounting for the beam and noise effect \\
$F_{\alpha \beta}$ & Fisher matrix element \\
$\fnl$ & primordial NG amplitude parameter \\
$F^s$ & second order perturbation theory kernel \\
$f_\mr{SKY}$ & fraction of sky \\
$G_{123}$, $G_{\lu \ld \lt}^{m_1 m_2 m_3}$ & Gaunt coefficient \\
$g_\nu$ & tSZ frequency dependence \\
$\vec{k}$ & 2D Fourier mode \\
$\kk$ & 3D Fourier mode \\
$\hn$ & direction, a unit vector \\
$n_\mr{config}$ & number of bispectrum configurations \\
$N_{\lu \ld \lt}$ or $N_{123}$ & normalisation of the bispectrum estimator, $\sim$ number of triangles \\
$\mathcal{P}^{(n)}$ & $n$-th order polyspectrum \\
$r(z)$ & comoving distance \\
$Y_i$, $Y_{\ell_i m_i}$ & spherical harmonic function \\
$\Delta_{\lu\ld\lt}$ & $\left\{\begin{array}{cc} 6 & \mr{equilateral \ triangles} \\ 2 & \mr{isosceles \ triangles} \\ 1 & \mr{general \ triangles} \end{array}\right.$\\
$\kappa_n$ & cumulant of order $n$ \\
$\mu_n$ & centered moment of order $n$ \\
$\nu$ & frequency \\
$\sigma_8$ & amplitude parameter for the density perturbations \\
$\Omega_i$ & normalised energy density of fluid $i$ \\
$\Omega_\mr{pix}$ & pixel surface area \\
\hline
\end{tabular}
\end{center}
\label{glossary:symbol}
\end{table}

%{\bf Abbreviations}
\begin{table}[htdp]
\begin{center}
\begin{tabular}{|c|c|}
\hline
{\bf Abbreviation} & {\bf Signification} \\
\hline
ACT & Atacama Cosmology Telescope \\
AGN & Active Galactic Nucleus \\
AME & Anomalous Microwave Emission \\
CDM & cold dark matter \\
c.f. & correlation function \\
c.g.f. & cumulant generating function \\
CIB & Cosmic Infrared Background \\
CMB & Cosmic Microwave Background \\
DM & dark matter \\
d.o.f. & degree(s) of freedom \\
DSFG & dusty star-forming galaxy \\
ERCSC & Early Release Compact Source Catalog (2011) \\
FFP6 & Full Focal Plane version 6 (simulations) \\
FL & Friedman-Lemaitre \\
FWHM & Full Width to Half Maximum \\
GMCA & General Morphological Component Analysis \\
GR & General Relativity \\
GUT & Grand Unification Theory \\
HFI & High Frequency Instrument \\
HOD & Halo Occupation Distribution \\
iff & if and only if \\
i.i.d. & independent identically distributed \\
IR & infrared \\
IRAS & Infrared Astronomical Satellite \\
ISW & integrated Sachs-Wolfe \\
KSW & Komatsu-Spergel-Wandelt (estimator) \\
kSZ & kinetic Sunyaev-Zel'dovich (effect) \\
LFI & Low Frequency Instrument \\
LSS & Large Scale Structure \\
m.g.f. & moment generating function \\
MILCA & Modified Internal Linear Combination Algorithm \\
NFW & Navarro-Frenk-White (profile) \\
NILC & Needlet Internal Linear Combination  \\
NG & non-Gaussianity \\
PCCS & Planck Catalogue of Compact Sources (2013) \\
p.d.f. & probability density function \\
PNG & primordial non-Gaussianity\\
ps or p.s. & point-source(s) \\
PSM & \planck Sky Model \\
rms & root mean squared \\
SNR & signal-to-noise ratio \\
SPT & South Pole Telescope \\
SW & Sachs-Wolfe (effect) \\
SZ & Sunyaev-Zel'dovich (effect) \\
tSZ & thermal Sunyaev-Zel'dovich (effect) \\
WMAP & Wilkinson Microwave Anisotropy Probe \\
\hline
\end{tabular}
\end{center}
\label{glossary:abbrev}
\end{table}

 %%%%%%%%%%%%%%%%%%%%%%%%%%%%%%%%%%%%%%%%%%%%%%
 %Bibliography
%\nocite{*}
\bibliographystyle{aa}
\bibliography{bibliography}
 
  %%%%%%%%%%%%%%%%%%%%%%%%%%%%%%%%%%%%%%%%%%%%%%
 
\end{document}